\documentclass[%
 reprint,
superscriptaddress,
 amsmath,
 amssymb,
prl
]{revtex4-2}

\usepackage{graphicx}
\usepackage{dcolumn}
\usepackage{bm}
\usepackage[usenames,dvipsnames]{color}
\usepackage{bbold}
\usepackage[hidelinks]{hyperref}
\usepackage{mathtools}

\DeclareUnicodeCharacter{2061}{}

\newcommand{\nyuphysics}{Center for Soft Matter Research, Department of Physics, New York University, New York 10003, USA}

\newcommand{\nyusimons}{Simons Center for Computational Physical Chemistry, Department of Chemistry, New York University, New York 10003, USA}
\newcommand{\nyucourant}{Courant Institute of Mathematical Sciences, New York University, New York 10003, USA}
\newcommand{\nyucns}{Center for Neural Science, New York University, NY 10003}

\begin{document}
\preprint{APS/123-QED}

\title{Gyromorphs: a new class of functional disordered materials}

\author{Mathias Casiulis}
\thanks{Equal contribution.}
\affiliation{\nyuphysics}
\affiliation{\nyusimons}
\author{Aaron Shih}
\thanks{Equal contribution.}
\affiliation{\nyucourant}
\affiliation{\nyuphysics}
\author{Stefano Martiniani}
\email{sm7683@nyu.edu}
\affiliation{\nyucourant}
\affiliation{\nyuphysics}
\affiliation{\nyusimons}
\affiliation{\nyucns}

\date{\today}

\begin{abstract}
We introduce a new class of functional correlated disordered materials, termed Gyromorphs, which uniquely combine liquid-like translational disorder with quasi-long-range rotational order, induced by a ring of $G$ delta peaks in their structure factor.
We generate gyromorphs in $2d$ and $3d$ by spectral optimization methods, verifying that they display strong discrete rotational order but no long-range translational order, while maintaining rotational isotropy at short range for sufficiently large $G$.
Using a coupled dipoles approximation, we numerically show that these structures outperform quasicrystals, stealthy hyperuniformity, and Vogel spirals in the formation of low-index-contrast isotropic bandgaps in $2d$, for both scalar and vector waves, and open complete isotropic bandgaps in $3d$.
This claim is further supported by analytical effective-medium theory and by numerical estimates of scattering mean-free paths.
Finally, we introduce ``polygyromorphs'' with several rotational symmetries at different length scales (i.e., multiple rings of delta peaks), enabling the formation of multiple bandgaps in a single structure, thereby paving the way for fine control over optical properties.
\end{abstract}


\maketitle

A crucial aspect of the design and discovery of new materials is understanding the relationship between structure and properties. 
Crystals have proven to be a highly versatile platform for engineering functions, as the periodicity of their atomic arrangement greatly facilitates the prediction and optimization of their properties. 
However, not all properties can be realized with periodic structures. 
Aperiodic media can achieve transport properties unattainable in periodic systems, such as the formation of isotropic photonic bandgaps, which are highly desirable in optoelectronic applications such as freeform waveguides~\cite{Man2013a,Milosevic2019} and tunable-reflectance coatings~\cite{Piechulla2021}.
While isotropic bandgaps have been demonstrated in some deterministic aperiodic systems such as Vogel spirals~\cite{Vogel1979, Pollard2009,Liew2012,Trevino2012,Razo-Lopez2024}, and near-isotropic bandgaps have been observed in quasicrystals~\cite{Man2005, Rechtsman2008, Vardeny2013}, many recent works have sought this property in correlated disordered structures~\cite{Yu2021,Vynck2023}, \textit{i.e.} random point patterns that lack conventional long-range order but exhibit spatial correlations.

Unlike in periodic systems, the structural origins of photonic bandgaps -- or more precisely, ``pseudogaps'', characterized by a depletion in the density of states relative to vacuum, leading to reduced transmission~\cite{Froufe-Perez2017,Monsarrat2022} -- remain poorly understood in aperiodic systems.
Disordered stealthy hyperuniform (SHU) structures~\cite{Torquato2018}, which suppress density fluctuations over long distances, have been shown to exhibit isotropic bandgaps in both $2d$ and $3d$~\cite{Florescu2009, Man2013, Muller2014, Tsitrin2015,Froufe-Perez2017,Klatt2022, Monsarrat2022, Siedentop2024}, but only when strong short-range correlations are also present~\cite{Florescu2009, Florescu2009a, Froufe-Perez2017, Zheng2020, Monsarrat2022, Froufe-Perez2023}.
Further complicating the picture, photonic bandgaps have also been observed in non-SHU disordered structures, such as jammed packings~\cite{Yang2010,Liew2011, Djeghdi2022}, and even in systems with only short-range correlations, like equilibrium hard spheres~\cite{Froufe-Perez2016}, which are not hyperuniform.
The role of short range correlations in disordered systems is especially pronounced for vector waves, for which bandgaps have so far only been observed in networked materials, such as honeycomb or tetrahedrally bonded structures ~\cite{Rechtsman2008, Florescu2009, Liew2011, Sellers2017,Klatt2019}.
In these random networks, the emergence of bandgaps has been linked to the similarity of local geometry and topology across the network \cite{Florescu2009a,Liew2011,Sellers2017}.

\begin{figure}
    \centering
    \includegraphics[width=\columnwidth]{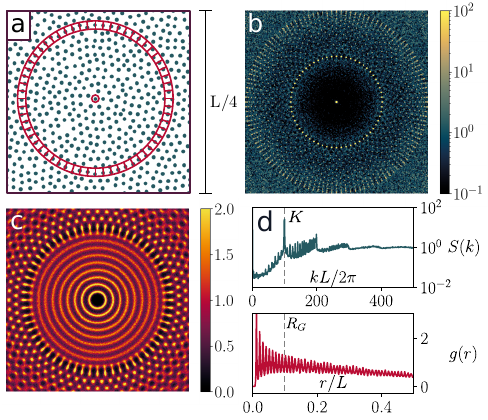}
    \caption{\textbf{Introducing: Gyromorphs.}
    $(a)$ Section of the point pattern of a $60$-fold gyromorph.
    We show the shortest distance displaying 60-fold order.
    $(b)$ Corresponding structure factor.
    $(c)$ Corresponding pair correlation function $g(\bm{r})$ near the origin.
    $(d)$ Radial structure factor $S(k)$ (top) and radial distribution function $g(r)$ (bottom) for the 60-fold gyromorph.
    }
    \label{fig:Gyromorphs}
\end{figure}
This raises the question: is there a unifying feature across these systems that promotes the formation of isotropic photonic bandgaps?
In the single-scattering regime, the emergence of a bandgap can be attributed to the presence of strong scattering at some characteristic frequency, and of weaker scattering at neighboring frequencies.
This argument has been invoked to justify the appearance of bandgaps in quasicrystals~\cite{Chan1998,Hagelstein1999,DellaVilla2005,Vardeny2013}, and explains why bandgaps also arise in systems with short-range correlations, such as equilibrium hard sphere liquids~\cite{Froufe-Perez2016,Shih2023}.
In fact, we observe that all previously cited aperiodic systems that exhibit isotropic bandgaps have an isotropic ring of high values in their structure factor, $S = |\hat{\rho}(\bm{k})|^2 / N$, where $\rho(\bm{r})$ is the density field defining the system.
This condition is also well approximated by quasicrystals, whose most intense structure factor peaks form regular polygons~\cite{Levine1984, [{See Supplemental Material at }][{ for a comparison between structure factors of known bandgap formers and gyromorphs, detailed numerical methods, additional discussions on the structure of gyromorphs, additional data on their optical behavior (including scalar-wave results and a discussion of our choice of definition for DOS), a reminder of the derivation of relevant quantities and equations within the coupled dipoles method, and a short derivation of the effective-medium theory.
}]supp}.

Bravely adhering to this single-scattering rationale, an ideal bandgap material should exhibit a ring of high values in $S(k)$, contrasting with low values around it.
Accordingly, we propose a correlated disordered structure of non-overlapping points whose structure factor displays one ring of $G \in 2 \mathbb{N}^\star$ delta peaks with intensities $\mathcal{O}(N)$ on a circle with radius $K$, but as little order as possible elsewhere.
In this work, we show that such structures, which we call \textit{gyromorphs}, are realizable, and we present an algorithm for their generation in $2d$ and $3d$.

In gyromorphs, liquid-like isotropic neighborhoods coexist with a ring of extensive $S(k)$ peaks, resulting in quasi-long-range rotational order.
These structures are thus fundamentally distinct from any previously known materials, as they reconcile seemingly contradictory features. 
We successfully generate finite gyromorphs with up to $G \sim 10^3$ peaks, whose heights surpass the peak intensities of finite quasicrystals obtained by usual deterministic methods~\cite{deBruijn1981,deBruijn1986}, see End Matter (EM).
Remarkably, we uncover a duality between the structure factor and pair correlation functions of gyromorphs and quasicrystals~\cite{supp}.
Using a coupled dipoles approximation~\cite{Lax1951,Lax1952,Carminati2021}, we thus demonstrate that gyromorphs outmatch previous candidate systems in $2d$ as low-index isotropic bandgap materials, for both vector and scalar waves~\cite{supp}.
Finally, we demonstrate that gyromorphs can be extended to $3d$ structures exhibiting \textit{complete} isotropic bandgaps, as well as to ``polygyromorphs'' with multiple rings of peaks, resulting in multiple bandgaps in a single system.

\textit{Generating gyromorphs --}
To generate gyromorphs, we use the Fast Reciprocal-Space Correlator (FReSCo)~\cite{Shih2023}.
We use the NUwNU (non-uniform real space with non-uniform k-space constraints) variant, which imposes constraints at continuous reciprocal space positions, assuming free boundary conditions for the point pattern.
Starting from an initially uncorrelated random point pattern with $N$ points, we optimize it to display $G$ $\mathcal{O}(N)$-high peaks regularly spaced on a ring while forbidding overlaps, thereby creating a $G$-fold gyromorph (see SM~\cite{supp}).

\begin{figure}
    \centering
    \includegraphics[height=0.47\columnwidth]{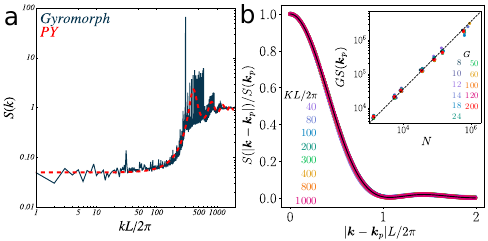}
    \caption{\textbf{Order in gyromorphs.}
    $(a)$ $S(k)$ of a $60$-fold gyromorph (blue) with $N\sim 10^5$ ($K L / 2\pi = 300$), compared to a Percus-Yevick $S(k)$ for $\phi = 0.57$ (dashed red line) in log scales.
    $(b)$ Corresponding radially averaged profile of peaks across $K$ values, rescaled by peak height (colored lines).
    A solid black line indicates the radial profile of $\text{sinc}^2(k_x L/2)\text{sinc}^2(k_y L/2)$.
    Inset: Rescaled peak height $G S(\bm{k}_p) $ against $N$, in log scales, across $G$ (colored symbols).
    A dashed black line shows $GS(\bm{k}_p) = 3.5 N$.
    }
    \label{fig:GyromorphOrder}
\end{figure}
\textit{Structure of gyromorphs --}
At high $G$, although the point pattern appears isotropic at short length scales, gyromorphs still exhibit strong $G$-fold order at large length scales, as shown in Fig.~\ref{fig:Gyromorphs}.
In Fig.~\ref{fig:Gyromorphs}$(a)$, we show a portion of an $N \sim 10^4$, $G=60$ gyromorph, highlighting in red the ``gear" with radius $R_G \approx G/K$~\cite{supp} onto which $60$-fold order is first achieved around an example point.
Note that this is different from bond-orientational order~\cite{Steinhardt1983}, which describes nearest-neighbor ordering only; at high $G$, gyromorphs have no such order, unlike quasicrystals~\cite{supp}.
The highlighted point in Fig.~\ref{fig:Gyromorphs}$(a)$ is arbitrary: by construction, gyromorphs are translation-invariant in the bulk.
In Fig.~\ref{fig:Gyromorphs}$(b)$, we show the corresponding structure factor, highlighting a strong $60$-fold ring of peaks, that are intense enough to display a few echoes, and low-intensity regions around the peaks.
In Fig.~\ref{fig:Gyromorphs}$(c)$, we show the central part of the corresponding pair correlation function $g(\bm{r})$.
This function illustrates that each particle is, on average, surrounded by a gear with the same orientation at the smallest ring of neighbors capable of accommodating $G$-fold order~\cite{supp}, with stronger fluctuations along the radial direction than the orthoradial direction around the gear.
Additionally, it reveals that, at shorter distances, the local neighborhoods are perfectly isotropic.
The short-range ``disorder'' is highlighted in Fig.~\ref{fig:Gyromorphs}$(d)$ by the radial plots of $S(k)$ (top) and $g(r)$ (bottom).
While $S$ is peaked at $K$ and shows a few nearby echoes, it eventually decays to $1$ at larger $k$, indicating the absence of short-range order.
Similarly, $g$ is notably less peaked than that of (quasi)crystalline structures, exhibiting no visible feature at $R_G$, and decaying with increasing distance.
Overall, Fig.~\ref{fig:Gyromorphs} highlights that $S-1$ is essentially just a $G$-fold ring of Dirac deltas, so that the $g(r)$ is well approximated by a sum of cosines akin to ``kaleidoscopic'' optical fields obtained by holographic techniques~\cite{Chen2011,Tuan2023}.
In contrast, quasicrystals display rings of peaked $g$ corresponding to a discrete set of allowed neighbor positions~\cite{deBruijn1981,deBruijn1986}, so that $S$ is well approximated by a sum of cosines at high $G$.
This ``Fourier duality'' between gyromorphs and quasicrystals is illustrated in SM~\cite{supp}.

To further investigate the nature of order in gyromorphs, we replot a structure factor in log-log scales in Fig.~\ref{fig:GyromorphOrder}$(a)$.
This panel shows that gyromorphs are \textit{not} hyperuniform, as the low-$k$ limit of $S$ saturates at a level similar to that of a Percus-Yevick model for equilibrium hard disk liquids~\cite{Rosenfeld1990}.
We further validate this observation by measuring local number fluctuations in real space over windows of increasing size~\cite{Torquato2018,Shih2023} (see SM), finding that the long-range density fluctuation scaling of gyromorphs is indistinguishable from that of hard disks.
Furthermore, at large $k$, the decay of $S(k)$ is also similar to that of a liquid.
These results suggest that, with respect to translational order, gyromorphs are more akin to liquids than to quasicrystals.

In Fig.~\ref{fig:GyromorphOrder}$(b)$, we analyze the peaks.
In the main panel, we show the profile of the peak obtained by plotting a radial projection of $S$ using the location of a peak, $\bm{k}_p = (K,0)$, as the origin for the polar coordinates.
Here, $S$ is normalized by the peak height and plotted across system sizes ($K \sim \sqrt{N}$) for $G = 60$. 
As $N$ grows, the profile converges to $\textrm{sinc}^2(k_x L/2) \textrm{sinc}^2(k_y L/2) $, implying that the linear width of the peaks in $k$-space decays like $1/L$.
The inset shows that, for different values of $G$, the peak heights grow like $\mathcal{O}(G N)$, indicating that the peaks are extensive for any fixed order symmetry.
Altogether, the peaks have extensive height while their area decays like $1/L^2 \sim 1/N$: they approach Dirac deltas as $N$ increases, like in quasicrystals, so that gyromorphs display quasi-long-range rotational order (see EM).

\textit{Coupled Dipoles Method --}
We now turn our attention to the optical properties of gyromorphs.
From an Ewald-sphere construction~\cite{Ewald1921, Kittel} one expects a ring of many high peaks at radius $K$ to generate strong backscattering at $K/2$~\cite{Hagelstein1999}.
That is why quasicrystals were deemed good candidates for isotropic bandgaps.
However, they display increasingly shallow bandgaps as their order of rotational symmetry grows~\cite{Rechtsman2008}, which can be attributed to weakening features in $S$ in de Bruijn's construction~\cite{deBruijn1981,deBruijn1986} as $G$ increases.
By contrast, gyromorphs are only constrained to display a finite set of high fine peaks, and they may retain a deep isotropic bandgap even at high $G$.

\begin{figure}
    \centering
    \includegraphics[width=\columnwidth]{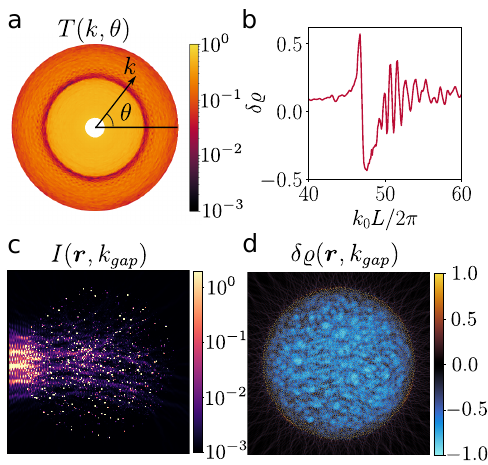}
    \caption{\textbf{Optical properties of gyromorphs}
    We use an example 60-fold gyromorph and focus on TE polarization.
    $(a)$ Intensity transmission for frequencies $40 \leq k_0L/2\pi \leq 60$ (radial direction) and $360$ incident angles (orthoradial direction) of a source Gaussian beam.
    $(b)$ Relative LDOS change with respect to vacuum, averaged over $1000$ random points, highlighting a dip close to $k_0 = K / 2$ (here $K L / 2\pi = 100$).
    $(c)$ Spatial map of the intensity at angle $0$ and $k_0 L / 2\pi = 47.6$, which corresponds to the minimum of both transmission and LDOS.
    $(d)$ Corresponding LDOS map.
    }
    \label{fig:TransmissionDOSExample}
\end{figure}
To verify this, we assume that each point in the gyromorph represents a non-magnetic scatterer with a dielectric contrast $\delta\varepsilon(\omega) \in \mathbb{C}$ from the surrounding medium at pulsation $\omega$, corresponding to a refractive index $n = \sqrt{1 + \delta\varepsilon}$, and a radius $a$.
Limiting our analysis to wave-vectors $k_0 = \omega / c$ (with $c$ the celerity of waves in the host medium) such that $k_0 a \ll 1$ (Rayleigh scattering~\cite{Carminati2021}), we employ a coupled dipoles approximation to describe the scattering of waves by the structure.
This standard method~\cite{Lax1951,Lax1952,Carminati2021} turns the problem of solving Maxwell's equations in the medium (a very strenuous task that involves a full discretization of space as well as very specific choices of boundary conditions~\cite{Oskooi2010,Yamilov2023}) into a linear system, see EM.
In short, setting only $a$ and $n$ for the scatterers, we measure transmitted intensities for any choice of monochromatic source field, as well as an integrated local density of optical states (LDOS).
Throughout the main text, we present results for vector waves (Transverse Electric, or TE, mode in $2d$, as opposed to the scalar-wave Transverse Magnetic, or TM, mode).
We report the scalar-wave results in SM, as scalar-wave bandgaps are less challenging to obtain in point patterns~\cite{Rechtsman2008,Florescu2009,Florescu2009a,Monsarrat2022}.
All optical properties are computed within the Materials Analysis via Green's Tensors (MAGreeTe) library, which we make publicly available~\cite{MAGreeTe}.

\textit{Existence of a bandgap --}
First, in Fig.~\ref{fig:TransmissionDOSExample}, we focus on a single $2d$ gyromorph with $G = 60$ and $N\sim 10^4$ to determine whether a bandgap is present (we discuss other $G$s in SM).
We first trim the system to a circular shape of radius $R = L/2$ to avoid any anisotropy due to the overall shape of the medium, like in many experimental tests~\cite{Man2005,Florescu2009,Man2013}.
We then set the radius $a$ of scatterers such that the filling fraction $\phi \equiv N a^2 / R^2$ equals $5\%$ and the dielectric contrast $\delta\varepsilon = 8$ ($n = 3$ for scatterers in vacuum), a typical value for metal oxides used in experiments~\cite{Siedentop2024}.
Following Refs.~\cite{Man2005,Florescu2009,Man2013,Siedentop2024}, we first perform transmission measurements.
We set the source to be a standard Gaussian beam focused at the center of the medium~\cite{Leseur2014,Leseur2016,Carminati2021}.
We solve the coupled dipoles system for $360$ regularly spaced orientations $\theta$ of the beam and for several frequencies $k_0$.
Then, following Refs.~\cite{Leseur2014}, we measure the transmitted intensity behind the system at a distance $D$ from its nearest edge~\footnote{The reason for not measuring the transmission in the far field is that the monochromatic Gaussian beam is not in fact a physical laser beam, in the sense that its intensity does not decay according to the Rayleigh-Sommerfeld boundary condition in the far field~\cite{Carminati2021}.}.
For each incoming beam orientation $\theta$, we define the transmission as the average intensity over 180 angles in $[\theta - \pi/2; \theta + \pi/2]$ on the half-circle with radius $D$, normalized by the value of the incident field at these points in the absence of the medium.
The results, shown in Fig.~\ref{fig:TransmissionDOSExample}$(a)$ in the plane of incident wave-vectors $(k_0,\theta)$, display a clear trough of low intensities at $k_{gap} \lesssim K/2$, as expected from single-scattering arguments~\cite{Hagelstein1999,Shih2023}.
An intensity map for $\theta = 0$ and $k_0 = k_{gap}$ is shown in Fig.~\ref{fig:TransmissionDOSExample}$(c)$, confirming that the field is strongly backscattered.
To confirm that this lower transmission is the sign of a bandgap, we measure the DOS of the system.
In Fig.~\ref{fig:TransmissionDOSExample}$(b)$, we show $\delta\varrho$, the relative change of DOS compared to vacuum averaged over $1000$ random points drawn randomly across the material but at least $2a$ away from scatterers, following Refs.~\cite{Pierrat2010,Froufe-Perez2017}.
We show strong mode depletion at $k_{gap} L /2\pi = 47.6$.
Finally, a map of the local $\delta\varrho$ across the system in Fig.~\ref{fig:TransmissionDOSExample}$(d)$ shows spatial fluctuations of $\delta\varrho$, coming both from finite-size and near-field effects, motivating our choice of a measure based on random points away from scatterers, rather than a single-point evaluation (see SM~\cite{supp}). 
The variability observed in Fig.~\ref{fig:TransmissionDOSExample}$(c)-(d)$ may indicate the presence of localized modes, similar to those reported in SHU systems~\cite{Monsarrat2022}.

\begin{figure}
    \centering
    \includegraphics[width=\columnwidth]{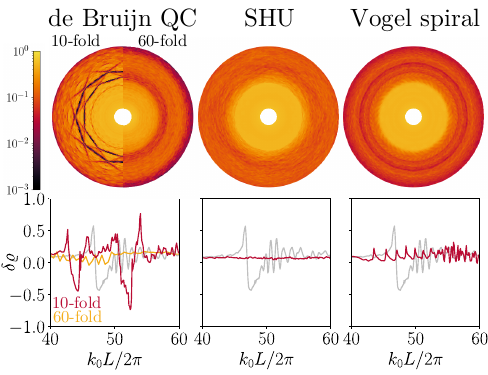}
    \caption{\textbf{Comparison with other systems}
    Top row: TE intensity transmissions as a function of $k_0$ (radial) and beam orientation (orthoradial) for (from left to right) $10$- and $60$-fold de Bruijn quasicrystals, an SHU structure, and a Vogel spiral.
    Bottom row: Corresponding $\delta\varrho$ (red lines), with the $60$-fold gyromorph shown as a light gray line.
    All results are obtained for $n=3$, $\phi = 0.05$.
    }
    \label{fig:Comparison}
\end{figure}
\textit{Comparison to other systems --}
We compare Fig.~\ref{fig:TransmissionDOSExample} to classic bandgap candidates, namely $10$- and $60$-fold deterministic quasicrystals obtained by the de Bruijn's construction~\cite{deBruijn1981,deBruijn1986, Lutfalla2021}, a deterministic Vogel spiral~\cite{Vogel1979}, and a random SHU structure generated by FReSCo~\cite{Shih2023} with a degree of stealthiness such that it displays a TM bandgap~\cite{Florescu2009,Man2013,Froufe-Perez2016,Froufe-Perez2017,Monsarrat2022,Klatt2022}.
Structures are adjusted to contain $N \sim 10^4$ points and a ring of higher values at the same $K$ as the gyromorphs, and we set all parameters to the same values as in Fig.~\ref{fig:TransmissionDOSExample} (see SM~\cite{supp} for $S(k)$).
In other words, the only difference between systems is the location of the $N$ scatterers.
We report transmissions in the top row of Fig.~\ref{fig:Comparison}, and DOS in the bottom row.
For this low $n$ and $\phi$, we show that Vogel and SHU systems display much shallower transmission troughs than the gyromorphs, which we attribute to weaker features in $S(k)$, and very little DOS change.
Furthermore, we show that quasicrystals may either display rather deep but anisotropic gaps at low symmetries, or isotropic but shallow gaps at high symmetry due to the broadening of their $S(k)$ features~\cite{Rechtsman2008,supp}.
In fact, the DOS depletion in gyromorphs are as deep as in low-order quasicrystals but more isotropic than them at comparable $\delta \varepsilon$.

\begin{figure}
    \centering
    \includegraphics[height=0.43\linewidth]{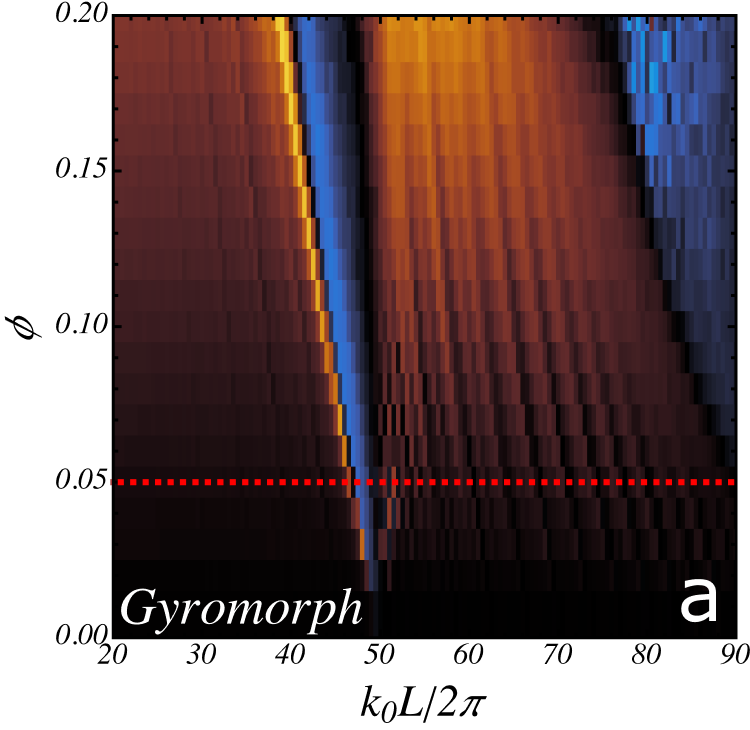}
    \includegraphics[height=0.43\linewidth]{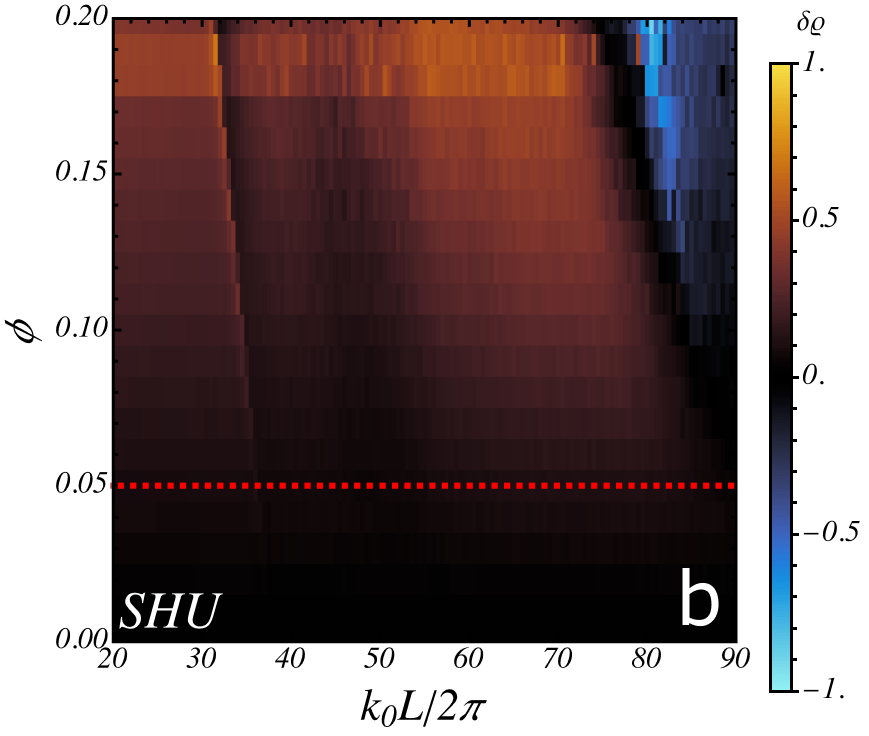} \\
    \includegraphics[height=0.43\linewidth]{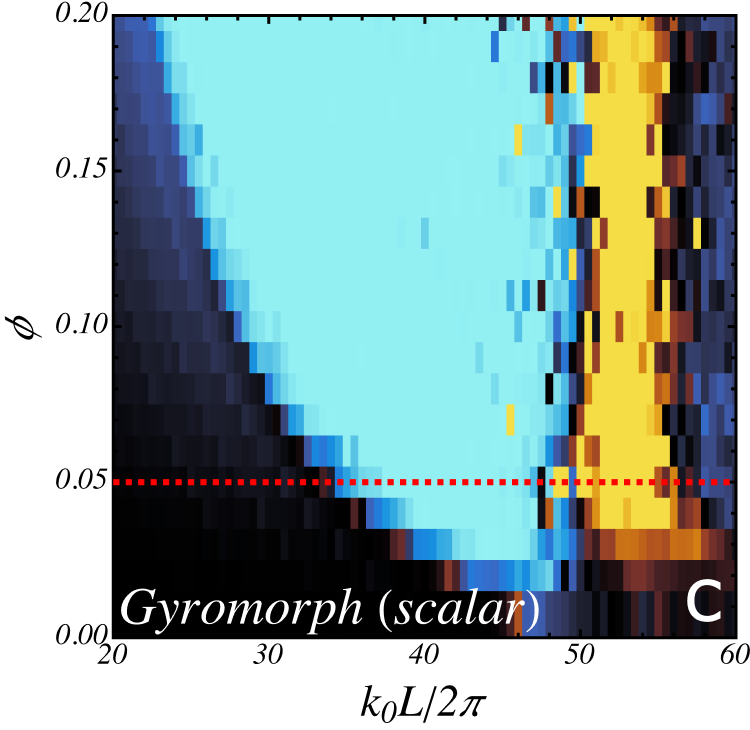}
    \includegraphics[height=0.43\linewidth]{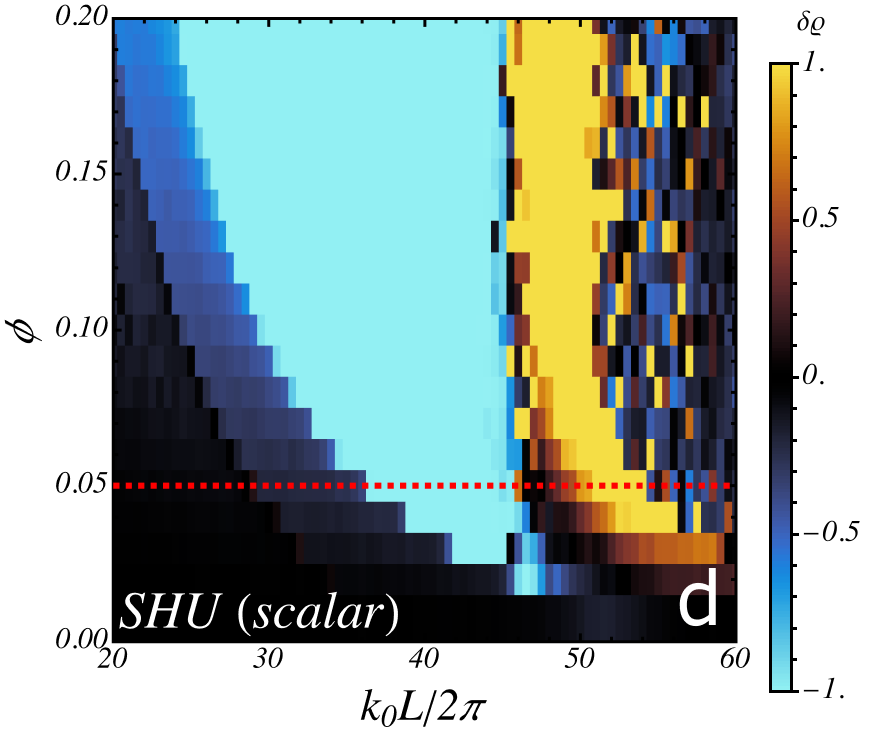}
    \caption{\textbf{Opening the gap.}
    Intensity maps of DOS against frequency and filling fraction at $n=3$, averaged over $30$ systems, for $(a)$ gyromorphs and $(b)$ SHU systems. Dotted red line: $\phi = 0.05$.
    $(c)-(d)$ Same plots for scalar waves.}
    \label{fig:maintext_phisweep}
\end{figure}

To further compare gyromorphs with SHU, Fig.~\ref{fig:maintext_phisweep}$(a)-(b)$ shows the DOS against $k_0L/2\pi$ and $\phi$ for both systems.
We confirm that gyromorphs develop a vector-wave bandgap that is absent in SHU.
In fact, the SHU bandgap reported in earlier works (\textit{e.g.}, Ref.~\cite{Monsarrat2022}) appears at larger $k_0$ and $\phi$, where gyromorphs display a similar feature, which can be recovered from the spectrum of the coupled dipoles matrix (see SM).
In the scalar waves case, Fig.~\ref{fig:maintext_phisweep}$(c)-(d)$, both systems display a bandgap at the same frequency but gyromorphs sustain it over a wider range of parameters.
We check that similar results are obtained when varying $n$ instead of $\phi$ (see SM), and that the depths of the DOS dip in gyromorphs increases with index contrast and system size (following a power law, like in defective crystals~\cite{Skipetrov2016, Skipetrov2020}), but does not strongly depend on $G$ (see SM~\cite{supp}).

\begin{figure}
    \centering
    \includegraphics[width=\columnwidth]{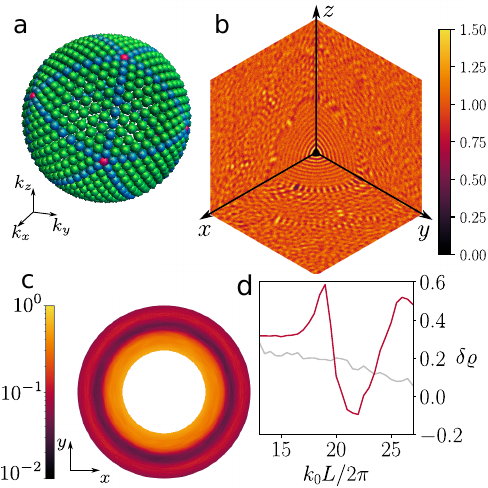}
    \caption{\textbf{Three-dimensional gyromorphs.}
    $(a)$ 3d-rendered orthographic view of the constrained peaks on the sphere.
    From an icosahedron (red), we subdivide edges (blue) and faces (green) so as to cover the whole sphere uniformly.
    $(b)$ Cross-sections (in the $x = 0$, $y = 0$, and $z=0$ planes) of the $g(\bm{r})$ of a gyromorph with $G = 1212$, $N = 83762$.
    $(c)$ Vector-wave transmission averaged over the forward half-sphere, for a beam source rotating in the $z = 0$ plane.
    $(d)$ Relative DOS change for the gyromorph (red line), compared to that of a Poisson point pattern (gray line).
    In $(c),(d)$, we use a ball of $N \approx 15000$ points from the structure and set $\phi = 0.1$, $n = 6.5$.
    }
    \label{fig:3d}
\end{figure}
\textit{Extension to $3d$ --}
We now extend gyromorphs to $3d$ structures.
We optimize structures to display $G$ regularly spaced peaks on a sphere with radius $K$ in their structure factor.
As there is no standardized way to populate points evenly spaced on a sphere while preserving centrosymmetry, we choose the location of peaks by building upon icosahedral geometry.
Starting from an icosahedron of peaks, we subdivide each icosahedral face into multiple triangular faces and project those additional vertices onto the sphere.
An example of such a set of peaks is shown in Fig.~\ref{fig:3d}$(a)$.
We show in Fig.~\ref{fig:3d}$(b)$ cross-sections of the $g(\bm{r})$ for a gyromorph thus obtained with $G \approx 10^3$.
Finally, in Fig.~\ref{fig:3d}$(c)-(d)$, we show optical properties measured for this system, namely a transmission plot and the averaged relative change of LDOS for vector waves.
We report an isotropic transmission gap, that is correlated to a depletion of DOS, which shows that $3d$ gyromorphs are promising bandgap formers.

\begin{figure}
    \centering
    \includegraphics[width=\columnwidth]{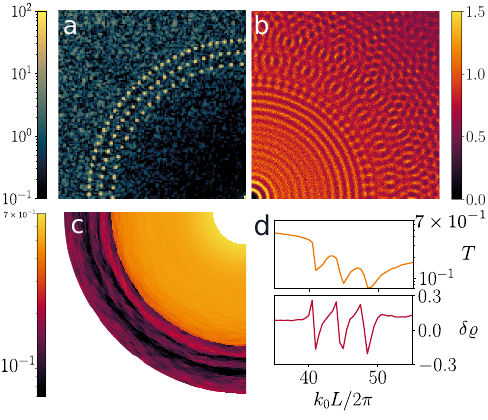}
    \caption{\textbf{Polygyromorphs.}
    $(a)$ Structure factor featuring $3$ rings with $82$, $106$, and $134$ peaks at $K L / 2\pi$ equal to $85$, $92.5$, and $100$ respectively.
    Corresponding
    $(b)$  $g(\bm{r})$, $(c)$ TE transmission, and $(d)$ Radially averaged transmission $T$ (top) and LDOS relative change (bottom).
    }
    \label{fig:Polygyromorphs}
\end{figure}
\textit{Beyond one ring --}
Finally, we extend the concept of gyromorphs to disordered systems displaying several rings of peaks in $S(k)$.
An example system is shown in Fig.~\ref{fig:Polygyromorphs}, where we impose $3$ rings of mutually prime (up to parity) rotational orders, at distances with irrational ratios to ensure that the structure is not trivially compatible with a single (quasi)crystal.
The resulting structure factor and pair correlation function are shown in Fig.~\ref{fig:Polygyromorphs}$(a)$ and $(b)$, respectively.
This structure is expected to display $3$ consecutive dips in transmission and DOS, which we indeed report in Fig.~\ref{fig:Polygyromorphs}$(c)$ and $(d)$.
This example highlights that gyromorphs are much more general than ``dual quasicrystals'' and can be used to obtain arbitrary transmission gaps in an aperiodic medium.

\textit{Conclusion --}
We introduced a novel class of disordered materials, gyromorphs, which are translationally disordered but exhibit one (or several) ring of delta peaks, resulting in quasi-long-range discrete rotational order.
We showed that gyromorphs with high rotational order exhibit deeper isotropic bandgaps than systems previously considered in the literature, in both 2d and 3d, and that several such features can be imposed at once.
This advance is significant both fundamentally and practically.
Fundamentally, we conceived and verified the existence of a new class of structures that reconciles seemingly contradictory features and stands apart from all known material types.
Practically, we showed that gyromorphs display isotropic photonic bandgaps that are either absent in other systems or persist over wider parameter ranges.
Manufactured gyromorphs therefore hold promise for applications including freeform waveguides~\cite{Man2013a,Milosevic2019} and lightweight coatings with strong backscattering, such as highly reflective ``white'' structures~\cite{Burresi2014,Wilts2018}.

\begin{acknowledgments}
\textit{Acknowledgments --}
The authors would like to thank Weining Man, Paul Chaikin, Dov Levine, Rémi Carminati, Luís Froufe-Pérez, Frank Scheffold, and Werner Krauth for insightful comments on this work.
A.S., M.C., and S.M. acknowledge the Simons Center for Computational Physical Chemistry for financial support.
M.C. and S.M. acknowledge financial support from the AFOSR Young Investigator Program under award FA9550-25-1-0359.
This work was supported in part through the NYU IT High Performance Computing resources, services, and staff expertise.
Parts of the results in this work make use of the colormaps in the CMasher package~\cite{VanderVelden2020}.
Correlations are computed using the rusted\_core library.
The FReSCo, MAGreeTe, and rusted-core libraries are publically available~\cite{FReSCo,MAGreeTe,rusted_core}.
\end{acknowledgments}

\bibliography{PostDoc-StefanoMartiniani,supp}

\appendix
\setcounter{figure}{0}
\setcounter{equation}{0}
\renewcommand{\figurename}{FIG.}
\renewcommand{\thefigure}{A\arabic{figure}}
\renewcommand{\thetable}{A\arabic{table}}
\renewcommand{\theequation}{A\arabic{equation}}
\newpage
\hphantom{a}
\newpage

\begin{figure}
    \centering
    \includegraphics[width=\columnwidth]{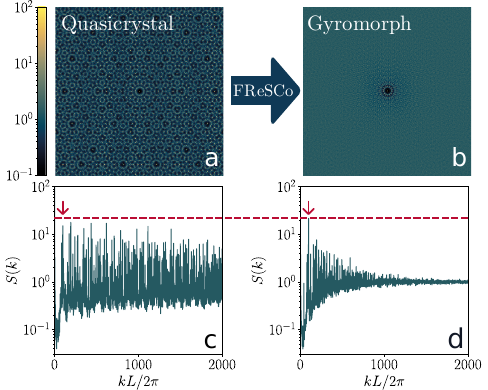}
    \caption{\textbf{Gyromorph generated from a Quasicrystal.}
    $2d$ structure factors of $(a)$ a 14-fold quasicrystal and $(b)$ a 14-fold gyromorph generated using the quasicrystal as an initial condition.
    $(c), (d)$ depict radial structure factors of the quasicrystal and gyromorph respectively.
    }
    \label{fig:QCtoGyro}
\end{figure}
\textit{Gyromorphs and Quasicrystals --}
Here, we further discuss the differences and commonalities between gyromorphs and quasicrystals.
First, we produce a $14$-fold quasicrystal by de Bruijn's method~\cite{deBruijn1981,deBruijn1986}, and report its $S(k)$ in Fig.~\ref{fig:QCtoGyro}$(a),(c)$, respectively as an intensity map in $k$-space and as a radially-averaged function of $k$.
The structure factor is patterned up to very large $k$, resulting in peaks with near-constant (but slowly decaying) intensity in the radial plot.
We use this structure as an initial condition for our algorithm, aiming to enhance the height of the existing first ring of high quasicrystalline peaks (indicated by the arrow in Fig.~\ref{fig:QCtoGyro}$(b)$).
The result, whose $S(k)$ is shown in Fig.~\ref{fig:QCtoGyro}$(b),(d)$ is indistinguishable from what we obtain from random initial conditions.
In particular, note that the large-$k$ structure in $S$ is completely destroyed: the far-away pattern has been ``pumped'' into the peaks to make them (about 3 times) higher, as highlighted by a red dashed line.
This experiment shows that gyromorphs, the minima of our loss function, are not simply noisy quasicrystals but do in fact display stronger peaks even at small $G$.
Furthermore, gyromorphs are not simply entropically more favorable than quasicrystals, they actually better minimize our loss function.
This difference becomes starker at larger-order rotational symmetries, as the height of the $S(k)$ peaks of de Bruijn quasicrystals decays fast with $G$~\cite{Rechtsman2008}.

\begin{figure}
    \centering
    \includegraphics[height=0.46\columnwidth]{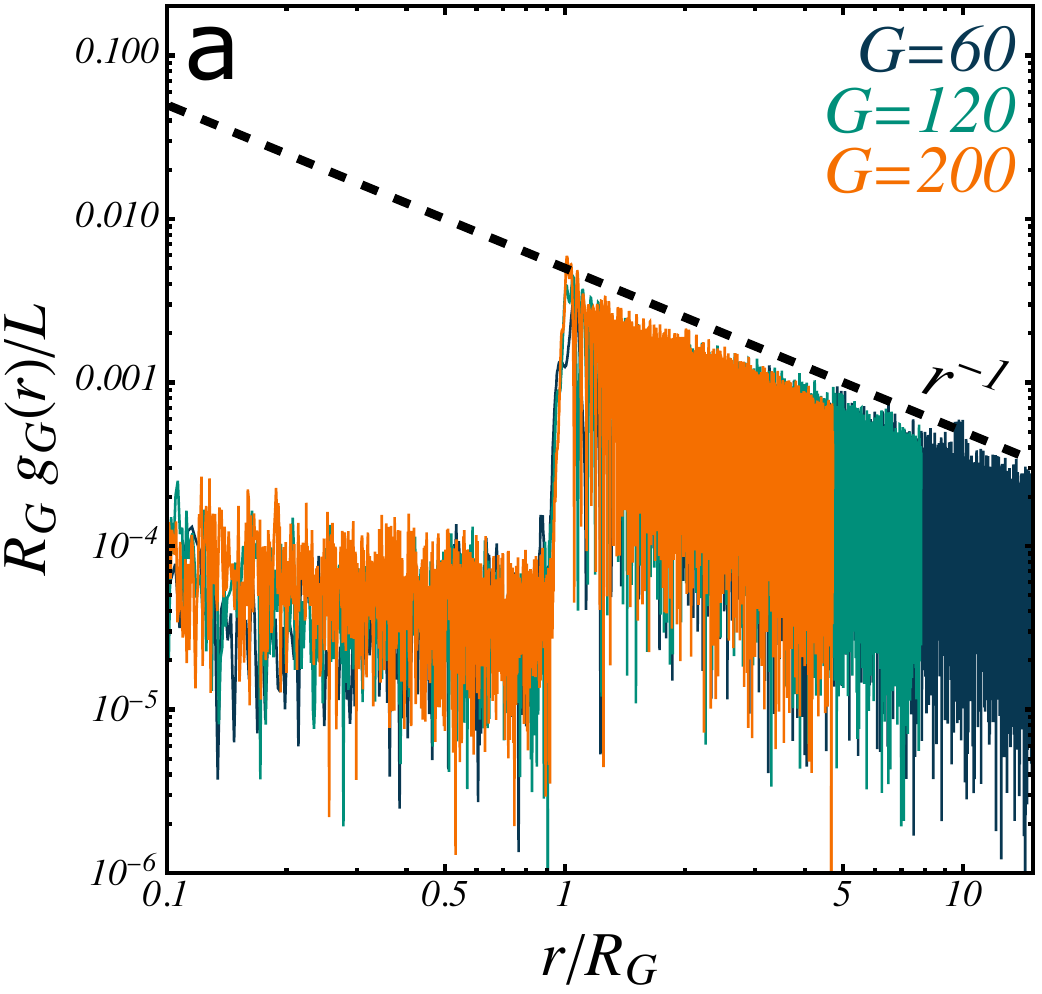}
    \includegraphics[height=0.46\columnwidth]{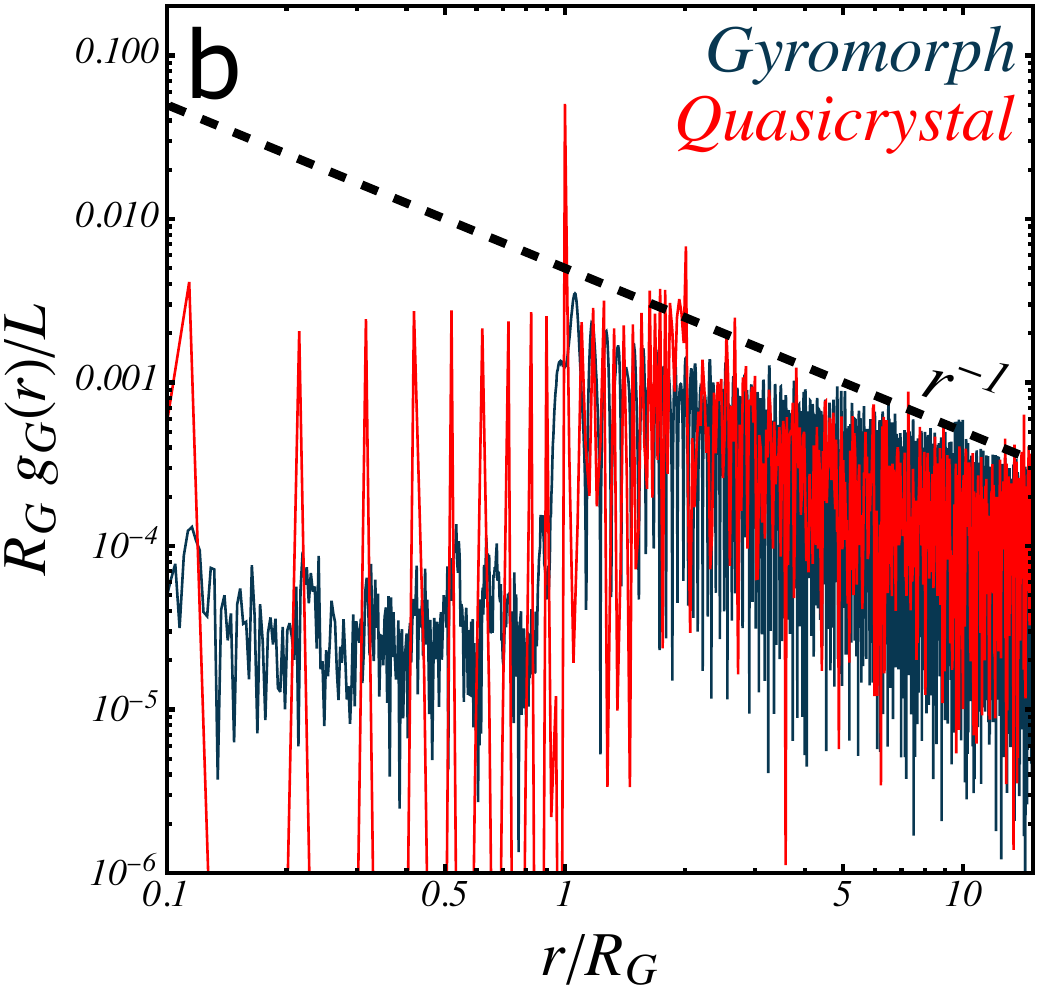}
    \caption{\textbf{Rotational order in gyromorphs.}
    Gyromorphic correlation function $g_G$, rescaled by $R_G/L$, against $r/R_G$, for $G = 60$ (dark blue), $G=120$ (green), $G = 200$ (orange).
    A dashed black line shows a $r^{-1}$ power law as a guide for the eye.
    Here, $KL/2\pi = 300$.
    }
    \label{fig:GyromorphicCorrelation}
\end{figure}

To highlight that gyromorphs display quasi-long-range rotational order, in Fig.~\ref{fig:GyromorphicCorrelation}$(a)$, we plot a gyromorphic correlation function, $g_G$, defined as 
\begin{align}
g_{G}(r) &\equiv \left|\frac{L^d}{N^2} \frac{1}{2\pi} \sum_{p \neq q} e^{i G \theta_{pq}} \delta(r - r_{pq})\right|,
\end{align}
where the sum runs over all pairs of particles, that are separated by a distance vector $\bm{r}_{pq}$ with polar components $(r_{pq}, \theta_{pq})$.
$g_G$ is essentially the modulus of the $G$-th Fourier mode of a shell of $g(r)$ at distance $r$.
When suitably rescaled by $R_G$, this function collapses across $G$'s, and displays an inverse-power decay.
We show in Fig.~\ref{fig:GyromorphicCorrelation}$(b)$ a comparison of that function between a $G=60$ gyromorph and the corresponding quasicrystal, with $R_G$ taken as the location of the strongest peak.
The difference is mainly that quasicrystals display strong rotational order with nearest neighbors, as they lie on a discrete set of orientations along a rhombic tiling, while gyromorphs are isotropic at short range.
The long-range behavior, however, is nearly the same.
Thus, gyromorphs display quasi-long-range rotational order~\cite{Shih2023,supp}.

\textit{Coupled dipoles in a nutshell --}
If $\bm{r}_i$ represents the position of scatterer $i$, $\bm{E}_{inc}(\bm{r}_i; \omega)$ is the $\omega$-component of the Fourier representation of an arbitrary incident field at $\bm{r}_i$ and $\bm{E}(\bm{r}_i; \omega)$ is the total exciting field at scatterer $i$, this system can be written as
\begin{align}
    \mathcal{M}(\left\{\bm{r}_{ij}\right\}; \omega)\cdot\mathcal{E}(\left\{\bm{r}_j\right\}; \omega) = \mathcal{E}_{inc}(\left\{\bm{r}_i\right\}; \omega),\label{eq:Linear_System}
\end{align}
where we introduced rank-$2$ tensor notations for both fields so that for instance $\mathcal{E}_{i,a} = \bm{E}(\bm{r}_i; \omega)\cdot \hat{\bm{e}}_a$ with $a \in \{x,y,z\}$, and $\mathcal{M}(\left\{\bm{r}_{ij}\right\}; \omega)$ is rank-$4$ with elements
\begin{align}
\mathcal{M}_{ijab} &= 
\begin{cases}
\delta_{ab}(1 - k_0^2 \alpha_0(\omega) \Sigma)     & \text{if } i = j \\
- k_0^2 \alpha_0(\omega) \hat{\bm{e}}_a\cdot\overline{\overline{G}}_0(\bm{r_i},\bm{r}_j; \omega)\cdot \hat{\bm{e}}_b & \text{otherwise}.
\end{cases} 
\label{eq:Mdef}
\end{align}
In this last expression, $\alpha_0(\omega) \equiv V \delta\varepsilon$ is the bare polarizability of a scatterer with volume $V$, $\Sigma$ is a self-interaction that ensures energy conservation and $\overline{\overline{G}}_0$ is the Green tensor of free propagation with the Rayleigh-Sommerfeld boundary conditions~\cite{Carminati2021}.
Solving Eq.~\ref{eq:Linear_System} for vector waves with $p$ components reduces to a $pN\times pN$ linear solve yielding $\mathcal{E}= \mathcal{W}\cdot\mathcal{E}_{inc}$ where $\mathcal{W} = \mathcal{M}^{-1}$.
The full electromagnetic field at any position $\bm{r}$ outside of scatterers can then be computed as
\begin{align}
    \bm{E}(\bm{r}; \omega) &= \bm{E}_{inc}(\bm{r}; \omega) + k_0^2 \alpha_0(\omega)\mathcal{G}_0(\bm{r},\{\bm{r}_i\}; \omega) \cdot \mathcal{E}(\{\bm{r}_i\}; \omega), \label{eq:outsideE} 
\end{align}
where $\mathcal{G}_0$ is a rank-$3$ tensor with elements $\mathcal{G}_{0,iab} = \hat{\bm{e}}_a \cdot \overline{\overline{G}}_0(\bm{r},\bm{r}_i;\omega)\cdot\hat{\bm{e}}_b$.
In practice, we use
\begin{align}
    \bm{E}_{inc}(\bm{r};\omega) &\equiv \bm{E}_0 \exp\left[- i \bm{k}_0 \cdot \bm{r}_{\parallel} - \frac{r_{\perp}^2}{w^2} \right],
\end{align}
with $\bm{r}_{\parallel}, \bm{r}_{\perp}$ the components of $\bm{r}$ parallel and perpendicular to a beam with width $w = L/5$, respectively.

Finally, to assert the presence of bandgaps, we compute the local density of optical states (LDOS)
\begin{align}
    \varrho(\bm{r};\omega) &\equiv \frac{2\omega}{\pi c^2} \text{Im}\text{Tr}_p \overline{\overline{G}}(\bm{r}, \bm{r}; \omega). \label{eq:LDOS_definition}
\end{align}
The coupled dipoles method lets us compute the relative change of LDOS compared to the vacuum value, $\delta\varrho = \varrho/\varrho_0 - 1$, as the trace of a tensor product.
\begin{align}
    \delta\varrho(\bm{r};\omega) &= C_p \textrm{Im}\left[ \alpha_0 \textrm{Tr}_p \left[\mathcal{G}_0(\bm{r}; \left\{\bm{r}_{i}\right\})\cdot\mathcal{W}\cdot {^t}\mathcal{G}_0(\bm{r}; \left\{\bm{r}_{j}\right\}) \right]\right], \label{eq:LDOS_CD}
\end{align}
where we dropped the $\omega$ dependence from the tensor notations for compactness and $C_p = (2 \omega^3)/[\pi c^4 \varrho_0(\omega)]$.
Following past works~\cite{Pierrat2010,Caze2013,Froufe-Perez2017}, we estimate the DOS of the medium at points at least $2a$ away from scatterer centers, and average the result over random points in the medium.
We remind the reader of the standard~\cite{Carminati2021} derivation of these equations in $2d$ and $3d$ in SM~\cite{supp}.

\textit{Effective medium theory --}
We now turn to effective medium properties.
In disordered media, the two most commonly used optical descriptors are the scattering mean-free path $\ell_s$ (the typical distance between two scattering events) and the anisotropy factor $g \in [-1;1]$ which quantifies how isotropic scattering events are~\cite{Carminati2021} -- $g=0$ corresponds to isotropic scattering, while $g= -1$ indicates purely backward scattering.
By expanding the medium’s Green tensor in powers of the density and retaining only terms that depend on pair correlations~\cite{VanTiggelen1994,Leseur2014,Conley2014,Leseur2016b,Cherroret2016,Carminati2021,Monsarrat2022,Vynck2023} one may link both quantities to $S(k)$.
In $d=2$, the final expressions read
\begin{align}
    \frac{1}{\ell_s} &= \frac{\rho_0}{4\pi k_R^2} \int\limits_{0}^{2 k_R} dq \frac{| t(q)|^2 \widetilde{S}(q)}{{\sqrt{1 - (q/2k_R)^2}}}, \nonumber \\
    g &= \frac{\rho_0 \sigma_s \ell_s}{4\pi k_R} \int\limits_{0}^{2k_R} dq \frac{1 - q^2/2k_R^2}{\sqrt{1 - (q/2k_R)^2}}\widetilde{S}(q), \label{eq:effective_medium_2d}
\end{align}
where $\widetilde{S}(k)$ is the structure factor, excluding the Dirac delta at $k = 0$, $t = k_0^2 \alpha_d$ is the amplitude of the $t$-matrix of one scatterer, $\rho_0 = N / L^d$ is the number density and $k_R \approx k_0 \sqrt{1 + \rho_0 \text{Re} [\alpha_d(\omega)]}$ is the real part of the effective wavenumber in the medium.
These expressions (derived in SM), are valid for both scalar and vector waves within this approximation, and are known~\cite{Conley2014,Leseur2016b}, although some works have mistakenly used the $3d$ expressions.

\begin{figure}[b]
    \centering
    \includegraphics[height=0.45\linewidth]{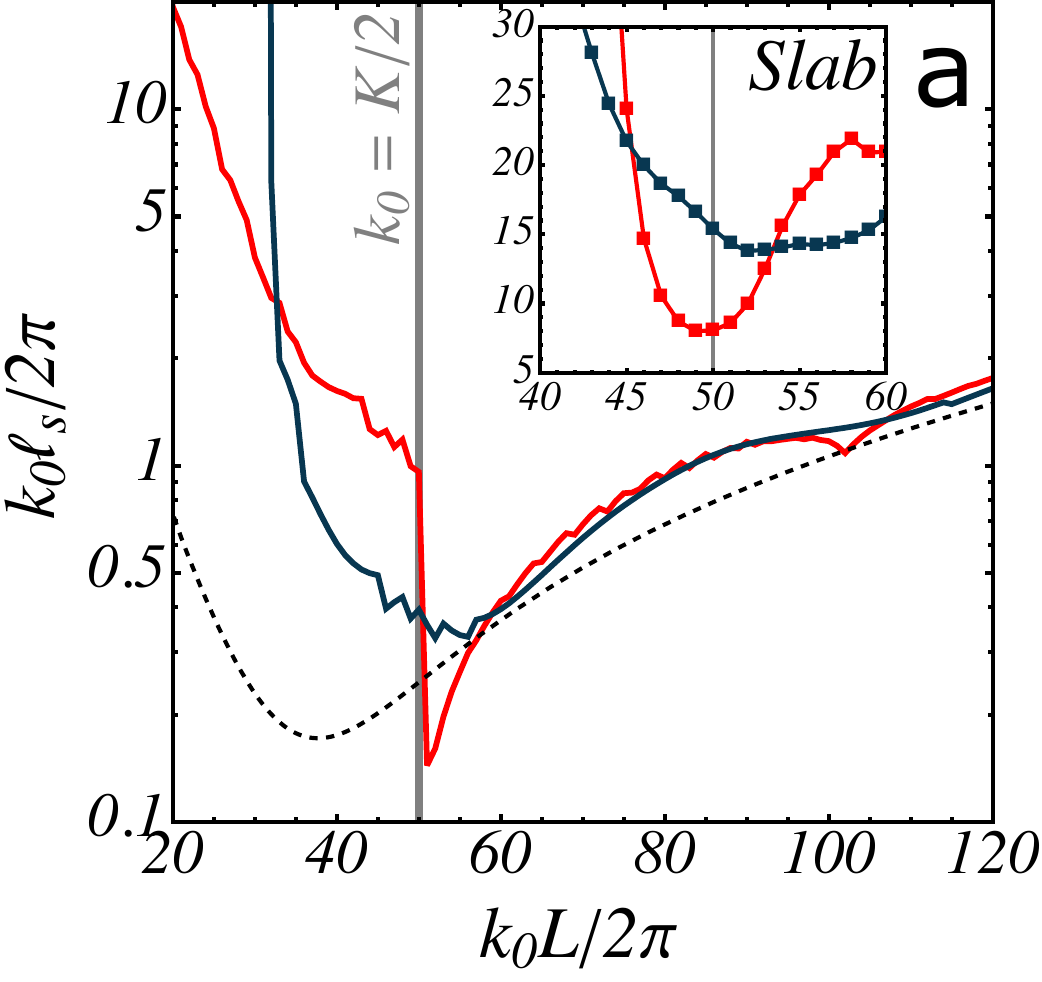}
    \includegraphics[height=0.45\linewidth]{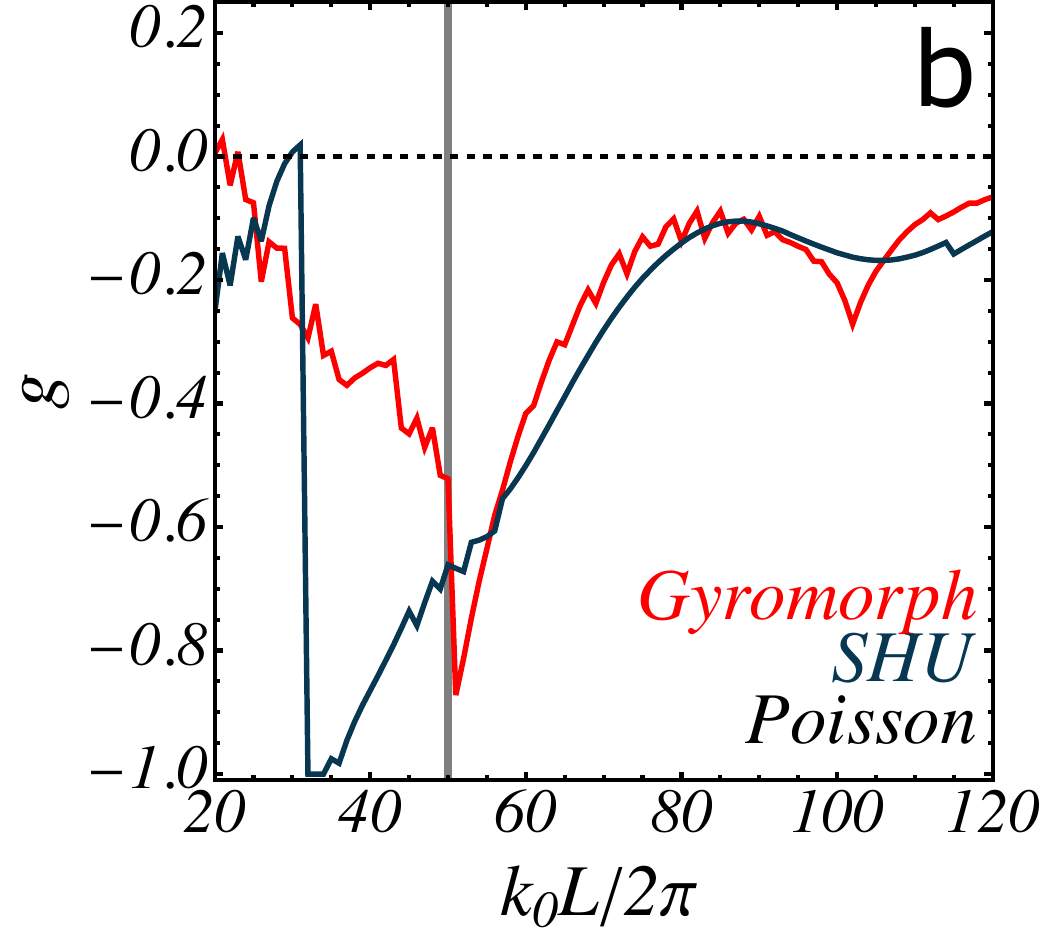}
    \caption{\textbf{Effective medium properties.}
    $(a)$ Rescaled scattering mean-free path $k_0 \ell_s / 2\pi$ (inset: numerical estimate in slab geometry) and $(b)$ anisotropy factor $g$ against $k_0 L / 2\pi$, for gyromorphs (red), SHU (dark blue) and Poisson point patterns (dashed black line), computed for $n=3$, $\phi = 0.05$ for scalar $2d$ waves.
    A gray line indicates $k_0 = K/2$.
    }
    \label{fig:Effective_Medium}
\end{figure}
We compute the $\ell_s$ and $g$ of gyromorphs and SHU systems, using the average $S(k)$ over $30$ realizations of each system, and show them in Fig.~\ref{fig:Effective_Medium}.
In Fig.~\ref{fig:Effective_Medium}$(a)$ we show that near the feature at $k_0L/2\pi = 50$, this theory predicts a shorter $\ell_s$ (stronger scattering) in gyromorphs.
We check qualitative agreement with measurements of $\ell_s$ (see inset), obtained by a standard slab-geometry measurement of the attenuation of the intensity of the disorder-averaged field~\cite{Carminati2021,Monsarrat2022} (see SM for numerical details).
As shown in Fig.~\ref{fig:Effective_Medium}$(b)$, this feature in $\ell_s$ coincides with more negative values of the anisotropy factor $g$, indicating more pronounced backward scattering.
Altogether, we report a lower effective transport length $\ell_t = \ell_s / (1-g)$.
This suggests that, at the level of pair correlations, gyromorphs are more likely than SHU systems to form bandgaps, where transport lengths should be short.
Due to this strong scattering, we expect gyromorphs to display transmission gaps, possibly bandgaps, at lower index contrasts and filling fractions compared to other isotropic bandgap candidate systems.

\end{document}


\preprint{APS/123-QED}

\title{Supplementary Material for: ``Gyromorphs: a new class of disordered functional materials''}

\author{Mathias Casiulis}
\thanks{Equal contribution.}
\affiliation{\nyuphysics}
\affiliation{\nyusimons}
\author{Aaron Shih}
\thanks{Equal contribution.}
\affiliation{\nyucourant}
\affiliation{\nyuphysics}
\author{Stefano Martiniani}
\email{sm7683@nyu.edu}
\affiliation{\nyucourant}
\affiliation{\nyuphysics}
\affiliation{\nyusimons}

\date{\today}


\maketitle

\renewcommand{\figurename}{FIG.}
\renewcommand{\thefigure}{S\arabic{figure}}
\renewcommand{\thetable}{S\arabic{figure}}
\newtagform{S}{(S})
\usetagform{S}

\section{Structure factors of bandgap formers}

\begin{figure}
    \centering
    \includegraphics[width=\linewidth]{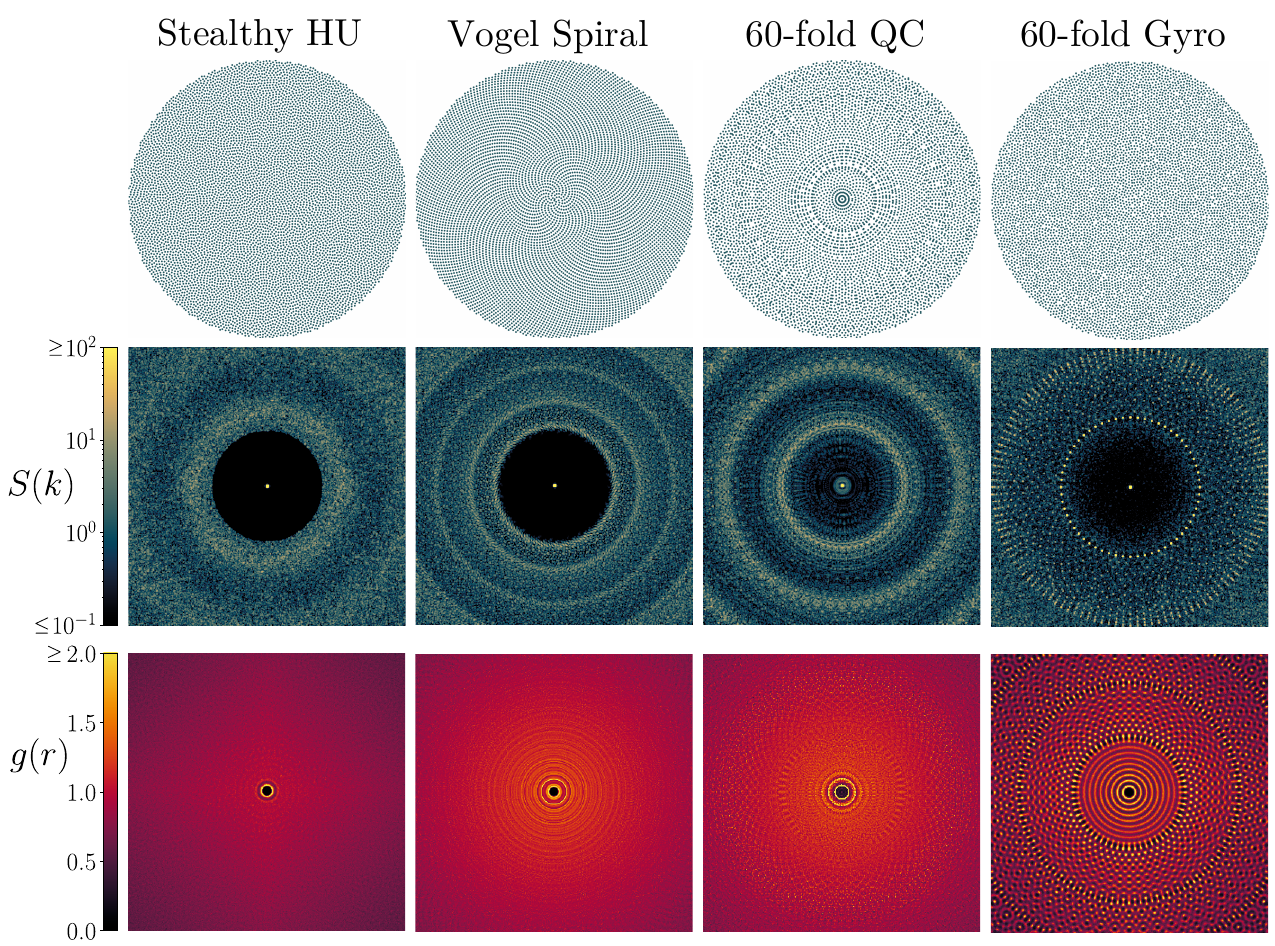}
    \caption{\textbf{Structure factors of bandgap formers.}
    Point patterns $\rho(\bm{r})$ (top row), structure factors $S(\bm{k})$(middle row) and $2d$ pair correlation functions $g(\bm{r})$ (bottom row) of the systems considered in the main text, from left to right, a stealthy hyperuniform point pattern, a Vogel spiral, a $60$-fold de Bruijn quasicrystal, and a $60$-fold gyromorph.
    }
    \label{fig:BandgapStructures}
\end{figure}

We argue in main text that all previously proposed aperiodic bandgap formers had one specific commonality: the presence of a feature in $S(k)$ at $k = 2 k_{gap}$, where $k_{gap}$ is the wavevector of the bandgap, such that $S(k) > 1$ surrounded by lower values.
In Fig.~\ref{fig:BandgapStructures} we show complete evidence supporting this fact.
The middle row shows intensity maps of the $2d$ structure factors, and the bottom row intensity maps of the $2d$ $g(\bm{r})$ of the systems studied in main text: a stealthy hyperuniform system, a Vogel spiral, a $60$-fold de Bruijn quasicrystal, and a $60$-fold gyromorph.
In this figure, all systems contain $N \sim 10^4$ points, filling up a circular volume with diameter $L$ homogeneously (top row).
The stealthy hyperuniform configuration was prepared using FReSCo~\cite{Shih2023}, at a ``degree of stealthiness'' $\chi \equiv \mathcal{N}_K / (2 d (N - 1)) = 0.5$, where $\mathcal{N}_K$ is the number of modes constrained to have low structure factor values.
This value was reported to display a pseudogap in past works~\cite{Froufe-Perez2017,Monsarrat2022,Froufe-Perez2023}.

The structure factor, both here and in the main text, is computed using a type-I transform from the FINUFFT~\cite{Barnett2019,Barnett2020} library, which avoids explicitly discretizing the point pattern to compute its FFT (and periodicizes the system for these plots).
The transform is here computed for $200$ modes on either side of the origin and on each axis.
Note that for the sake of comparison, we cap the log-scale intensity map on both ends.
The $g(\bm{r})$, both here and in the main text, is computed independently in real space by binning distance vectors between particles, using square bins with sidelength $\delta \equiv 0.05 * L / (2 \sqrt{N})$ ($1/20$ of the typical distance between points).
The result is here plotted in the range $[-L/4; L/4]^2$.
To better show the structure, the top of the intensity map is capped.

The figure clearly shows commonalities between all considered systems in both real and reciprocal spaces.
In $S(\bm{k})$, all these systems display a near-isotropic low-high-low feature in their intensity when moving away from the center radially.
The ring of high values assumes different maximal intensities, degrees of isotropy, and typical widths depending on the system under consideration: gyromorphs are a proposed way to maximize intensity, remain close to isotropic, and minimize width for that feature.
Note that the values of the $S(\bm{k})$ near zero are generally smaller, but not generically stealthy.
Quasicrystals, while they are stealthy in the limit of infinite system size, have a dense set of finite-width peaks in finite size, leading to many non-zero values in the measured $S(\bm{k})$, and gyromorphs achieve low but clearly non-zero values near the center, as discussed in the main text.

Note that the ring of higher values in the stealthy hyperuniform system is not a consequence of being stealthy in general, but is known to systematically appear at high $\chi$ $(\chi \gtrsim 0.5)$.
It has been proposed that this ripple could be understood as a Fourier-space analogue to the appearance of secondary peaks in the $g(r)$ of hard spheres~\cite{Torquato2015}.
Coincidentally, bandgaps in stealthy hyperuniform systems prepared using reciprocal-space optimization like FReSCo~\cite{Uche2004,Uche2006,Morse2023} only ever display bandgaps at $\chi \approx 0.5$; the ripple is likely the feature responsible for it.
This is also why the bandgap in stealthy hyperuniform systems is in fact very similar to those seen in equilibrium hard sphere configurations~\cite{Froufe-Perez2016,Shih2023}, or in networked structures with only short-range order~\cite{Liew2011, Sellers2017}.

The pair correlation functions, in the bottom row of Fig.~\ref{fig:BandgapStructures}, also show some similarities (as they should, since they are simply the Fourier transforms of $S(k)$).
The first similarity is that all systems display an isotropic exclusion radius near $\bm{r} = \bm{0}$, a feature not trivially seen from the structure factor, but argued to be important for bandgaps in the past~\cite{Sellers2017}.
Next, notice that there is a typical lengthscale between rings of neighbors appearing in fine, often weak features of the pair correlation.
In fact, this feature is particularly strong in gyromorphs: points have extremely regular spacings up to large distances.
Note that a shape reminiscent of the gyromorph's $g(\bm{r})$ structure is also present, but much weaker, in the de Bruijn quasicrystal.
This is because the quasicrystal has many peaks of comparable heights, at very different length scales, creating a more complicated superposition of cosines in the $g(\bm{r})$.
Gyromorphs, on the other hand, are by definition structures that amplify just one ring of peaks in $S(\bm{k})$, and as few other features as possible, resulting in a much less ``harmonic-rich'' $g(\bm{r})$.

Note that the comparison between the $S(\bm{k})$ of quasicrystals and the $g(\bm{r})$ of gyromorphs (and vice-versa) presented in the main text is performed at very different scales: the rings shown in the $S(\bm{k})$ of the quasicrystal in Fig.~\ref{fig:BandgapStructures} are the innermost rings in the $g(\bm{r})$ of the gyromorph.
Likewise, it is the nearest-neighbor ring in the $g(\bm{r})$ of the quasicrystal that matches the ring of peaks of the gyromorph.

\section{Numerical methods}

\subsection{Gyromorph generation}

\begin{figure}
    \centering
    \includegraphics[width=\columnwidth]{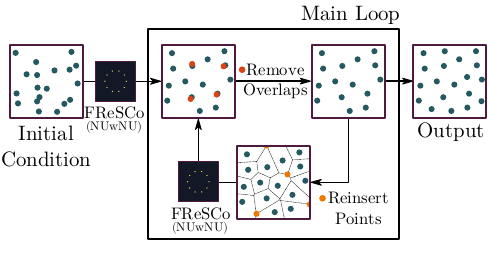}
    \caption{\textbf{Algorithm: generation of a gyromorph.}
    Sketch of the generation procedure for a $10$-fold gyromorph.
    }
    \label{fig:Algorithm}
\end{figure}

To generate gyromorphs, we use the Fast Reciprocal-Space Correlator (FReSCo)~\cite{Shih2023}, an optimization method that imposes arbitrary features into the structure factor of a point pattern.
Specifically, we use a variant of the NUwNU (non-uniform real space with non-uniform k-space constraints) routine, which imposes constraints at continuous reciprocal space positions, assuming free boundary conditions for the point pattern.
The usual NUwNU amounts to minimizing a loss function, defined as
\begin{align}
    L_{\text{NUwNU}}(\bm{r}_1, \ldots, \bm{r}_N) \equiv \sum\limits_{p = 1}^{G/2} (S(\bm{k}_p) - S_0 (\bm{k}_p))^2, \label{eq:FReSCoLoss}
\end{align}
where the sum runs over $G/2$ peaks placed at locations $\bm{k}_p$ of a continuum Fourier space, and $S_0$ is a target value at these specific locations.
Assuming that all $\bm{k}_p$ are distinct and belong to the same half-space, the $G/2$ peaks at $-\bm{k}_p$ are implicitly constrained since the point pattern is a real-valued field~\cite{Friedel1913,Schwarzenbach1996,Shih2023}.
Within this paper, we set $\bm{k}_p$ to be regularly spaced on a circle in $2d$, and on a sphere in $3d$, with a modulus $K$.
Furthermore, we set the target to height $S_0(\bm{k}_p) = N \sim K^2$.
To ensure that peaks are about equal heights throughout the minimization, which prevents the emergence of lower-symmetry configurations, we also add an extra term to the loss that controls the variance across peaks,
\begin{align}
    L_{var} \equiv \frac{C_{var}}{G}\sum\limits_{p=1}^{G} \left(S(\bm{k}_p) - \overline{S_G} \right)^2,
\end{align}
where $C_{var}$ is an arbitrary scaling factor that weighs this term against the main FReSCo term, and 
\begin{align}
    \overline{S_G} \equiv  \frac{1}{G}\sum\limits_{p=1}^{G} S(\bm{k}_p).
\end{align}
Using that, for $\rho(\bm{r}) = \sum_{n=1}^N c_n \delta(\bm{r} - \bm{r}_n)$ with $c_n \in \mathbb{C}$,
\begin{align}
    \frac{\partial S}{\partial \bm{r}_n}(\bm{k}) = \frac{2}{N} \text{Re}\left[ \widehat{\rho}(\bm{k}) \frac{\partial \widehat{\rho}^\dagger}{\partial \bm{r}_n}(\bm{k}) \right] = 2 Re \left[ - \frac{i\bm{k}c^\dagger_n}{N}  \widehat{\rho}\,(\bm{k})e^{-i\bm{k}\cdot\bm{r}_n} \right]. \label{eq:Sgradient}
\end{align}
The gradient of the variance component of the loss may then be written explicitly as
\begin{align}
    \frac{\partial L_{var}}{\partial \bm{r}_n} = -\frac{2 C_{var}}{NG} \sum\limits_{p=1}^{G}\left[\frac{\partial S}{\partial \bm{r}_n}(\bm{k}_p) - \frac{\partial \overline{S_G}}{\partial \bm{r}_n}  \right] \left(S(\bm{k}_p) - \overline{S_G} \right).
\end{align}
which, noticing that 
\begin{align}
    \sum\limits_{p=1}^{G}\frac{\partial \overline{S_G}}{\partial \bm{r}_n}  \left(S(\bm{k}_p) - \overline{S_G} \right) = \bm{0}
\end{align}
eventually yields
\begin{align}
    \frac{\partial L_{var}}{\partial \bm{r}_n} = -\frac{2 C_{var}}{NG} \text{Re}\left[\sum\limits_{p=1}^{G} \left(S(\bm{k}_p) - \overline{S_G} \right) i \bm{k}_p \hat{\rho}(\bm{k}_p) c_n^{\dagger}e^{-i \bm{k}_p \cdot \bm{r}_n}  \right].
\end{align}
Like the FReSCo loss~\cite{Shih2023}, this can be read as an inverse Fourier transform, $\frac{\partial L_{var}}{\partial \bm{r}_n} = \text{Re}[\widehat{g}_{var}(\bm{r}_n)]$ with $g_{var}(\bm{k})$ defined as
\begin{align}
    g_{var}(\bm{k}) = -\frac{2 C_{var}}{NG} \left(S(\bm{k}) - \overline{S_G} \right) i \bm{k} \hat{\rho}(\bm{k}) c_n^{\dagger}.
\end{align}
This expression conveniently uses only values that are already computed to evaluate the usual FReSCo gradient~\cite{Shih2023}: this extra loss and its gradient may both be computed within the same Fourier transforms as those used to compute the FReSCo loss and gradient.
Starting from an initially uncorrelated random point pattern, we minimize the total loss $L = L_{NUwNU} + L_{var}$.
In practice, throughout this work, we set $C_{var} = 1$.

Interestingly, we show in main text that after minimizing, $S(\bm{k}_p) \propto N/G$ (the largest achievable peak height scaling, see argument in Sec.~\ref{sec:RealSpaceStructure}), so that $S(\bm{k}_p) \ll S_0 (\bm{k}_p)$ at the imposed peaks.
As a result, the loss in Eq.~\ref{eq:FReSCoLoss} is well approximated by:
\begin{align}
    L(\bm{r}_1, \ldots, \bm{r}_N) \sim - \sum\limits_{p=1}^{G/2} S(\bm{k}_p). \label{eq:SimplerLoss}
\end{align}
This simpler asymptotic expression is useful to get intuition about the effective interaction between particles.
Indeed, recalling the definition of $S(k)$ from the Fourier transform of the density field, and introducing $\bm{r}_{mn} \equiv \bm{r}_n - \bm{r}_m$ one may write
\begin{align}
    S(\bm{k}) &= \frac{1}{N} \sum\limits_{m,n = 1}^{N} e^{i \bm{k}\cdot \bm{r}_{mn}} \\
    &= 1 + \frac{2}{N} \sum\limits_{1 \leq m<n \leq N} \cos \bm{k} \cdot \bm{r}_{mn}.
\end{align}
As a result, for $G \gg 1$, gyromorphs can be thought of as minima of an energy landscape, built up by pairwise interactions given by a potential
\begin{align}
    V_{G}(\bm{r}) &\propto - \sum\limits_{p = 1}^{G/2} \cos (K \hat{\bm{e}}(\theta_p) \cdot \bm{r}), \label{eq:Sum_of_cosines}
\end{align}
where $\theta_p = 2\pi p/ G$.
This effective pairwise interaction is long-ranged, and very peculiar in that it depends on externally defined directions.
In fact, it assumes a shape similar to the $g(\bm{r})$ we report in the main text and in the gyromorph column of Fig.~\ref{fig:BandgapStructures}, as obtained from an inverse Fourier transform of $S(k)$.
Note however that, unlike in holographic techniques used to trap colloids along a similar function~\cite{Jagannathan2014} or in kaleidoscopic fields~\cite{Chen2011,Tuan2023} used to etch into materials~\cite{Zoorob2000,Lubin2012}, this function here acts as a pair potential between points, not just as an external field, making our gyromorphs distinct from both categories of systems.
Also note that our full loss also contains a $4$-point potential term due to the $S(k)^2$ term, which is not captured by this approximation, and the variance component of the loss that also affects interactions.

The optimization is performed by feeding the configuration and gradient to L-BFGS~\cite{Liu1989}, a quasi-Newton method, with a maximal step size and a backtracking line-search~\cite{Nocedal1999}.
During minimization, points are neither precluded from exiting the box where they started, nor from overlapping exactly.
We thus periodically remove points that drifted outside of the box, as well as points closer than $\pi/K$, the first maximum of the effective potential in Eq.~\ref{eq:Sum_of_cosines}, and reinsert as many points as we removed, then minimize again.
To limit the number of overlaps at the next cycle, the reinsertion sites are chosen as the Voronoi sites lying furthest from their neighboring points (farthest-first batch insertion~\cite{Rosenkrantz1977}).
Each minimization step is run either until $10^4$ steps have been computed, or until the gradient of the loss becomes smaller than $10^{-20}$, whichever comes first.
We thus create a $G$-fold gyromorph.

\subsection{Structure factor calculations}

Throughout the paper, we plot a number of structure factors.
To plot them in spite of the finite size of systems and of the fact that they have free boundary conditions, we perform the following operations.

First, the point pattern is rescaled so that its points are in the support $[-L/2; L/2]$.
Then, we cut the system into a disk with radius $L/2$, the largest disk inscribed in the square, which is the system we use for optical measurements in practice.
Then, we window the density function using a Hamming window, meaning that the effective density field fed into Fourier transforms is 
\begin{align}
    \rho_w (\bm{r}) = \sum\limits_{n = 1}^{N} c_n(r) \delta(\bm{r} - \bm{r}_n),
\end{align}
where
\begin{align}
    c_n(r) \equiv \frac{1}{\mathcal{N}}\left(a_0 + a_1 \cos \frac{2 \pi r}{ \sqrt{2} L}\right)
\end{align}
and $a_0 = 0.54$, $a_1 = 0.46$, following standard conventions for the coefficients of the Hamming window~\cite{Mallat1999}, and setting the zero at the maximal distance in the initial square.
The normalization $\mathcal{N}$ is set so as to preserve the total mass in the field, $\sum_{n=1}^N c_n = N$.

We emphasize that these operations are strictly performed to avoid periodicization artifacts in $2d$ intensity maps, which are computed with non-uniform to uniform Fourier transforms (Type 1 transforms in fiNUFFT~\cite{Barnett2019,Barnett2020}), and are \textit{not} affecting the gyromorph generation algorithm, which only relies on non-uniform to non-uniform transforms (Type 3 transforms in fiNUFFT~\cite{Barnett2019,Barnett2020}).
Likewise, they do not affect any $g(\bm{r})$, as these are computed in real space by binning pair distances.
Likewise, when reporting the profile and height of peaks, we use Type 3 transforms to evaluate points around the peak.

\subsection{Measurement of scattering mean-free paths}

To numerically measure scattering mean-free paths $\ell_s$, we rely on a well-known~\cite{Carminati2021} measurement strategy using slab-geometry transmission measurements, following Ref.~\cite{Monsarrat2022} for instance.
Starting from a square chunk of a $2d$ system with sidelength $L$, we subselect a slice (or ``slab'') with depth $D < L$.
We then shine the system with a Gaussian beam with pulsation $\omega = k_0 c$ and beam waist $w < L$ (in practice $w = L/5$), orthogonal to the long direction of the slab and centered at the center of the slab.
We solve the coupled dipoles linear system for that source.
We then evaluate the field $\bm{E}(\bm{r}; \omega)$ on a plane parallel to the long direction of the slab, a distance $X \gg \lambda = 2\pi/k_0$ away from the slab to avoid strong near-field effects (in practice $X/\lambda = 10$.
Within that plane, we spatially average the field over a line segment with sidelength $d = X$ symmetric around the center of the slab, yielding $\overline{\bm{E}}$.
We finally average that field over $30$ realizations of the disorder (either gyromorphs or SHU systems prepared with FReSCo), and compute the ballistic part of the intensity through the average medium,
\begin{align}
    I_{b}( D; \omega) \equiv \left|\langle \overline{\bm{E}}(D; \omega)\rangle \right|^2.
\end{align}
One may show~\cite{Carminati2021} that this ballistic component of the intensity is expected to decay exponentially with the depth of the slab, with a rate of decay given by the scattering mean-free path $\ell_s$, $I_b(D; \omega) \propto e^{-D/\ell_s(\omega)}$.
We thus perform the measurement of $I_b$ across values of $D$, here $L/100 \leq D \leq L/10$ by steps of $L/100$.
We then perform a linear fit on $\ln I_b$ against $D$ to evaluate $\ell_s$.

\section{Structure of Gyromorphs}

\subsection{Gyromorphs are not Hyperuniform}

\begin{figure}[b]
    \centering
    \includegraphics[width=0.96\columnwidth]{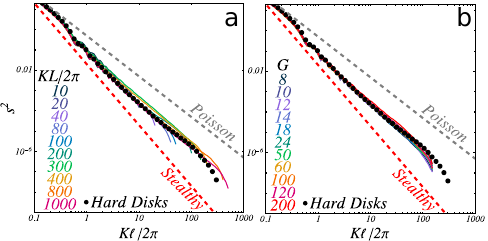}
    \caption{\textbf{Gyromorphs are not hyperuniform.}
    $(a)$ Reduced number variance from randomly drawn disks with radii $\ell$ in $60$-fold gyromorphs with various $KL/2\pi$ from $10$ to $1000$ (solid colored lines), against $K \ell/2\pi$, in log-log scale.
    A dashed gray line indicates Poisson scaling, $s^2 \sim \ell^{-d}$, and a dashed red line the stealthy scaling $s^2 \sim \ell^{-d-1}$.
    Black disks indicate data from equilibrium hard disks with a $\phi = 0.6$ filling fraction, for which the $x$ axis has been rescaled using the location of the first peak as a $K$ value.
    $(b)$ Similar plot but varying the order of rotational symmetry $G$ for $KL/2\pi =300$.
    }
    \label{fig:hyperuniformity}
\end{figure}

In the main text, we show that gyromorphs are not hyperuniform based on their structure factor.
We present further evidence that gyromorphs are not hyperuniform in Fig.~\ref{fig:hyperuniformity}, where we perform discrepancy measurements for disks~\cite{Pilleboue2015,Torquato2018, Shih2023}.
We draw random disks with radius $\ell$ at uniform positions within $60$-fold gyromorphs and count the number of points that each disk contains.
We then measure the mean number $\mu$ and the variance $\sigma^2$ of this number using $10,000$ disks, and compute the reduced variance $s^2 \equiv \sigma^2 / \mu^2$ for each $\ell$.
The output is shown when varying $K$ for $G = 60$ in Fig.~\ref{fig:hyperuniformity}$(a)$, where $s^2$ is plotted against $K \ell / 2\pi$ in logarithmic scales.
We show that gyromorphs display number fluctuations close to Poissonian ($s^2 \sim \ell^{-d}$) at small length scales, then feature sub-Poissonian fluctuations at ranges larger than $1/K$.
While the behaviour is non-Poissonian, the slope is much weaker than that of stealthy hyperuniform systems ($s^2 \sim \ell^{-d-1}$).
In fact, we show that gyromorphs are near-indistinguishable from equilibrium hard disk liquids, using data from an $N\sim 10^5$ equilibrium configuration obtained by Event-Chain Monte Carlo~\cite{Bernard2009}.
This conclusively shows that gyromorphs are \textit{not} hyperuniform, even in the largest systems we test, $N \sim 10^6$.
We show in Fig.~\ref{fig:hyperuniformity}$(b)$ that this result holds when scanning values of $G$ at a fixed $K$, even at rotational symmetries of order as low as $8$-fold, where gyromorphs may more easily be confused~\cite{Shih2023} with quasicrystals (which follow the stealthy scaling~\cite{Torquato2018}).

\subsection{``Duality'' between Gyromorphs and Quasicrystals}

\begin{figure}[b]
    \centering
    \includegraphics[width=\columnwidth]{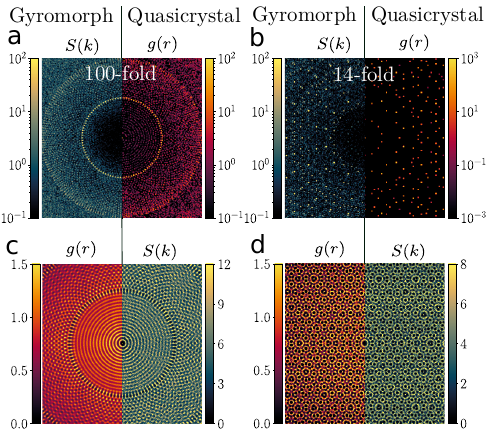}
    \caption{\textbf{Gyromorphs as duals of Quasicrystals.}
    Comparisons between the $S(k)$ (resp. $g(r)$) of $2d$ gyromorphs and the $g(r)$ (resp. $S(k)$) of quasicrystals for $(a), (c)$ $G=100$ and $(b), (d)$ $G = 14$.
    The plot ranges and intensity scales are adjusted to match images.
    }
    \label{fig:gScomparison}
\end{figure}
We here highlight an interesting link between gyromorphs and quasicrystals.
In Fig.~\ref{fig:gScomparison} we show side-by-side comparisons of the $g(\bm{r})$ of a gyromorph and the $S(\bm{k})$ of a quasicrystal (and vice-versa) for $G=100$ in panels Fig.~\ref{fig:gScomparison}$(a)$,$(c)$, and $G=14$ in panels Fig.~\ref{fig:gScomparison}$(b)$,$(d)$.
All systems were made using $N \approx 10^4$ points.
The quasicrystals are constructed using de Bruijn's multigrid and dual method~\cite{deBruijn1981,deBruijn1986,Lutfalla2021}.
The panels show near-perfect agreement between the real (resp. reciprocal) features of gyromorphs and the reciprocal (resp. real) features of quasicrystals.
In other words, gyromorphs look like duals of quasicrystals.
This is because the $S(k)$ of gyromorphs (resp. the $g(r)$ of quasicrystals), as further discussed in Sec.~\ref{sec:RealSpaceStructure}, is dominated by one ring of high peaks, so that its Fourier transform is well approximated by a sum of cosines -- although $g(r) = 0$ at short range in gyromorphs by design.

\subsection{Scalings of $3d$ gyromorphs' peaks}

\begin{figure}
    \centering
    \includegraphics[width=0.5\linewidth]{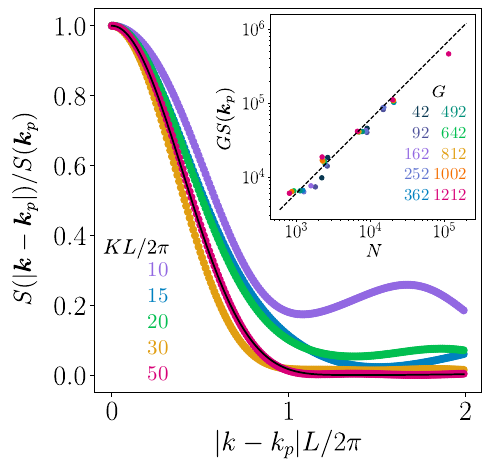}
    \caption{\textbf{Scaling of peak height in $3d$ gyromorphs.}
    Radially averaged profile of peaks across $K$ values, rescaled by peak height (colored lines) a solid black line indicates the radial profile of $\text{sinc}^2(k_x L/2)\text{sinc}^2(k_y L/2)\text{sinc}^2(k_z L/2)$.
    Inset: Rescaled peak height $G S(\bm{k}_p) $ against $N$, in log scales, across $G$ (colored symbols).
    A dashed black line shows $GS(\bm{k}_p) = N$.
    }
    \label{fig:3dpeakscaling}
\end{figure}

In the main text, we show evidence that peak heights scale like $S(\bm{k}_p) \sim N/G$ at peak locations $\bm{k}_p$ in $2d$.
Here, we show that the same result is recovered in $3d$ gyromorphs.
We generate a set of $3d$ gyromorphs with various $K$ and $G$ values, and report results in Fig.~\ref{fig:3dpeakscaling}.
In the main panel, we show the profile of the peak at $(KL/2\pi,0)$, normalized by its height, as a function of the distance $|k-k_p|L/2\pi$ to the maximum, across system sizes for $G = 1212$. 
As $N$ grows, the profile converges to $\prod_{a \in {x,y,z}}\textrm{sinc}^2(k_a L/2)$, implying that the linear width of the peaks in $k$-space decays like $1/L$.
Note that the convergence is slower with $N$ than in $2d$, as it depends on the number of particles per sidelength, $N^{1/d}$, rather than the number of points.
The inset shows that, for different values of $G$, the peak heights grow like $\mathcal{O}(G N)$, indicating that the peaks are extensive for any fixed order symmetry.
Altogether, the peaks have extensive height while their area decays like $1/L^3 \sim 1/N$: they approach Dirac deltas as $N$ increases, like in the $2d$ case discussed in main text.

\subsection{Real-space structure \label{sec:RealSpaceStructure}}

\begin{figure}
    \centering
    \includegraphics[width=0.9\linewidth]{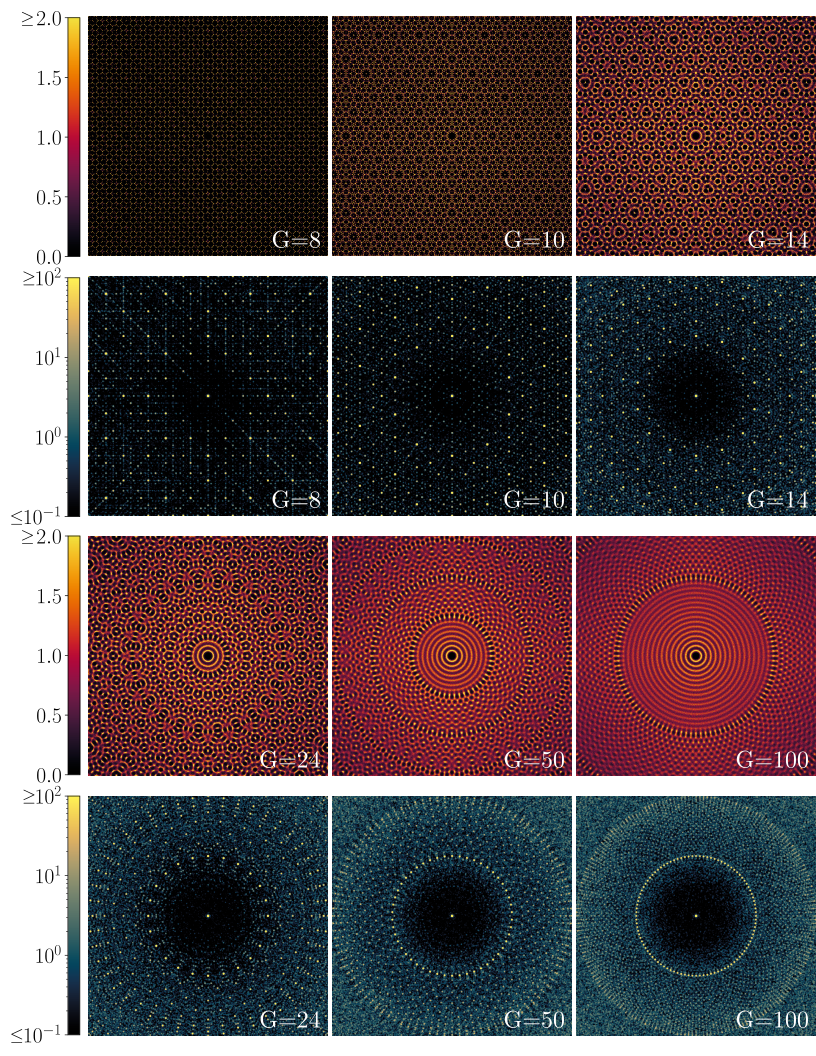}
    \caption{\textbf{Structure of gyromorphs.}
    $g(\bm{r})$ (first and third rows) and $S(\bm{k})$ (second and fourth rows) of gyromorphs for $G = 8$, $10$, $14$, $24$, $50$, $100$ and $KL/2\pi = 100$.
    $S(\bm{k})$ are plotted up to $2 K$ in both directions, and $g(\bm{r})$ up to $L/4$.
    }
    \label{fig:GyroGallery}
\end{figure}

In this section, we discuss the real-space structure of gyromorphs in more detail, in  particular the radius at which a gear forms and the relation between gyromorphs, Moiré patterns, and quasicrystals.

First, we give a simple argument to evaluate the value of $R_G$ and the number of rings in the isotropic part of the $g(r)$ of gyromorphs.
In a nutshell, we note that $G$-fold order can only be supported on a ring large enough that it may contain $G$ points.
In other words, since the distance between two neighboring points on the ring is the nearest-neighbor distance $a=2\pi/K$, the perimeter of the gear has to be such that $2\pi R_G \geq G a$.
Thus, $R_G \gtrsim G/K$.
Furthermore, since rings are located at radii multiple of $a$, $R_G = N_G a$ and the ring carrying the rotational order is the $N_G \equiv \lceil G/(2\pi) \rceil$-th one.
For instance, in the example $G=60$ that we focus on in the main text, $N_G = 10$.
Note that for small $G$, we recover the square ($G=4$) and triangular ($G=6$) lattices ($N_G = 1$)~\cite{Shih2023}.

We now turn our attention to the structure of gyromorphs as we tune the value of $G$.
To do so, in Fig.~\ref{fig:GyroGallery}, we show intensity maps of the $S(\bm{k})$ and $g(\bm{r})$ for a few different values of $G$, namely $8, 10, 14, 24, 50, 100$, all of them for $K = 100$.
The intensity maps highlight that as $G$ grows, the extent of peaked $S(\bm{k})$ features decreases.
While at low $G$ we find structure factors with many peaks other than the ones we imposed, in a way that is reminiscent of quasicrystals, at high $G$ only a small region has peaks.
Furthermore, even these peaks have intensities that are noticeably lower than on the ring we constrained.
In real space, at low $G$ the $g(\bm{r})$ displays very sharp anisotropic features (high values surrounded by very low values) up to large distances.
However, at large $G$, the long-distance region becomes less contrasted, with broader features.
Furthermore, note that the number of isotropic rings near the center grows following $N_G \equiv \lceil G/(2\pi) \rceil$ as we proposed: $N_8 = N_{10} = 1$, $N_{14} = 2$, $N_{24} = 4$, $N_{50} = 8$ and $N_{100} = 16$.

As the structure factor becomes better and better approximated by one ring of peaks surrounded by a flat background, the $g(\bm{r})$ actually approaches a Moiré pattern (save near $0$, as $g(0) = 0$ by construction).
Indeed, if $S$ approaches the shape
\begin{align}
    S(\bm{k}) = 1 + (2\pi)^d \rho_0 \delta(\bm{k}) + C (2\pi)^d \rho_0 \sum\limits_{p=1}^{G} \delta(\bm{k} - \bm{k}_p)
\end{align}
with $C>0$ a real constant and $\rho_0 = N / L^2$ the average number density (the constant multiplying $C$ being chosen in the same way as for usual Bragg peaks~\cite{Schwarzenbach1996}), and the central peak ensures that the point pattern contains $N$ points.
One may write $g(\bm{r})$ as the Fourier transform~\cite{Hansen2006},
\begin{align}
g(\bm{r}) = \frac{1}{ \rho_0} \int_\mathcal{F} \frac{d^d \bm{k}}{(2\pi)^d} (S(\bm{k}) - 1) e^{-i \bm{k}\cdot \bm{r}}, \label{eq:gfromS}
\end{align}
which immediately yields
\begin{align}
g(\bm{r}) = 1 + C \sum\limits_{p=1}^{G} \cos \bm{k}_p\cdot \bm{r}.
\end{align}
Note in particular that non-negativity of $g$ already suggests $C \propto 1/G$ at most, as empirically observed in gyromorphs.

\begin{figure}
    \centering
    \includegraphics[height=0.30\linewidth]{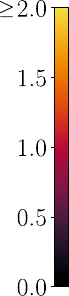}
    \includegraphics[height=0.30\linewidth]{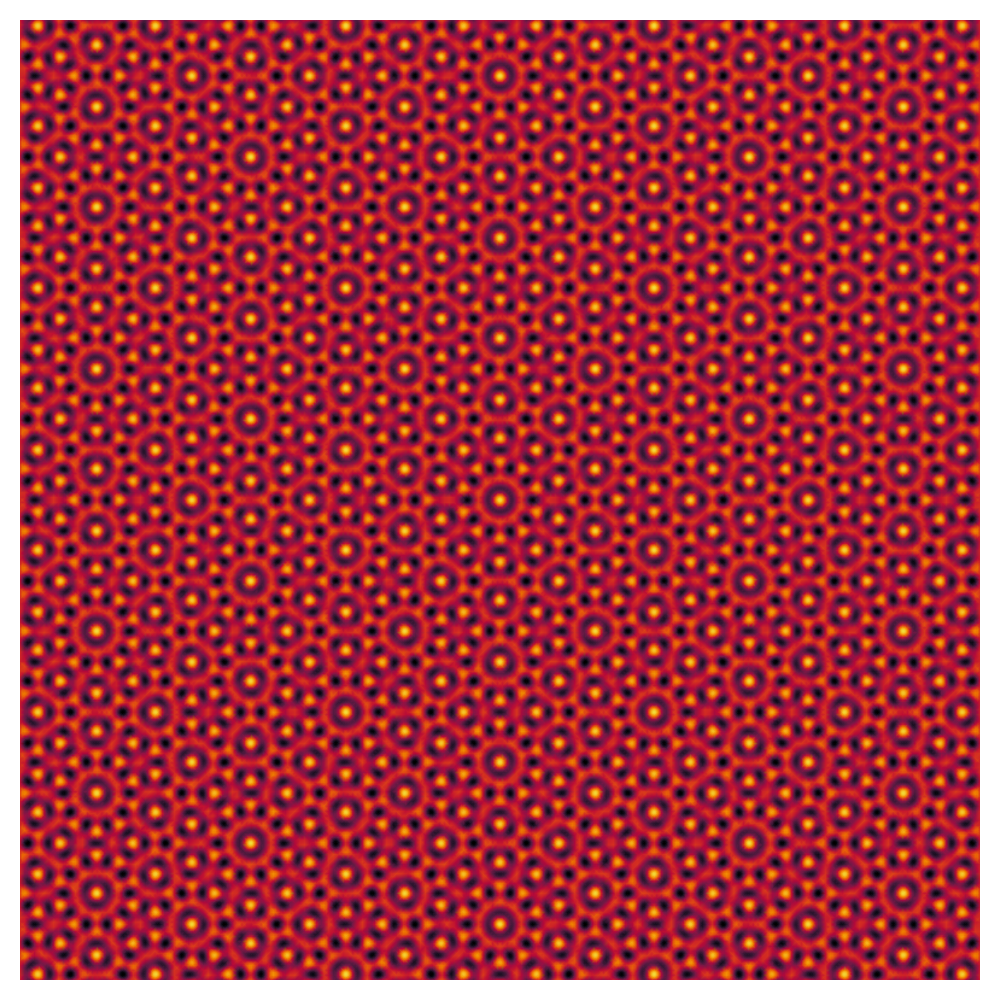}
    \includegraphics[height=0.30\linewidth]{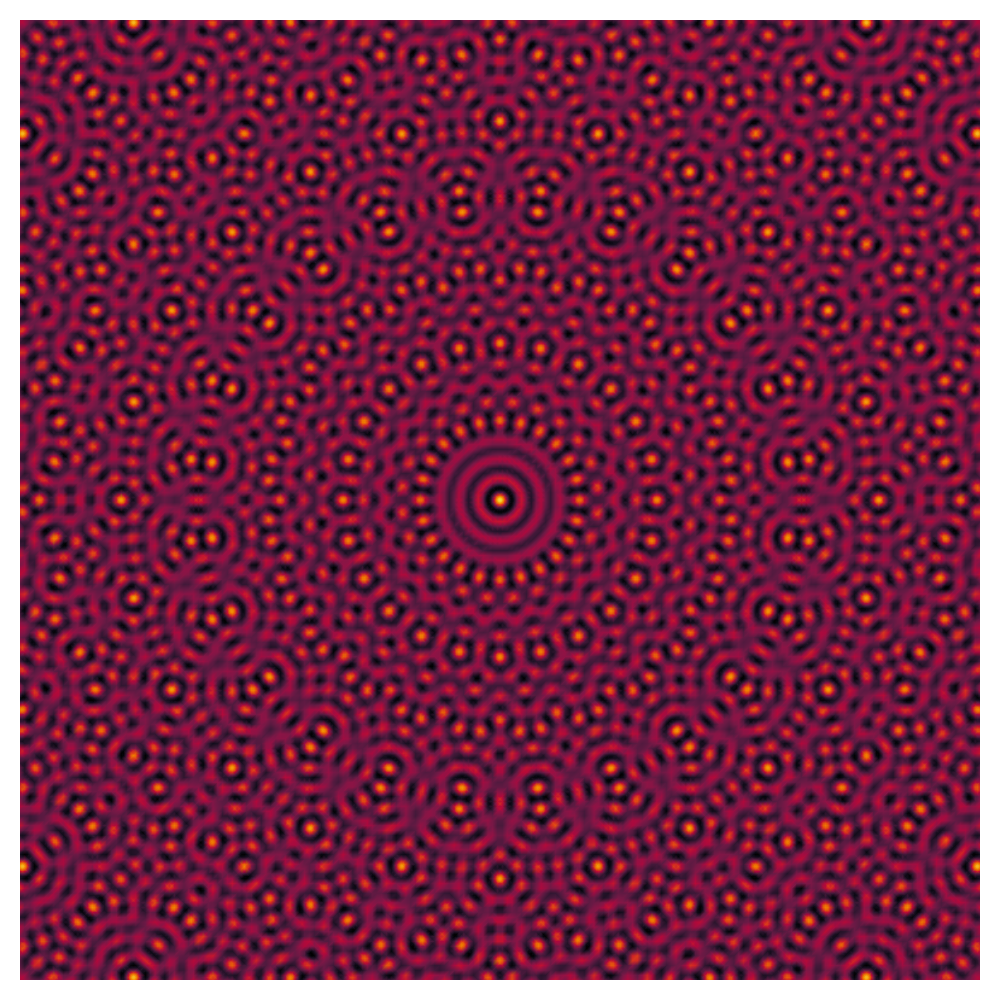}
    \includegraphics[height=0.30\linewidth]{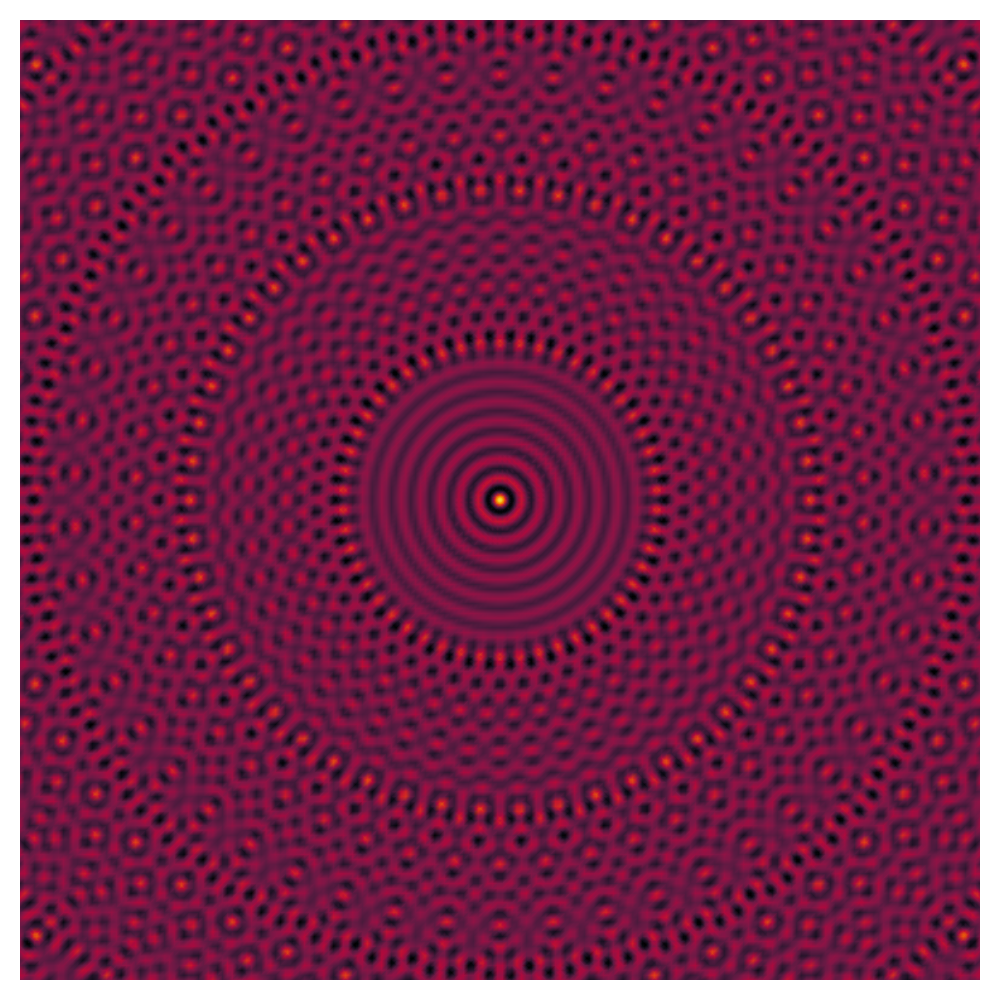}
    \caption{\textbf{Moiré patterns.}
    Example intensity plots of analytical Moiré patterns for $G = 10$, $G = 24$, and $G = 50$ from left to right.
    We choose $C = 1/G$, $K = 2\pi$ and the plotting range is $[-50;50]^2$.
    }
    \label{fig:Moiré}
\end{figure}

Examples of this function are shown in Fig.~\ref{fig:Moiré} for a subset of the symmetries of Fig.~\ref{fig:GyroGallery}.
Notice how close to the observed $g(\bm{r})$ of gyromorph this function becomes as $G$ becomes large -- the main difference being that due to the exclusion of perfect overlaps the $g$ of gyromorphs does not feature a central peak.
This is an indication that gyromorphs actually do approach the maximal possible height for the constrained peaks.
Note that this Moiré shape is however distinct from the point pattern itself: one may not simply subselect all local maxima of the $g(\bm{r})$ and recover the gyromorph.
Finally, note that the replacement of the central peak by hardcore exclusion in gyromorphs qualitatively explains the presence of a hard-sphere-like smooth variation of $S$ at small $k$.

\begin{figure}
    \centering
    \includegraphics[width=0.30\linewidth]{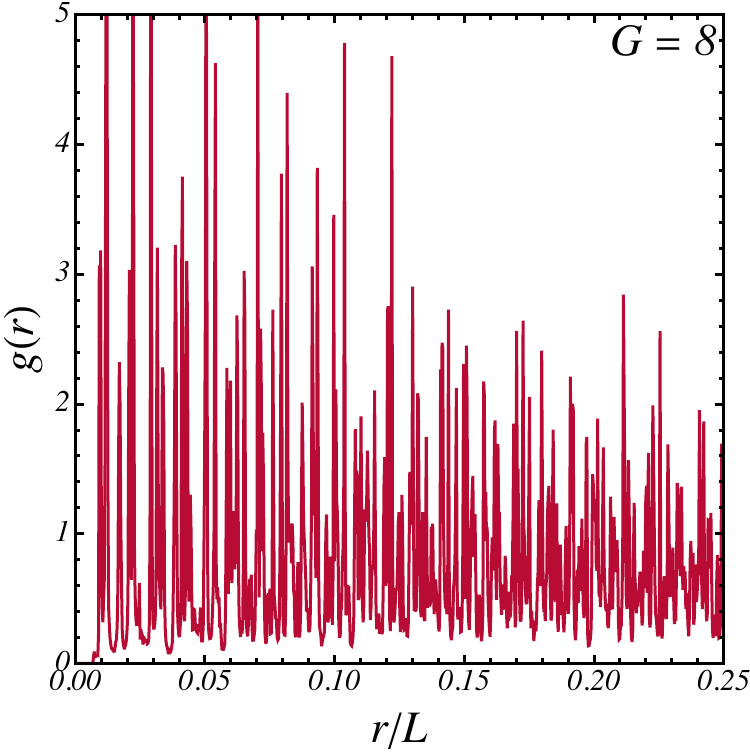}
    \includegraphics[width=0.30\linewidth]{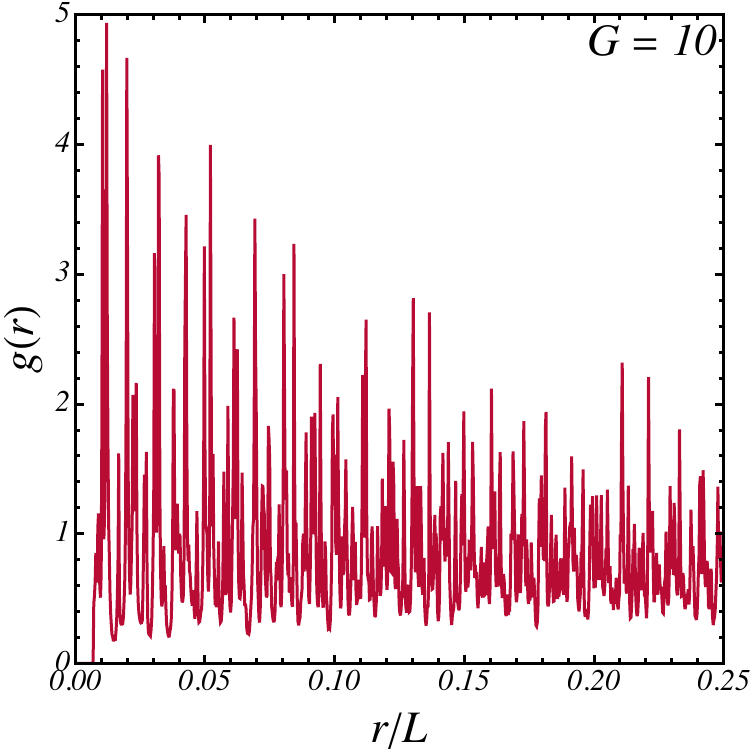}
    \includegraphics[width=0.30\linewidth]{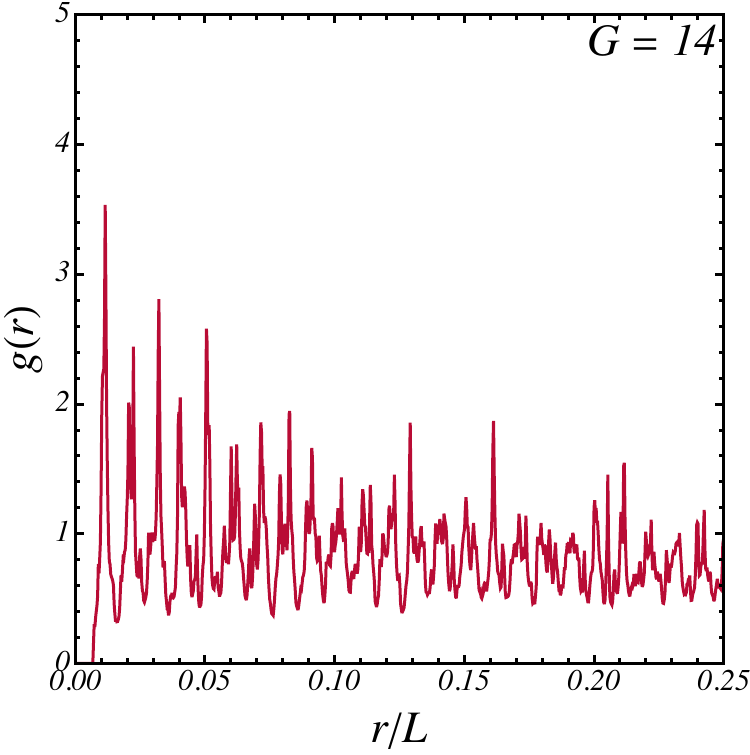}\\
    \includegraphics[width=0.30\linewidth]{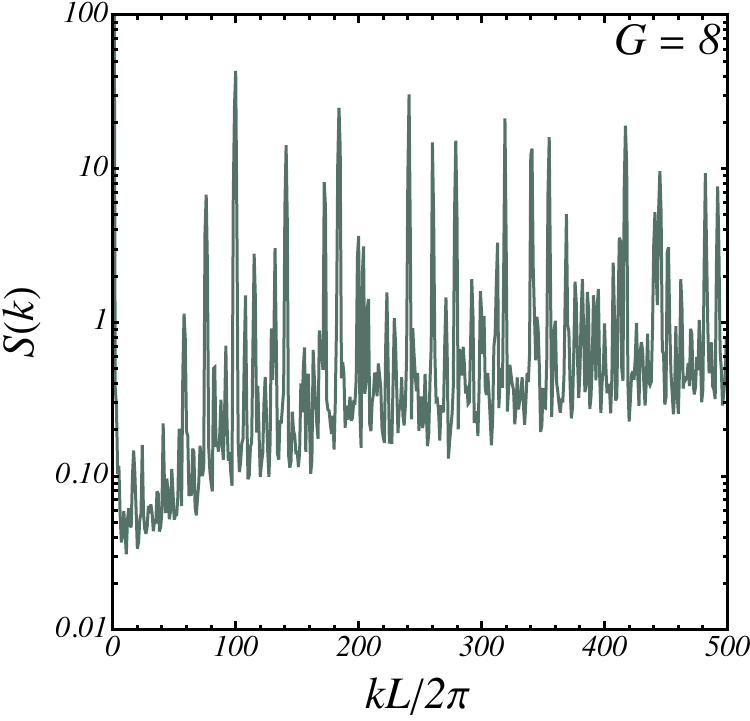}
    \includegraphics[width=0.30\linewidth]{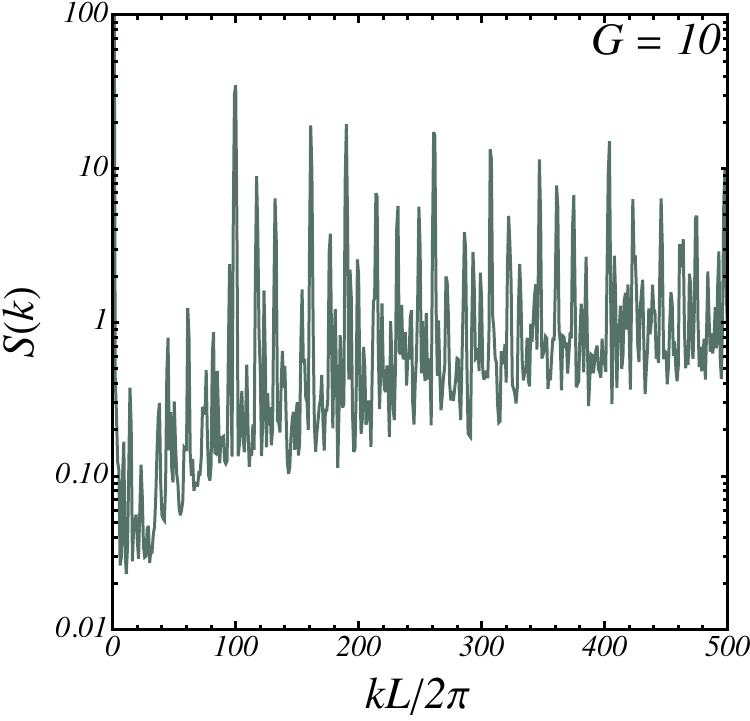}
    \includegraphics[width=0.30\linewidth]{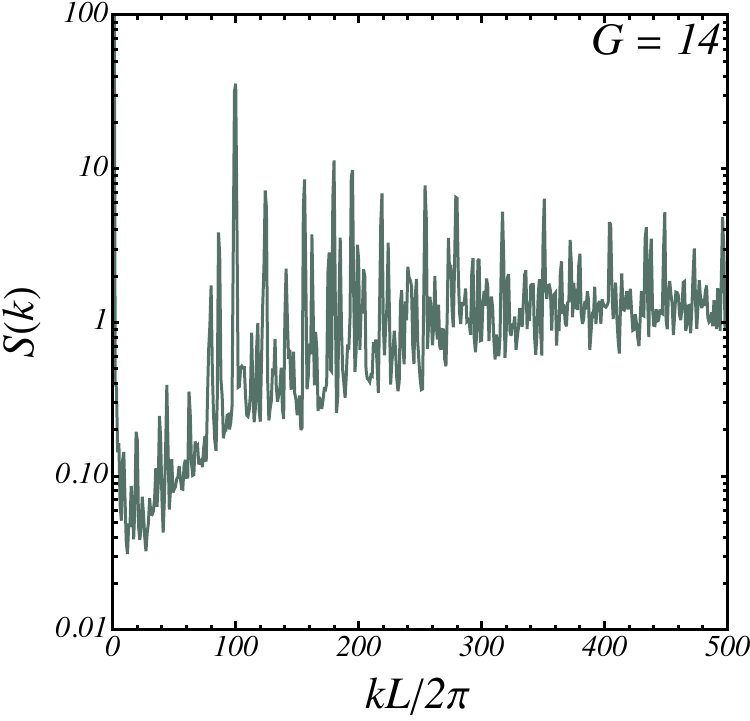} \\
    \includegraphics[width=0.30\linewidth]{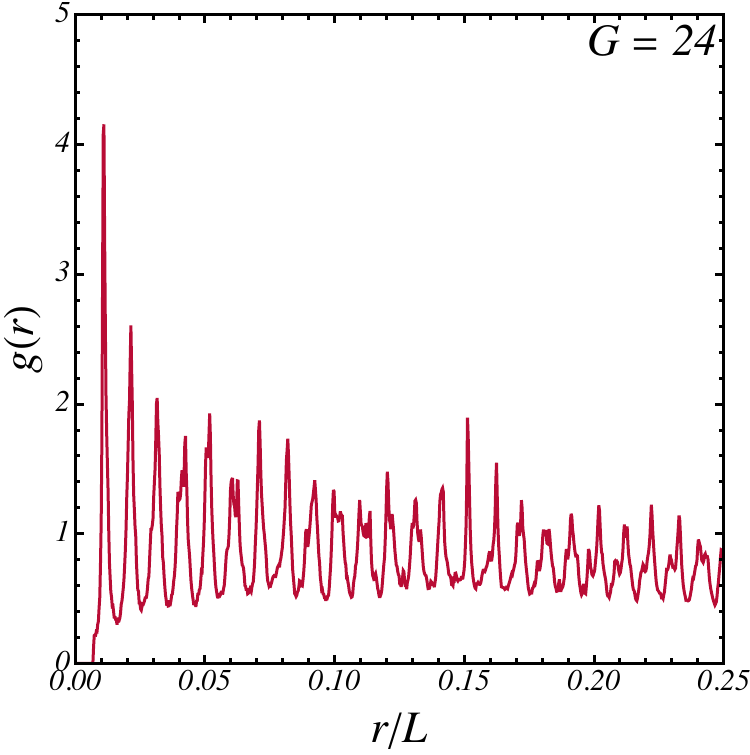}
    \includegraphics[width=0.30\linewidth]{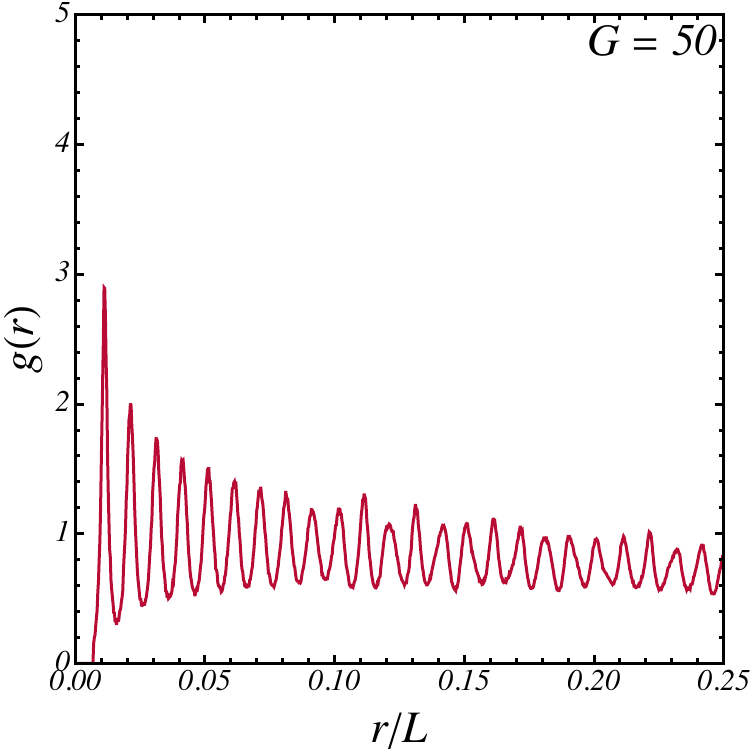}
    \includegraphics[width=0.30\linewidth]{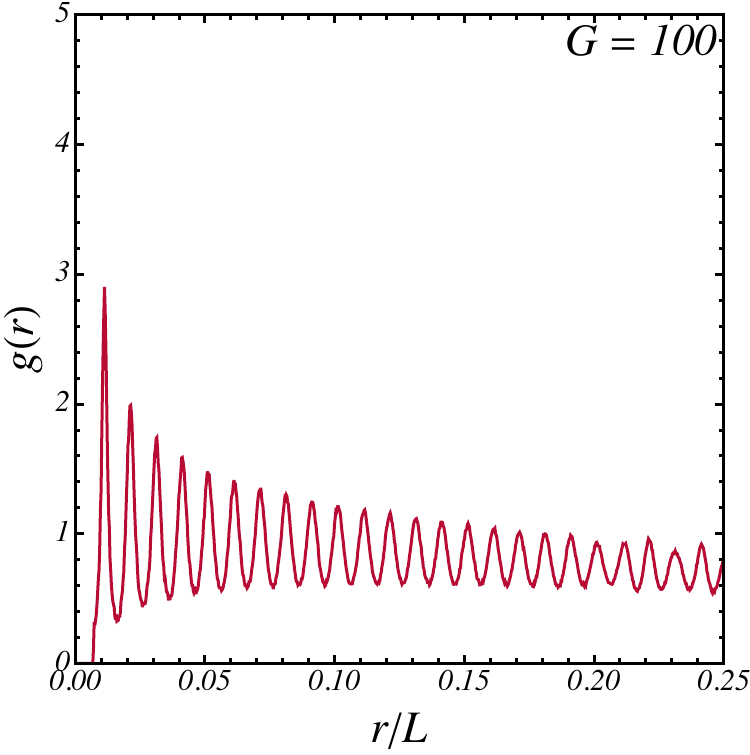} \\
    \includegraphics[width=0.30\linewidth]{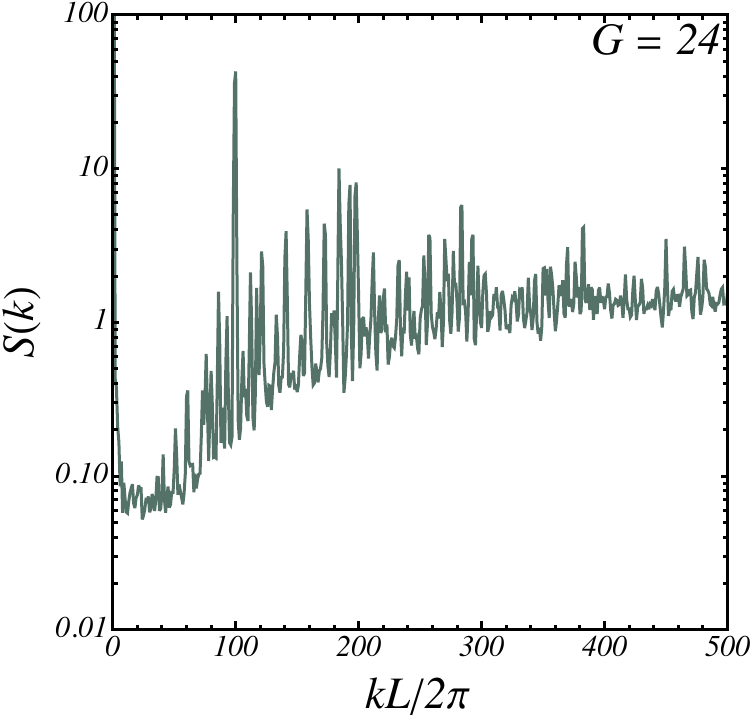}
    \includegraphics[width=0.30\linewidth]{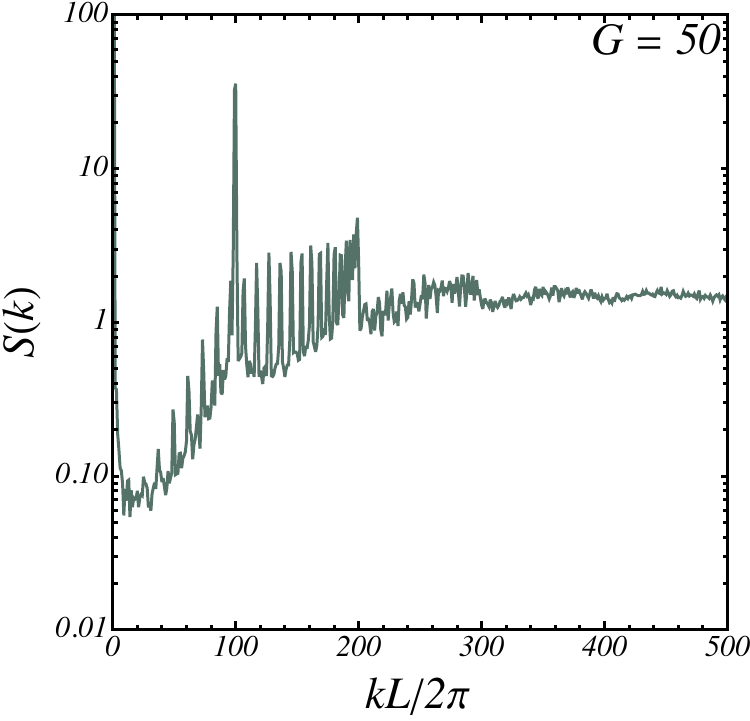}
    \includegraphics[width=0.30\linewidth]{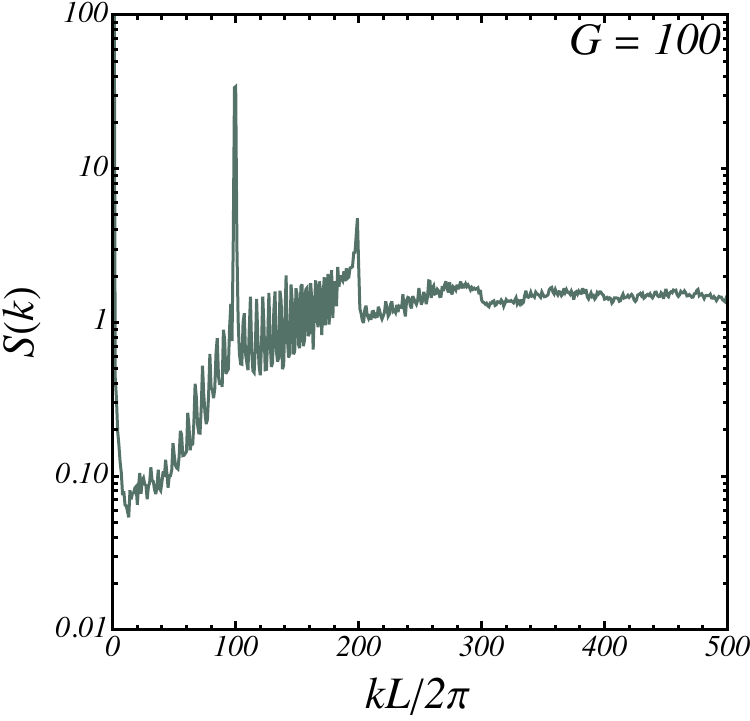} \\
    \caption{\textbf{Angularly averaged structure of gyromorphs.}
    $g(r)$ (first and third rows, in linear scales) and $S(k)$ (second and fourth rows, in lin-log scale) of gyromorphs for $G = 8$, $10$, $14$, $24$, $50$, $100$ and $KL/2\pi = 100$, same as in Fig.~\ref{fig:GyroGallery}.
    }
    \label{fig:GyroRadialGallery}
\end{figure}

To complete the picture, we compute the angularly averaged versions of both the structure factors and pair correlation functions of Fig.~\ref{fig:GyroGallery}.
The results are shown in Fig.~\ref{fig:GyroRadialGallery}, for the same set of systems, showing the structure factor in lin-log scales and the radial correlation function in linear scale.
For these plots, as well as the angularly averaged curves in the main text: $S(\bm{k})$ is computed like in the intensity map of Fig.~\ref{fig:GyroGallery} then pixel distances to the origin are binned with a binsize equal to the pixel size, while the RDF is directly binned into radial distances.
These curves show, as argued in the main text, and in the discussion of Fig.~\ref{fig:GyroGallery}, that at low $G$ the $S(k)$ retains peaks of high intensity reminiscent of a quasicrystal up to large values of $k$.
Note however that these peaks are attenuated much faster than in a finite quasicrystal, and that the first peak is systematically higher (see Appendix of the main text).
As $G$ increases, fewer and fewer extra peaks are visible in $S(k)$, to the point that the system essentially becomes radially Poisson-like for $k \gtrsim 2K$ at $G = 100$.
As for $g(r)$, the low-$G$ structures display very sharp and high angularly averaged peaks but, as $G$ increases, the oscillations of $g(r)$ become smoother and smoother.

\begin{figure}
    \centering
    \includegraphics[height=0.30\linewidth]{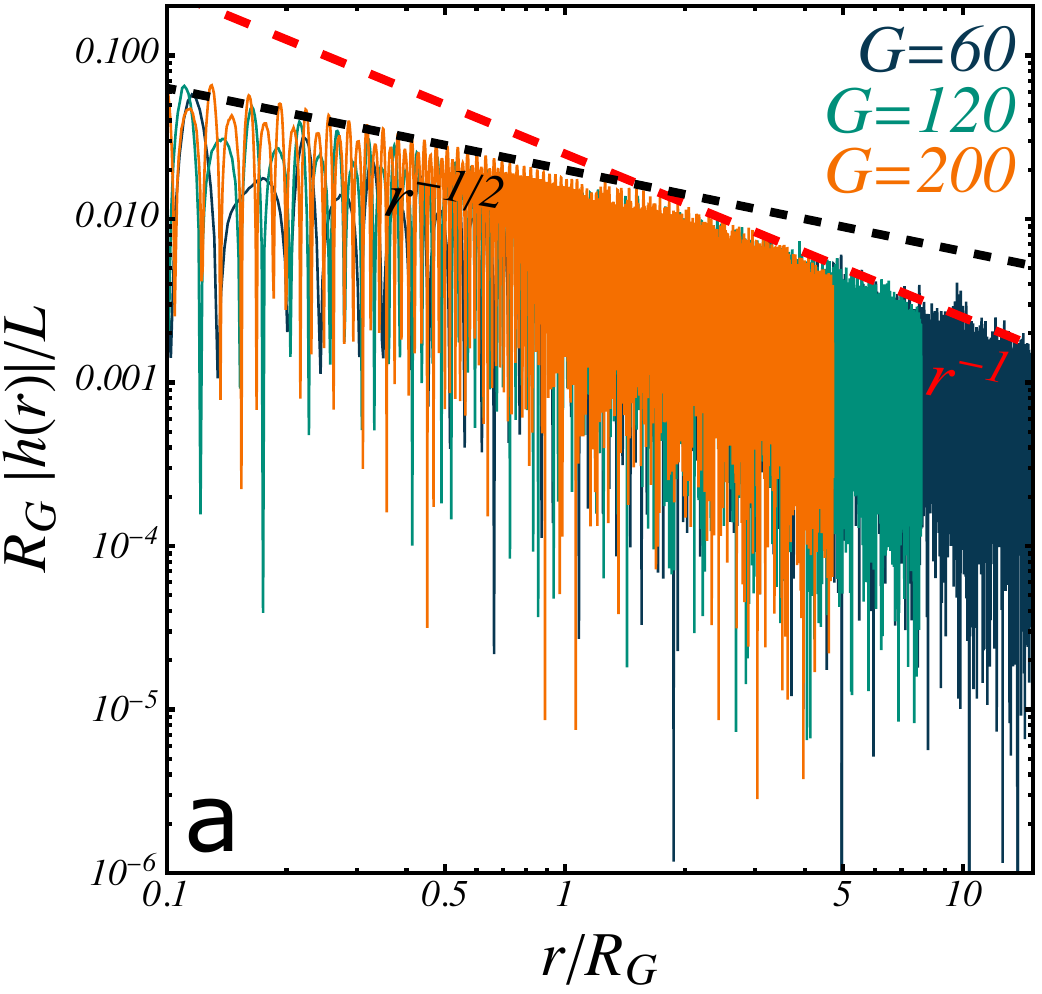}
    \includegraphics[height=0.30\linewidth]{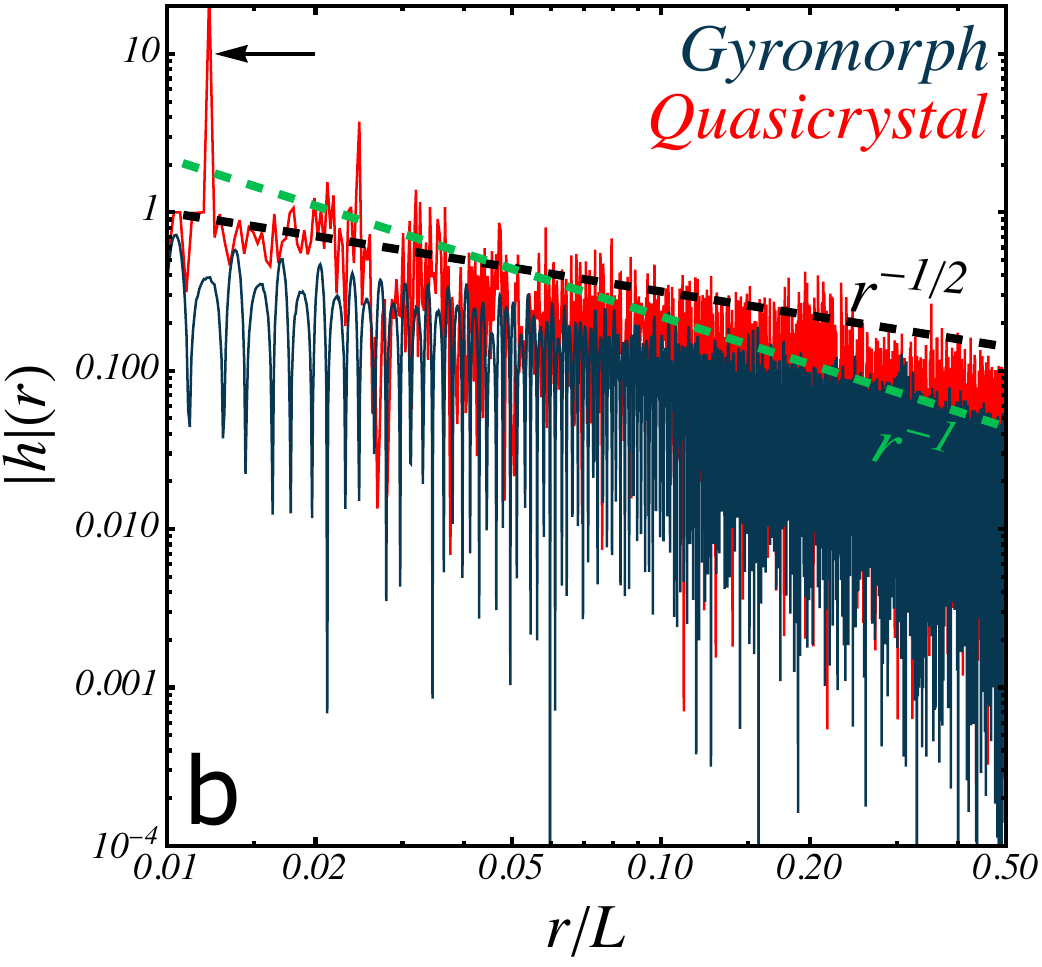}
    \includegraphics[height=0.30\linewidth]{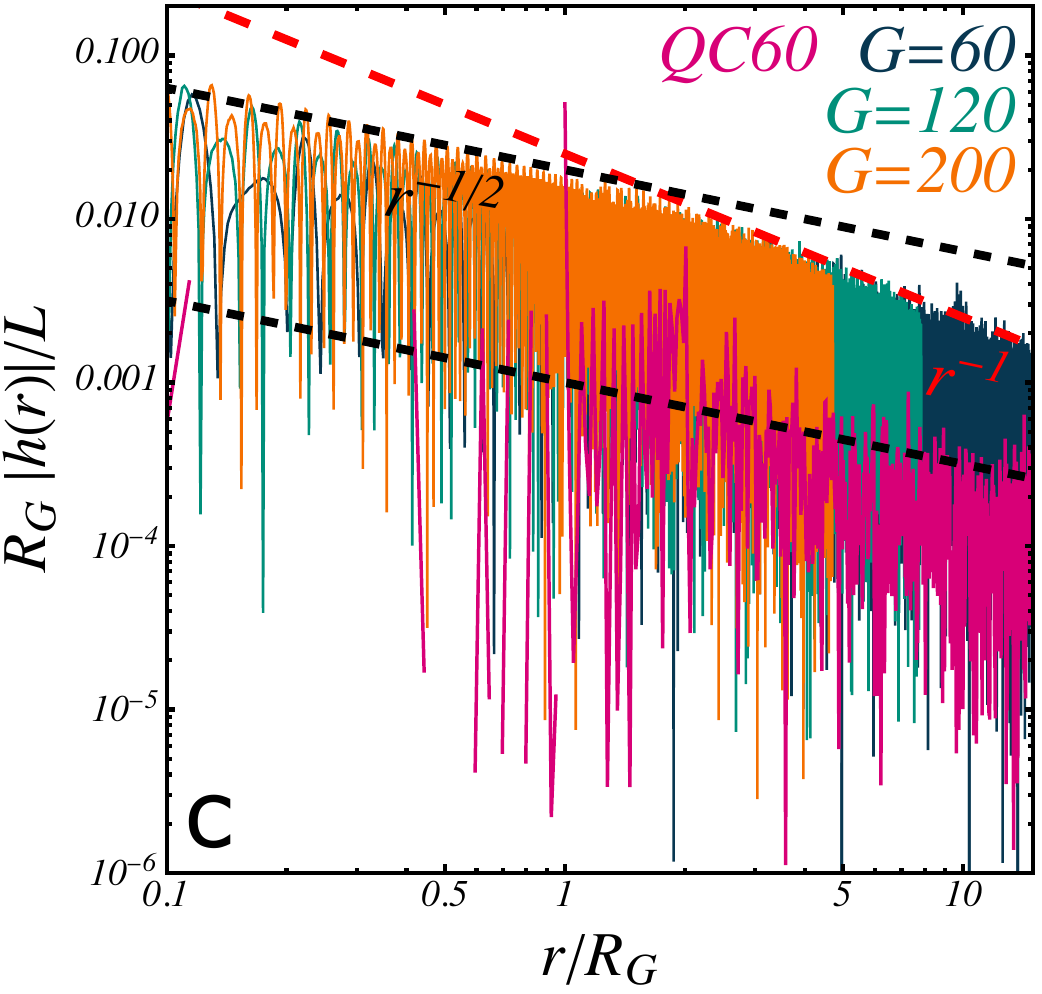}
    \caption{\textbf{Decay of pair correlations in gyromorphs.}
    $(a)$ Log-log plot of the absolute value of the connected radial pair correlation $|h|$ against distance, rescaled by $R_G$ so as to collapse curves for a few $G$.
    Dashed lines are guides for the eye, showing powers $-0.6$ (black dashed line) and power $-1$ (red long-dashed line).
    $(b)$ Comparison between the $|h|$ of a 60-fold gyromorph (blue) and a 60-fold quasicrystal (red).
    An arrow indicates the $60$-fold (nearest-neighbor) peak of the quasicrystal.
    The $1/r$ trend is reproduced in green instead of red for readability in this panel.
    $(c)$ Quasicrystal from panel $(b)$, rescaled and replotted in the same way (and along with the same data as) panel $(a)$, in magenta.
    To highlight the scaling of its decay off-peak, we copy the $1/\sqrt{r}$ trend lower on the graph.
    }
    \label{fig:PowerLawRDF}
\end{figure}

Importantly, the rate of decay of $g(r)$ can be characterized from such radial measurements.
As noted in main text, the decay is power-law at large distances like that of $g_G(r)$.
We show additional evidence of that fact in Fig.~\ref{fig:PowerLawRDF}.
In Fig.~\ref{fig:PowerLawRDF}$(a)$, we plot rescaled versions of $|h|$ in log-log so as to collapse a few values of $G$ together.
The curve shows that for $r > R_G$, the decay of the envelope is close to a $1/r$ behavior, much like the $g_G$ shown in main text.
For $r< R_G$, there is also a slower decay.
While it is not clear whether the envelope actually follows a power-law due to the variability of peak heights within that domain, the decay lies close to a $1/\sqrt{r}$ power.
Without lending too much credence to that value, it is at the very least clearly slower than the long-range behavior.
To go further, in Fig.~\ref{fig:PowerLawRDF}$(b)$, we compare the decay in a gyromorph to that in a quasicrystal with the same typical number of points.
While the quasicrystal, as usual, features an extremely high peak at the edge length of its rhombus tiling, it also displays some slow decay.
Interestingly, it is closer to the slow, low-$r$ decay of gyromorphs than to their large-$r$ decay, as shown by the dashed line indicating a power $1/\sqrt{r}$.
For the sake of completeness in the comparison, in Fig.~\ref{fig:PowerLawRDF}$(c)$, we also reproduce the quasicrystal data, rescaled using the location of its highest peak in $g(r)$ as an $R_G$, along with the rest of Fig.~\ref{fig:PowerLawRDF}$(a)$.
This comparison shows in a clearer way that the quasicrystalline $g(r)$ has a different decay behavior than the gyromorphic ones.
Thus, the long-range decay of correlations in gyromorphs, while it appears to be power-law, is notably faster than in quasicrystals.

\begin{figure}
    \centering
    \includegraphics[width=0.5\linewidth]{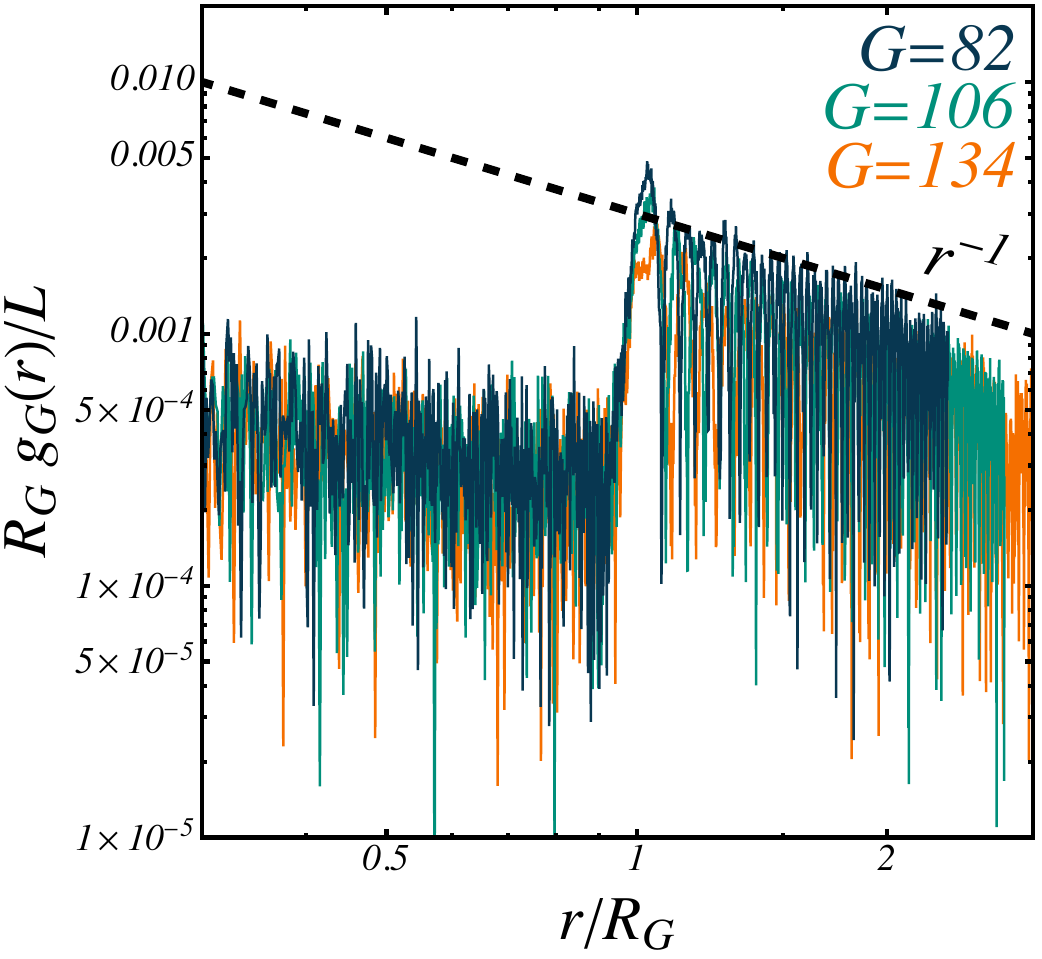}
    \caption{\textbf{Decay of gyromorphic correlations in polygyromorphs.}
    Log-log plot of the gyromorphic correlation for all 3 imposed rotational orders in the polygyromorph presented in main text.
    Each correlation is rescaled by the relevant $R_G$.
    A dashed line indicates $1/r$ decay.
    }
    \label{fig:polygyro_decay}
\end{figure}

The decay of rotational order can also be characterized in polygyromorphs.
We mention in the main text that they display the same decay of each of their rotational orders as gyromorphs that have a single ring of peaks.
We show evidence of that fact for the example polygyromorph used in the main text in Fig.~\ref{fig:polygyro_decay}.
We plot the gyromorphic correlation for each of the imposed rotational orders, rescaling each time both axes by the corresponding $R_G$.
The result is similar to that obtained across $G$ values in distinct gyromorphs: the correlation is low for $r<R_G$, then decays with an exponent close to $-1$ for $r>R_G$.

\begin{figure}
    \centering
    \includegraphics[width=0.9\linewidth]{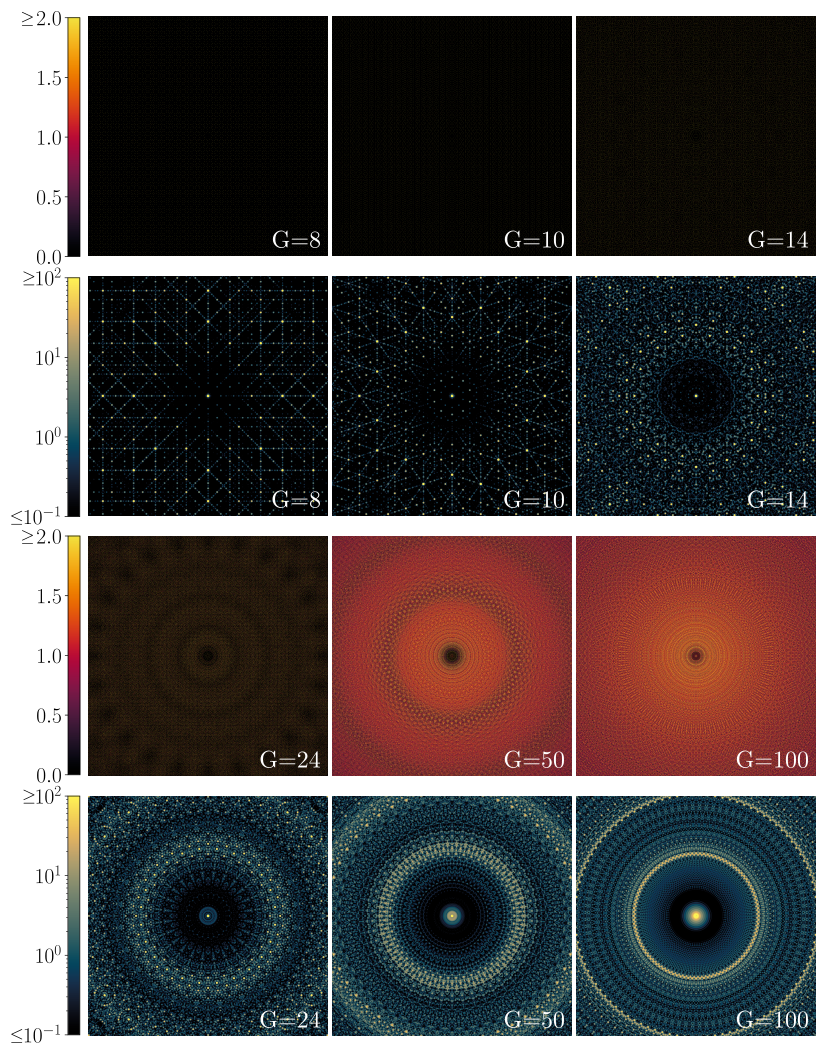}
    \caption{\textbf{Structure of quasicrystals.}
    $g(\bm{r})$ (first and third rows) and $S(\bm{k})$ (second and fourth rows) of de Bruijn quasicrystals for $G = 8$, $10$, $14$, $24$, $50$, $100$ and $KL/2\pi = 100$.
    $S(\bm{k})$ are plotted up to $2 K$ in both directions, and $g(\bm{r})$ up to $L/4$.
    }
    \label{fig:dBQCGallery}
\end{figure}

It is natural to compare the order of gyromorphs as reported in Figs.~\ref{fig:GyroGallery} and~\ref{fig:GyroRadialGallery} to that of de Bruijn quasicrystals.
To do so, we generate quasicrystals with a similar number of points, $N \sim 10^4$ at the same symmetries, and compute the same quantities \textit{over the same ranges of $k$ and using the same intensity scales}, although these are not the most adapted scales to characterize quasicrystalline order.
The results for $2d$ intensity maps in Fig.~\ref{fig:dBQCGallery}.
The main take-away is that it is now the $g(r)$ that is Dirac-delta peaked (with peaks one pixel wide in the figure), due to the underlying construction relying on a finite deterministic set of rhombi with a finite number of allowed orientations at each site.
As a result, the $g(r)$ contains a ring of Dirac deltas forming a regular $G$-gon at its center, so that the $S(k)$ features a Moiré pattern structure, as pointed out in the Appendix of the main text.
However, note that the correspondence established in that same appendix between $(g(\bm{r}),S(\bm{k}))$ of quasicrystals and $(S(\bm{k}),g(\bm{r}))$ of gyromorphs is only visible by changing the plotrange of these intensity maps.
Indeed, the ring of peaks in the $S(\bm{k})$ of the gyromorphs corresponds to the nearest-neighbor peaks of the $g(\bm{r})$ of quasicrystals (a very small part of the intensity maps of Fig.~\ref{fig:dBQCGallery}), and consequently the Moiré pattern in the $g(\bm{r})$ of gyromorphs maps to a very zoomed out version of the $S(\bm{k})$ shown in Fig.~\ref{fig:dBQCGallery}.
At the largest symmetry, the rings visible in the $S(\bm{k})$ in Fig.~\ref{fig:dBQCGallery} are the equivalent of the isotropic rings at the center of the $g(\bm{r})$ of gyromorphs.

\begin{figure}
    \centering
    \includegraphics[width=0.30\linewidth]{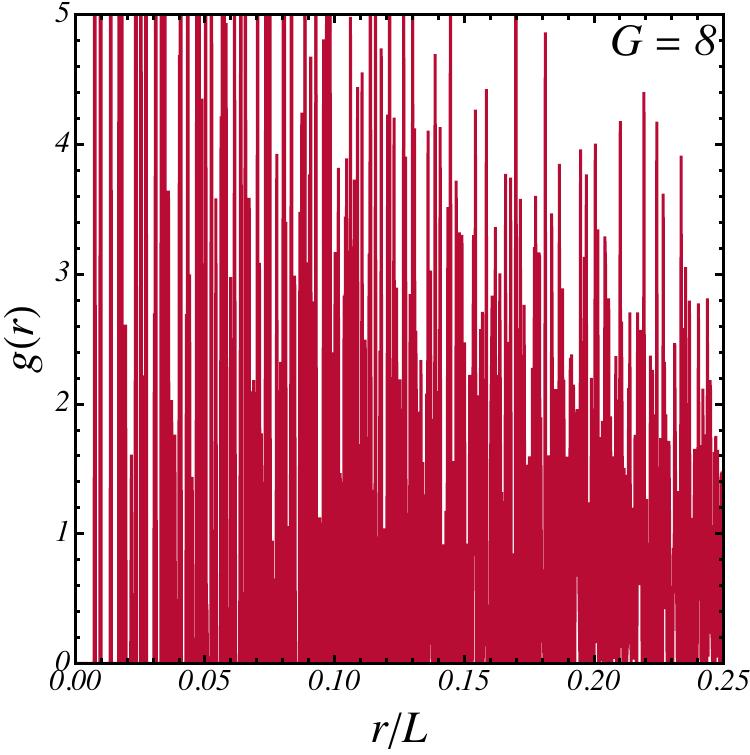}
    \includegraphics[width=0.30\linewidth]{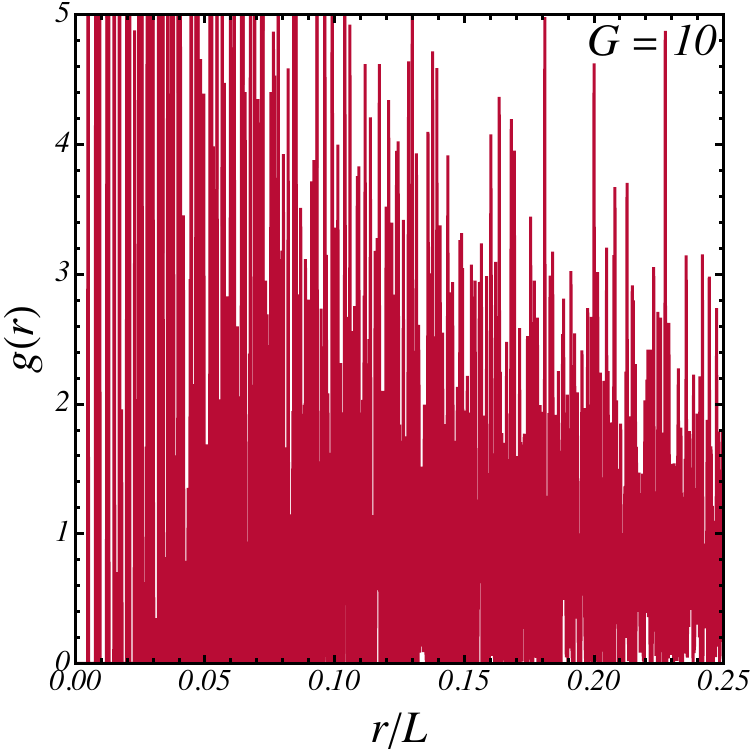}
    \includegraphics[width=0.30\linewidth]{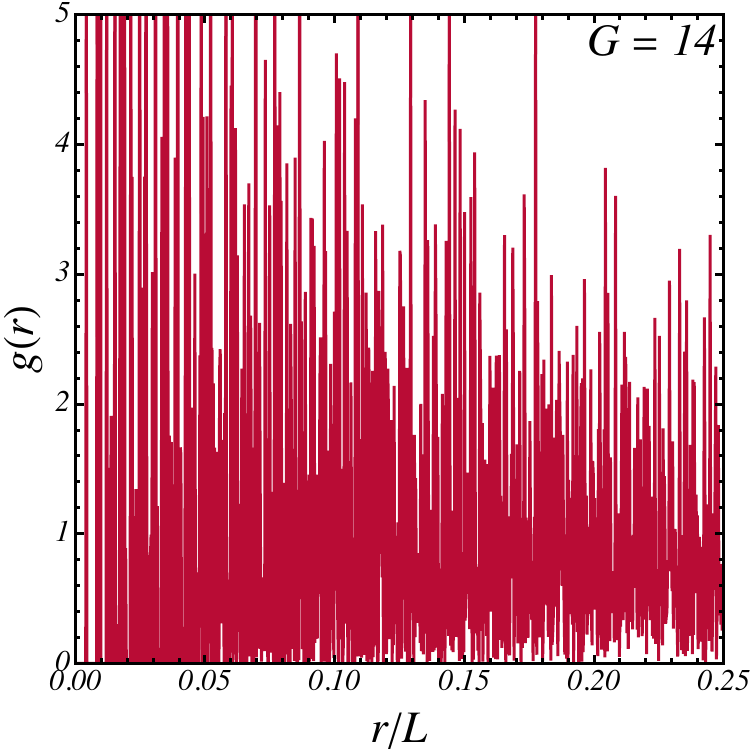}\\
    \includegraphics[width=0.30\linewidth]{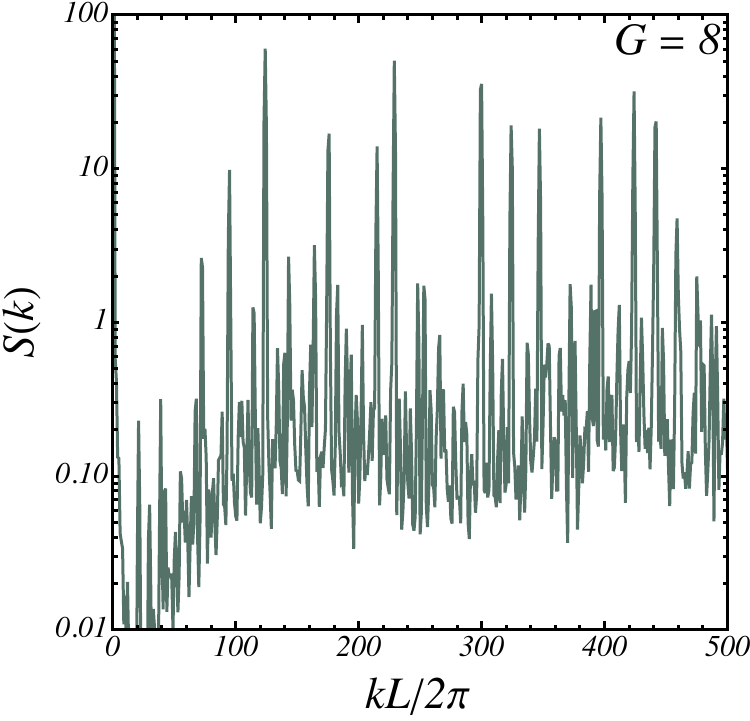}
    \includegraphics[width=0.30\linewidth]{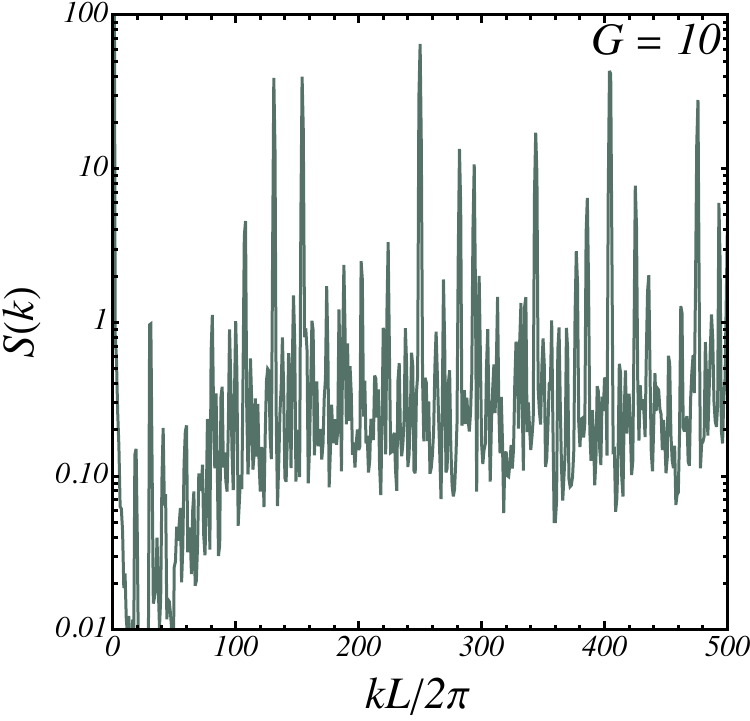}
    \includegraphics[width=0.30\linewidth]{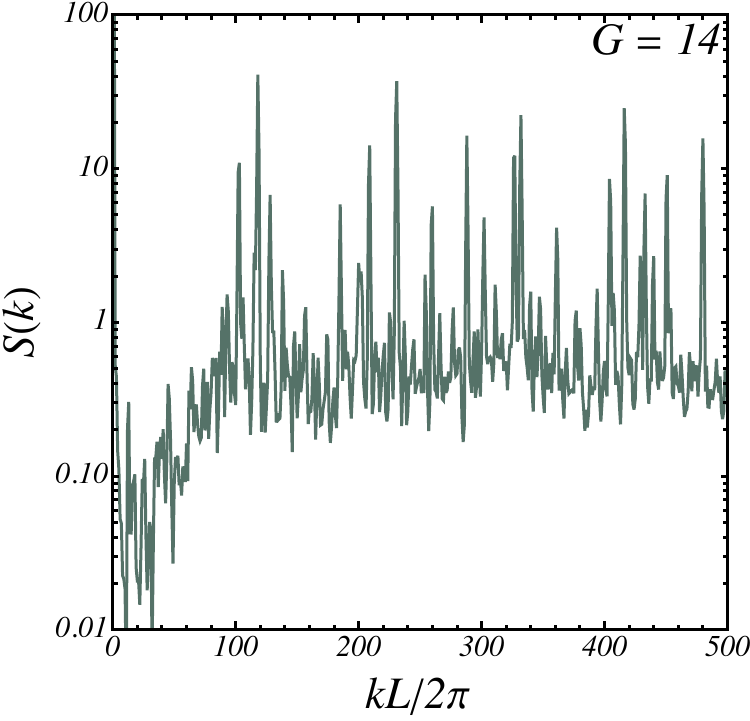} \\
    \includegraphics[width=0.30\linewidth]{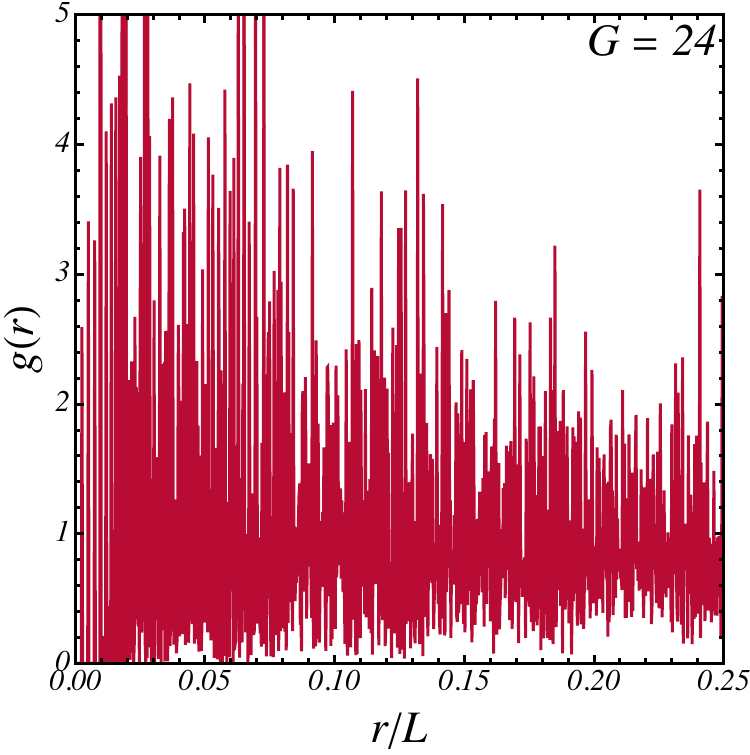}
    \includegraphics[width=0.30\linewidth]{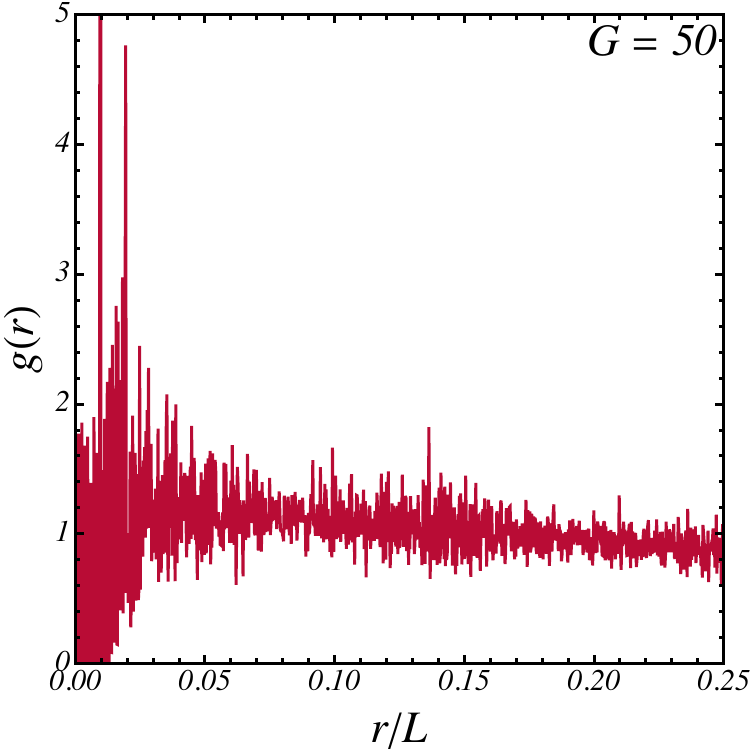}
    \includegraphics[width=0.30\linewidth]{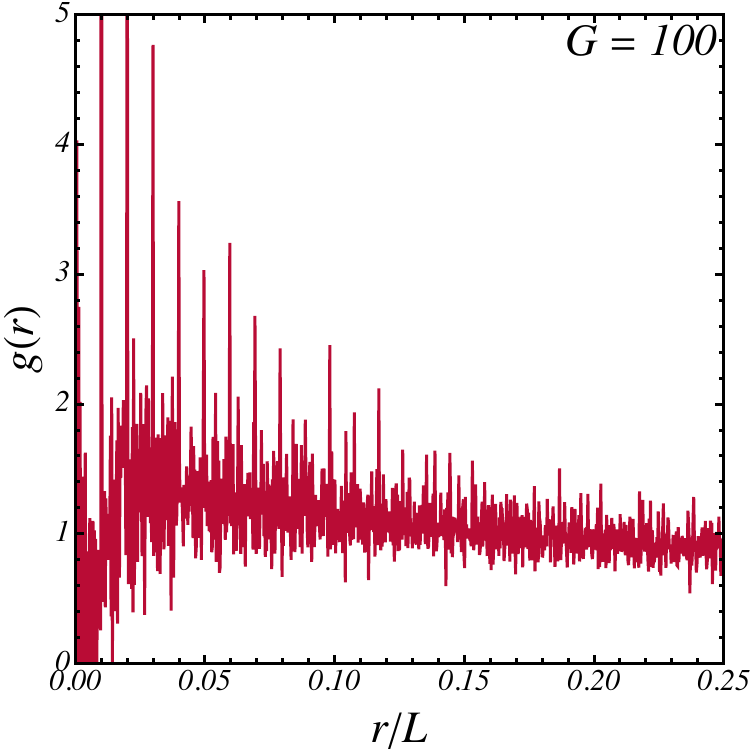} \\
    \includegraphics[width=0.30\linewidth]{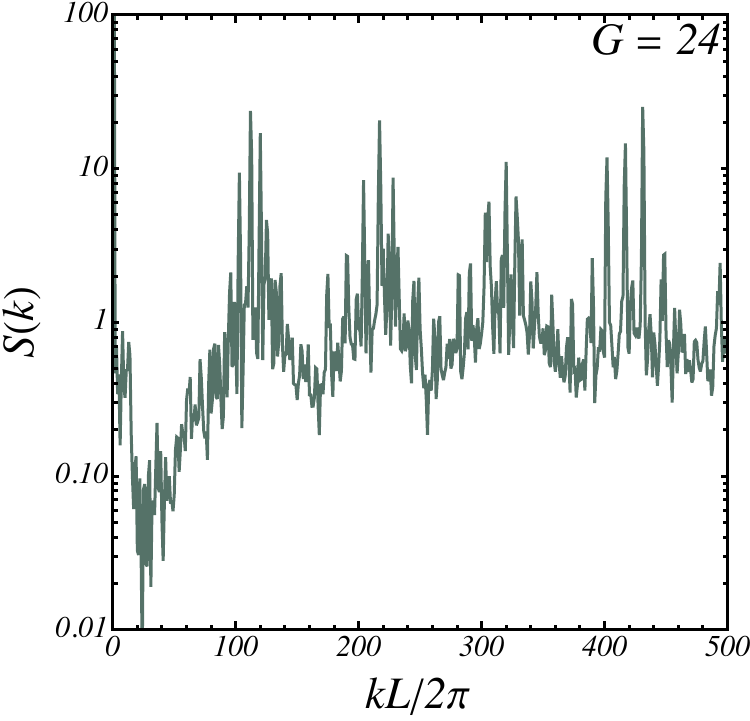}
    \includegraphics[width=0.30\linewidth]{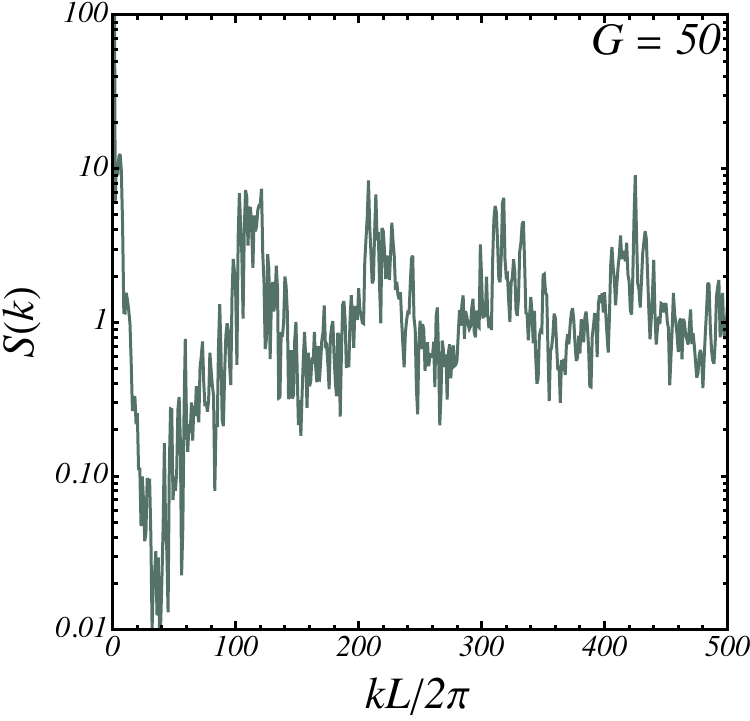}
    \includegraphics[width=0.30\linewidth]{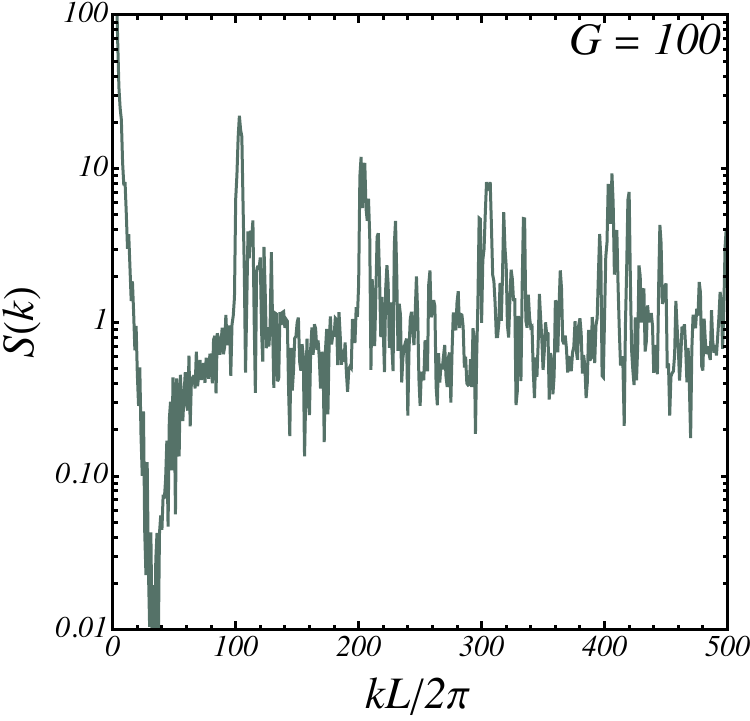}
    \caption{\textbf{Angularly averaged structure of quasicrystals.}
    $g(r)$ (first and third rows, in linear scales) and $S(k)$ (second and fourth rows, in lin-log scale) of quasicrystals for $G = 8$, $10$, $14$, $24$, $50$, $100$ and $KL/2\pi = 100$, same as in Fig.~\ref{fig:dBQCGallery}.
    }
    \label{fig:dBQCRadialGallery}
\end{figure}

Likewise, we show in Fig.~\ref{fig:dBQCRadialGallery} the corresponding radial profiles of the $g(\bm{r})$ and $S(\bm{k})$.
Again, these plots are shown using the same plot ranges and axis ranges as Fig.~\ref{fig:GyroRadialGallery}, at the cost of losing some information about the height of the peaks in the $g(r)$.
We show that the quasicrystalline $S(k)$ are clearly distinct from the gyromorphic ones, especially at high $G$, where quasicrystals have smooth, broad angularly-averaged features, as opposed to the very peaked features in gyromorphs.
As a result, as also justified in the appendix of the main text, the peaks in gyromorphs are higher than in quasicrystals, even at the angularly-averaged level.
The real-space structures are also clearly different: while the $g(r)$ of gyromorphs get consistently smoother as $G$ grows, quasicrystals retain very peaked structures by construction.

\begin{figure}
    \centering
    \includegraphics[width=\linewidth]{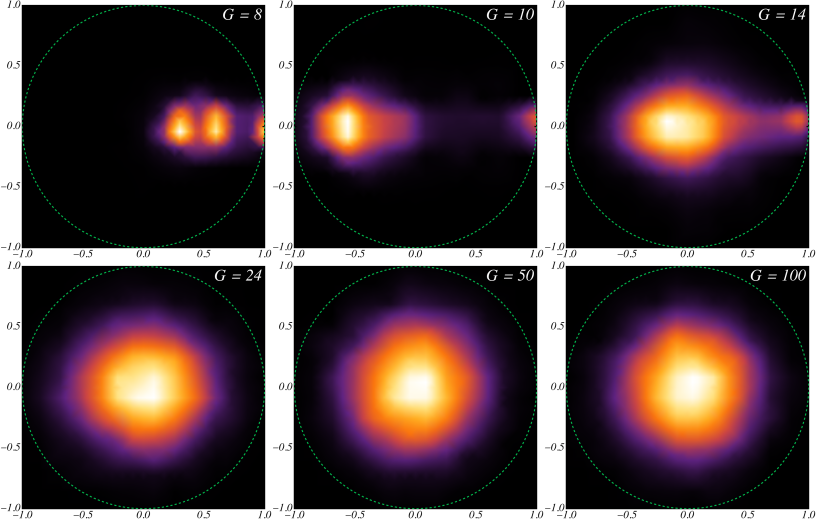}
    \caption{\textbf{BOOPs in Gyromorphs.}
    Intensity maps of the distribution of $\psi_{G,n}$ in the gyromorphs of Fig.~\ref{fig:GyroGallery}.
    A dashed green circle indicates the border of the unit disk.}
    \label{fig:GyroBOOPs}
\end{figure}

\begin{figure}
    \centering
    \includegraphics[width=\linewidth]{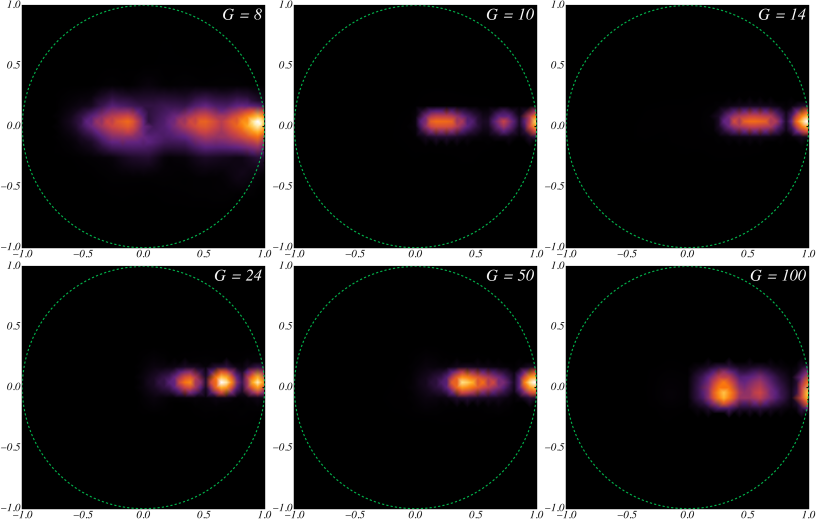}
    \caption{\textbf{BOOPs in Quasicrystals.}
    Intensity maps of the distribution of $\psi_{G,n}$ in the quasicrystals of Fig.~\ref{fig:dBQCGallery}.
    A dashed green circle indicates the border of the unit disk.}
    \label{fig:dBQCBOOPs}
\end{figure}

Finally, as a last indication that gyromorphs and quasicrystals are clearly different structures, we perform an analysis of the relevant Bond-Orientational Order Parameters~\cite{Steinhardt1983}.
For point $n$ in the point pattern, at a given $G$, we compute the microscopic observable $\psi_{G,n} \in \mathbb{C}$ defined as
\begin{align}
    \psi_{G,n} \equiv \frac{1}{z}\sum\limits_{m \in \partial n} e^{i G \theta_{mn}},
\end{align}
with $\theta_{mn} = \measuredangle (\hat{\bm{e}}_x, \bm{r}_m - \bm{r}_n)$ the angle formed between the horizontal and the vector linking point $n$ to one of its $z$ neighbors, here defined by a Delaunay neighborhood.
The complex modulus of $\psi_{G,n}$ is close to $1$ if a point has neighbors lying on a subset of the regular $G$-gon, and a lower value otherwise.
The argument of $\psi_{G,n}$ encodes for the orientation $\vartheta$ of the $G$-gon on the $2\pi / G$-wide interval it can explore, $\arg \psi_{G,n} = G \vartheta$.
In Figs.~\ref{fig:GyroBOOPs} (gyromorphs) and~\ref{fig:dBQCBOOPs} (quasicrystals), we plot for each $G$ the $2d$ histograms of these quantities in the unit disk, which encode the full variety of local Delaunay environments.
In gyromorphs, for small $G$ these histograms show strong directionality along the $x$ axis.
This is expected as for these $G$ the orientational order sets in at the level of nearest neighbors, and we impose one of the peaks at $K \hat{\bm{e}}_x$.
The moduli of the microscopic BOOPs shows that there are a few non-equivalent environments (different peaks), and the peaks have finite widths in both the radial and orthoradial directions, indicating some variety in environments.
As $G$ grows, however, the orientational order gets imposed at an $R_G$ increasingly far from the nearest-neighbor distance, so that the local BOOPs distribution become isotropic peaks around $0$ like in a fluid.
This is a stronger indication of local isotropy than $g(\bm{r})$.
Indeed, one could have clear unit-modulus BOOPs but a perfectly isotropic first ring in $g(\bm{r})$: this is for instance the expected scenario for very large equilibrium hexatic liquids~\cite{Nelson1979} -- in that case, while all local environments are hexatic, they also span all possible angles due to soft modes, so that the expected distribution of hexatic vectors in hard disks is an annulus close to $1$.
Here, instead, we find no BOOP at the relevant $G$ in gyromorphs.
By contrast, in quasicrystals, at all orders the distribution of BOOPs is a fine line with finite moduli along positive $\hat{bm{e}}_x$, indicating strong orientational order with the prescribed symmetry at the level of nearest neighbors.

\subsection{Stability against noise\label{sec:NoiseStability}}

A common source of concern for photonic structures is how resilient they are to fabrication errors, and to random modifications.
In particular, it has been argued that disordered structures could be more advantageous than ordered ones because, by virtue of being disordered, they would be less sensitive to extra disorder.
We here assess the effect of noise on gyromorphs, as compared to other systems like stealthy hyperuniform patterns.
There are essentially 3 kinds of noise that could affect a point pattern: random displacements on the positions, random deletions of points, and random additions of points.
The effect of all three has been formalized in the context of stealthy hyperuniform systems~\cite{Kim2018}, we here recall the main results and their implications for gyromorphs.

\begin{figure}
    \centering
    \includegraphics[width=0.32\linewidth]{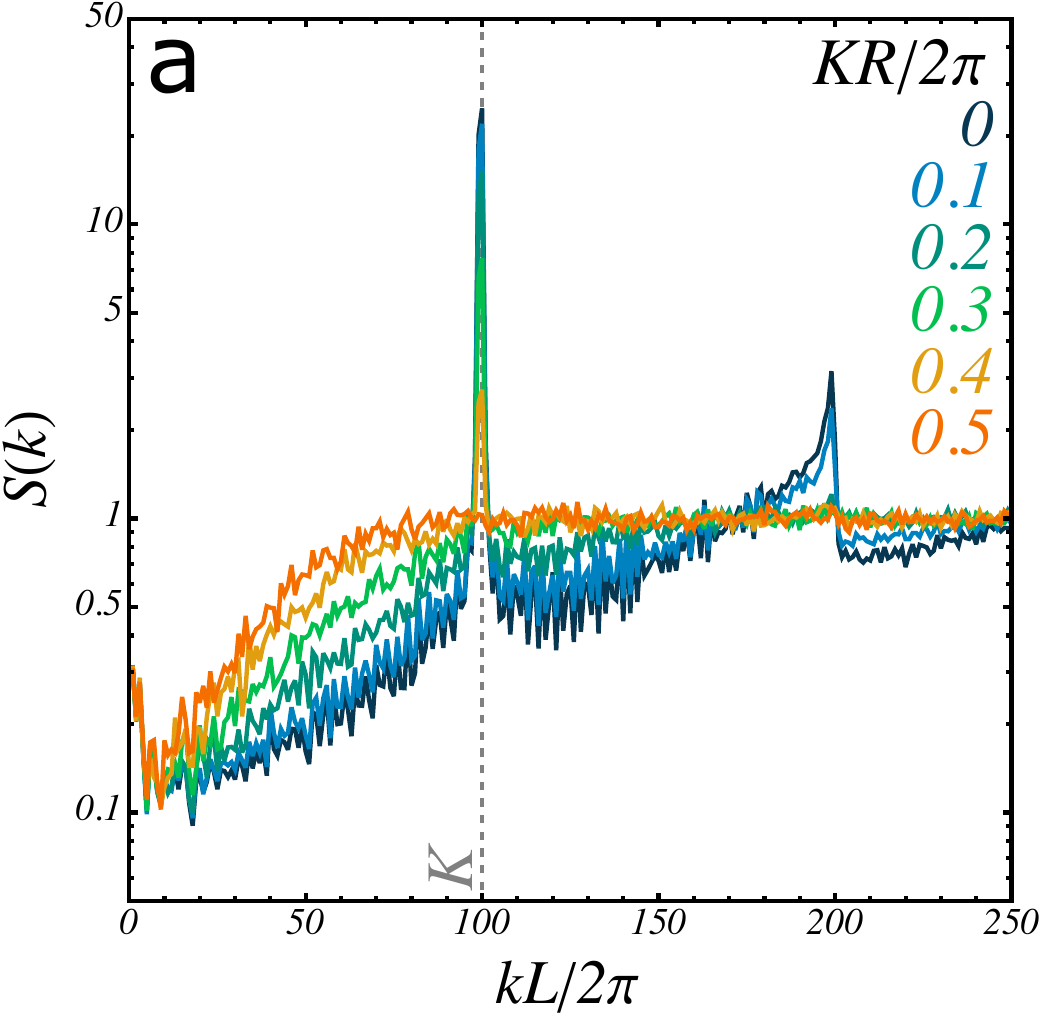}
    \includegraphics[width=0.32\linewidth]{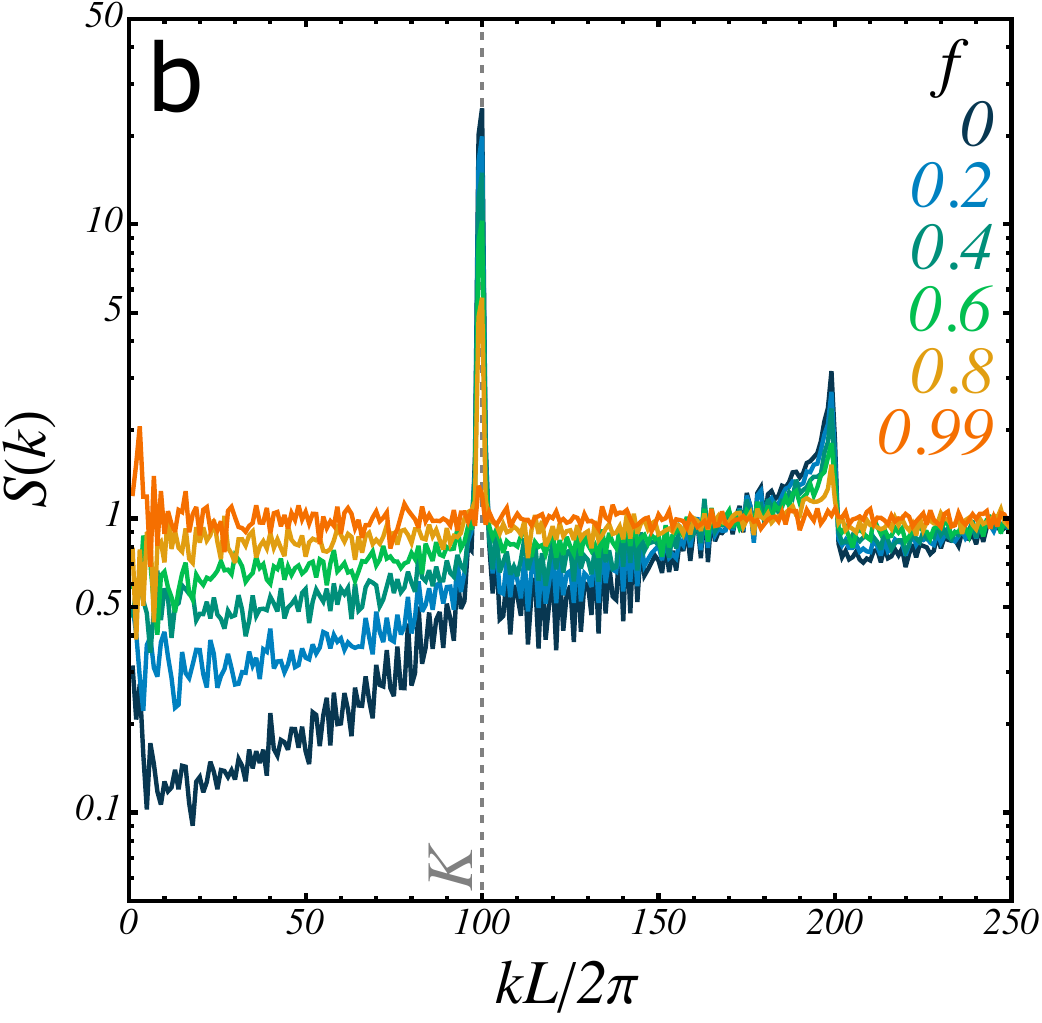}
    \includegraphics[width=0.32\linewidth]{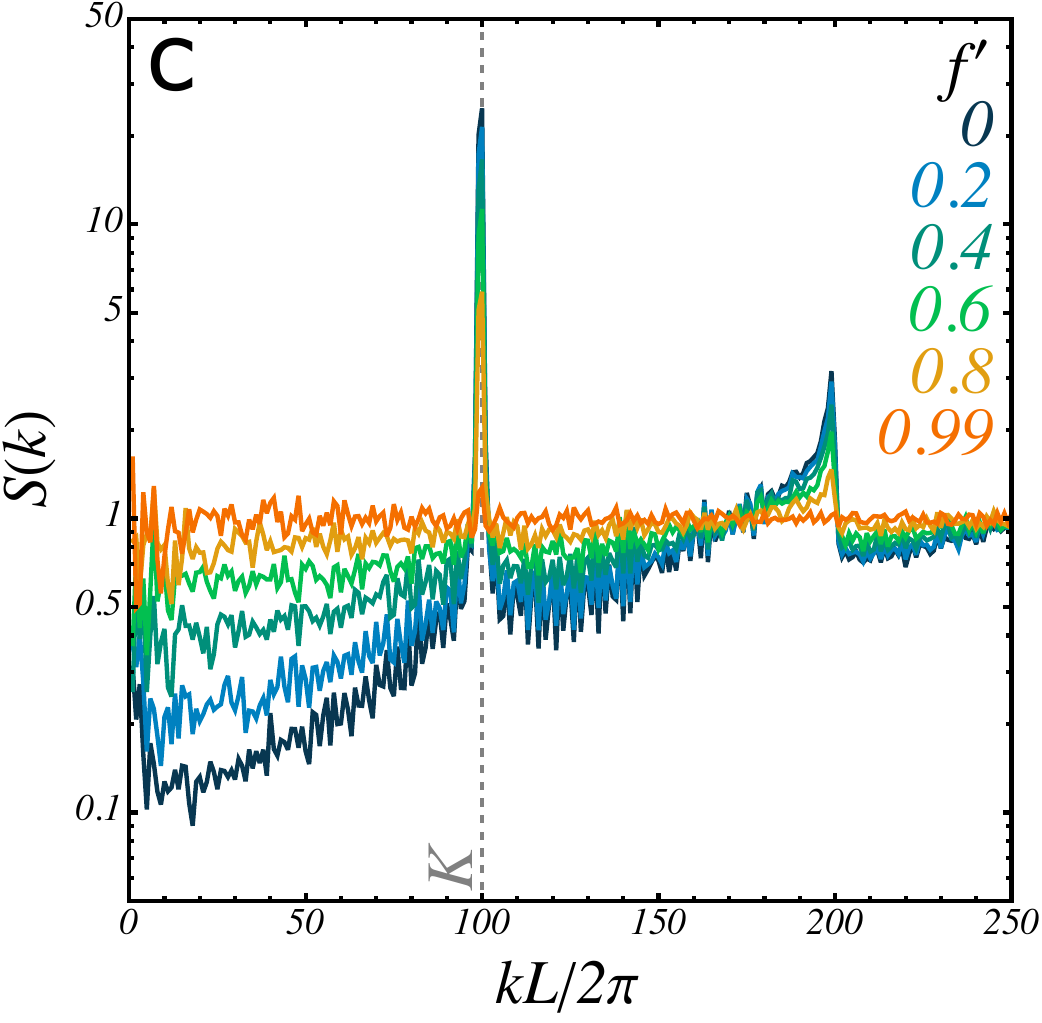} \\
    \includegraphics[width=0.32\linewidth]{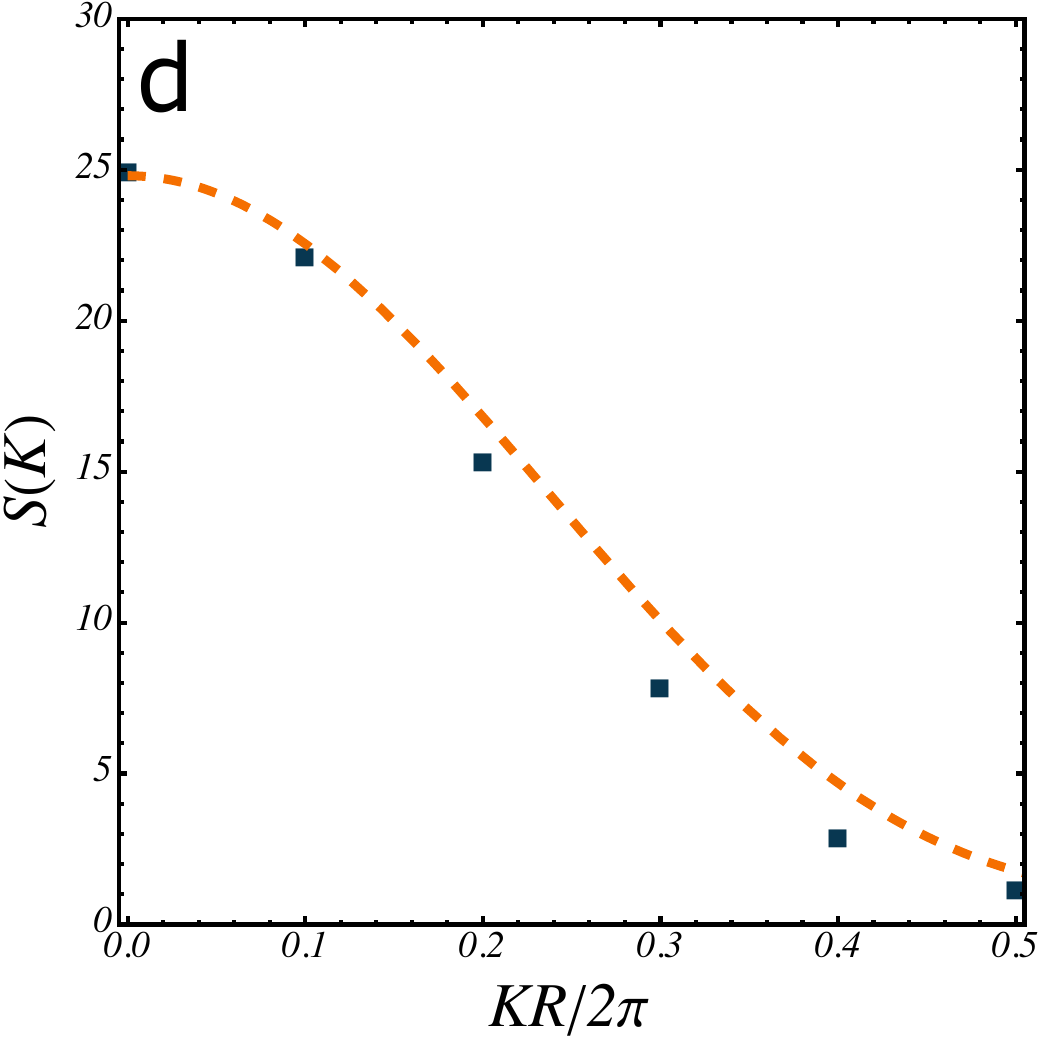}
    \includegraphics[width=0.32\linewidth]{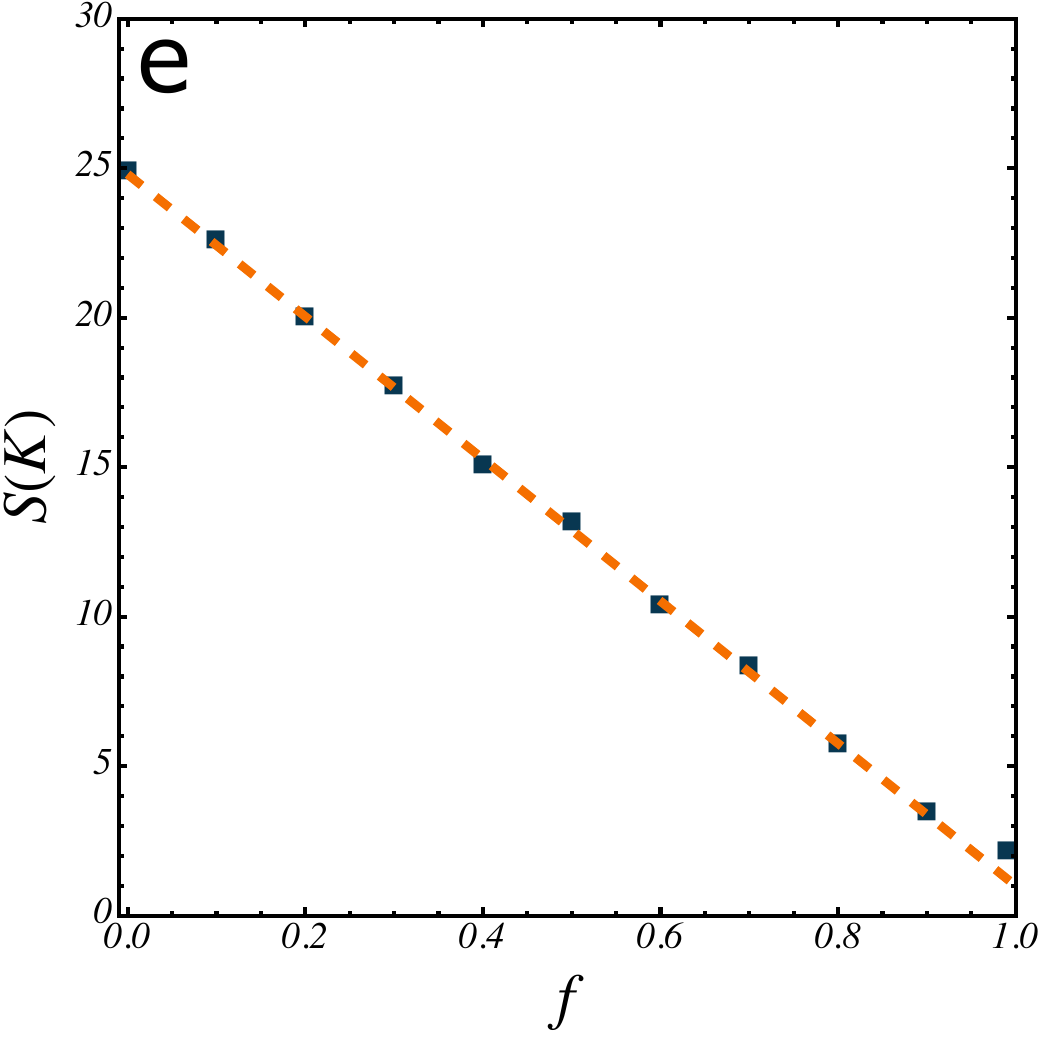}
    \includegraphics[width=0.32\linewidth]{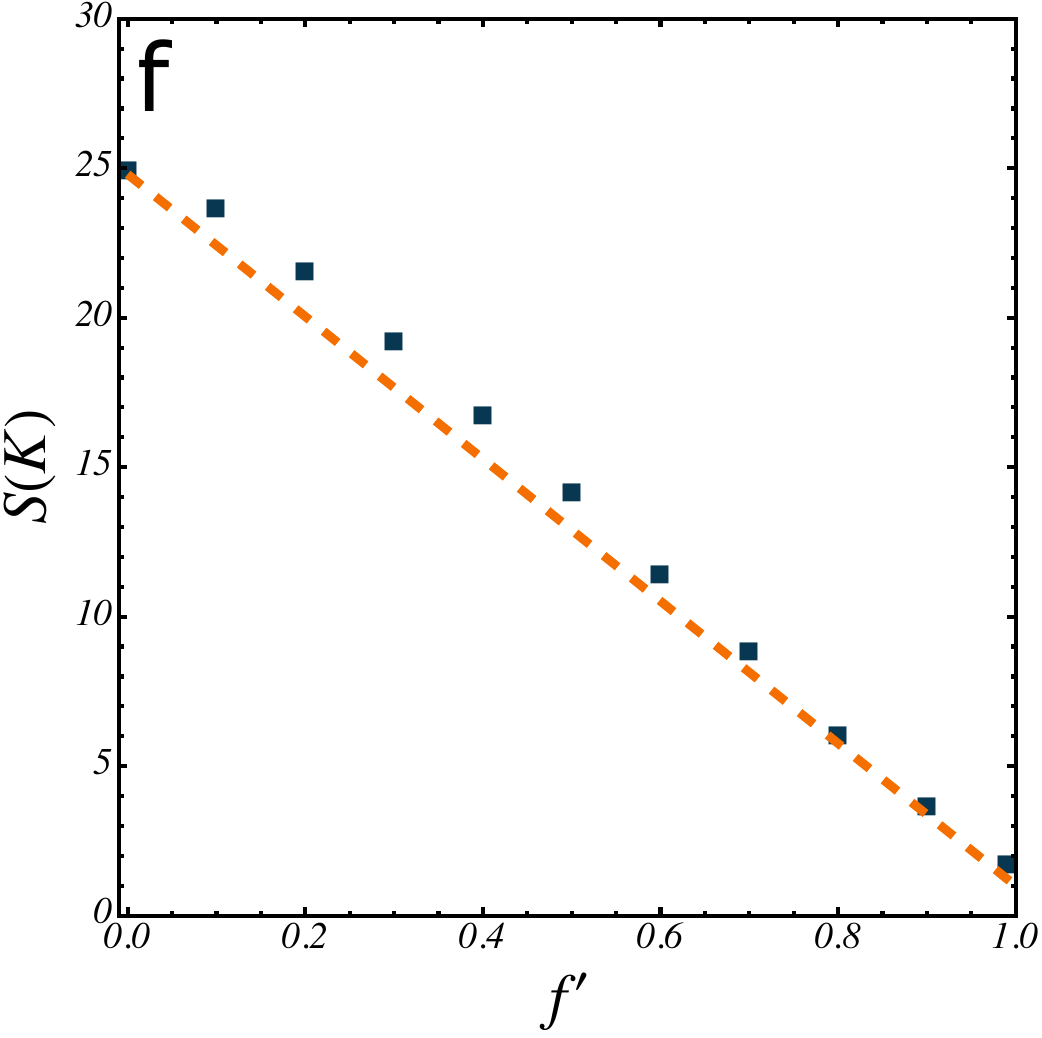}
    \caption{\textbf{Noisy gyromorphs.}
    Evolution of the structure factor of a $200$-fold gyromorph with $K = 100$ ($N \sim 10^4$) when subjected to noise.
    $(a)$ Points are kicked at random uniformly within a disk with radius $R$.
    The kick size grows from $0$ (dark blue) to $R = K/2$ (orange), corresponding to a disk with a diameter equal to the typical interparticle spacing.
    $(b)$ Poisson-distributed points are added to the same gyromorph, with a final fraction $f$ of added points.
    $(c)$ Random uncorrelated deletions are performed, with a fraction of deletions $f'$
    $(d)-(f)$ Corresponding evolutions of the peak height.
    Measured values in single realizations (dark blue squares) are compared to predictions for the average (dashed orange lines).}
    \label{fig:NoisyGyromorphs}
\end{figure}

First, assume that a point pattern is perturbed by independent random kicks of the location of each point.
The resulting noised density field reads
\begin{align}
    \rho_{kicked}(\bm{r}) = \sum\limits_{n=1}^{N} \delta(\bm{r} - \bm{r}_n - \bm{d}_n),
\end{align}
with $\bm{d}_n$ independent random vector drawn from some distribution.
By definition, the Fourier transform of the density field can be written as
\begin{align}
    \widehat{\rho}_{kicked}(\bm{k}) = \sum\limits_{n=1}^{N} e^{i \bm{k} \cdot (\bm{r}_n +\bm{d}_n)},
\end{align}
and the ensemble averaged structure factor as
\begin{align}
    \left\langle S_{kicked}(\bm{k}) \right\rangle = \frac{1}{N} \left\langle\left| \sum\limits_{n=1}^{N} e^{i \bm{k} \cdot (\bm{r}_n +\bm{d}_n)} \right|^2 \right\rangle,
\end{align}
where $\langle \cdot \rangle$ represents an average over the distribution of all $N$ displacements.
This last expression can be rewritten as
\begin{align}
    \left\langle S_{kicked}(\bm{k}) \right\rangle = \frac{1}{N} \sum\limits_{m,n=1}^{N} e^{i \bm{k} \cdot \bm{r}_{mn}} \left\langle  e^{i \bm{k} \cdot\bm{d}_{mn}}  \right\rangle,
\end{align}
where $\bm{r}_{mn} = \bm{r}_n - \bm{r}_m$ and  $\bm{d}_{mn} = \bm{d}_n - \bm{d}_m$.
Assuming that kicks are are independent random variables, one may write
\begin{align}
    \left\langle  e^{i \bm{k} \cdot\bm{d}_{mn}}  \right\rangle &= \left\langle  e^{i \bm{k} \cdot\bm{d}_{n}}  \right\rangle \left\langle  e^{-i \bm{k} \cdot\bm{d}_{m}}  \right\rangle = \delta_{m,n} + (1 - \delta_{m,n})\left| F(\bm{k}) \right|^2.
\end{align}
in which we identified the square modulus of the characteristic function $F$ of the distribution of displacements, and introduced a Kronecker delta notation.
For instance, considering uniform kicks within a $2d$ disk with radius $R$,
\begin{align}
    F(\bm{k}) &= \frac{1}{\pi R^2}\int\limits_{r = 0}^{R} dr \int\limits_{\theta = -\pi}^{\pi} d\theta \, r e^{i k r \cos\theta}, \\
    &= 2 \frac{J_1(k R)}{k R},
\end{align}
with $J_1$ a Bessel function of the first kind.
Thus, for this example,
\begin{align}
    \left\langle S_{kicked}(\bm{k}) \right\rangle = 1 + \left| F(\bm{k})\right|^2 (S_0(\bm{k}) -1 ) = 1 + 4 \left(\frac{J_1(k R)}{k R}\right)^2(S_0(\bm{k}) -1 ),
\end{align}
with $S_0$ the structure factor before kicking points.
As expected, $\left\langle S_{kicked}(\bm{k}) \right\rangle \to 1$ at all $k$ as $R \to \infty$.
Furthermore, at finite values, the long-$k$ regime of the structure factor (short range order) is most attenuated.
At small $kR$, one may expand the Bessel function to yield
\begin{align}
    \left\langle S_{kicked}(\bm{k}) \right\rangle &\approx 1 + \left( 1 - \frac{k^2 R^2}{4}\right)(S_0(\bm{k}) -1 ), \\
    &= \frac{k^2 R^2}{4} + \left( 1 - \frac{k^2 R^2}{4}\right) S_0(\bm{k}).
\end{align}
This last result is actually fairly generic: for small noise amplitudes, and with most usual noise distributions, a kicked point pattern gains a quadratic term at small $k$.
Now, how does this sort of noise affect the features of the structures we study in this paper?
As already noted in past work~\cite{Kim2018,Klatt2020,Shih2023}, random kicks quickly deteriorates low features, $S(k) \to 0$, near $k \to 0$, so that stealthiness essentially doesn't survive any amount of noise.
As for other features, the larger $|S_0(\bm{k}) - 1|$, the better they survive.
For instance, Bragg peaks in crystals remain visible as long as kicks are not comparable to the lattice spacing~\cite{Klatt2020}.
In the example of gyromorphs, we illustrate this in Fig.~\ref{fig:NoisyGyromorphs}$(a)$, where we show the structure factor of a $200$-fold gyromorph where each point has been kicked by independent, random displacements drawn uniformly in a disk with radius $R$, for a few values of $R$.
The peak remains visible even for a diameter $80\%$ of the typical nearest neighbor distance, but disappears at $100\%$.
Notice that the low-$k$ behavior goes from a plateau to a growing function, as expected.

Next, we consider random additions of uncorrelated points to an existing point pattern.
Consider a system starting with $N-p$ points, to which $p$ points are added.
The structure factor of that one realization may be written as
\begin{align}
    S_N(\bm{k}) = \frac{1}{N} \left| \widehat{\rho}_{0}(\bm{k}) + \widehat{\rho}_p(\bm{k}) \right|^2,
\end{align}
where $\widehat{\rho}_{0}$ and $\widehat{\rho}_p$ represent the Fourier densities of the initial and additional points, respectively.
This expression may be rewritten as
\begin{align}
    S_N(\bm{k}) &= \frac{1}{N} \left| \widehat{\rho}_{0}(\bm{k})\right|^2  + \frac{1}{N} \left| \widehat{\rho}_p(\bm{k}) \right|^2 + \frac{2}{N} \text{Re}\left[ \widehat{\rho}_{0}(\bm{k}) \widehat{\rho}_p^\dagger(\bm{k}) \right], \\
    &= (1 - f) S_0(\bm{k})  + f S_p(\bm{k})+ \frac{2}{N} \text{Re}\left[ \widehat{\rho}_{0}(\bm{k}) \widehat{\rho}_p^\dagger(\bm{k}) \right].
\end{align}
In the last expression, we introduced $S_0$ and $S_p$ the structure factors of the initial and additional point patterns taken separately, as well as the fraction of added points $f = p / N$.
Assuming that the additional points are a Poisson point pattern, the average modified structure factor reads, for $\bm{k} \neq \bm{0}$,
\begin{align}
    \left\langle S_N(\bm{k}) \right\rangle &= (1 - f) S_0(\bm{k})  + f,\\
    &= 1 + (1-f)(S_0(\bm{k}) - 1),
\end{align}
as $\langle S_p \rangle = 1$ in a Poisson point pattern, and $\langle \widehat{\rho}_p (\bm{k})\rangle \propto \delta(\bm{k})$.

Before commenting on this result, we establish the expression of the final structure factor when performing random uncorrelated removals from a point pattern.
Consider a point pattern initially containing $N$ points, and assume that a random point is removed, with all points being equiprobably likely to be selected.
The new structure factor for one realization of this removal reads
\begin{align}
    S_{N-1}(\bm{k}) = \frac{1}{N-1} \left| \widehat{\rho}_0(\bm{k}) - e^{i \bm{k}\cdot\bm{r}} \right|^2,
\end{align}
with $\bm{r}$ the location of the point that was removed.
Once again, the square modulus may be expanded as
\begin{align}
    S_{N-1}(\bm{k}) = \frac{1}{N-1} + \frac{N}{N-1} S_N(\bm{k}) - \frac{2}{N-1} \text{Re}\left[\widehat{\rho}_0(\bm{k}) e^{-i \bm{k}\cdot\bm{r}} \right].
\end{align}
To average this result over possible realizations of the removal, only the last term needs to be averaged, and for equiprobable removals
\begin{align}
    \left\langle \widehat{\rho}_0(\bm{k}) e^{-i \bm{k}\cdot\bm{r}} \right\rangle = \frac{1}{N} \widehat{\rho}_0(\bm{k}) \sum\limits_{p = 1}^{N} e^{-i \bm{k}\cdot\bm{r}_p} = \frac{1}{N}\left|\widehat{\rho}_0(\bm{k}) \right|^2 = S_0(\bm{k}).
\end{align}
All in all, for a single removed point,
\begin{align}
    \left\langle S_{N-1}(\bm{k})\right\rangle = \frac{1}{N-1} + \frac{N-2}{N-1} S_N(\bm{k}).
\end{align}
Treating this expression as a recurrence between average structure factors, one can approximate the structure factor after $p$ removals as
\begin{align}
    \left\langle S_{N-p}(\bm{k})\right\rangle &\approx \frac{p}{N-1} + \left(1 - \frac{p}{N-1}\right)S_{N}(\bm{k}) \\
    &= 1 + \left(1 - f'\right)(S_{N}(\bm{k}) - 1),
\end{align}
where $f'\equiv \frac{p}{N-1}$ is close to the fraction of removed points.
Note that $S_{N-p} = 1$ for $p = N-1$ as a single point remains, which is why the fraction is offset by $1$.
This expression should be understood as a sort of mean-field level approximation since it was obtained by treating the relation between the average after one removal and the starting point pattern, rather than the average of a realization with $p$ removals.
Nevertheless, this expression is interesting as it assumes nearly the same form as that obtained for the addition of points, and yields the same intuition.

Indeed, for both additions and removals, the result we obtained is that the new structure factor is the weighted arithmetic average of the starting structure factor and that of a Poisson pattern.
As a result, features that survive noise best are, again, those that are as distant as possible from $1$.
This again means that low features like stealthiness are very affected by usual metrics of their quality: if one prepares an $N = 10^6$ point pattern with $S(k) = 10^{-50}$ at small $k$~\cite{Morse2023}, on average, losing or adding a single point will bounce that value up to $f \approx f' \approx 10^{-6}$, a jump by over 40 orders of magnitude!
This illustrates that it is in practice unrealistic to achieve features lower than $1/N$ in $S(\bm{k})$ if structures are expected to be defective.
High values, on the other hand, are qualitatively less affected: a peak that is $N$-high will remain roughly $N$-high after a deletion or addition of a fraction $f \ll 1$ of random points.
This is illustrated with the example of gyromorphs in Fig.~\ref{fig:NoisyGyromorphs} $(b)-(c)$, where the gyromorphic peak remains visible even at large fractions of added or removed points ($f \approx 0.8$), and only asymptotically disappears as $f \to 1$.
Note that the low-$k$ behavior also becomes flatter and closer to $1$ as expected.

For all 3 types of noise, the observed peak height in an example realization is compared to the corresponding analytical prediction in Fig.~\ref{fig:NoisyGyromorphs}$(d)-(f)$, highlighting good agreement throughout.

All in all, for all forms of possible uncorrelated defects on the point pattern, we show that the features that survive best are features with $S(k) \gg 1$.
As a result, gyromorphs are, in a sense, the best possible structures to resist noise.
Furthermore, note that all types of uncorrelated noises level the structure factor back to the Poisson value.
This is an extra indication that gyromorphs are not, as one might fear from their large-$k$ behavior, ``just'' noisy quasicrystals: if they were, their peaks would necessarily be lower than those of quasicrystals, and we showed in main text that they are in fact higher.

As a follow-up to this section, we show in Sec.~\ref{sec:KickedDOS} the effect of random kicks on point poisitions on the DOS of gyromorphs.

\subsection{Diversity of solutions}

In this section, we discuss the differences between solutions reached by our algorithm. 
To quantify the similarity between point patterns, we introduce an overlap metric
\begin{align}
    O[\rho_1, \rho_2] \equiv \frac{\langle \rho_1 | \rho_2 \rangle}{\sqrt{\langle \rho_1 | \rho_1 \rangle \langle \rho_2 | \rho_2 \rangle}},
\end{align}
where we introduced a scalar product
\begin{align}
    \langle \rho_1 | \rho_2 \rangle \equiv \int\limits_{\mathbb{R}^d} d^d\bm{r} \widetilde{\rho}_1 (\bm{r}) \widetilde{\rho}_2 (\bm{r}),
\end{align}
that involves the density fields convolved by a narrow Gaussian
\begin{align}
    \widetilde{\rho} (\bm{r}) &\equiv \sum\limits_{n=1}^{N} c_n e^{(\bm{r} - \bm{r}_n)^2 / 2 \sigma^2}.
\end{align}
The overlap $O$ can be evaluated efficiently by performing the convolution of the point pattern by a Gaussian in Fourier space.
By construction, $O$ is non-negative and becomes asymptotically $0$ for point patterns with no overlapping points, and $1$ for identical point patterns.
In practice, we use FINUFFT and, for a gyromorph with peaks at $K$, compute the FFT up for $N_k = 50 * KL/2\pi$ modes, and set $\sigma$ such that $ \sigma = 1/(10K)$.

\begin{figure}
    \centering
    \includegraphics[width=0.48\linewidth]{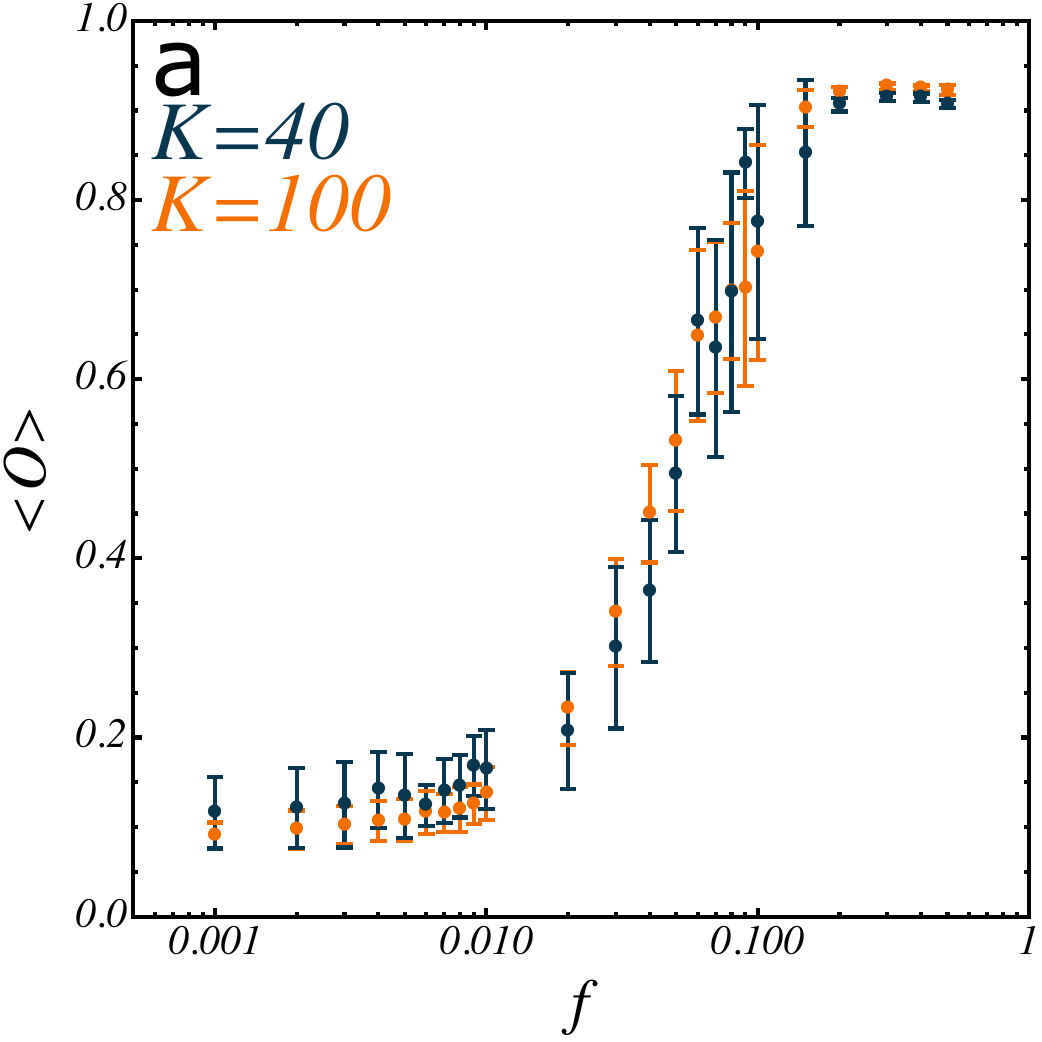}
    \includegraphics[width=0.48\linewidth]{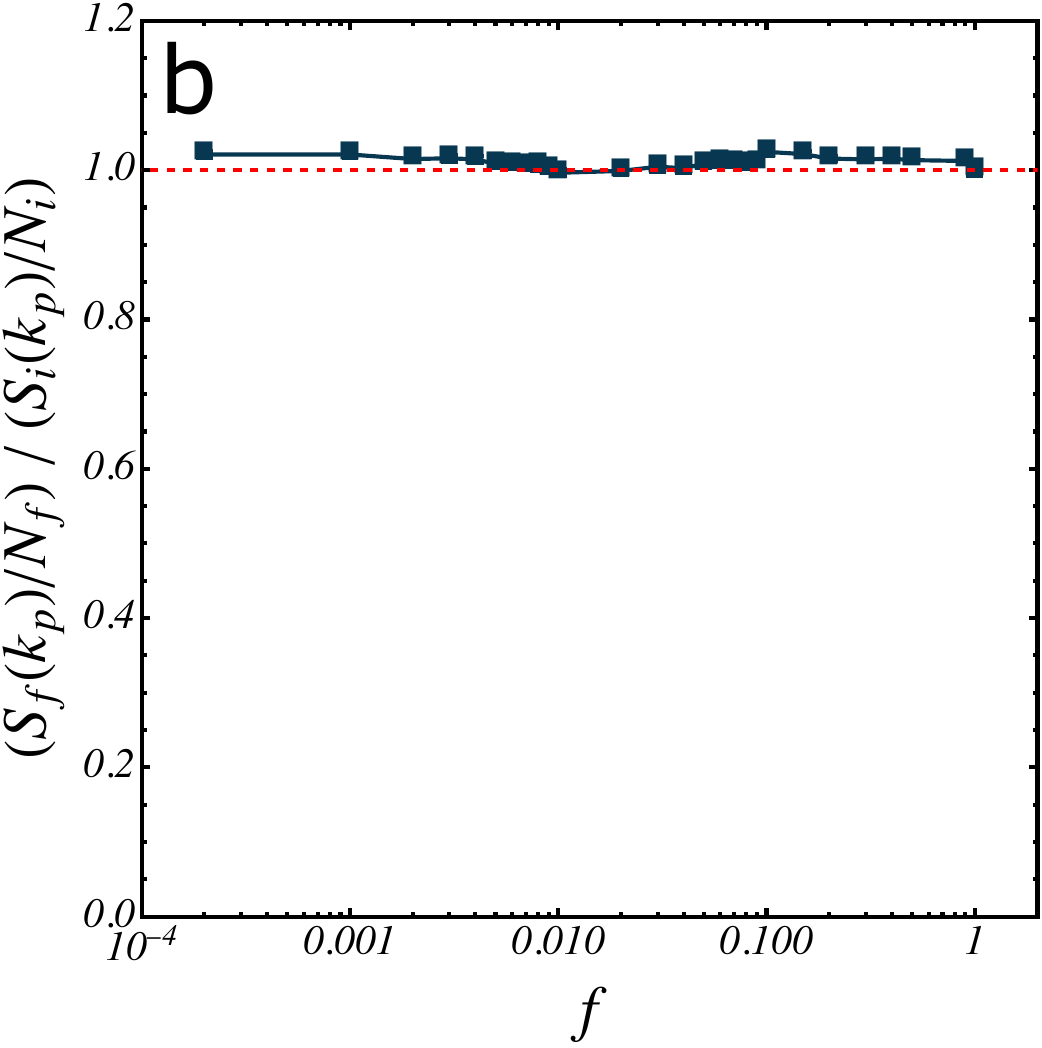} \\
    \includegraphics[width=0.46\linewidth]{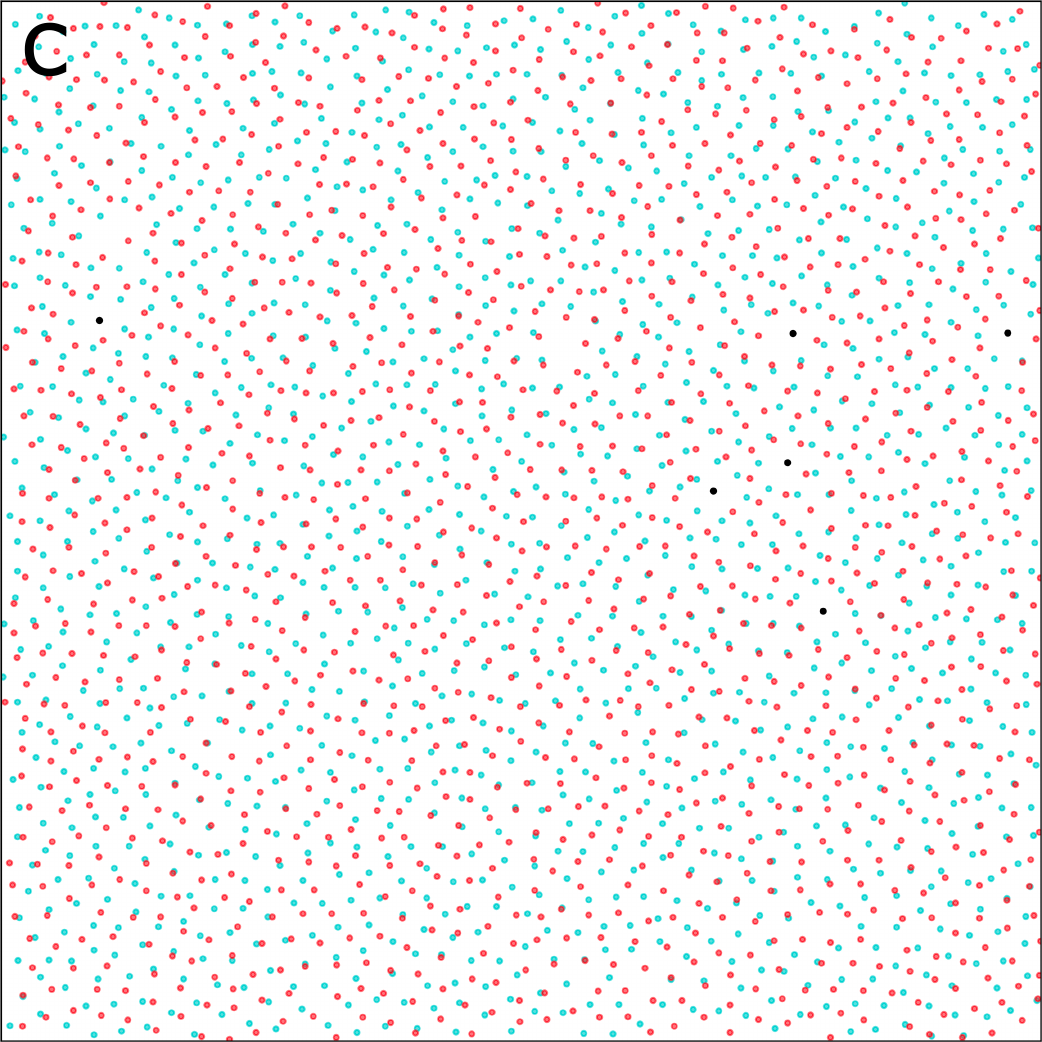} \hspace{0.03\linewidth}
    \includegraphics[width=0.46\linewidth]{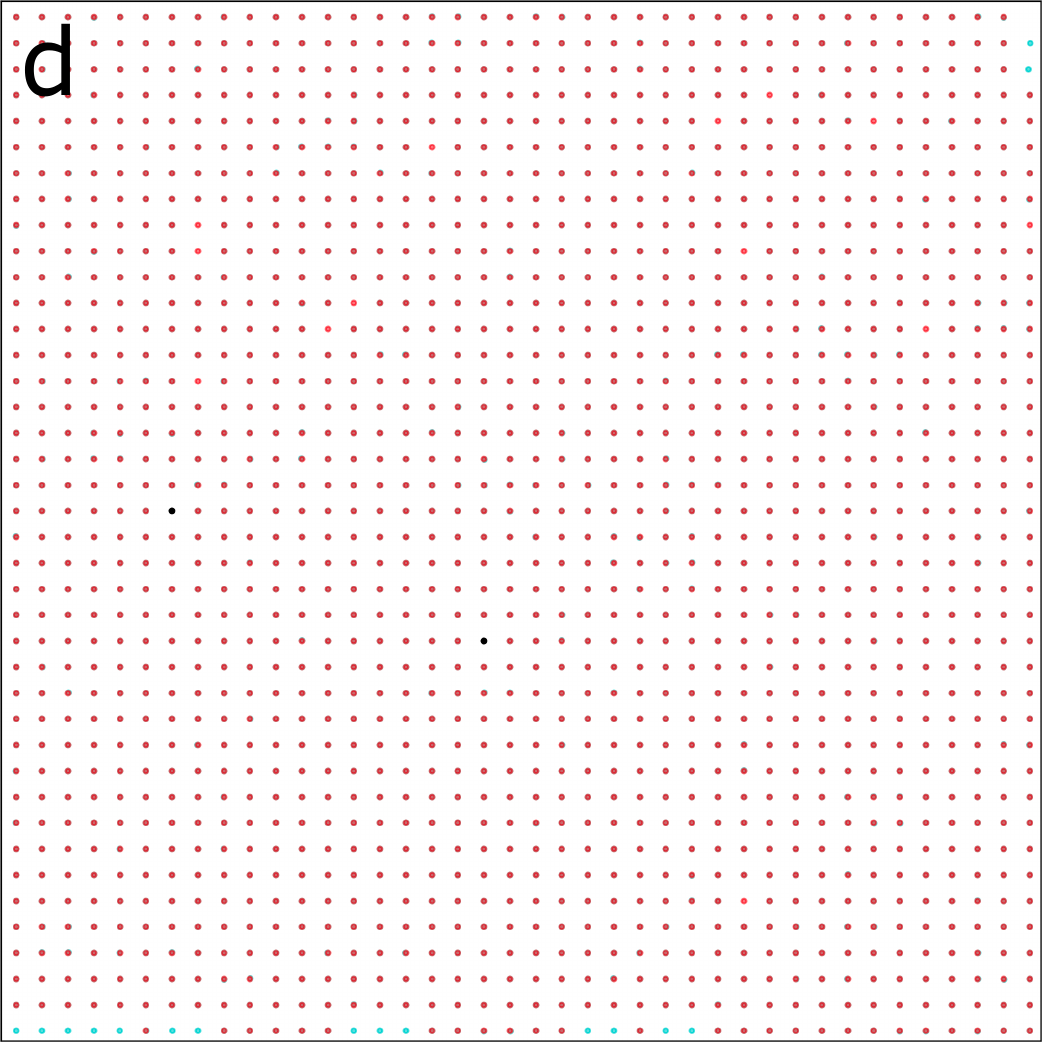}
    \caption{\textbf{Similarity across minima.}
    $(a)$ Averaged overlap across minimizations against fraction of pinned points, in log-linear scales, for $G = 60$.
    The overlap is averaged over 10 realizations using the same initial configuration but random sets of pinned points.
    The same operation is performed for $K=40$ ($N\sim 1000$) and $K = 100$ ($N \sim 10^4$).
    Error bars are Student $95\%$ confidence intervals.
    $(b)$ Ratio of the final peak heights to initial peak heights, normalized by the ratio of number of points before and after repopulation.
    A dashed red line indicates 1.
    $(c)$ Example realization for $K = 40$, $G=60$, pinning $f = 5\times 10^{-3}$ points.
    The pinned points are shown in black, the starting gyromorph in blue and the final gyromorph in red.
    $(d)$ Example realization following the same recipe for $G=4$ (square lattice) and pinning only $2$ random points.
    }
    \label{fig:overlap}
\end{figure}
In order to obtain comparable point patterns and measure a meaningful overlap, we consider the following experiment: we generate a gyromorph for a given $G$ and $K$.
Then, we pin a randomly selected fraction $f$ of the points and freeze their positions, but redraw the complementary fraction $1-f$ uniformly at random in the system and reminimize loss by adjusting the positions of these points only, using the same algorithm.
We thus obtain collections of different gyromorphs with a set of fixed points, and we compute $O$ using only the density field $\rho_{f}(\bm{r})$ containing the set of $(1-f)N$ points that were free to move.
The reported quantity therefore quantifies how similar the final reminimized positions of the free points are across realizations, in the spirit of point-to-set measurements of configurational entropy in glasses~\cite{Guiselin2022}.
We show the results in Fig.~\ref{fig:overlap}$(a)$ for $60$-fold gyromorphs at two sizes, $K = 40$ ($N\sim 10^3$) and $K = 100$ ($N \sim 10^4$).
Starting from a unique gyromorph, for each value of $f$ the overlap is averaged over $10$ realizations with different random pinned points.
In panel $(b)$, we show the ratios of the final peak heights $S_f$ to initial peak heights $S_i$, normalized by the ratio of number of points before and after repopulation, $N_i$ and $N_f$ respectively.
We show that the peak height is essentially not affected throughout, so that every repopulated configuration is a gyromorph as good as the starting one.
An example realization for $K = 40$, $f = 5\times 10^{-3}$ is shown in Fig.~\ref{fig:overlap}$(c)$, highlighting the differences between the old (blue) and new (red) configurations for a set of pinned (black) points.
The curves show that the overlap is low at small fractions of pinned particles, and grows to values close to $1$ later on.
Note that they do not finish strictly at $1$ as there can be equivalent sites after point removals even at rather large $f$.
The overlap value at a given $f$ does not seem to change significantly when the system size is changed, meaning that one needs to fix a subset $\mathcal{O}(N)$ of the points to obtain a similar configuration, as opposed to a fixed number of points independent of $L$.
This is in contrast with lattices, as illustrated in Fig.~\ref{fig:overlap}$(d)$ by using our algorithm in the case $G=4$, which yields perfect square lattices -- by pinning only 2 points, thus removing translational invariance from the problem, the whole lattice is perfectly recovered up to boundary effects.

All in all, this experiment shows that gyromorphs are not simply deterministic patterns like crystals or de Bruijn quasicrystals, as these would be perfectly determined by a fixed number of points.
Instead, a variety of gyromorphs can be obtained even when setting an extensive fraction of the points to the same locations.

\subsection{Gyromorphs do not form continuous families}

Looking at Fig.~\ref{fig:overlap}, one may wonder whether the configurations found by pinning subsets of the particles could be linked by continuous deformations of the point patterns that would not affect the peak heights.
In other words, considering the minimization part of our algorithm only, one may wonder whether the Hessian of the loss has non-trivial flat directions at minima.
In order to check, we first derive the expression of the Hessian of the FReSCo part of the loss, Eq.~\ref{eq:FReSCoLoss}.
Each $d\times d$ block $H_{mn}$ of the Hessian matrix is defined by
\begin{align}
    \overline{\overline{H}}_{mn}(\bm{r_1}, \ldots, \bm{r}_N) \equiv \frac{\partial^2 L_{\text{NUwNU}}}{\partial \bm{r}_m \otimes \partial \bm{r}_n} (\bm{r_1}, \ldots, \bm{r}_N),
\end{align}
where $\otimes$ represents an outer product and we introduce a short-hand notation for the double outer derivative.
Using the expression of the loss, a $d\times d$ tensor element of the Hessian can be written at any point as
\begin{align}
    \overline{\overline{H}}_{mn} &= 2 \sum\limits_{p=1}^{G/2}\left[ \frac{\partial^2 S(\bm{k}_p)}{\partial \bm{r}_m  \otimes \partial \bm{r}_n} (S(\bm{k}_p) - S_0 (\bm{k}_p)) +  \frac{\partial S(\bm{k}_p)}{\partial \bm{r}_m } \otimes \frac{\partial S(\bm{k}_p)}{\partial \bm{r}_n}\right].
\end{align}
The relevant derivatives of $S$ can be written in terms of the Fourier transform of the density field, $\widehat{\rho}$ and its complex conjugate $\widehat{\rho}^\dagger$, recalling that $S \equiv |\widehat{\rho}|^2/N$,
\begin{align}
    \frac{\partial S(\bm{k})}{\partial \bm{r}_n} &= \frac{2}{N} \text{Re}\left[ \widehat{\rho}(\bm{k}) \frac{\partial \widehat{\rho}^\dagger}{\partial \bm{r}_n}(\bm{k}) \right],\\
    \frac{\partial^2 S(\bm{k})}{\partial \bm{r}_m \otimes \partial \bm{r}_n} &= \frac{2}{N} \text{Re}\left[ \frac{\partial \widehat{\rho}}{\partial \bm{r}_m}(\bm{k}) \otimes \frac{\partial \widehat{\rho}^\dagger}{\partial \bm{r}_n}(\bm{k}) + \widehat{\rho}(\bm{k})\frac{\partial^2 \widehat{\rho}^\dagger}{\partial \bm{r}_m \otimes \partial\bm{r}_n}(\bm{k}) \right].
\end{align}
Finally, we introduce a general form for the density field like in Ref.~\onlinecite{Shih2023},
\begin{align}
    \rho(\bm{r}) \equiv \sum\limits_{n=1}^{N} c_n \delta(\bm{r} - \bm{r}_n)
\end{align}
with $c_n \in \mathbb{C}$, and define our Fourier transform such that
\begin{align}
    \widehat{\rho}(\bm{k}) = \sum\limits_{n=1}^{N} c_n e^{i \bm{k} \cdot \bm{r}_n}.
\end{align}
Injecting the expression of $\widehat{\rho}$ into the derivatives of $S$ leads to
\begin{align}
     \frac{\partial S(\bm{k})}{\partial \bm{r}_m} \otimes \frac{\partial S(\bm{k})}{\partial \bm{r}_n} &= \frac{-2 \bm{k}\otimes\bm{k}}{N^2} \text{Re}\left[ \widehat{\rho}^{\,2} c_m^\dagger c_n^\dagger e^{- i \bm{k}\cdot(\bm{r}_m + \bm{r}_n)} - |\widehat{\rho}|^2 c_m^\dagger c_n e^{- i \bm{k}\cdot(\bm{r}_m - \bm{r}_n)} \right],\\
    \frac{\partial^2 S(\bm{k})}{\partial \bm{r}_m \otimes \partial \bm{r}_n} &= \frac{2 \bm{k} \otimes \bm{k}}{N} \text{Re}\left[c_m^\dagger c_n e^{-i \bm{k}\cdot (\bm{r}_m - \bm{r}_n)} - \delta_{m,n} \widehat{\rho} \, c_n^\dagger e^{-i\bm{k}\cdot\bm{r}_n} \right],
\end{align}
where we dropped the argument of $\widehat{\rho}$, $S$, and $S_0$ to avoid clutter.
Just like the gradient in FReSCo~\cite{Shih2023}, the Hessian may then be written as Fourier series,
\begin{align}
    \overline{\overline{H}}_{mn} &= \text{Re}\left[ \delta_{m,n} \widehat{D}_n(\bm{r}_n) + \widehat{S}_{mn}(\bm{r}_m + \bm{r}_n) + \widehat{A}_{mn}(\bm{r}_m - \bm{r}_n) \right],
\end{align}
with $\widehat{D}_n$ a ``diagonal'' term, $\widehat{S}_{mn}$ a term that depends only on a symmetric combination of $\bm{r}_m$ and $\bm{r}_n$, and $\widehat{A}_{mn}$ one that depends only on their difference, and the hat indicates that any of the three terms is of the form
\begin{align}
    \widehat{T}(\bm{r}) = \sum\limits_{\bm{k}_p} \overline{\overline{T}}(\bm{k}_p) e^{- i \bm{k}_p \cdot \bm{r}}.
\end{align}
The expressions of the corresponding Fourier coefficents (analogues of $T(\bm{k}_p)$) read
\begin{align}
    \overline{\overline{D}}_n(\bm{k}) &\equiv -\frac{4 \bm{k}\otimes\bm{k}}{N}\widehat{\rho}(\bm{k}) c_n^\dagger (S(\bm{k}) - S_0 (\bm{k})),  \\
    \overline{\overline{S}}_{mn}(\bm{k}) &\equiv -\frac{4 \bm{k}\otimes\bm{k}}{N^2}\widehat{\rho}^{\,2}(\bm{k}) c_m^\dagger c_n^\dagger, \\
    \overline{\overline{A}}_{mn}(\bm{k}) &\equiv \frac{4 \bm{k}\otimes\bm{k}}{N}c_m^\dagger c_n (2S(\bm{k}) - S_0(\bm{k})).
\end{align}

\begin{figure}
    \centering
    \includegraphics[width=0.46\linewidth]{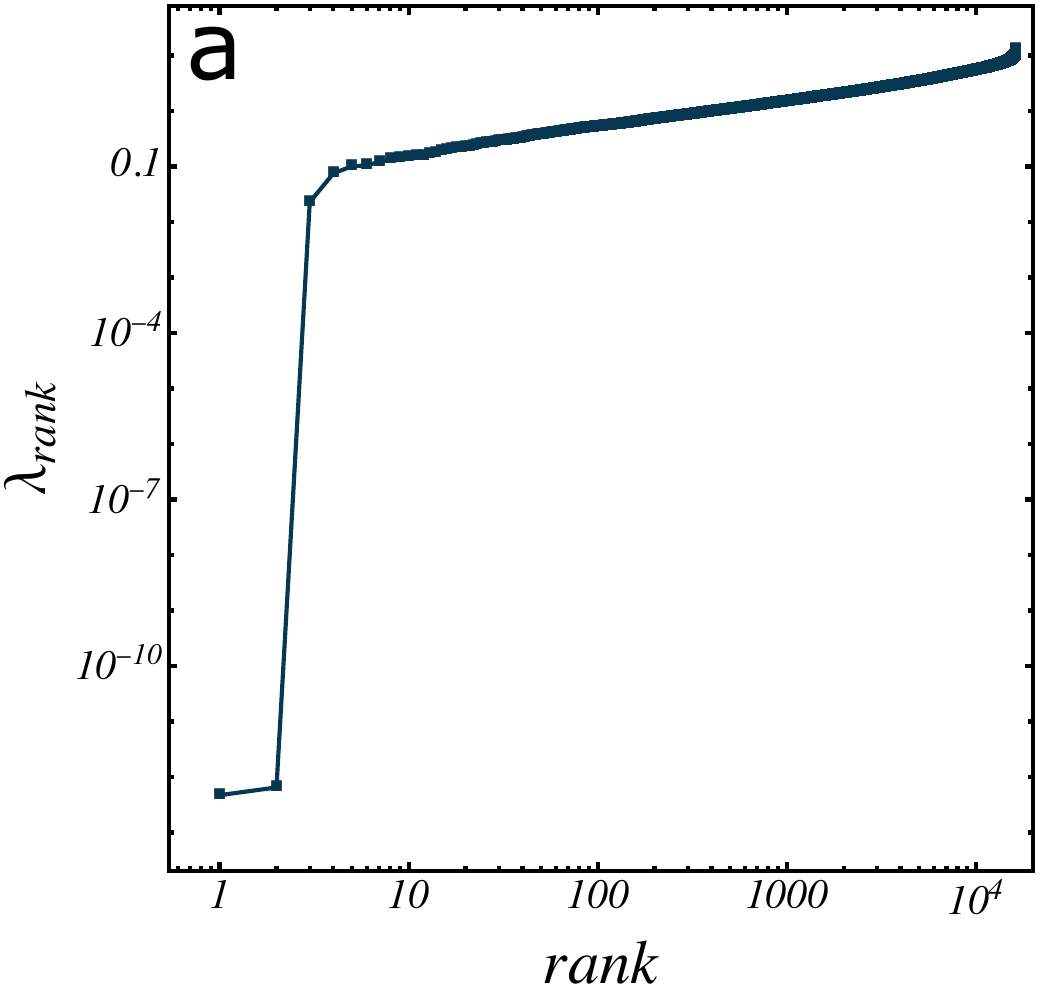}
    \includegraphics[width=0.46\linewidth]{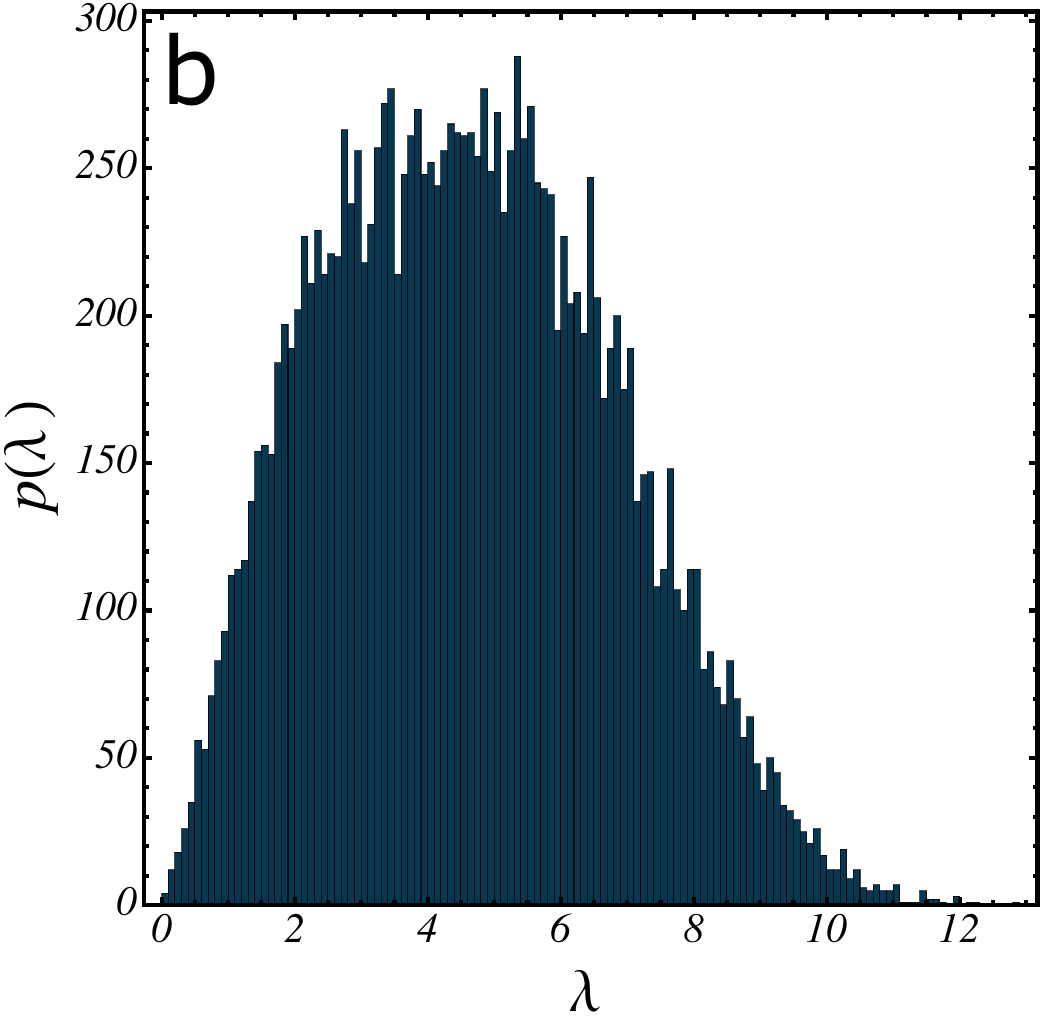}
    \caption{\textbf{Spectrum of the FReSCo Hessian at a minimum.}
    $(a)$ Value of sorted eigenvalues against their rank, in log-log.
    $(b)$ Histogram of eigenvalues.
    Both plots are obtained for a $G = 60$ gyromorph with $N \sim 8000$ points.
    }
    \label{fig:HessianSpectrum}
\end{figure}

The algorithm of Fig.~\ref{fig:Algorithm} is not just a minimization, as it contains cycles of particle removals and reinsertions.
Thus, to compute the Hessian at a meaningful point, we start from a configuration generated by our algorithm, then we reminimize the FReSCo part of the loss alone (without the variance part), remove overlapping points, and repeat the procedure until $N$ stabilizes, then evaluate the Hessian there.
Note that this is \textit{not} our usual recipe as this one, while it generates true minima of a loss function, also has more short-range density fluctuations, which is often argued to be bad for optics~\cite{Sellers2017}.
The relevant part of the obtained Hessian is its spectrum.
We perform an eigendecomposition and report the eigenvalues for $G = 60$ and $K = 100$ in Fig.~\ref{fig:HessianSpectrum}$(a)$, where they are sorted by increasing order.
The relevant information is that 2 eigenvalues are much smaller than the rest -- we check that they correspond to global translations, which leave the loss invariant.
The next eigenvalues are orders of magnitude larger, and form the first bin of a smooth distribution of positive values, whose histogram is shown in Fig.~\ref{fig:HessianSpectrum}$(b)$.
This shows that gyromorphs may not be obtained from each other via smooth deformations without making the gyromorph worse in intermediate stages.

\section{Additional data on optical behavior}

\begin{figure}
    \centering
    \includegraphics[width=\columnwidth] {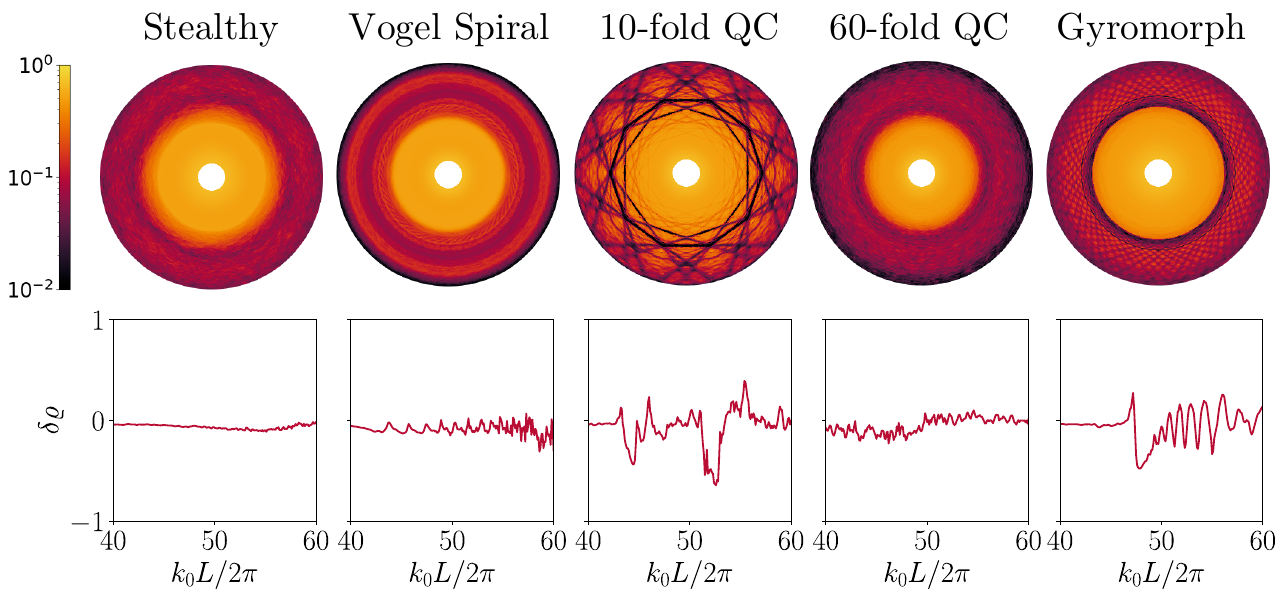}
    \caption{\textbf{Scalar-wave results: low contrast.}
    Transmission plots (top) and density of states (bottom) of incident scalar-waves for five 2$d$ systems.
    In the whole figure, $\phi = 5\%$, $n=1.5$. 
    }
    \label{fig:ScalarWaves}
\end{figure}

\begin{figure}
    \centering
    \includegraphics[width=\columnwidth] {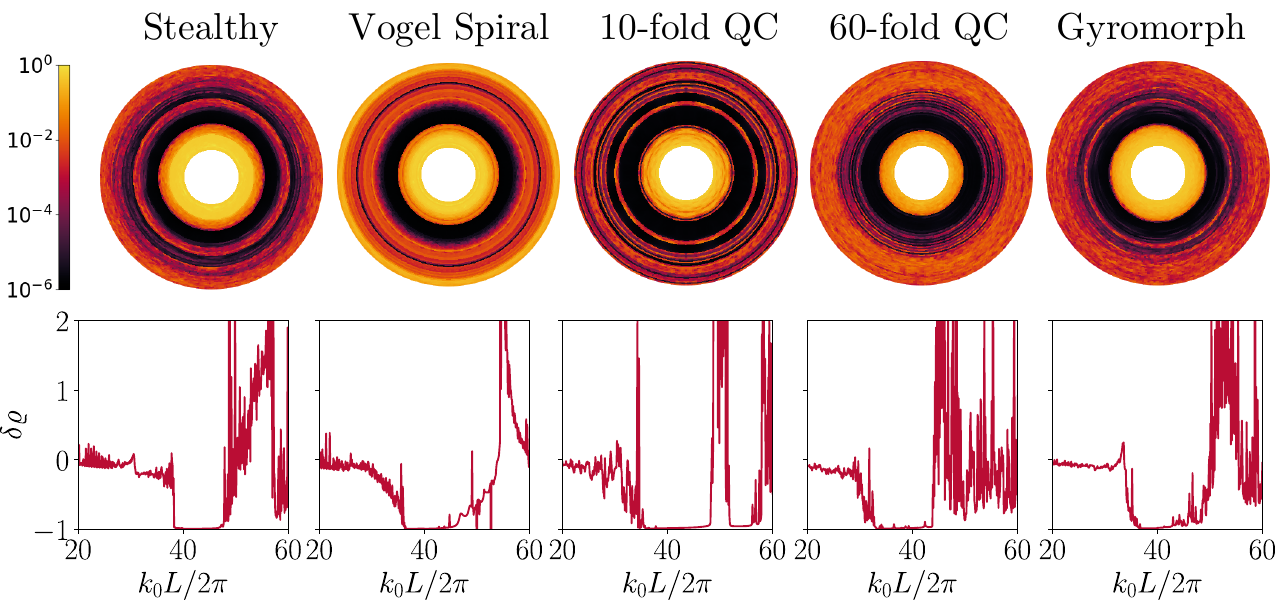}
    \caption{\textbf{Scalar-wave results: high contrast.}
    Transmission plots (top) and density of states (bottom) of incident scalar-waves for five 2$d$ systems.
    In the whole figure, $\phi = 5\%$, $n=3$.
    }
    \label{fig:ScalarWaves_n3}
\end{figure}

\subsection{Scalar-wave results}

In the main text, we mostly focus on vector wave results.
Here, we present the corresponding results for scalar waves, which have been the subject of most previous studies, as bandgaps are generally easier to achieve in the scalar case~\cite{Florescu2009,Froufe-Perez2016,Froufe-Perez2017,Monsarrat2022}.
First, in Fig.~\ref{fig:ScalarWaves}, we show results for $2d$ scalar waves (``TM polarization'') for parameters leading to curves with a similar appearance to those obtained for vector waves (``TE polarization'') of the main text, namely $\phi = 5\%$ and $n = 1.5$, for the same systems examined in the main text.
The results closely resemble the low-contrast TE ones: the deeper features in transmission and LDOS variation appear in gyromorphs and low-order quasicrystals, with gyromorphs being more isotropic than the latter.
In Fig,~\ref{fig:ScalarWaves_n3}, we show the same set of quantities, now using the same parameters as in the main text, namely $\phi = 5\%$ and $n=3$.
As the TM response is typically stronger than the TE one, a clear gap emerges for all systems, consistent with previous studies.
The main transmission gap in gyromorphs is as wide or wider than in its counterparts, and there are fewer features (local maxima and minima) of transmission as a function of frequencies than in other systems like quasicrystals.
The DOS displays strong depletion to near-zero number of modes, like other systems.
Finally, note that this measurement indicates ``completeness''  with respect to the polarization of mode depletion in gyromorphs.

\begin{figure}
    \centering
    \includegraphics[width=\columnwidth] {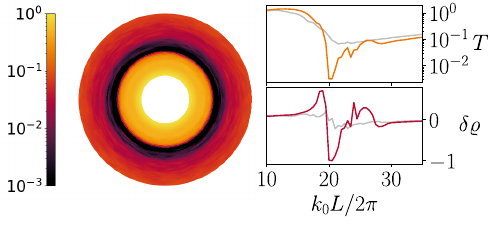}
    \caption{\textbf{Scalar-wave results for 3d gyromorph}
    Transmission plot (left) for the 3$d$ gyromorph. Radially averaged transmission (upper right) and density of states (lower right) of incident scalar-waves for the 3$d$ gyromorph (orange, red lines) and a 3d poisson point pattern (gray lines).
    In the whole figure, $n=3$ and $\phi = 10\%$.
    }
    \label{fig:ScalarWaves3d}
\end{figure}

In Fig.~\ref{fig:ScalarWaves3d}, we show scalar-wave results for the $3d$ gyromorph from the main text, again choosing the same packing fraction, but a lower index contrast than in the main text (here, $n=3$).
This response is again much stronger than in the vector-wave case.

\subsection{Additional parameter sweeps}

\begin{figure}
    \centering
    \includegraphics[height=0.45\linewidth]{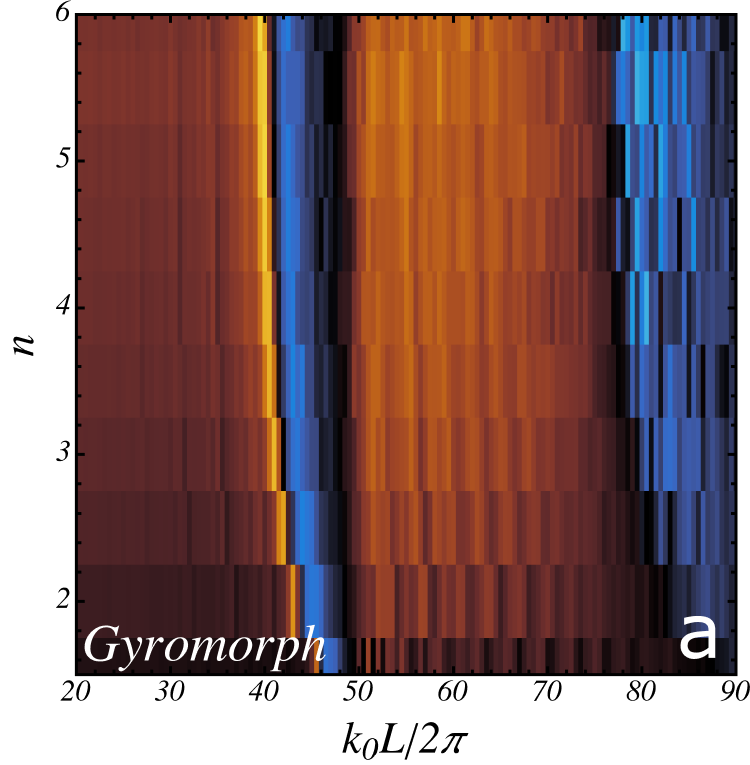}
    \includegraphics[height=0.45\linewidth]{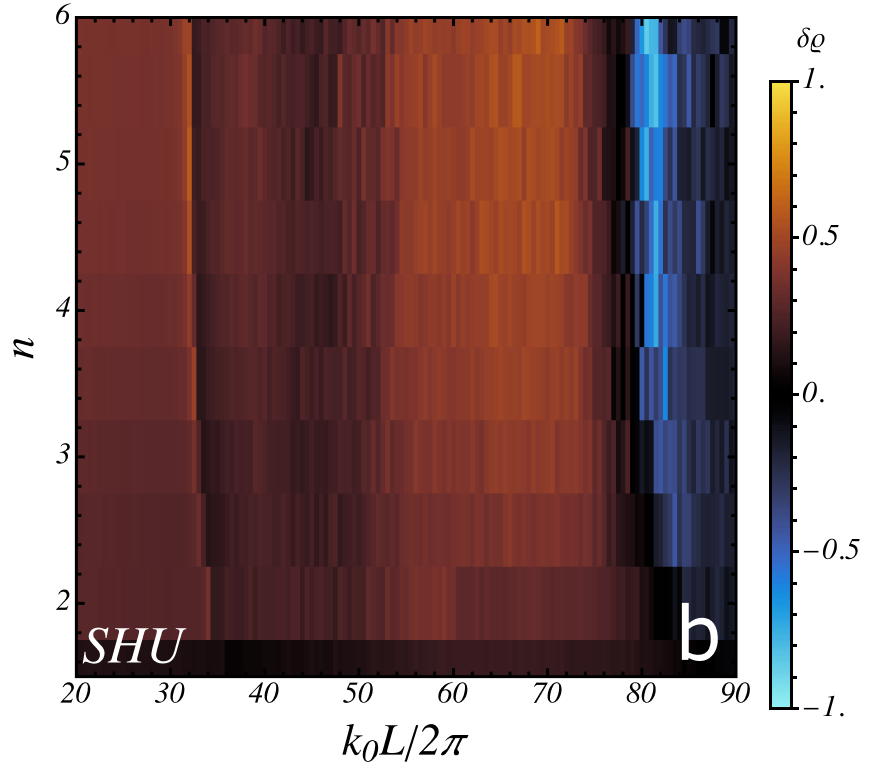} \\

    \includegraphics[height=0.45\linewidth]{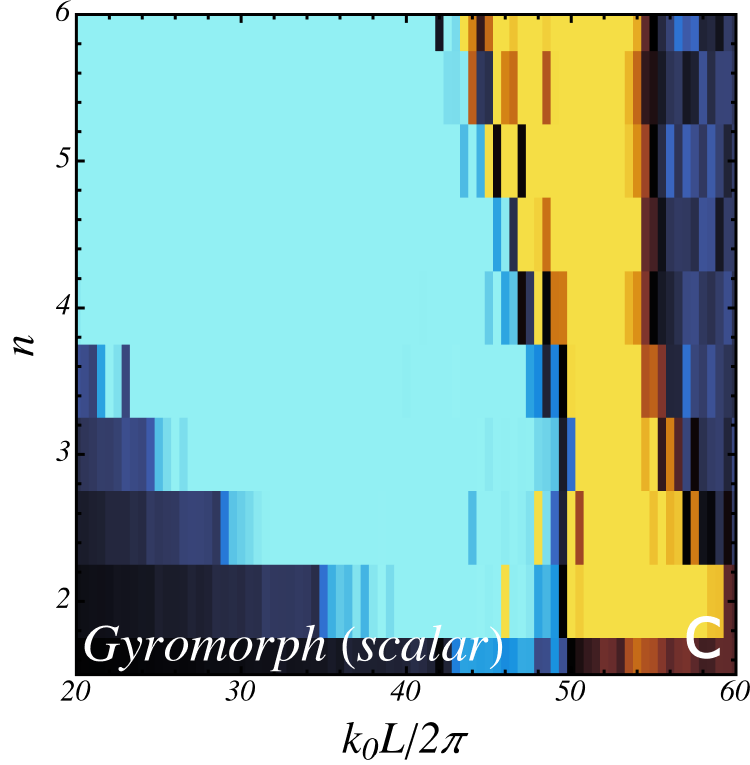}
    \includegraphics[height=0.45\linewidth]{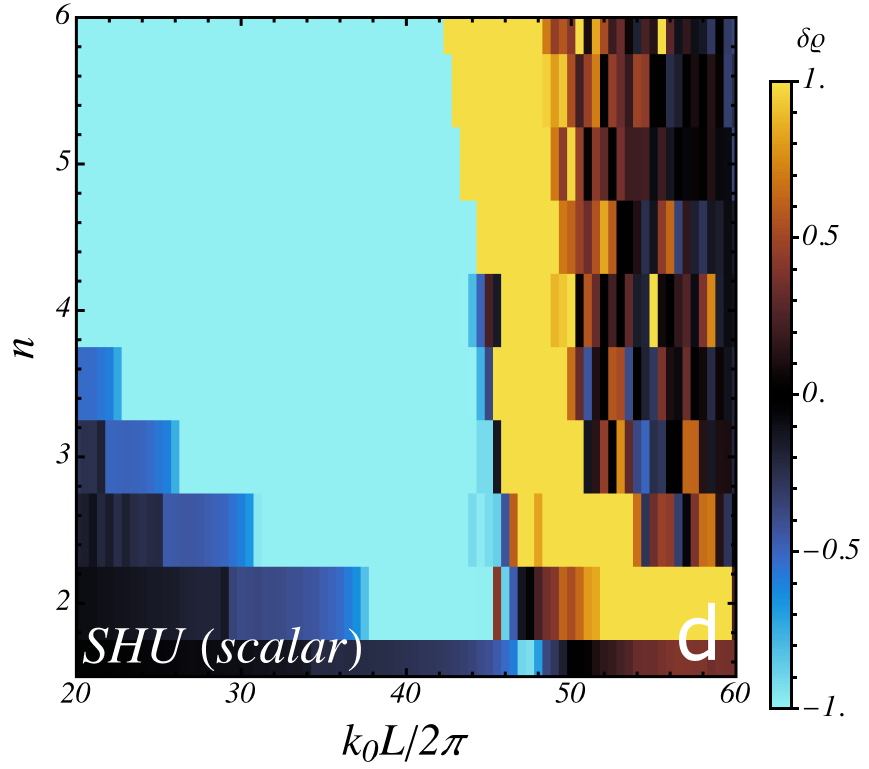}
    
    \caption{\textbf{Refractive index sweep.}
    Intensity maps of DOS against frequency and refractive index at $\phi = 0.15$, averaged over $30$ systems, for $(a)$ gyromorphs and $(b)$ SHU systems. Dotted red line: $\phi = 0.05$.
    $(c)-(d)$ Same plots for scalar waves.}
    \label{fig:idos_nsweep_0p15}
\end{figure}

We provide additional data on how optical parameters influence the DOS of the correlated disordered point patterns examined in this paper.
Like in the main text, we compute the relative change in the LDOS, $\delta\varrho$, averaged over $30$ independent realizations of each system and over positions within a single realization (here 1000 independent random points within the medium, each located at least $2a$ from any scatterer).
In Fig.~\ref{fig:idos_nsweep_0p15}$(a)-(b)$, we report DOS maps for vector waves at $\phi = 0.15$, where SHU systems display a bandgap at large $k_0$.
Notably, only the gyromorph displays a mode depletion at $k_0 L / 2 \pi \approx 50$, which is remarkably stable across $n$.
The fact that the DOS in gyromorphs does not saturate at $-1$ is due to two effects: fluctuations of the DOS within one system, and fluctuations of the precise location of the depletion across systems.
We illustrate this with further data from individual systems in Sec.~\ref{sec:CDOSvsIDOS}.
In Fig.~\ref{fig:idos_nsweep_0p15}$(c)-(d)$, we report results for scalar waves.
Again, both systems display a bandgap but it is clear that the DOS is depleted over a wider range of parameters in the gyromorph.

\subsection{$G=60$ is not special}

\begin{figure}
    \centering
    \includegraphics[width=0.5\linewidth]{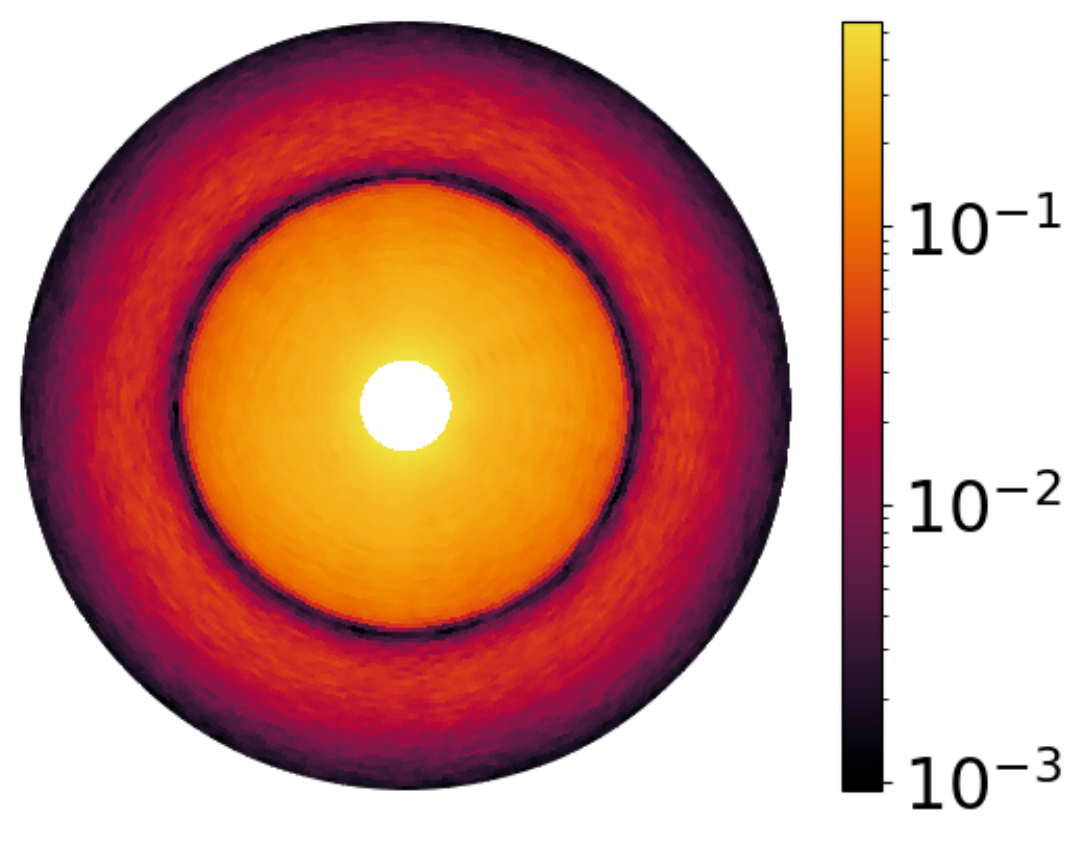}
    \caption{\textbf{Transmission spectrum of a $G=200$ gyromorph.}
    Intensity transmission for frequencies $40 \leq k_0L/2\pi \leq 60$ (radial direction) and $360$ incident angles (orthoradial direction) of a source Gaussian beam, like in Fig.~3 of the main text, for a system of $N\approx 9000$ points in a $G=200$ gyromorph, for $n = 3$ and $\phi = 0.05$ and vector waves.
    The center of the intensity drop lies at $k_0L/2\pi \approx 50$.
    }
    \label{fig:200-fold}
\end{figure}

In the main text and in this document, we overwhelmingly focus on the case of $G=60$.
We make this choice not because $G=60$ is special, but because it is already sufficiently large compared to $1$ that the system displays isotropic transmission gaps and bandgaps.
To minimally illustrate that one may choose a different $G$, in Fig.~\ref{fig:200-fold}, we show the vector-wave transmission for a $G=200$ gyromorph comprising $N \approx 9000$ points, for the same parameters as in the main text and for vector waves.
The isotropic intensity drop is qualitatively similar to that of the main text.

\subsection{Scaling of bandgap depth with physical parameters}

\begin{figure}
    \centering
    \includegraphics[width=\columnwidth] {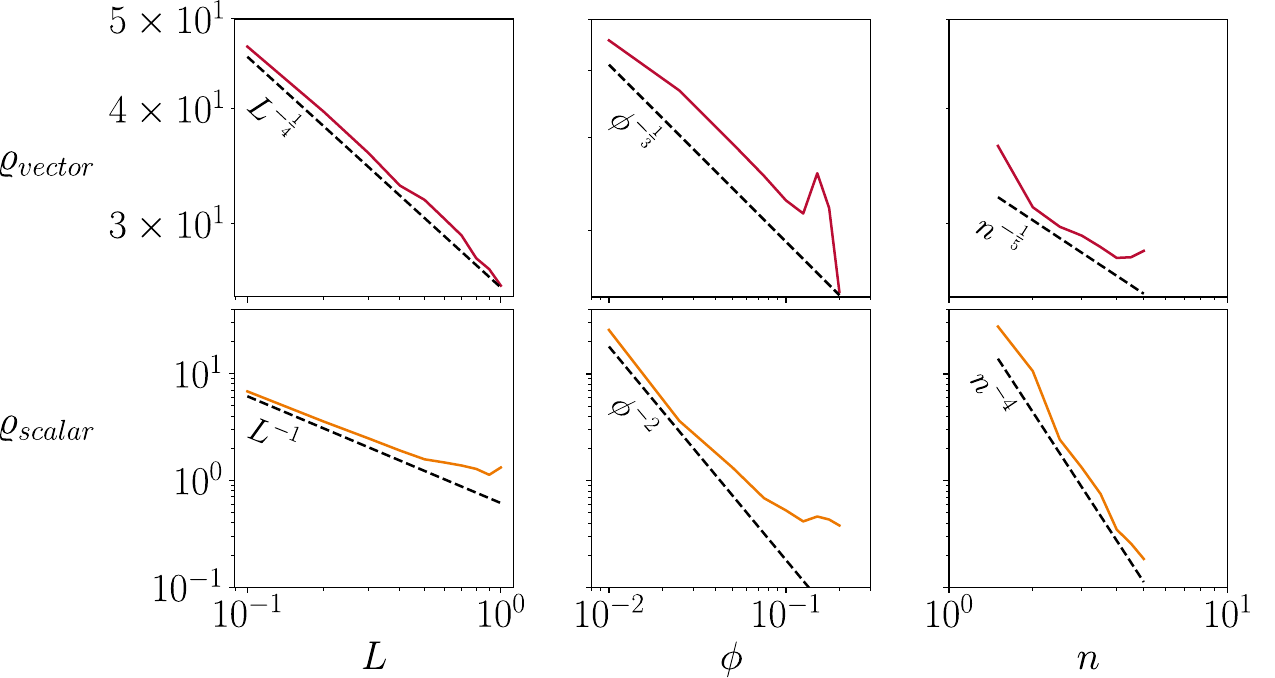}
    \caption{\textbf{Scalings of the bandgap in the 2d 60-fold gyromorph}
    Absolute DOS at the deepest point of the bandgap, $\varrho = \varrho_0 (\delta \varrho_{min}+ 1)$, against the diameter of the subsampled medium $L$, the filling fraction $\phi$, and the refractive index $n$, in log scales, for TE (red) and TM (orange) polarizations.
    Dashed lines indicate power-law tendencies.
    In the whole figure, $n = 3$ and $\phi = 5\%$ if not otherwise specified.
    }
    \label{fig:Scalings2d}
\end{figure}

To assess the quality of the bandgaps formed by gyromorphs, we measure the absolute DOS $\varrho = \varrho_0 (1 + \delta \varrho )$, averaged over points away from scatterers, for both vector and scalar waves, when varying physical parameters, $L$, $\phi$, and $n$, keeping the other parameters fixed to the values used in the main text.
To vary $L$, following Ref.~\onlinecite{Froufe-Perez2017}, we subsample from one bigger configuration with a radius that we tune.
The $2d$ results are shown in Fig.~\ref{fig:Scalings2d}.
We show that the mode depletion indeed gets stronger for both polarizations as the medium gets thicker.
In fact, following Refs~\cite{Skipetrov2016, Skipetrov2020}, we show that the decay with $L$ is consistent with the expected $\varrho \sim 1/L$ for scalar waves, as expected in crystalline bandgaps for the TM polarization.
We also report a (weaker) decay for the TE component, with an exponent $\varrho \sim 1/L^{1/4}$.
The changes with $\phi$ and $n$ are more subtle, but we do observe a deepening of the gap with both $\phi$ and $n$.
We show nearby power laws as guides for the eyes in Fig.~\ref{fig:Scalings2d}.

\begin{figure}
    \centering
    \includegraphics[width=\columnwidth] {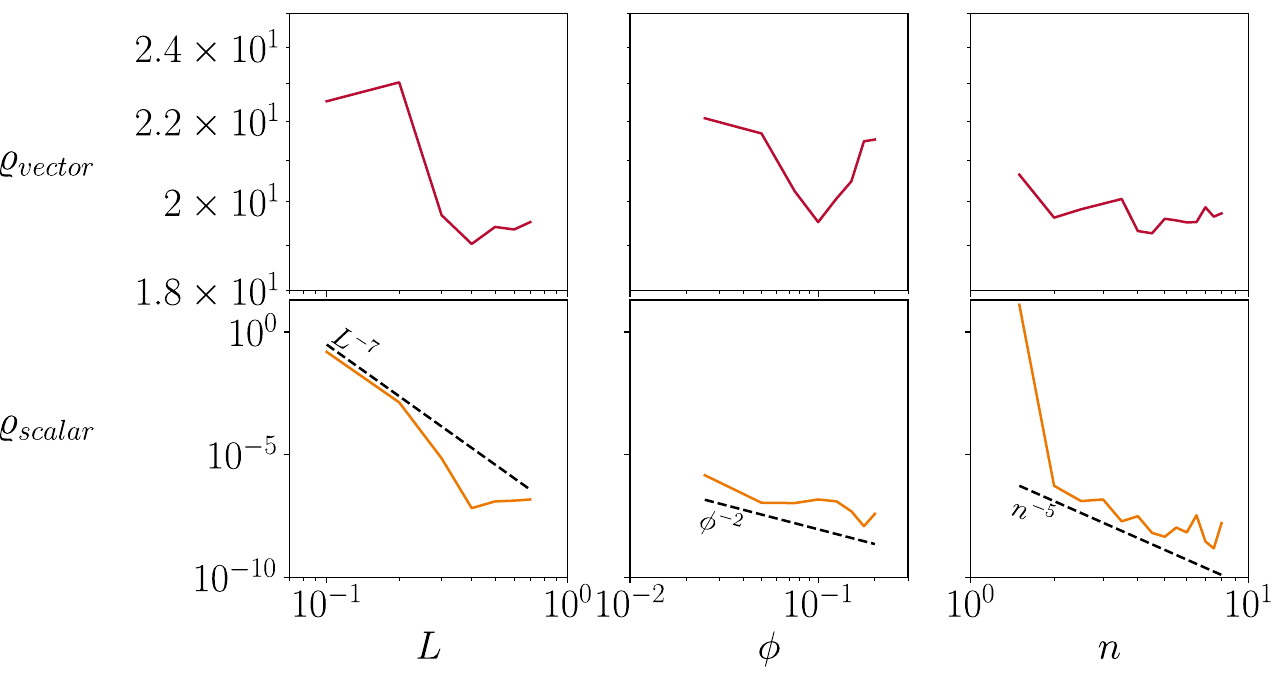}
    \caption{\textbf{Scalings of gap in the 3d gyromorph}
    Absolute DOS at the deepest point of the bandgap, $\varrho = \varrho_0 (\delta \varrho_{min}+ 1)$, against the diameter of the subsampled medium $L$, the filling fraction $\phi$, and the refractive index $n$, in log scales, for TE (red) and TM (orange) polarizations.
    Dashed lines indicate power-law tendencies.
    Dashed lines indicate results for typical bandgap forming candidates.
    In the whole figure, $n = 6.5$ for vector waves, $n=3$ for scalar waves and $\phi = 10\%$ if not otherwise specified.
    }
    \label{fig:Scalings3d}
\end{figure}

The same measurements are performed in $3d$ gyromorphs and reported in Fig.~\ref{fig:Scalings3d}.
We emphasize that the vector-wave results almost do not move, probably suggesting that a larger variety of $L$ would be needed to catch any clear tendency.
On the scalar side though, the DOS decreases extremely fast with $L$ (a nearby power is $L^{-7}$).
That fast decay could be a sign of an exponential decay, as suggested in similar measurements in a stealthy hyperuniform system in past work~\cite{Froufe-Perez2017}.

\subsection{DOS definition comparison\label{sec:CDOSvsIDOS}}

\begin{figure}
    \centering
    \includegraphics[width=0.48\linewidth]{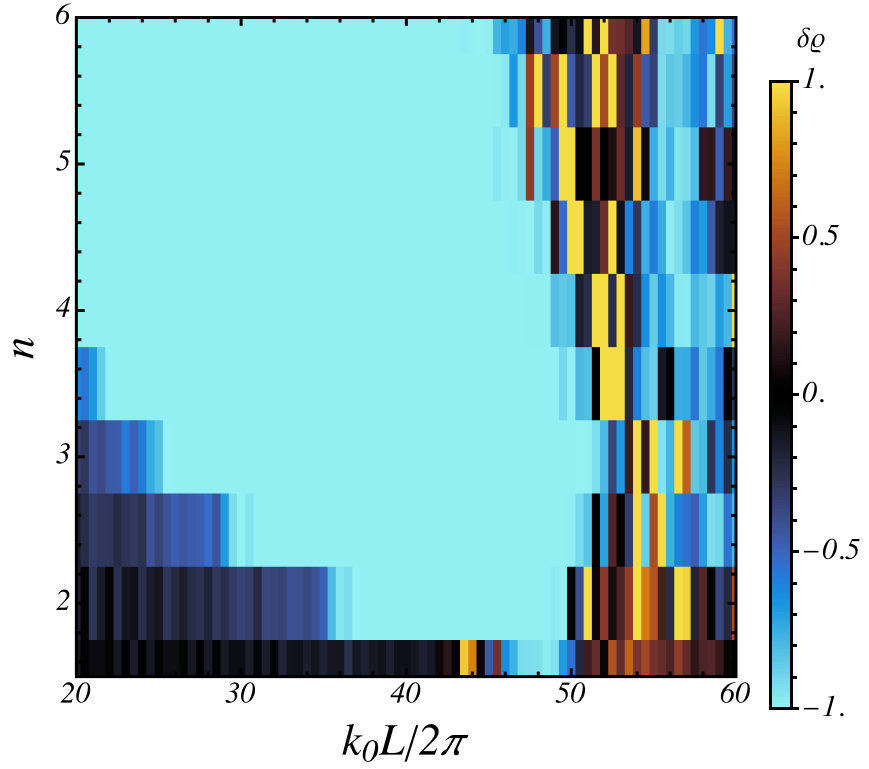}
    \includegraphics[width=0.48\linewidth]{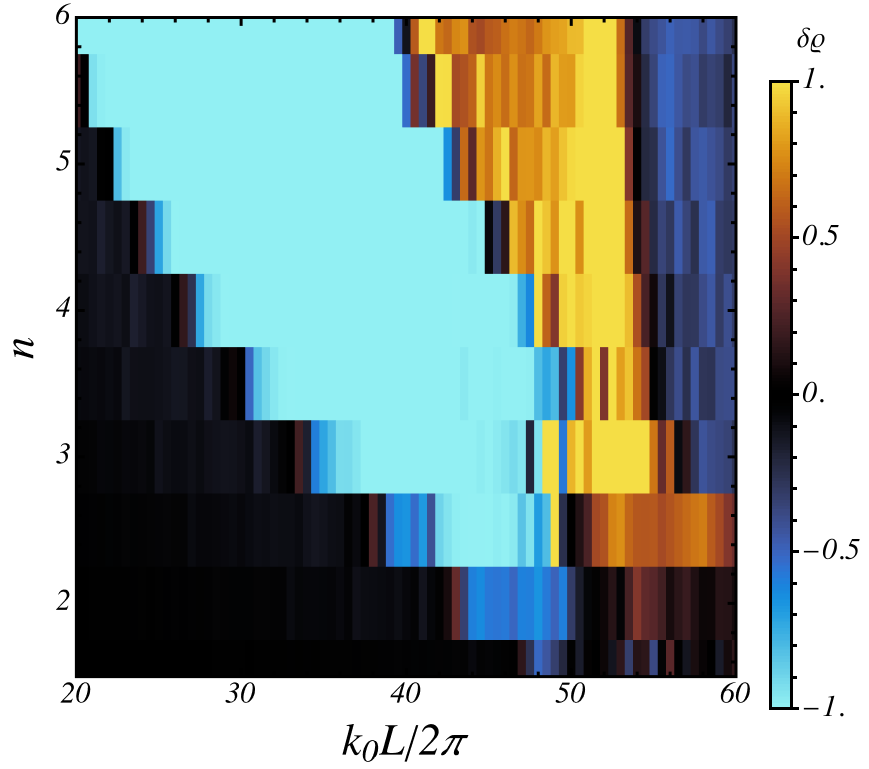}
    \caption{\textbf{CDOS vs IDOS.}
    CDOS (left) and IDOS (right) intensity maps in the $(k_0L/2\pi,n)$ plane for $\phi = 0.05$, averaged over $30$ gyromorphs with $G = 60$ and $N \approx 9000$, for scalar waves.}
    \label{fig:cdos_vs_idos}
\end{figure}
We compare to our preferred definition a commonly accepted definition of the DOS of a material (\textit{e.g.}, see~\cite{Pierrat2010}), which in this section and in MaGreeTe~\cite{MAGreeTe} we refer to as ``CDOS'', with C standing for cavity.
CDOS uses a single LDOS evaluation at the center of a finite system, but ensures that this point lies far enough away from scatterers that the LDOS does not diverge.
To do so, all points originally in the point pattern are deleted in the vicinity of the center (here with a radius $2a$ around the center) and the LDOS is evaluated in the newly created cavity.
The result may then be averaged over realizations of the system.
Due to the fact that only one point at the center is used, this definition disregards possible fluctuations of the LDOS across locations in a system.
Furthermore, since a single point is used in each system, this definition may take a large number of independent realizations of disorder to converge to a stable value.

By contrast, in our preferred definition (called ``IDOS'' in MaGreeTe, with I standing for interstitial), we draw $M$ measurement points uniformly in the system that lie at least $2a$ away from any scatterer (which is just achieved by rejection and resampling of points).
This definition takes into account spatial fluctuations of the LDOS in a single system, and also typically converge to stable values faster than CDOS.
However, it may also catch boundary effects in finite systems, and thus have artificially weaker features.

We illustrate the practical difference between the two approaches in Fig.~\ref{fig:cdos_vs_idos} for scalar waves in gyromorphs.
We show that the CDOS definition typically obscures local fluctuations of DOS that here overestimate the depth and width of the bandgap.
These local fluctuations are, within our work, particularly important to consider as we work exclusively with finite disks (balls in $3d$) of material in free boundary conditions, not periodic boundaries that artificially suppress boundary effects.
Thus, at the risk of making DOS suppressions look weaker than past works, but with the clear advantage that we include spatial fluctuations of the DOS into the averages, we prefer using IDOS throughout this work.

\begin{figure}
    \centering
    \includegraphics[width=0.94\linewidth]{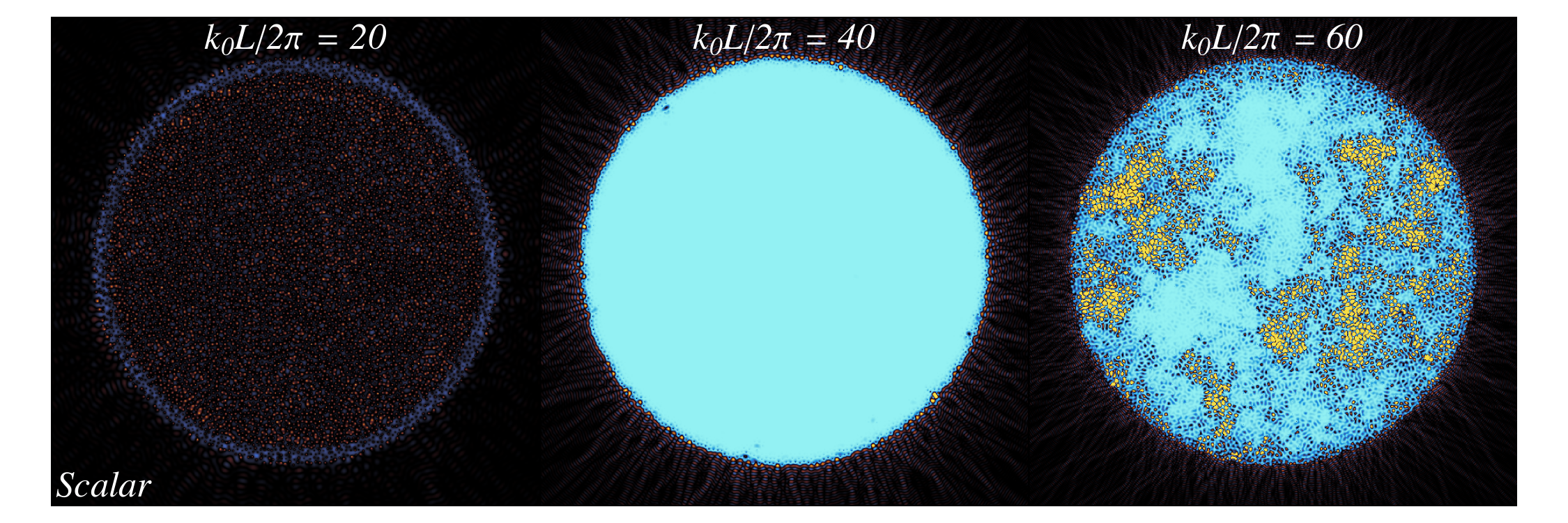}
    \includegraphics[height=0.28\linewidth]{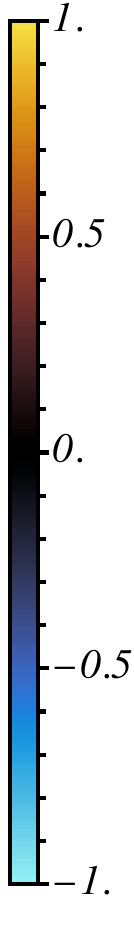}
    \caption{\textbf{LDOS maps: scalar waves in a gyromorph.}
    Maps of LDOS change relative to vacuum, $\delta\varrho(\bm{r}; \omega)$, obtained using a $1001\times1001$ regular grid of pixels in a gyromorph with $N \approx 9000$, $\phi = 0.05$ and $n=3$, for scalar waves.
    From left to right, $k_0L /2\pi =20$ (below gap), $40$ (in gap), $60$ (above gap)}
    \label{fig:LDOS_Gyro_Scalar}
\end{figure}

These fluctuations are particularly important for vector waves, as LDOS is suppressed much more homogeneously in the case of scalar waves.
This is illustrated by the LDOS maps of Fig.~\ref{fig:LDOS_Gyro_Scalar}, where we show three LDOS maps: one below the gap ($k_0L/2\pi = 20$), one inside the gap ($k_0L/2\pi = 40$), and one above the gap ($k_0L/2\pi = 60$).
The map inside the gap displays very little fluctuations in the bulk of the system, only small ones at the boundary.
Notice that while the map below the gap is unambiguous, the map above the gap could have been mistakenly classified as gapped when using a single point near the center -- which indeed happens in Fig.~\ref{fig:cdos_vs_idos}.

\begin{figure}
    \centering
    \includegraphics[width=0.94\linewidth]{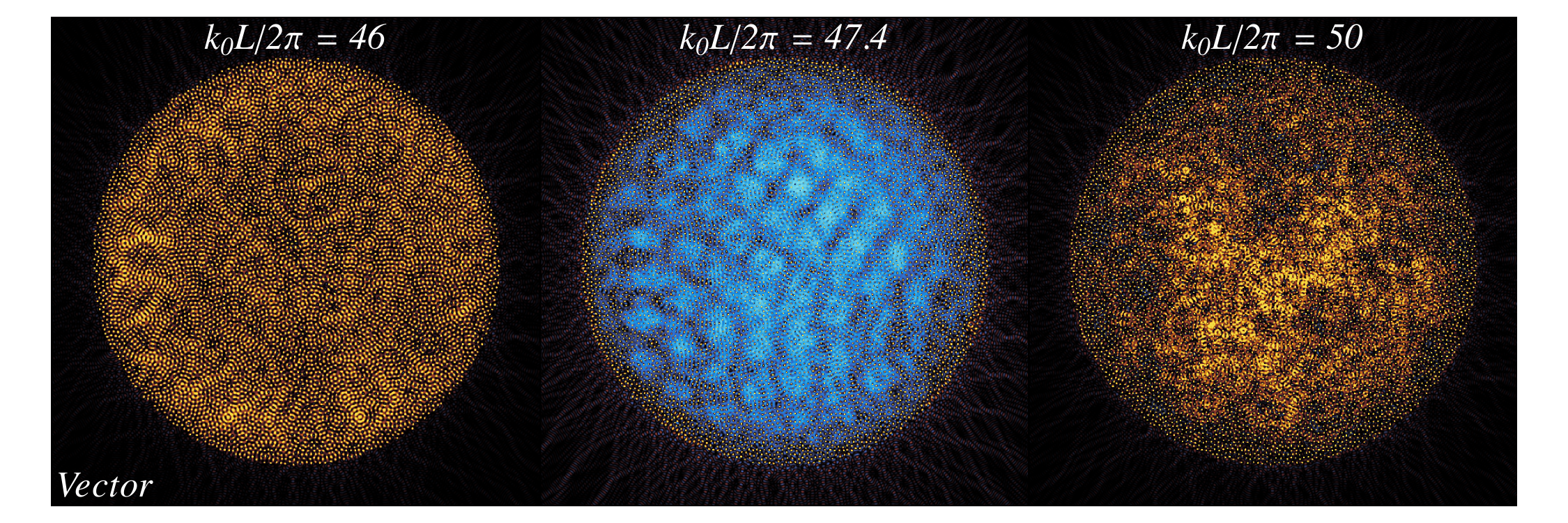}
    \includegraphics[height=0.28\linewidth]{Figures/LDOS_maps/DOSbar.pdf}
    \caption{\textbf{LDOS maps: vector waves in a gyromorph.}
    Maps of LDOS change relative to vacuum, $\delta\varrho(\bm{r}; \omega)$, obtained using a $1001\times1001$ regular grid of pixels in a gyromorph with $N \approx 9000$, $\phi = 0.05$ and $n=3$, with vector waves.
    From left to right, $k_0L /2\pi =46$ (below gap), $47.4$ (in gap), $50$ (above gap)}
    \label{fig:LDOS_Gyro_Vector}
\end{figure}

This is to be compared to the vector-wave version of these maps, shown in Fig.~\ref{fig:LDOS_Gyro_Vector}.
First, notice that scatterers are typically always locally associated with an enhanced LDOS, as expected for a single nanoparticle with vector waves~\cite{Carminati2006}.
This is why it is so important to exclude the immediate vicinity of scatterers from the estimate, as taking into account the near-field effects would shift the average DOS up.
Furthermore, even inside the gap, the map is associated to larger fluctuations than in the scalar case: using a single point is then a particularly bad representation of the bulk.

\begin{figure}
    \centering
    \includegraphics[width=0.48\linewidth]{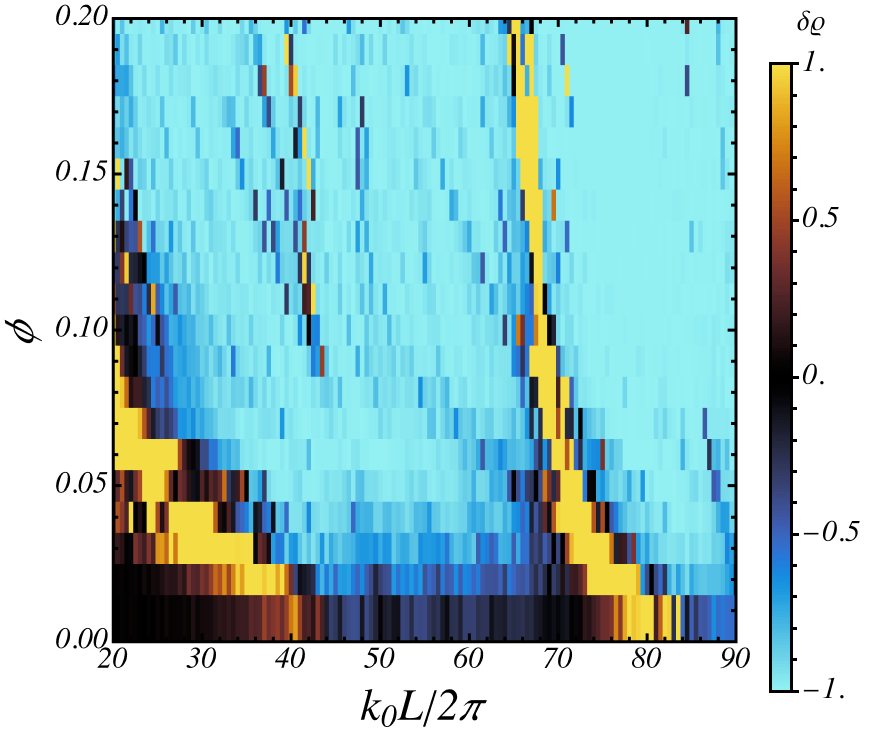}
    \includegraphics[width=0.48\linewidth]{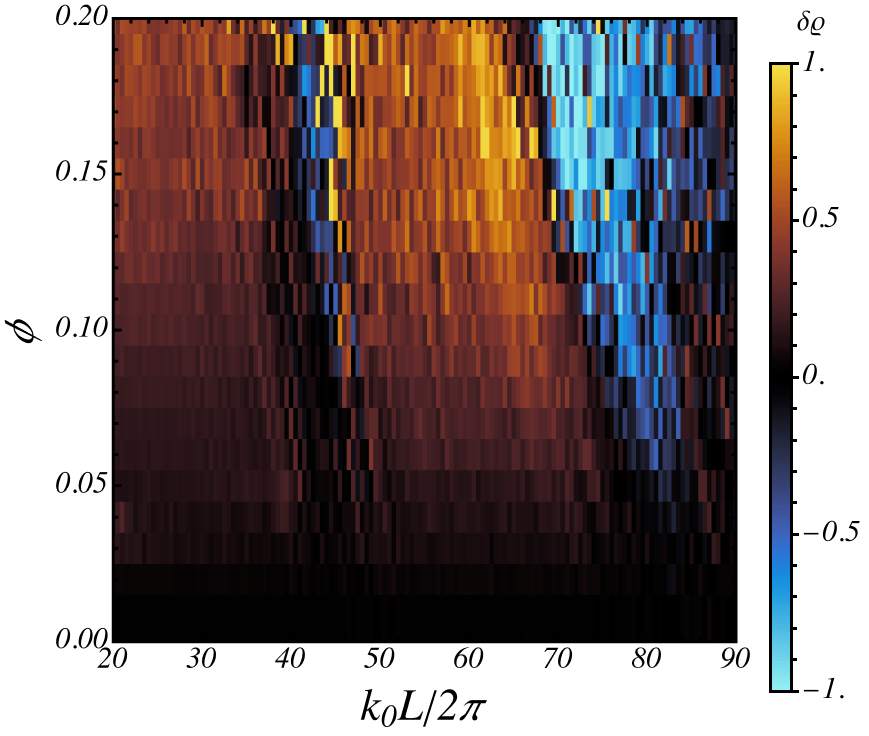}
    \caption{\textbf{CDOS vs IDOS in a quasicrystal.}
    Intensity maps of the relative change of DOS with respect to vacuum as computed from the CDOS definition (left) and IDOS definition (right), against the frequency $k_0 L / 2\pi$ and the filling fraction $\phi$, for a 60-fold quasicrystal of $N \approx 9000$ scatterers of refractive index $n =3$, for vector waves.}
    \label{fig:QC60_CDOS_IDOS}
\end{figure}

Note that this bad representativity of the center point is, as one may expect, worse in structures with a clear radial symmetry around a center.
This is for instance a particularly strong effect in quasicrystals when they are obtained by the standard de Bruijn construction~\cite{deBruijn1981,deBruijn1986}, like the ones illustrated in Fig.~\ref{fig:BandgapStructures}.
This is illustrated for vector waves in Fig.~\ref{fig:QC60_CDOS_IDOS}.
The CDOS map, on the left, displays an impressively wide region of strong depletion compared to all other maps obtained for vector waves.
However, the illusion is lifted when averaging over spatial fluctuations and looking at the IDOS map, on the right.
The region of depleted DOS is then much narrower, and has lost much of its depth.
In particular, the depth of the feature near $k_0 L /2\pi = 50$ is greatly reduced.

\begin{figure}
    \centering
    \includegraphics[width=0.94\linewidth]{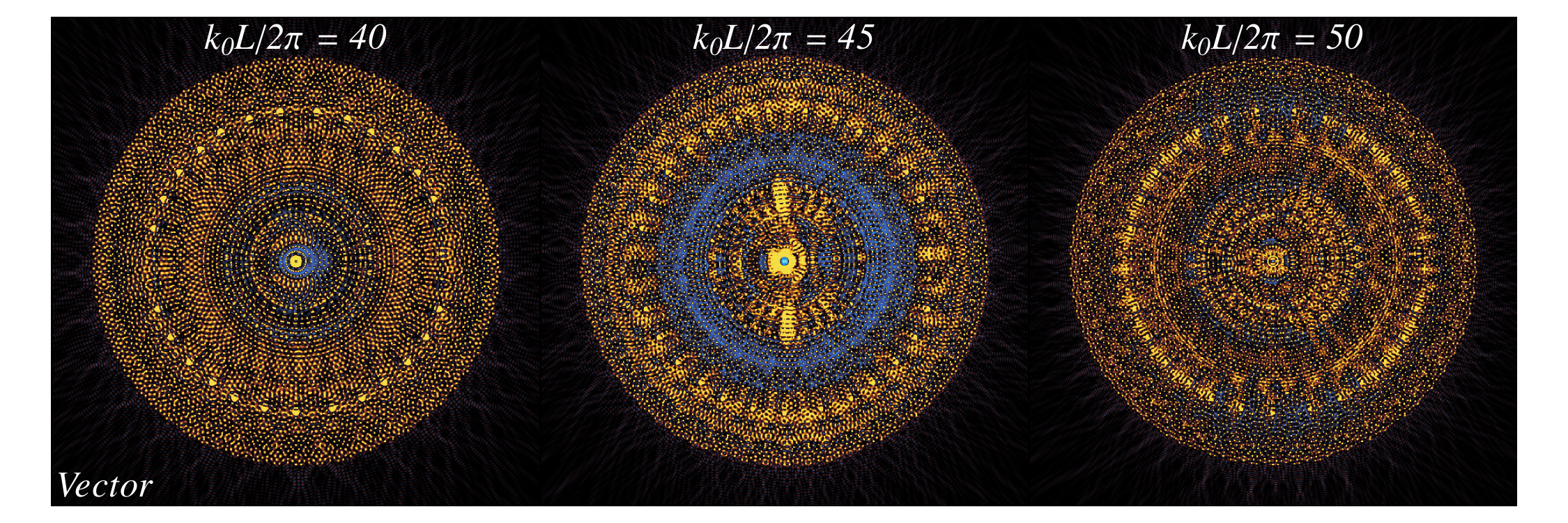}
    \includegraphics[height=0.28\linewidth]{Figures/LDOS_maps/DOSbar.pdf}
    \caption{\textbf{LDOS maps: vector waves in a quasicrystal.}
    Maps of LDOS change relative to vacuum, $\delta\varrho(\bm{r}; \omega)$, obtained using a $1001\times1001$ regular grid of pixels in a 60-fold quasicrystal with $N \approx 9000$, $\phi = 0.05$ and $n=3$, for vector waves.
    From left to right, $k_0L /2\pi =40$, $45$, $50$.}
    \label{fig:QC60_LDOS_maps}
\end{figure}

The origin of this discrepancy is clear when looking at maps of the relative change of LDOS within the quasicrystal.
We show a few examples in Fig.~\ref{fig:QC60_LDOS_maps}.
In all three examples, the central point displays a strong LDOS depletion due to the radial structure around it.
Yet, and in spite of the differences in the precise structure of the LDOS across the system, the overall IDOS is enhanced compared to vacuum.

\begin{figure}
    \centering
    \includegraphics[width=0.48\linewidth]{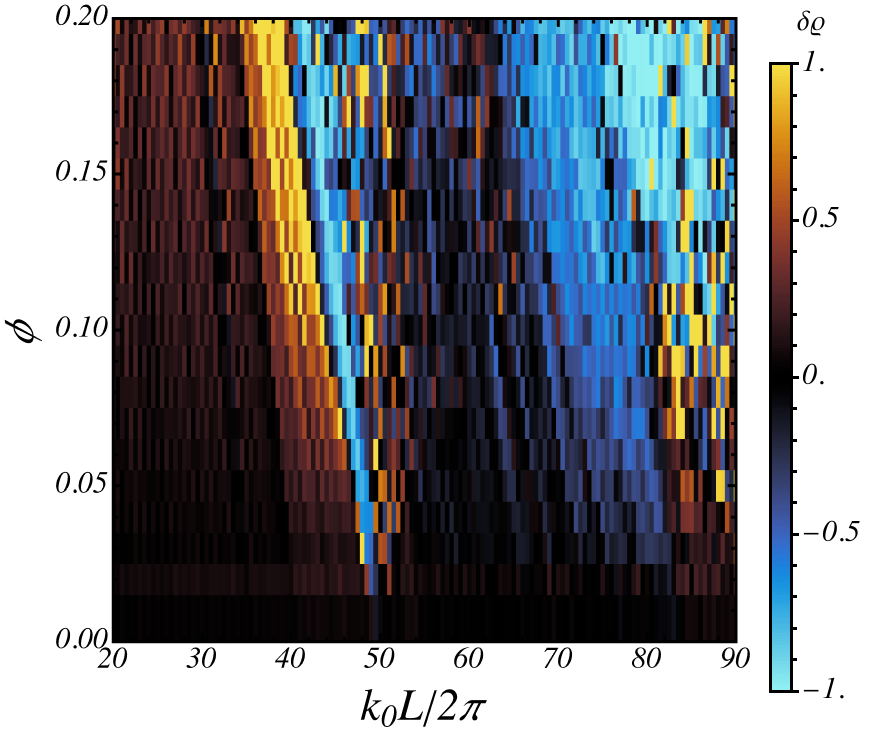}
    \includegraphics[width=0.48\linewidth]{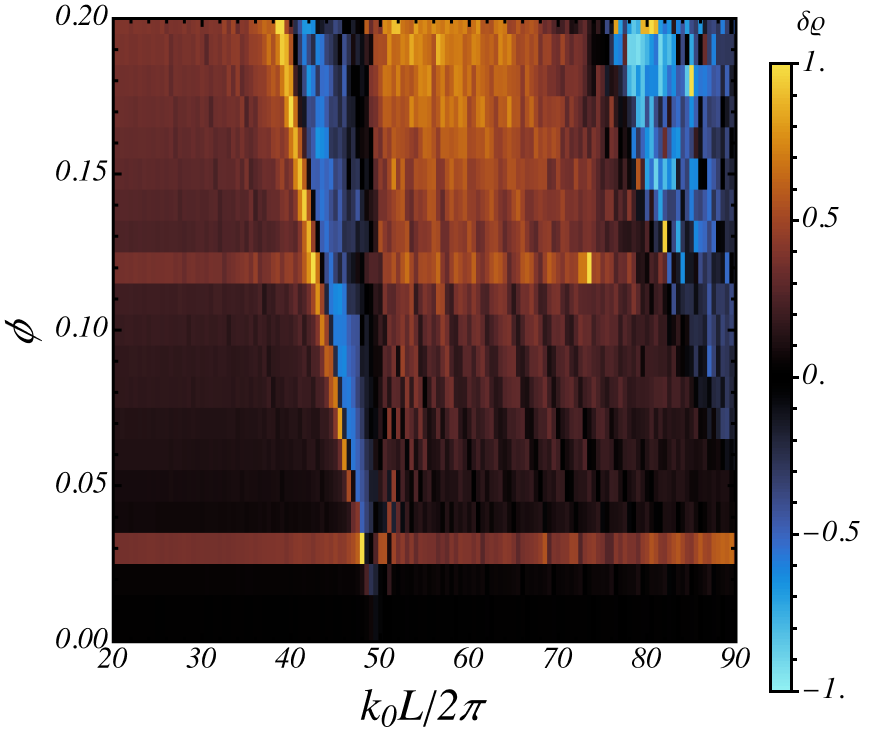}
    
    \caption{\textbf{Single-system CDOS and IDOS sweep.}
    CDOS (left) and IDOS (right) sweeps for a single gyromorph with $N \approx 9000$ and $G = 60$.}
    \label{fig:cdos_sweep_vector}
\end{figure}

\begin{figure}
    \centering
    \includegraphics[height=0.48\linewidth]{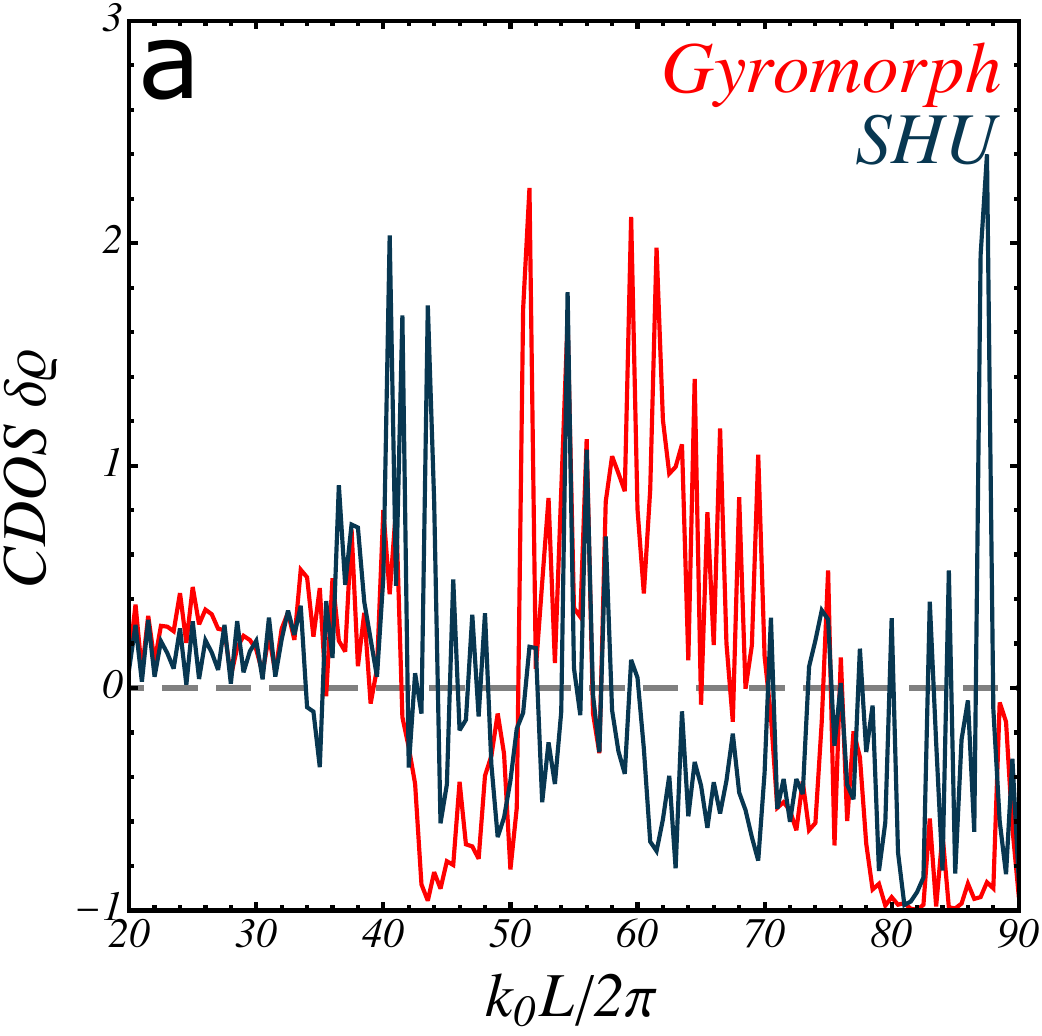}
    \includegraphics[height=0.48\linewidth]{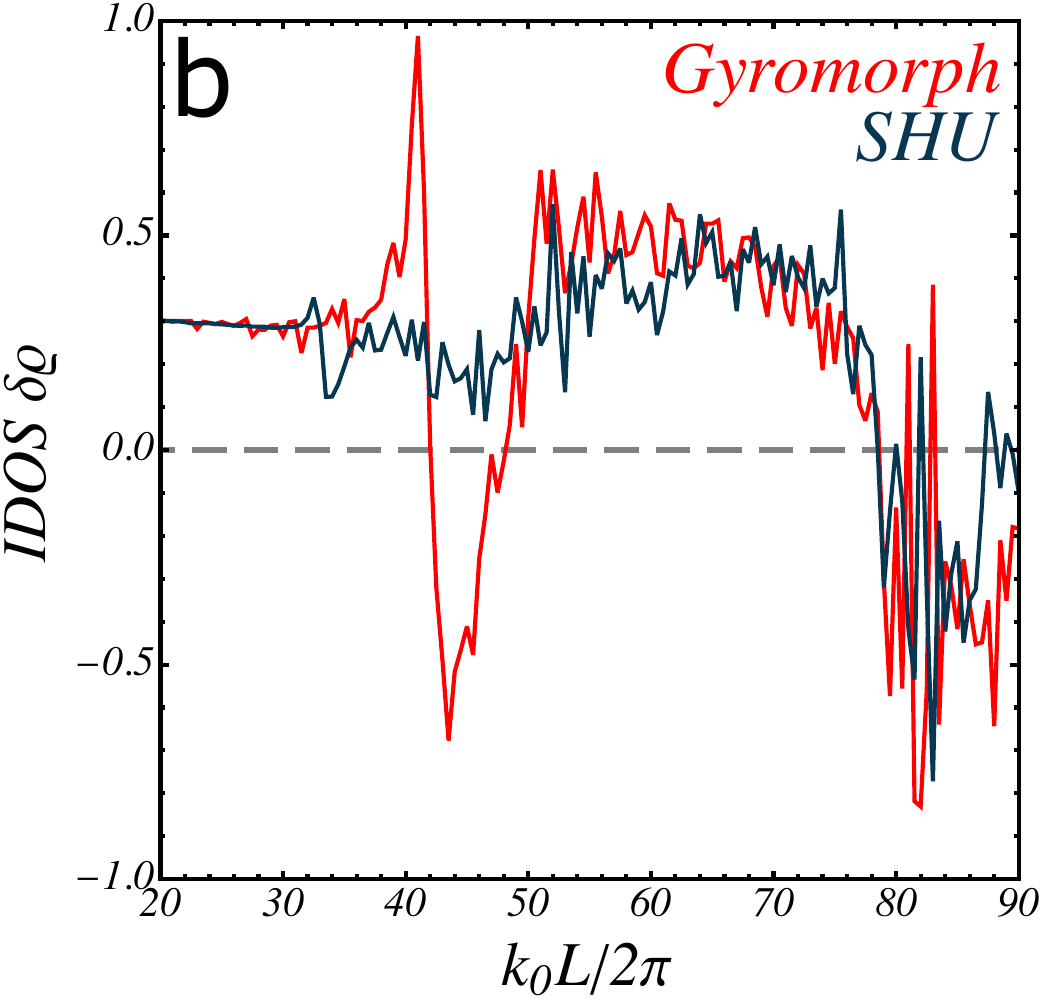}
    \caption{\textbf{CDOS vs IDOS: lineplots.}
    Relative change compared to vacuum of $(a)$ CDOS and $(b)$ IDOS against $k_0 L /2 \pi$ for gyromorphs (red) and SHU systems (blue), for vector waves, at $\phi = 0.15$ and $n = 3$, for the configuration of Fig.~\ref{fig:cdos_sweep_vector}.
    A dashed gray line indicates 0.
    }
    \label{fig:CDOS_IDOS_lineplots}
\end{figure}

Finally, we highlight that inter-sample fluctuations may slightly displace the minimum of the DOS in frequency, leading to less contrasted averaged DOS curves.
To be more concrete, in the vector-wave case, we show in Fig.~\ref{fig:cdos_sweep_vector} the CDOS and IDOS parameter sweeps shown in EM for a single gyromorph.
The CDOS value is locally very close to $-1$, indicating a seemingly much more convincing vector bandgap.
The IDOS map, however, highlights spatial fluctuations and thus seemingly weakens features.
We highlight, however, that these features are weakened regardless of the considered system -- SHU systems also display less deep features in DOS when taking into account spatial fluctuations.
To illustrate this last point quantitatively, in Fig.~\ref{fig:CDOS_IDOS_lineplots}, we show line plots of both the CDOS $(a)$ and IDOS $(b)$ measurements of the relative change of DOS $\delta\varrho$ for the gyromorph of Fig.~\ref{fig:cdos_sweep_vector} and an SHU system, both with $N \approx 9000$ and for $\phi = 0.15$ and $n =3$.
The lineplots highlight that the CDOS of both systems goes close to $\delta\varrho = -1$ near $k_0 L /2\pi = 80$ (the gyromorph dips to lower values), and that only the CDOS of the gyromorph goes close to $\delta\varrho = -1$ near $k_0 L /2\pi = 50$.
However, when averaging over spatial fluctuations in the IDOS measurement, the lowest point of both systems is brought higher up, to values around $\delta\varrho \approx -0.8$ at best.
In other words, the IDOS measurement, by incorporating spatial fluctuations, attenuates DOS depletion in both systems.
Furthermore, the deepest points of both CDOS and IDOS for the gyromorph reach depths comparable to those of the SHU system in what has been shown to be a bandgap, making our measurement in gyromorphs credible as the sign of an underlying bandgap.

\subsection{Hamiltonian DOS}

\begin{figure}
    \centering
    \includegraphics[height=0.45\linewidth]{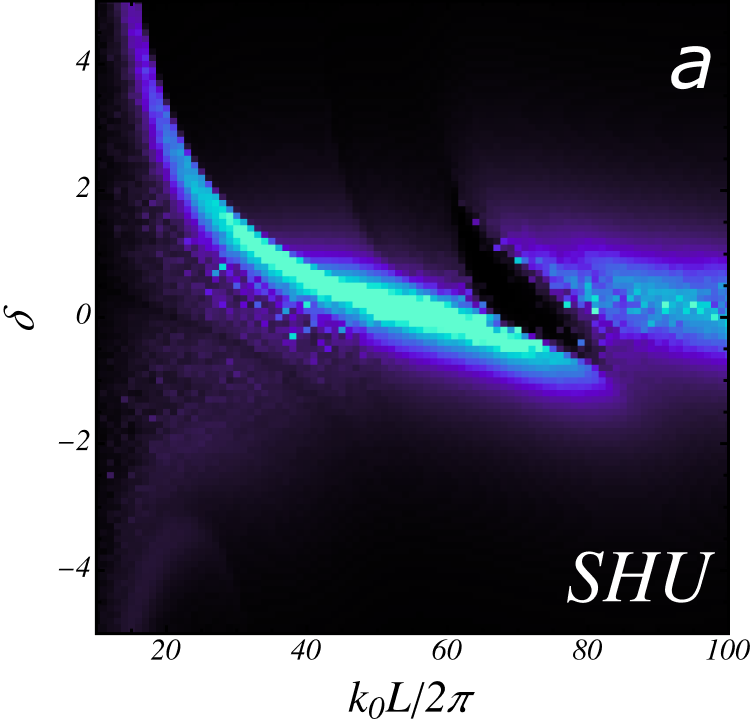}
    \includegraphics[height=0.45\linewidth]{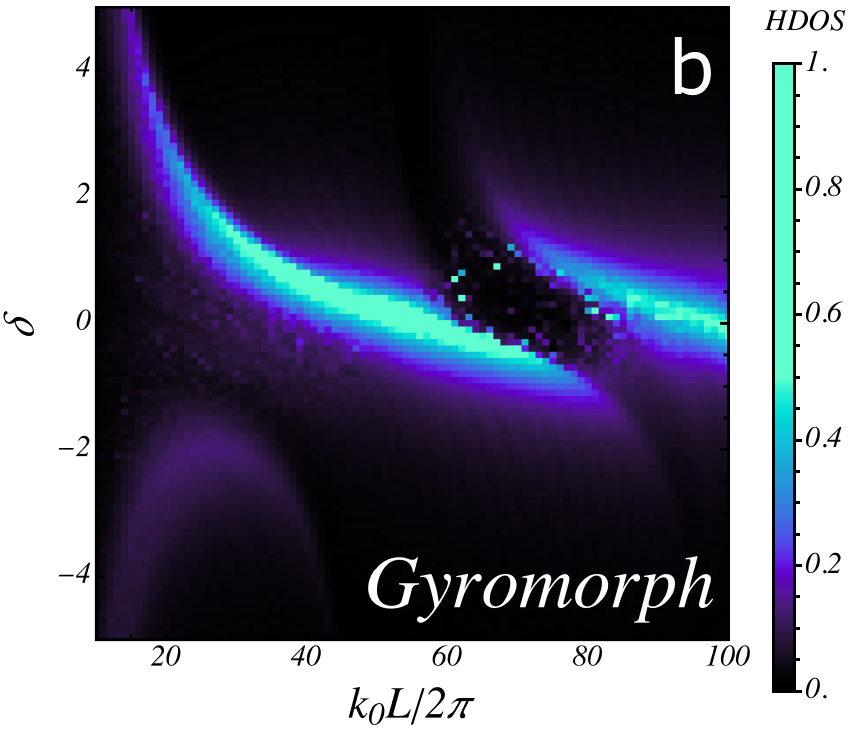} \\
    \includegraphics[width=0.48\linewidth]{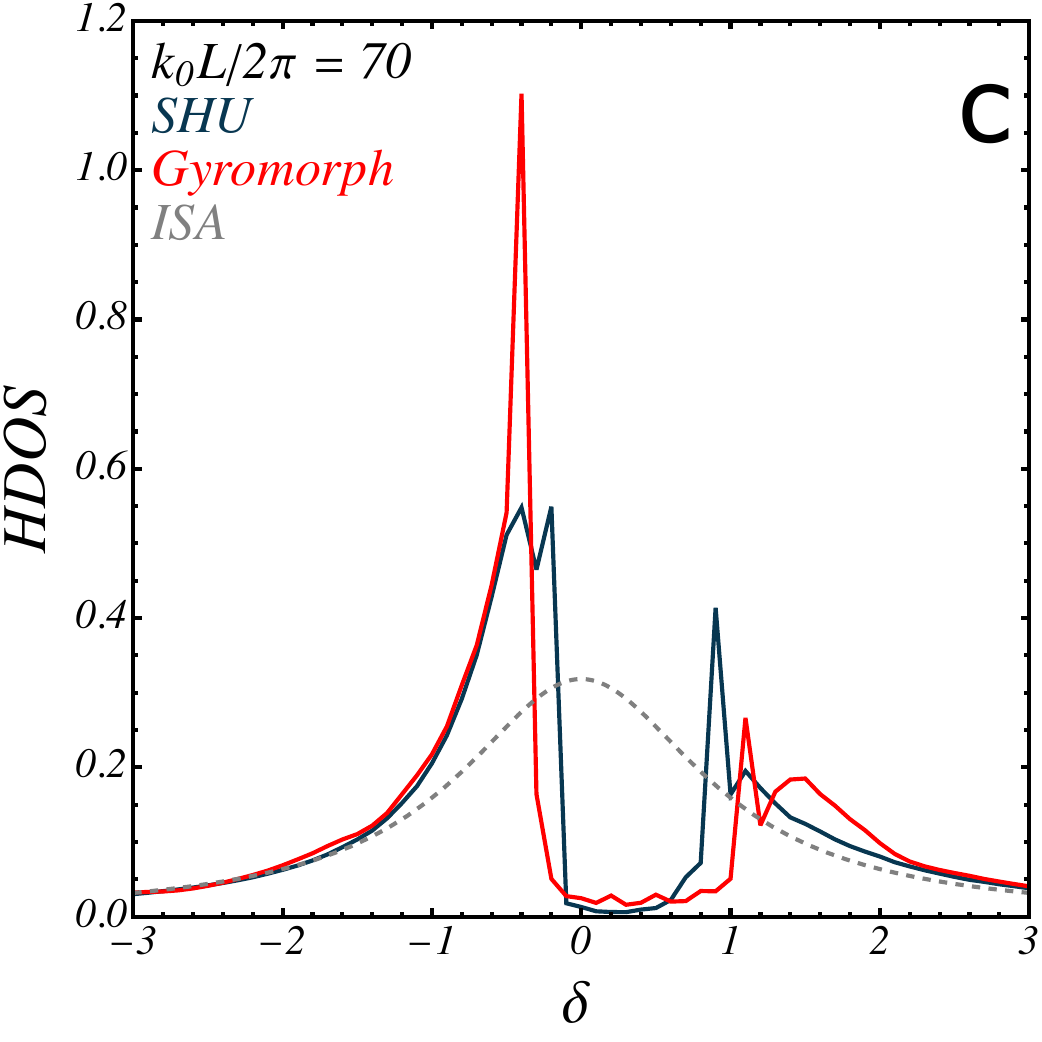}
    \includegraphics[width=0.48\linewidth]{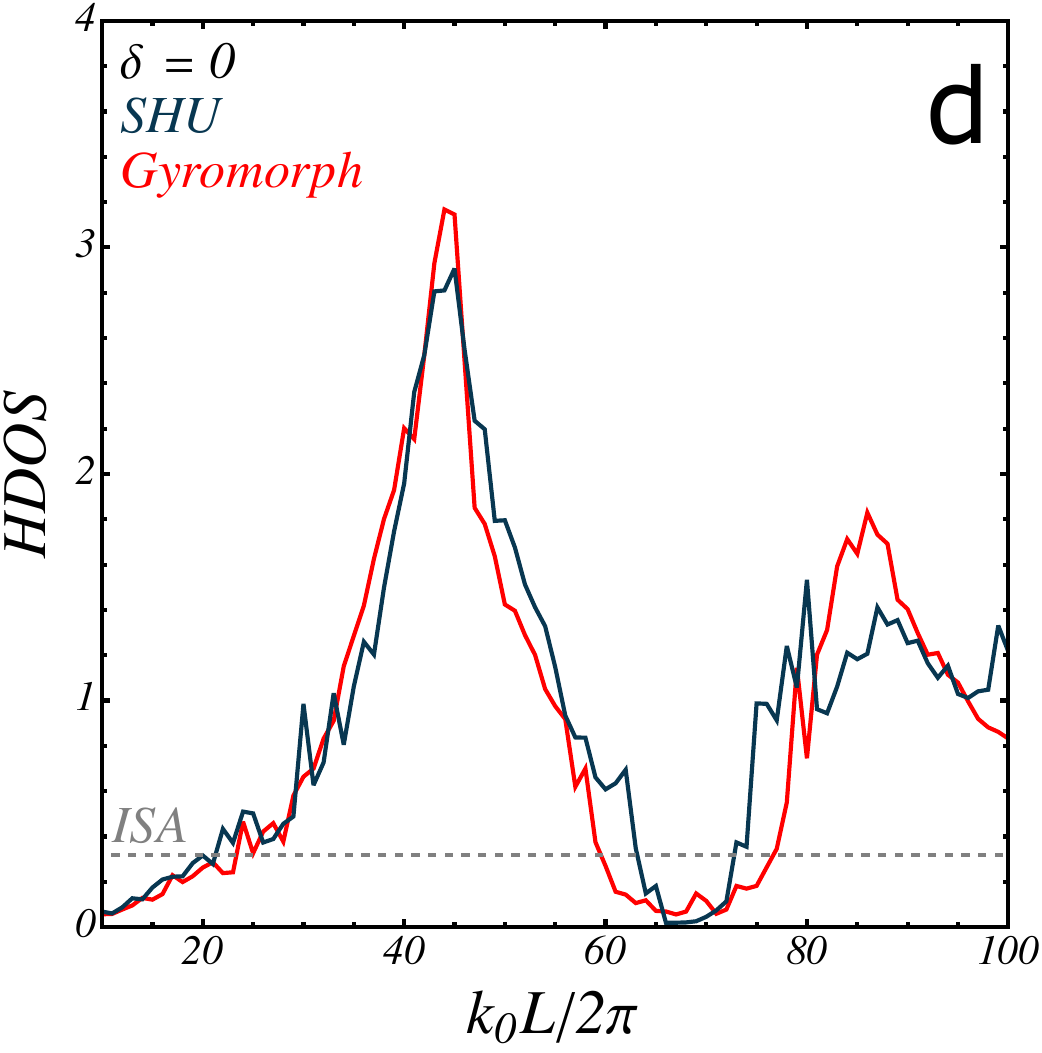}
    \caption{\textbf{Hamiltonian DOS.}
    $(a)$ HDOS for $2d$ vector waves in SHU systems similar to that of Ref.~\cite{Monsarrat2022} and $(b)$ for $G = 60$ gyromorphs.
    Line plots of the HDOS at $(c)$ $k_0 L /2\pi = 70$ and $(d)$ $\delta = 0$ are shown for SHU (dark blue) and gyromorph (red) systems, and compared to the ISA prediction (dashed gray).
    For all panels, $N \approx 7000$ points in a finite disk like those used in transmission measurements, and we use $30$ independent copies of each systems within the average.
    }
    \label{fig:HDOS}
\end{figure}

The LDOS computed elsewhere in this paper is an experimentally accessible quantity (see, \textit{e.g.}, Ref.~\cite{Razo-Lopez2024}) that measures how much power a source current would actually emit at a given location, using optical and geometrical properties of the surrounding medium.
Some works (\textit{e.g.} Ref.~\cite{Monsarrat2022}) however prefer introducing a definition of a DOS that is more agnostic to optical properties.
Its definition relies on the assumption that the system is probed near a resonance of $\alpha_d$, so that it may be approximated as $\alpha_d(\omega) \propto k_0^{-d}/(\delta + i)$ with $\delta$ a dimensionless detuning from the resonance, that in practice indirectly encodes for optical properties.
Using this expression, one may show that finding collective resonances of the system is tantamount to studying the spectrum of an effective Hamiltonian in which the potential energy term is proportional to the Green's tensor $\mathcal{G}_0$~\cite{Monsarrat2022}.
As a result, one may write the density of states of the system in a resolvent form involving the eigenvalues $(\Lambda_1, \ldots,\Lambda_{pN})$ of $\mathcal{G}_0$ with $p$ the dimension of the wave polarization vector ($1$ for scalar waves, $d$ for vector waves).
More concretely, the ``Hamiltonian'' DOS, or HDOS, reads
\begin{align}
    HDOS(k_0,\delta) = \frac{1}{\pi}\text{Im}\left[g(k_0, \delta)\right]
\end{align}
with $g$ the resolvent
\begin{align}
    g(k_0, \delta) = \left\langle\frac{1}{\delta - i -\Lambda_n} \right\rangle,
\end{align}
where the average is over eigenvalues (for translationally invariant systems, since it is performed over the spectrum of an Euclidean Random Matrix, the average can be improved either by taking $N \to \infty$ of averaging over copies at finite $N$).
We show intensity maps of HDOS for vector waves in Fig.~\ref{fig:HDOS}.
Fig.~\ref{fig:HDOS}$(a)$ is a map obtained using SHU systems with $\chi = 0.5$ like in Ref.~\cite{Monsarrat2022} while Fig.~\ref{fig:HDOS}$(b)$ is obtained for $G=60$ gyromorphs.
Note that we here perform the measurement using finite disks of the point patterns with free boundaries, as opposed to periodic squares in Ref.~\cite{Monsarrat2022}.
We show that the HDOS is attenuated to near-zero levels in the vicinity of $k_0L/2\pi = 70$ and $\delta = 0$ in both systems, revealing a bandgap opening.
The depletion for the SHU system is, as expected, similar to that of Ref.~\cite{Monsarrat2022} (down to its precise location: here $N=9000$ in the original periodic square so that the gap is near $k_0 \rho_0^{-1/d} \approx 4.5 $).
By comparison, the gyromorph yields a similar depletion over a comparatively larger range of $\delta$ (optical parameters) and frequencies.
This is confirmed by line plots at a constant $k_0$ in Fig.~\ref{fig:HDOS}$(c)$ and at a constant $\delta$ in Fig.~\ref{fig:HDOS}$(d)$, where we also report the prediction of the independent scattering approximation (ISA) to highlight the effects of correlations.
Across the board, we find that gyromorphs deplete the HDOS about as much as SHU systems but over a wider range of parameters.
In both systems, isolated modes still appear in the suppressed region, likely due to the effects of finite size and free boundaries.
This measurement, which depends less on practical details of the implementation of optical behavior, thus confirms that gyromorphs open bandgaps, like SHU systems.
It however predicts mode depletion around $k_0L/2\pi \approx 80$ -- this depletion is indeed seen across all DOS measurements in both systems (\textit{e.g.} Fig.~\ref{fig:idos_nsweep_0p15}).
The DOS depletion reported in gyromorphs at $k_0L/2\pi \approx 50$ for vector waves is not captured by this approach.
This is hardly surprising because the HDOS is \textit{more} approximate than the DOS measurement of the main text, which is a direct equivalent of experimentally accessible quantities~\cite{Razo-Lopez2024}.
In particular, HDOS assumes near-resonance, and thus neglects the $\omega$ dependence of $\alpha_d$.

\subsection{Effect of noise on optical properties \label{sec:KickedDOS}}

\begin{figure}
    \centering
    \includegraphics[width=0.48\linewidth]{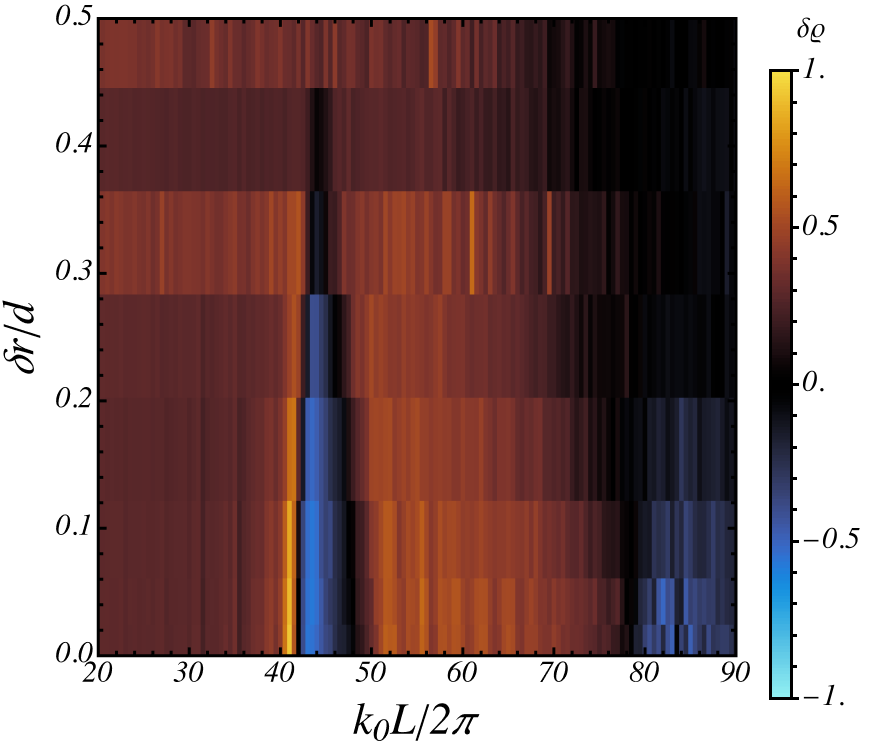}
    \includegraphics[width = 0.48\linewidth]{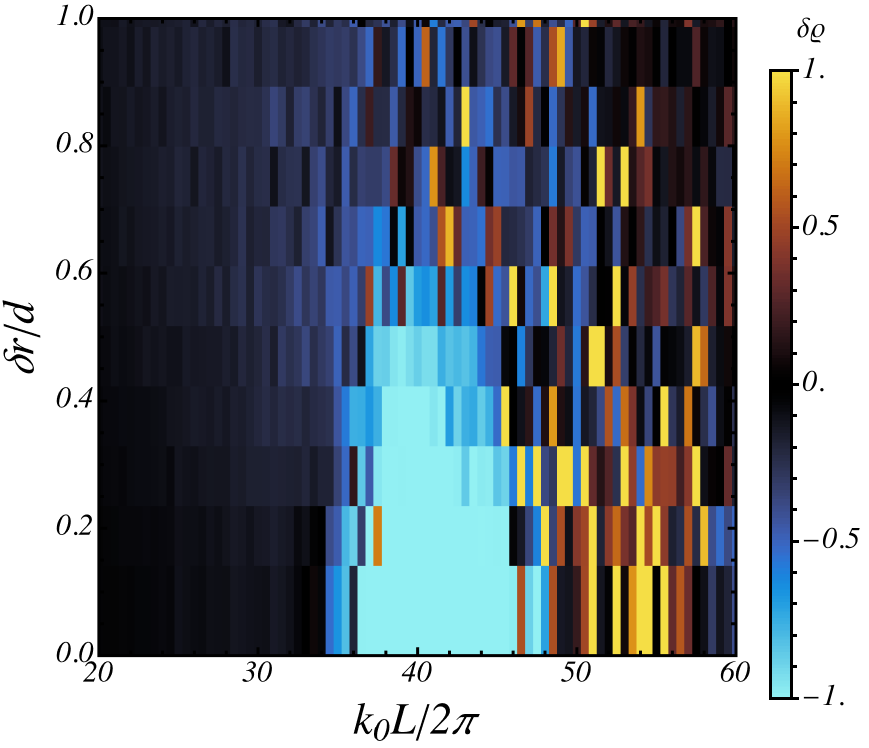}
    \caption{\textbf{Effect of noise on DOS in gyromorphs.}
    Intensity map of the spatially-averaged DOS, averaged over $30$ gyromorphs with $G = 60$, against the rescaled frequency $k_0L/2\pi$ and the ratio $\delta r/d$ between the standard deviations of random kicks applied to scatterers and the typical inter-scatterer distance.}
    \label{fig:KickedDOS}
\end{figure}

As a follow-up to Sec.~\ref{sec:NoiseStability}, we here briefly discuss the effect of random kicks on scatterer positions on the depletion of the DOS in gyromorphs, to model manufacturing errors.
In Fig.~\ref{fig:KickedDOS} we show intensity maps of the DOS computed over $1000$ random points at least $2a$ away from any scatterer and averaged over $30$ systems, when moving scatterers by independent identically distributed Gaussian kicks with mean $0$ and standard deviation $\delta r$ for vector (left) and scalar (right) waves.
The maps are computed for $n=3$ and using $\phi = 0.15$ (vector waves) and $\phi = 0.05$ (scalar waves) so as to make the features easier to read.
We show that the DOS depletion survives kicks reaching up to tens of percent of the typical distance between scatterers $d = 1/\sqrt{\rho}$, which suggests that gyromorphs would retain their properties within reasonable ranges of manufacturing errors on the positions of scatterers.

\section{Coupled Dipoles Method}

\subsection{General Definitions}

In the main text, we rely on the Coupled Dipoles Method (CDM) to describe the optical properties of gyromorphs.
As the paper is aimed at a general audience that may not be familiar with light scattering, we give a pedagogical reminder of the theory and definitions below.
The starting point of the calculation is the $3d$ monochromatic Maxwell-Helmholtz equation for the electric field $\bm{E}$, at pulsation $\omega$ in a non-magnetic heterogeneous system characterized by an isotropic relative dielectric constant field $\varepsilon(\bm{r}; \omega)$~\cite{MorseFeshbachI,Jackson}
\begin{align}
    \bm{\nabla}\times\bm{\nabla}\times \bm{E}(\bm{r}; \omega) - \frac{\omega^2}{c^2} \varepsilon(\bm{r}; \omega) \bm{E}(\bm{r};\omega) = i \mu_0 \omega \bm{j}_{ext}(\bm{r};\omega),
\end{align}
where $c$ is the speed of light in vacuum, $\mu_0$ is the magnetic permeability of vacuum and $\bm{j}_{ext}$ is the externally imposed charge current density, that results in light sources.
Following usual conventions~\cite{Carminati2021}, we define the incident field $\bm{E}_{inc}(\bm{r})$ as the solution of the wave equation for the same sources, but in vacuum,
\begin{align}
    \bm{\nabla}\times\bm{\nabla}\times \bm{E}_{inc}(\bm{r}; \omega) - \frac{\omega^2}{c^2}\bm{E}_{inc}(\bm{r};\omega) = i \mu_0 \omega \bm{j}_{ext}(\bm{r};\omega),\label{eq:incidentE}
\end{align}
as well as the scattered field $\bm{E}_s = \bm{E} - \bm{E}_{inc}$.
The latter obeys the equation
\begin{align}
    \bm{\nabla}\times\bm{\nabla}\times \bm{E}_s(\bm{r}; \omega) - \frac{\omega^2}{c^2} \bm{E}_s(\bm{r};\omega) = \frac{\omega^2}{c^2} \delta\varepsilon(\bm{r}; \omega) \bm{E}(\bm{r};\omega), \label{eq:scatteredE}
\end{align}
in which we introduced the relative dielectric contrast $\delta\varepsilon = \varepsilon - 1$.
To go further, we introduce the \textit{dyadic Green's function}~\cite{Economou2006,Carminati2021}, $\overline{\overline{G}}_0(\bm{r},\bm{r'}; \omega)$, associated to propagation in free space.
This $3\times 3$ rank-$2$ tensor is defined as the solution of the free-space Maxwell-Helmholtz equation when the source is replaced by a Dirac delta in each direction,
\begin{align}
    \bm{\nabla}\times\bm{\nabla}\times \overline{\overline{G}}_0(\bm{r},\bm{r'}; \omega) - \frac{\omega^2}{c^2}\overline{\overline{G}}_0(\bm{r},\bm{r'}; \omega) = \delta(\bm{r} - \bm{r}') \overline{\overline{I}},
\end{align}
with $\overline{\overline{I}}$ the identity tensor.
Defining the Green tensor enables to write any field propagating in vacuum as an integral equation over the source term.
For instance, the incident field defined in Eq.~\ref{eq:incidentE} verifies
\begin{align}
    \bm{E}_{inc}(\bm{r}; \omega) = i \mu_0 \omega \int d^3\bm{r}'\, \overline{\overline{G}}_0(\bm{r},\bm{r'}; \omega) \bm{j}_{ext}(\bm{r}'; \omega).
\end{align}
More importantly, writing the analogue equation for the scattered field, Eq.~\ref{eq:scatteredE}, leads to the Lippman-Schwinger equation~\cite{Carminati2021},
\begin{align}
    \bm{E}(\bm{r}; \omega) = \bm{E}_{inc}(\bm{r}; \omega) + \frac{\omega^2}{c^2} \int d^3\bm{r}'\, \overline{\overline{G}}_0(\bm{r},\bm{r'}; \omega) \delta\varepsilon(\bm{r}'; \omega) \bm{E}(\bm{r}'; \omega).
    \label{eq:Lippman-Schwinger}
\end{align}

In this paper, we study propagation through media composed of $N$ discrete scatterers, each with homogeneous dielectric contrasts $\delta\varepsilon_i$, placed at positions $\bm{r}_i$, in a homogeneous medium that can be assumed to be vacuum, leading to
\begin{align}
    \bm{E}(\bm{r}; \omega) = \bm{E}_{inc}(\bm{r}; \omega) + \frac{\omega^2}{c^2} \sum\limits_{i=1}^N \delta\varepsilon_i(\omega) \int_{V_i} d^3\bm{r}'\, \overline{\overline{G}}_0(\bm{r},\bm{r'}; \omega)  \bm{E}(\bm{r}'; \omega),
\end{align}
with $V_i$ the volume of scatterer $i$.
Up until this point, all equations were exact.
We henceforth introduce a version of the discrete dipole approximation (DDA)~\cite{Mishchenko2000}: we assume that each scatterer is smaller than a wavelength, $\omega a /(2 \pi c) \ll 1$ with $a$ the typical size of scatterers.
In this regime, the field component at $\omega$ is well approximated by the field at the center of the scatterer, so that
\begin{align}
    \bm{E}(\bm{r}; \omega) \approx \bm{E}_{inc}(\bm{r}; \omega) + \frac{\omega^2}{c^2}  \sum\limits_{i=1}^N \left[ \int_{V_i} d^3\bm{r}'\, \overline{\overline{G}}_0(\bm{r},\bm{r'}; \omega)\right] \delta\varepsilon_i(\omega)  \bm{E}(\bm{r}_i; \omega)  .
\end{align}
This equation is useful in two scenarios~\cite{Lax1951,Lax1952,Leseur2014,Leseur2016,Carminati2021}.
First, it can be used to evaluate the field outside of scatterers, which only requires the knowledge of the incident field and of the field at every scatterer.
In that case, the expression above can be further approximated by taking only the leading contribution of the integral in the limit of small scatterers,
\begin{align}
    \bm{E}(\bm{r}; \omega) \approx \bm{E}_{inc}(\bm{r}; \omega) + \frac{\omega^2}{c^2} \sum\limits_{i=1}^N V_i \delta\varepsilon_i(\omega) \overline{\overline{G}}_0(\bm{r},\bm{r_i}; \omega) \bm{E}(\bm{r}_i; \omega). \label{eq:outsideE}
\end{align}
Second, the same equation can be applied at the center of a scatterer, yielding a set of equations of the form
\begin{align}
    \bm{E}(\bm{r}_j; \omega) \approx \bm{E}_{inc}(\bm{r_j}; \omega) + \frac{\omega^2}{c^2} \sum\limits_{j\neq i}^N V_i \delta\varepsilon_i(\omega) \overline{\overline{G}}_0(\bm{r}_j,\bm{r_i}; \omega) \bm{E}(\bm{r}_i; \omega) + \frac{\omega^2}{c^2} \delta\varepsilon_j(\omega) \left[\int_{V_j} d^3\bm{r}'\, \overline{\overline{G}}_0(\bm{r_j},\bm{r'}; \omega)\right] \bm{E}(\bm{r}_j; \omega), \label{eq:linearsystem}
\end{align}
where the last term, that encodes a self-induction effect at scatterer $j$, has to be computed differently from other terms in the sum due to the singularity of the Green's function of free space at $\bm{r} = \bm{r}'$.
Once this integral is computed, one finds a system of $N$ linear equations between the fields at the centers of scatterers, which can be solved frequency by frequency for any choice of incident field.
For future convenience, we introduce $k_0 = \omega/c$ the wave number in free space, $\alpha_i(\omega) = V_i \delta\epsilon_i(\omega)$ the bare polarizability of scatterers, $\mathcal{E}(\left\{\bm{r}_i\right\}; \omega) = (\bm{E}(\bm{r}_1;\omega), \ldots, \bm{E}(\bm{r}_N;\omega))$ the vector of electric fields at scatterers, $\mathcal{E}_{inc}(\left\{\bm{r}_i\right\}; \omega)$ the corresponding vector of incident fields, and $\mathcal{M}(\left\{\bm{r}_{ij}\right\}; \omega)$ the matrix that verifies
\begin{align}
    \mathcal{M}(\left\{\bm{r}_{ij}\right\}; \omega)\cdot\mathcal{E}(\left\{\bm{r}_j\right\}; \omega) = \mathcal{E}_{inc}(\left\{\bm{r}_i\right\}; \omega),\label{eq:M_definition}
\end{align}
with $\cdot$ indicating a matrix multiplication.
Notice that in our notations, $\mathcal{M}(\left\{\bm{r}_{ij}\right\}; \omega)$ is an $N\times N$ matrix of $3\times 3$ tensors, $\mathcal{M}_{ij} \equiv \overline{\overline{M}}(\bm{r}_i, \bm{r_j}; \omega)$.
In the simple cases of identical scatterers, $\alpha_i(\omega) = \alpha(\omega)$, the matrix above can be decomposed by introducing the \textit{Green matrix}~\cite{DalNegro2016,Skipetrov2020} $\mathcal{G}$, which verifies
\begin{align}
    \mathcal{M}(\left\{\bm{r}_{ij}\right\}; \omega) = \mathcal{I} - k_0^2 \alpha(\omega)\mathcal{G}(\left\{\bm{r}_{ij}\right\}; \omega), \label{eq:Mmatrix}
\end{align}
with $\mathcal{I}$ a diagonal $N\times N$ matrix of $3\times 3$ identity tensors.
With these notations, it clearly appears that solving the linear system in Eq.~\ref{eq:linearsystem} is tantamount to computing the inverse $\mathcal{W} \equiv \mathcal{M}^{-1}$, a task that can be achieved numerically by encoding it as a $3N\times 3N$ matrix.
Through Eq.~\ref{eq:outsideE}, this solution is enough to reconstruct the full electric field in all of space.
Note that this approach is not strictly limited to small scatterers: one can approximate any large scatterer by a collection of voxels with a size smaller than the considered lengthscale, and solve the system above~\cite{Mishchenko2000}.

Let us now focus on self-interaction.
Consider a single small scatterer centered at $\bm{0}$ in an incident field, 
\begin{align}
    \bm{E}(\bm{0}; \omega) = \bm{E}_{inc}(\bm{0}; \omega) + \frac{\omega^2}{c^2} \delta\varepsilon(\omega)\overline{\overline{S}}(\omega) \bm{E}(\bm{0}; \omega), \label{eq:singlescatterer}
\end{align}
where $\overline{\overline{S}}$ denotes the self-interaction term,
\begin{align}
    \overline{\overline{S}} \equiv \left[\int_{V} d^3\bm{r}'\, \overline{\overline{G}}_0(\bm{r_j},\bm{r'}; \omega)\right]
\end{align}
For scalar dielectric constants such as the ones considered here (\textit{i.e.} non-birefringent media), the self-interaction is a diagonal tensor $\overline{\overline{S}} = S \overline{\overline{I}}$.
One may write the value of the polarization field at the center of the particle, defined as $\alpha_0$ multiplied by the total field there, and express it as a response to the incident field instead,
\begin{align}
    \bm{P}(\bm{0}; \omega ) \equiv \alpha_0(\omega) \bm{E}(\bm{0}; \omega) = \alpha_{d}(\omega) \bm{E}_{inc}(\bm{0}; \omega) \label{eq:dressed_bare_alpha}
\end{align}
with $\alpha_d$ a \textit{dressed} polarizability that accounts for self-interaction.
Using the diagonal nature of $\overline{\overline{S}}$, this yields an expression for the dressed polarizability,
\begin{align}
    \alpha_d(\omega) = \frac{\alpha_0(\omega)}{1 - k_0^2 \alpha_0(\omega) S / V}. \label{eq:generic_dressed}
\end{align}
This polarizability describes the true response of the scatterer to an incident field, which takes into account its finite size and leads to resonant behaviour.
More importantly, neglecting self-interaction, which is equivalent to considering $\alpha_d = \alpha_0$, leads to violations of the optical theorem, which encodes energy conservation at scattering events~\cite{Carminati2021}.
In some works~\cite{Pierrat2010,Leseur2014,Leseur2016,Leseur2016b,Skipetrov2014,Monsarrat2022}, the self-interaction is not written explicitly into the linear problem that is being solved.
It is because the linear problem can be rewritten in terms of dressed polarizabilities (\textit{i.e.} the vertices of the diagrams in the Born expansion are renormalized), 
\begin{align}
    \bm{E}(\bm{r}_j; \omega) \approx \bm{E}_{inc}(\bm{r_j}; \omega) + k_0^2 \sum\limits_{j\neq i}^N  \alpha_{d,i}(\omega)\overline{\overline{G}}_0(\bm{r}_j,\bm{r_i}; \omega) \bm{E}(\bm{r}_i; \omega), \label{eq:linearsystem_renorm}
\end{align}
so that the matrix in Eq.~\ref{eq:Mmatrix} is replaced by
\begin{align}
    \mathcal{M}_d(\left\{\bm{r}_{ij}\right\}; \omega) = \mathcal{I} - k_0^2 \alpha_d(\omega)\mathcal{G}_d(\left\{\bm{r}_{ij}\right\}; \omega), \label{eq:dressed_system}
\end{align}
with $\mathcal{G}_d^{iiab} = 0$, for all particle indices $i$ and coordinate pairs $a,b \in \{x,y,z \}^2$.
As noted in the End Matter, this formulation is completely equivalent as far as solving the Coupled Dipoles system is concerned.
Concretely, at any scatterer position $\bm{r}_s$, one may introduce the auxiliary field $\bm{E}_{\text{aux}}$ such that 
\begin{align}
    \alpha_d (\omega) \bm{E}_{\text{aux}}(\bm{r}_s) = \alpha_0 (\omega) \bm{E}(\bm{r}_s), \label{eq:AuxiliaryField}
\end{align}
then solve the linear system
\begin{align}
    \mathcal{M}_d(\left\{\bm{r}_{ij}\right\}; \omega)\cdot\mathcal{E}_\text{aux}(\left\{\bm{r}_j\right\}; \omega) = \mathcal{E}_{inc}(\left\{\bm{r}_i\right\}; \omega),\label{eq:Linear_System_Tmatrix}
\end{align}
where the dressed linear operator $\mathcal{M}^d$ can be written explicitly as
\begin{align}
\mathcal{M}_{d,ijab} &= 
\begin{cases}
\delta_{ab}     & \text{if } i = j \\
- k_0^2 \alpha_d(\omega) \hat{\bm{e}}_a\cdot\overline{\overline{G}}_0(\bm{r_i},\bm{r}_j; \omega)\cdot \hat{\bm{e}}_b & \text{otherwise}.
\end{cases} 
\label{eq:Mdef_Tmatrix}
\end{align}
One may then either solve the coupled dipoles equation for $\mathcal{E}$ using Eq.~\ref{eq:M_definition}, or for the auxiliary field $\mathcal{E}_{\text{aux}}$ using Eq.~\ref{eq:Linear_System_Tmatrix}.
The two solutions are, by definition, linked through the proportionality relation introduced in Eq.~\ref{eq:AuxiliaryField}.
Thus, the full field at any position outside of scatterers can be obtained by propagating the field from scatterers to any point, using either solution
\begin{align}
    \bm{E}(\bm{r}; \omega) &= \bm{E}_{inc}(\bm{r}; \omega) + k_0^2 \alpha_0(\omega)\mathcal{G}_0(\bm{r},\{\bm{r}_i\}; \omega) \cdot \mathcal{E}(\{\bm{r}_i\}; \omega), \label{eq:outsideEbis}  \\
    = &\bm{E}_{inc}(\bm{r}; \omega) + k_0^2 \alpha_d(\omega)\mathcal{G}_0(\bm{r},\{\bm{r}_i\}; \omega) \cdot \mathcal{E}_{\text{aux}}(\{\bm{r}_i\}; \omega), 
\end{align}
so that it is \textit{completely equivalent} to solve either version of the problem.

It is customary to characterize structured dielectric media by a \textit{density of states} (DOS) for the medium, which quantifies the number of electromagnetic modes that are allowed to propagate through it at a given frequency~\cite{Florescu2009,Man2013,Tsitrin2015,Froufe-Perez2017}.
The quantity that is easily accessible when using a Green's function formalism is the \textit{local} density of states (LDOS), $\varrho(\bm{r}, \omega)$ at a given position $\bm{r}$ and around pulsation $\omega$.
While its expression is not original to this paper, past works typically resort to a different set of approximations than us to model cold atomic vapors rather than soft materials~\cite{Economou2006,Prosentsov2007,Caze2013,Leseur2014,Skipetrov2020,Carminati2021,Monsarrat2022}, so that we here reproduce our computation to avoid possible confusions.
The starting point of the calculation is that the local density of states of a lossless material placed in a cavity can be written as~\cite{Carminati2021} 
\begin{align}
    \varrho(\bm{r}, \omega) = \frac{2 \omega}{\pi c^2} \textrm{Im}\left[\textrm{Tr}_3 \overline{\overline{G}}(\bm{r},\bm{r}; \omega)\right], \label{eq:LDOS_def}
\end{align}
where $\textrm{Im}$ denotes the imaginary part, $\textrm{Tr}_3$ the trace over the 3 diagonal elements of a rank-2 tensor, and $\overline{\overline{G}}$ is the Green's function of the medium, defined through
\begin{align}
    \bm{\nabla}\times\bm{\nabla}\times \overline{\overline{G}}(\bm{r},\bm{r'}; \omega) - \frac{\omega^2}{c^2} \varepsilon(\bm{r},\omega)\overline{\overline{G}}(\bm{r},\bm{r'}; \omega) = \delta(\bm{r} - \bm{r}') \overline{\overline{I}}.
\end{align}
This definition ensures that the \textit{full} field obeys the equation
\begin{align}
    \bm{E}(\bm{r}; \omega) = i \mu_0 \omega \int d^3\bm{r}'\, \overline{\overline{G}}(\bm{r},\bm{r'}; \omega) \bm{j}_{ext}(\bm{r}'; \omega). \label{eq:fullE}
\end{align}
In particular, one can choose, a point-electric dipole source with moment $\bm{p}$ placed at $\bm{r}_d$, corresponding to 
\begin{align}
\bm{j}_{ext}(\bm{r};\omega) \equiv - i \omega \bm{p} \delta(\bm{r} - \bm{r}_{d}). \label{eq:pointdipole_current}
\end{align}
For this specific choice of source (but without any loss of generality), using Eqs.~\ref{eq:incidentE}, ~\ref{eq:fullE}, ~\ref{eq:outsideE}, and $\mu_0 \varepsilon_0 c^2 = 1$, one can write $\overline{\overline{G}}$ when the observation point $\bm{r}$ is not inside a scatterer as
\begin{align}
    \overline{\overline{G}}(\bm{r},\bm{r}'; \omega) = \overline{\overline{G}}_0(\bm{r},\bm{r}'; \omega) + k_0^2 \alpha(\omega)\sum\limits_{j,l=1}^{N} \overline{\overline{G}}_0(\bm{r},\bm{r}_j; \omega) \overline{\overline{W}}_{jl} \overline{\overline{G}}_0(\bm{r}_l,\bm{r}'; \omega), \label{eq:explicit_Dyson}
\end{align}
where we introduced $\overline{\overline{W}}_{jl}$ the $jl$-th ($d\times d$)-tensor element of matrix $\mathcal{W} \equiv \mathcal{M}^{-1}$.
The second term of this equation, which encodes the effect of scatterers on propagation, assumes the shape of a matrix product: defining the $1\times N$ matrix $\mathcal{G}_0^{(1)}(\bm{r};\left\{\bm{r}_{i}\right\}; \omega) = (\overline{\overline{G}}_0(\bm{r},\bm{r}_1; \omega), \ldots, \overline{\overline{G}}_0(\bm{r},\bm{r}_N; \omega))$, and the transpose operator ${^t}$, one can write
\begin{align}
    \overline{\overline{G}}(\bm{r},\bm{r}'; \omega) = \overline{\overline{G}}_0(\bm{r},\bm{r}'; \omega) + k_0^2 \alpha(\omega) \, \mathcal{G}_0^{(1)}(\bm{r}; \left\{\bm{r}_{i}\right\}; \omega)\cdot\mathcal{W}(\left\{\bm{r}_{ij}\right\};\omega)\cdot {^t}\mathcal{G}_0^{(1)}(\bm{r'}; \left\{\bm{r}_{j}\right\}; \omega).
\end{align}
As a result, the LDOS of a system of small discrete scatterers, evaluated outside of scatterers, reads
\begin{align}
    \varrho(\bm{r}, \omega) = \varrho_0(\omega) + \frac{2 \omega^3}{\pi c^4} \textrm{Im}\left[ \alpha(\omega) \textrm{Tr}_3 \left[\mathcal{G}_0^{(1)}(\bm{r}; \left\{\bm{r}_{i}\right\}; \omega)\cdot\mathcal{W}(\left\{\bm{r}_{ij}\right\};\omega)\cdot {^t}\mathcal{G}_0^{(1)}(\bm{r}; \left\{\bm{r}_{j}\right\}; \omega) \right]\right], \label{eq:LDOS_explicit}
\end{align}
where $\varrho_0(\omega)$ is the LDOS of vacuum, which does not depend on location~\cite{Prosentsov2007,Carminati2021}.

One may also define a density of state for the medium as a whole.
A straightforward definition from the LDOS is~\cite{Economou2006}
\begin{align}
\varrho_{med}(\omega) = \int_{V_{med}} d^3\bm{r} \varrho(\bm{r}, \omega),
\end{align}
with $V_{med}$ the volume of the region that contains scatterers.
In practice, this integral can be evaluated numerically: by defining $M$ measurement points at positions $\left\{\bm{r}^{m}_n\right\}$, uniformly spaced within the medium but distinct from the positions of scatterers, and $\mathcal{G}_0^{(M)}(\left\{\bm{r}^{m}_n\right\}; \left\{\bm{r}_{i}\right\}; \omega)$ the $M\times N$ matrix of propagators linking the measurement points to the scatterers, one finds
\begin{align}
    \varrho_{med}(\omega) \approx \varrho_0(\omega) V_{med} +  \frac{2 \omega^3 V_{med}}{M \pi c^4} \textrm{Im}\left[ \alpha(\omega) \textrm{Tr}_{3M} \left[\mathcal{G}_0^{(M)}(\left\{\bm{r}^{m}_n\right\}; \left\{\bm{r}_{i}\right\}; \omega)\cdot\mathcal{W}(\left\{\bm{r}_{ij}\right\};\omega)\cdot {^t}\mathcal{G}_0^{(M)}(\left\{\bm{r}^{m}_n\right\}; \left\{\bm{r}_{j}\right\}; \omega) \right]\right],
\end{align}
where the trace is now computed over all $3M$ diagonal elements of the matrix product.
In practice, we compute the relative corrections to the density of states of vacuum, $\delta \varrho(\bm{r}, \omega) = (\varrho(\bm{r}, \omega) - \varrho_0( \omega))/\varrho_0( \omega)$ and $\delta \varrho_{med}( \omega) = (\varrho_{med}( \omega) - \varrho_0( \omega) V_{med})/(\varrho_0( \omega) V_{med})$, which are numbers in $\left[-1;\infty\right[$, the sign of which indicate depletion or enrichment in number of states.
Note that it is common~\cite{Carminati2021} to define the generalized Purcell factor $F(\omega, \bm{r})$ such that $\delta \varrho(\bm{r}, \omega) = F(\omega,\bm{r}) - 1$.

\subsection{Specific geometries}

The previous subsection introduced every quantity in a way that did not specify the geometry of the problem, encoded by propagators, explicitly.
We here remind the expressions of every quantity of interest in the 4 cases considered in the paper: $2d$ vector, $2d$ scalar, $3d$ vector, and $3d$ scalar waves, all in free space.

\subsubsection{Rayleigh scatterers for vector waves in $3d$ space}

The most usual context for light propagation is that of free $3d$ space with finite scatterers.
In this case, the Green's function associated to free propagation can be written as~\cite{MorseFeshbachI,Economou2006,Carminati2021}
\begin{align}
    \overline{\overline{G}}_0(\bm{r},\bm{r}'; \omega) = PV\left(\frac{e^{i k_0 R}}{4 \pi R}\left[ \overline{\overline{I}} - \hat{\bm{R}}\otimes\hat{\bm{R}} - \left(\frac{1}{i k_0 R} + \frac{1}{k_0^2 R^2}\right) \left( \overline{\overline{I}} - 3 \hat{\bm{R}}\otimes\hat{\bm{R}}\right)\right]\right) - \frac{\delta(\bm{R})}{3 k_0^2}\overline{\overline{I}}, \label{eq:G03d}
\end{align}
where $PV$ indicates a Cauchy principal value, $\bm{R} \equiv \bm{r} - \bm{r}' = R \hat{\bm{R}}$, $k_0 = \omega/c$, $\overline{\overline{I}}$ is the identity tensor, and $\otimes$ is the outer product.
Note that the Green's function should be understood as a distribution, and consists of a regular part that diverges as $R\to 0$ together with a singular part at $\bm{R} = \bm{0}$.

From this expression, one may write the expression of the self-interaction for small spherical scatterers.
The relevant integral to compute, following Eq.~\ref{eq:linearsystem}, is
\begin{align}
    \overline{\overline{S}}_{3d}(\omega) = \int_{\mathcal{B}(\bm{0},a)} d^3\bm{r}'\, \overline{\overline{G}}_0(\bm{0},\bm{r'}; \omega),
\end{align}
where the integral runs over the volume of the scatterer, the ball ${\mathcal{B}(\bm{0},a)}$ centered at $\bm{0}$ and with radius $a$.
This integral can be rewritten in spherical coordinates as
\begin{align}
    \overline{\overline{S}}_{3d}(\omega) = \int\limits_{0}^{a} dr' \int\limits_{-\pi}^{\pi} d\varphi  \int\limits_{0}^{\pi} d\theta\,  r'^2 \sin\theta \overline{\overline{G}}_0(\bm{0},\bm{r'}; \omega). \label{eq:S3dDef}
\end{align}
To compute this integral, it is useful to notice that
\begin{align}
    \int\limits_{-\pi}^{\pi} d\varphi  \int\limits_{0}^{\pi} d\theta\, \sin\theta \hat{\bm{R}}\otimes\hat{\bm{R}} = \frac{4 \pi}{3} \overline{\overline{I}} = \frac{1}{3}  \int\limits_{-\pi}^{\pi} d\varphi  \int\limits_{0}^{\pi} d\theta\, \sin\theta \overline{\overline{I}},
\end{align}
so that substituting Eq.~\ref{eq:G03d} into Eq.~\ref{eq:S3dDef} yields
\begin{align}
    \overline{\overline{S}}_{3d}(\omega) &= - \frac{1}{3 k_0^2} \overline{\overline{I}} + \frac{2}{3} \overline{\overline{I}} \int\limits_{0}^{a} dr' \, r' e^{i k_0 r'}, \\
    &= - \frac{1}{3 k_0^2} \overline{\overline{I}} + \frac{2}{3} \overline{\overline{I}} \left( \frac{e^{i k_0 a}}{k_0^2}(1 - i k_0 a) - \frac{1}{k_0^2} \right) \\
    &=  \frac{1}{k_0^2}\left( \frac{2}{3}  e^{i k_0 a}(1 - i k_0 a) - 1\right) \overline{\overline{I}} \label{eq:exactself3d}
\end{align}
In this result, the first term is kept separate as it comes from the singular part of the Green's function.
Interestingly, this full form is not trivially consistent with energy conservation in the CDM.
Indeed, to ensure energy conservation, the renormalization of polarizability by the self-interaction must ensure that
\begin{align}
    \sigma_e = \sigma_a + \sigma_s
\end{align}
where $\sigma_{e,a,s}$ are respectively the extinction, absorption, and scattering cross-sections which, if a scatterer is illuminated by a plane-wave with intensity $I_0$, yield the extinguished, absorbed, and scattered powers via $P_{e,a,s} = \sigma_{e,a,s} I_0$.
One may show that, in $3d$ space and in the (Rayleigh) limit $a \to 0$~\cite{Carminati2021}
\begin{align}
    \sigma_e = k_0 \textrm{Im}\left[ \alpha_d(\omega) \right], \label{eq:Rayleigh_extCS}\\
    \sigma_s = \frac{k_0^4}{6\pi} \left| \alpha_d(\omega) \right|^2. \label{eq:Rayleigh_scatCS}
\end{align}
For a non-absorbing scatterer, $\alpha_0 \in \mathbb{R}$, Eq.~\ref{eq:generic_dressed} can be rewritten in a way that makes the real and imaginary parts more apparent, and introducing the notation $\overline{\overline{S}}_{3d} = S_{3d} \overline{\overline{I}}$,
\begin{align}
    \alpha_d = \frac{\alpha_0(\omega)}{\left|1 - k_0^2 \alpha_0(\omega) S_{3d} / V \right|^2} \left(1 - k_0^2 \alpha_0(\omega) S_{3d}^{\dagger} / V \right).
\end{align}
As a result, the extinction cross-section reads
\begin{align}
    \sigma_e = \frac{k_0^3 \alpha_0^2(\omega)}{\left|1 - k_0^2 \alpha_0(\omega) S_{3d} / V \right|^2} \textrm{Im}\left[S_{3d}\right] / V.
\end{align}
On the other hand, the scattering cross-section reads
\begin{align}
    \sigma_s = \frac{k_0^4}{6\pi} \frac{\alpha_0^2(\omega)}{\left|1 - k_0^2 \alpha_0(\omega) S_{3d} / V \right|^2}.
\end{align}
These two values are equal if:
\begin{align}
     \textrm{Im}\left[S_{3d}\right] = \frac{k_0 V}{6\pi}.
\end{align}

To meet this condition, one can consider an approximate form~\cite{Carminati2021}, which is obtained by Taylor-expanding the regular part of the integral to leading order in $k_0 a$, so that
\begin{align}
    \overline{\overline{S}}_{3d}(\omega) &=  \left(- \frac{1}{3 k_0^2} + \frac{a^2}{3} + \frac{2}{9}i k_0 a^3 \right) \overline{\overline{I}}  + \mathcal{O}(k_0^2 a^4). \label{eq:approxself3d}
\end{align}
For a single scatterer with radius $a$ and volume $V = 4 \pi a^3 / 3$, this last approximate equation leads to a well-known result,
\begin{align}
    \alpha_d &\approx \frac{\alpha_r(\omega)}{1 - \frac{i k_0^3}{6\pi} \alpha_r(\omega)} \label{eq:Tayloralpha3d}
\end{align}
where the second term in the denominator is often called the \textit{radiative correction} to the polarizability, as it ensures conservation of the energy of the optical field, and we defined a rescaled polarizability,
\begin{align}
    \alpha_r(\omega) \equiv 3V \frac{\delta\varepsilon(\omega)}{ 3 + \delta\varepsilon(\omega)(1 - k_0^2 a^2)}.
\end{align}
Note that this last expression is also often used in its so-called quasistatic limit,
\begin{align}
    \alpha_r(\omega) \underset{k_0 a\to 0}{\to} 3V \frac{\varepsilon - 1}{\varepsilon + 2}, \label{eq:ClausiusMossotti3d}
\end{align}
which is known as the Clausius-Mossotti, or Lorentz-Lorenz, equation.
Note that, assuming non-absorbing scatterers with a constant dielectric contrast, the real part of the denominator of Eq.~\ref{eq:Tayloralpha3d} diverges at a resonant frequency,
\begin{align}
    \omega_r^2 = \frac{c^2}{a^2} \frac{3 + \delta\varepsilon_s}{ \delta\varepsilon_s}.
\end{align}

Finally, let us compute the LDOS of vacuum for $3d$ vector waves.
By definition,
\begin{align}
    \varrho_0(\omega) = \frac{2 \omega}{\pi c^2} \textrm{Im}\left[\textrm{Tr}_3 \overline{\overline{G}}_0(\bm{r},\bm{r}; \omega)\right]. \label{eq:LDOS_vacuum_def}
\end{align}
To compute it, it is useful to first write the trace of the Green's function for any pair of points. 
Making use of $\textrm{Tr}_3 \hat{\bm{R}}\otimes\hat{\bm{R}} = 1$, one gets
\begin{align}
    \textrm{Tr}_3 \overline{\overline{G}}_0(\bm{r},\bm{r}'; \omega) =  PV\left[\frac{e^{i k_0 R}}{2 \pi R}\right] - \frac{\delta(\bm{R})}{ k_0^2},
\end{align}
so that 
\begin{align}
    \textrm{Im}\left[\textrm{Tr}_3 \overline{\overline{G}}_0(\bm{r},\bm{r}'; \omega)\right] =  \frac{\sin( k_0 R)}{2 \pi R} \underset{R\to 0}{\to} \frac{k_0}{2\pi},
\end{align}
and 
\begin{align}
    \varrho_0(\omega) = \frac{\omega^2}{\pi^2 c^3}. \label{eq:LDOS_vacuum}
\end{align}
A consequence of this result for small scatterers in free $3d$ space is that the LDOS, Eq.~\ref{eq:LDOS_explicit}, of the system can be written with the LDOS of vacuum explicitly factored out:
\begin{align}
    \delta \varrho(\bm{r}, \omega) \equiv \frac{\varrho(\bm{r}, \omega)}{\varrho_0(\omega)} - 1 = 2 \pi k_0 \textrm{Im}\left[ \alpha(\omega) \textrm{Tr}_3 \left[\mathcal{G}_0^{(1)}(\bm{r}; \left\{\bm{r}_{i}\right\}; \omega)\cdot\mathcal{W}(\left\{\bm{r}_{ij}\right\};\omega)\cdot {^t}\mathcal{G}_0^{(1)}(\bm{r}; \left\{\bm{r}_{j}\right\}; \omega) \right]\right].\label{eq:LDOS_explicit_3d}
\end{align}

\subsubsection{$3d$ scalar waves}

In acoustics, but also in simplified optics models, it is common to consider scalar waves in $3d$, \textit{i.e.} a scalar field $E$ that verifies 
\begin{align}
\nabla^2 E(\bm{r};\omega) - \frac{\omega^2}{c^2} \varepsilon (\bm{r};\omega) E(\bm{r};\omega) = i \mu_0 \omega j_{ext}(\bm{r};\omega).
\end{align}
This is in spirit equivalent to a linearly polarized field that is unable to couple to other polarizations, like the TM component in the context of $2d$-like propagation, or the remaining component of $1d$-like propagation.
Defining $\bm{R} = \bm{r} - \bm{r}'$, one may associate a free-space Green's (scalar) function $G_0$ to this equation that reads~\cite{Carminati2021}
\begin{align}
G_0(\bm{r}, \bm{r}'; \omega) = \frac{e^{i k_0 R}}{4 \pi R}.
\end{align}

From this expression, we compute the other quantities derived in the previous subsection.
First, the self-interaction integral for a small spherical scatterer with radius $a$ is defined by
\begin{align}
S_{3d}^{scalar}(\omega) \equiv \int_{\mathcal{B}(\bm{0},a} d^3\bm{r}' G_0(\bm{0},\bm{r}';\omega).
\end{align}
Rewriting the integral in spherical coordinates, it may be rewritten as
\begin{align}
S_{3d}^{scalar}(\omega) = \int\limits_{0}^{a} d\rho \,  \rho e^{i k_0 \rho}
\end{align}
and finally
\begin{align}
S_{3d}^{scalar}(\omega) = \frac{1}{k_0^2}\left(\vphantom{\int} e^{i k_0 a}(1 - i k_0 a) - 1\right).
\end{align}

The first few terms of the Taylor expansion of this self-interaction read
\begin{align}
    S_{3d}^{scalar}(\omega) = \frac{a^2}{2} + i \frac{k_0 a^3}{3} + \mathcal{O}(k_0^2 a^3),
\end{align}
so that the dressed polarizability of a single scatterer can be written as
\begin{align}
    \alpha_d = \frac{\alpha_0(\omega)}{1 - k_0^2 \alpha_0(\omega) S_{3d}^{scalar}/V}
\end{align}
which, after injecting the Taylor expansion and some simple algebra, yields
\begin{align}
    \alpha_d \approx \frac{\alpha_r(\omega)}{1 - i \frac{k_0^3}{4\pi} \alpha_r(\omega)}
\end{align}
with 
\begin{align}
\alpha_r(\omega) \equiv \frac{\alpha_0}{1 - \frac{3 k_0^2 \alpha_0}{8\pi a}}.
\end{align}
Rewriting the bare polarizability as a function of volume and dielectric contrast, the expression becomes
\begin{align}
\alpha_r(\omega) = V \frac{\delta\varepsilon}{1 - \frac{k_0^2 a^2 \delta\varepsilon}{2}} \underset{k_0 \to 0}{\to} \alpha_0.
\end{align}
This is a well-known result~\cite{Carminati2021}: in particular, away from $k_0 = 0$ this expression contains radiative corrections that generate a resonance at 
\begin{align}
k_r a = \sqrt{\frac{8\pi a^3}{3 \textrm{Re}[\alpha_0]}} = \sqrt{\frac{2}{ \textrm{Re} [\delta\varepsilon]}}.
\end{align}

Note that for scalar waves the scattering cross-section in the limit $a \to 0$ is defined via
\begin{align}
    \sigma_s = \frac{k_0^4}{4\pi} \left| \alpha_d(\omega) \right|^2,
\end{align}
so that the optical theorem being valid requires
\begin{align}
    \textrm{Im}[S] = \frac{k_0}{4\pi} V = \frac{k_0 a^3}{3}.
\end{align}

Finally, the density of state of vacuum in this problem is defined as
\begin{align}
\rho_0^{3d, scalar} (\omega) \equiv \frac{2 \omega}{\pi c^2} \textrm{Im}\left[  G_0(\bm{r},\bm{r}; \omega)\right].
\end{align}
Here, the imaginary part of the propagator simply reads
\begin{align}
\textrm{Im}\left[  G_0(\bm{r},\bm{r}'; \omega)\right] = \frac{k_0}{4\pi}\textrm{sinc}(k_0 R) \underset{R \to 0}{\to} \frac{k_0}{4\pi}
\end{align}
so that
\begin{align}
\rho_0^{3d, scalar} (\omega) = \frac{\omega^2}{2\pi^2 c^3}.\label{eq:LDOS_vacuum_3d_scalar} 
\end{align}
Note that this is half of the vector wave value, Eq.~\ref{eq:LDOS_vacuum}.
As a result,
\begin{align}
    \delta \varrho(\bm{r}, \omega) \equiv \frac{\varrho(\bm{r}, \omega)}{\varrho_0(\omega)} - 1 = 4 \pi k_0 \textrm{Im}\left[ \alpha(\omega) \left[\mathcal{G}_0^{(1)}(\bm{r}; \left\{\bm{r}_{i}\right\}; \omega)\cdot\mathcal{W}(\left\{\bm{r}_{ij}\right\};\omega)\cdot {^t}\mathcal{G}_0^{(1)}(\bm{r}; \left\{\bm{r}_{j}\right\}; \omega) \right]\right].\label{eq:LDOS_explicit_3d_scalar}
\end{align}

\subsubsection{Infinite, parallel straight rods with small cross-sections}

The other case we consider at length is that of a system of scatterers that are well approximated by infinite right cylinders, all parallel to the $z$ axis, with small circular cross-sections.
Such a system is invariant in the $z$ direction, so that it is natural to rewrite the Green's function in cylindrical coordinates, $(\bm{\rho},z)$ where (for clarity), we now call $\bm{\rho}$ the polar position in the $xy$ plane.
Due to invariance in the $z$ direction, the source imposes the $z$ dependence of the field.
Suppose the dependence of the source current is simply plane in the $z$ direction,
\begin{align}
    \bm{j}_{ext}(\bm{r}) = \bm{j}_{ext}(\bm{\rho}) e^{ik_{z,inc} z}.
\end{align}
One may then define the $2d$ Green's function between any two polar coordinates in the $xy$ plane as the integral over $z$ of the full Green's function applied to the $z$-dependent part of the current:
\begin{align}
    \overline{\overline{G}}_0(\bm{\rho},\bm{\rho}'; \omega) = \int\limits_{-\infty}^{\infty} dz' \overline{\overline{G}}_0(\bm{r},\bm{r}'; \omega) e^{i k_{z,inc} z'}.
\end{align}
Using the expression in Eq.~\ref{eq:G03d} of the $3d$ Green's function of free space, rewritten in a convenient way as
\begin{align}
    \overline{\overline{G}}_0(\bm{r},\bm{r}'; \omega) = \left[ \overline{\overline{I}} + \frac{1}{k_0^2}\bm{\nabla\otimes\nabla}\right] \frac{e^{i k_0 R}}{4 \pi R}, \label{eq:G03d_nablas}
\end{align}
as well as the integral definition of the zeroth order Hankel's functions of the first kind $H_0^{(1)}$, here through the expression
\begin{align}
    \int\limits_{-\infty}^{\infty} dz' \frac{e^{i k_0 \sqrt{|\bm{\rho} - \bm{\rho}'|^2 + (z - z')^2}}}{\sqrt{|\bm{\rho} - \bm{\rho}'|^2 + (z - z')^2}} e^{i k_{z,inc} z'} = i \pi H_0^{(1)}(k_\rho |\bm{\rho} - \bm{\rho}'| )e^{i k_{z,inc} z},
\end{align}
where $k_{\rho}^2 = k_0^2 - k_{z,inc}^2$, and noticing that the nablas here reduce to $\nabla = (\partial_x,\partial_y, i k_{z,inc})$ one finds
\begin{align}
    \overline{\overline{G}}_0(\bm{\rho},\bm{\rho}'; \omega) =
    \begin{pmatrix}
    1 + \frac{\partial_x^2}{k_0^2} && \frac{\partial_x\partial_y}{k_0^2} && \frac{i k_{z,inc} \partial_x}{k_0^2} \\
    \frac{\partial_x\partial_y}{k_0^2} && 1 + \frac{\partial_y^2}{k_0^2} &&  \frac{i k_{z,inc} \partial_y}{k_0^2} \\
    \frac{i k_{z,inc} \partial_x}{k_0^2} && \frac{i k_{z,inc} \partial_y}{k_0^2} &&  \frac{k_\rho^2}{k_0^2}
    \end{pmatrix} \frac{i}{4} H_0^{(1)}(k_\rho |\bm{\rho} - \bm{\rho}'| )e^{i k_{z,inc} z}.
\end{align}
In particular, if the source is assumed to be invariant along $z$ as well, or at least such that $k_{z,inc} \to 0$, this matrix reduces to a block-diagonal one
\begin{align}
    \overline{\overline{G}}_0(\bm{\rho},\bm{\rho}') =
    \begin{pmatrix}
    \overline{\overline{G}}_0^{TE}(\bm{\rho},\bm{\rho}') && \begin{matrix}0 \\0 \end{matrix} \\
    \begin{matrix} 0 && 0 \end{matrix}  &&  G_{0}^{TM}(\bm{\rho},\bm{\rho}')
    \end{pmatrix},
\end{align}
where we introduced the so-called Transverse Electric (TE) and Transverse Magnetic (TM) propagators.
In practice, the fact that the matrix is block-diagonal means that the $z$ and $(x,y)$ components of a field, also called TM and TE polarizations, here propagate independently.
Using this expression and the relation between the first two Hankel functions, 
\begin{align}
    \frac{d}{dx}H_0^{(1)}(x) = - H_1^{(1)}(x),
\end{align}
and now calling $\bm{R} = \bm{\rho} - \bm{\rho}'$ one may write the explicit expressions of both propagators as
\begin{align}
    \overline{\overline{G}}_0^{TE}(\bm{\rho},\bm{\rho}'; \omega) &= \frac{i}{4} \left(\overline{\overline{I}} - \hat{\bm{R}}\otimes\hat{\bm{R}}\right) H_0^{(1)}(k_0 R) - \frac{i}{4} PV\left[ \left( \overline{\overline{I}} - 2 \hat{\bm{R}}\otimes\hat{\bm{R}}\right)\frac{ H_1^{(1)}(k_0 R) }{k_0 R}\right] - \frac{\delta(\bm{R})}{2 k_0^2}\overline{\overline{I}},  \label{eq:G02dTE} \\
    G_{0}^{TM}(\bm{\rho},\bm{\rho}'; \omega) &=  \frac{i}{4} H_0^{(1)}(k_0 R). \label{eq:G02dTM}
\end{align}
As a result of the propagator breaking down into two independent parts, all scattering properties can also be decomposed into two independent polarization components.

With this definition for the components of the Green's function, the whole $3d$ calculation can be adapted and the system of equations between scatterers written purely in-plane,
\begin{align}
    \bm{E}(\bm{\rho}_j; \omega) \approx \bm{E}_{inc}(\bm{\rho}_j; \omega) + \frac{\omega^2}{c^2} \sum\limits_{j\neq i}^N \Sigma_i \delta\varepsilon_i(\omega) \overline{\overline{G}}_0(\bm{\rho}_j,\bm{\rho}_i; \omega) \bm{E}(\bm{\rho}_i; \omega) + \frac{\omega^2}{c^2} \delta\varepsilon_j(\omega) \left[\int_{\Sigma_j} d^2\bm{\rho}'\, \overline{\overline{G}}_0(\bm{\rho}_j,\bm{\rho}'; \omega)\right] \bm{E}(\bm{\rho}_j; \omega), \label{eq:linearsystem_2d}
\end{align}
where $\Sigma_i$ is the cross-section area of cylinder $i$, leading to the $2d$ definition of the bare polarizability $\alpha_i(\omega) \equiv \Sigma_i \delta\epsilon_i(\omega)$.

We first establish the value of the self-interaction integral for both TE and TM polarizations.
In the TM case, 
\begin{align}
    S_{2d}^{TM}(\omega) = \int_{\mathcal{D}(\bm{0},a)} d^2\bm{r}'\, G_0^{TM}(\bm{0},\bm{r'}; \omega),
\end{align}
where $\mathcal{D}(\bm{0},a)$ is the disk centered at $\bm{0}$ with radius $a$.
This integral is best rewritten in polar coordinates, 
\begin{align}
    S_{2d}^{TM}(\omega) = \frac{i}{4}\int\limits_{0}^a d\rho \int\limits_{-\pi}^{\pi} d\theta \, \rho  H_0^{(1)}(k_0 \rho),
\end{align}
which yields
\begin{align}
    S_{2d}^{TM}(\omega) =  i \frac{\pi a^2}{2} \frac{H_1^{(1)}(k_0 a)}{k_0 a} - \frac{1}{k_0^2}.
\end{align}
Again, this expression can be approximated by its leading order terms in its Taylor expansion~\cite{Silveirinha2006},
\begin{align}
    S_{2d}^{TM}(\omega) =  - \frac{a^2}{4}\left(2 \gamma - 1 - i \pi + 2\log\frac{k_0 a}{2}\right) + \mathcal{O}(k_0^2a^2),
\end{align}
where $\gamma = 0.577216\ldots$ is Euler's gamma constant.
Like in $3d$, one may define a dressed TM polarizability,
\begin{align}
    \alpha_d^{TM} &= \frac{\alpha_0(\omega)}{1 - k_0^2 \alpha_0(\omega) S_{2d}^{TM} / \Sigma },
\end{align}
with $\Sigma = \pi a^2$ the cross section area of the scatterers.
Using the Taylor-expanded version of $S_{2d}$, one can find the analog of Eq.~\ref{eq:Tayloralpha3d},
\begin{align}
    \alpha_d^{TM} &\approx \frac{\alpha_r^{TM}(\omega)}{1 - \frac{i k_0^2}{4} \alpha_r^{TM}(\omega)}, \label{eq:Tayloralpha2d_TM}
\end{align}
with
\begin{align}
    \alpha_r^{TM}(\omega) &= \frac{\alpha_0 }{1 + \frac{\alpha_0 k_0^2}{4\pi}\left( 2\gamma -1 + 2 \log\frac{k_0 a}{2}\right)} \underset{k_0 \to 0}{\to} \alpha_0.
\end{align}

As for the TE polarization, the self-interaction integral reads
\begin{align}
    \overline{\overline{S}}_{2d}^{TE}(\omega) = \int_{\mathcal{D}(\bm{0},a)} d^2\bm{r}'\, \overline{\overline{G}}_0^{TE}(\bm{0},\bm{r'}; \omega).
\end{align}
It is useful to notice that
\begin{align}
    \int\limits_{-\pi}^{\pi}d\theta \, \hat{\bm{R}}\otimes\hat{\bm{R}} = \pi \overline{\overline{I}} = \frac{1}{2} \int\limits_{-\pi}^{\pi}d\theta \, \overline{\overline{I}},
\end{align}
so that
\begin{align}
    \overline{\overline{S}}_{2d}^{TE}(\omega) &= \frac{i \pi}{4} \int\limits_{0}^a d\rho \, \rho H_0^{(1)}(k_0 \rho) \overline{\overline{I}} - \frac{1}{2k_0^2} \overline{\overline{I}}, \\
    &= \left(i \frac{\pi a^2}{4} \frac{H_1^{(1)}(k_0 a)}{k_0 a} - \frac{1}{k_0^2}\right) \overline{\overline{I}}.
\end{align}
Rewriting the last expression as $\overline{\overline{S}}_{2d}^{TE}(\omega) = S_{2d}^{TE}(\omega) \overline{\overline{I}}$, and noticing that the Taylor expansion of the scalar prefactor reads
\begin{align}
    S_{2d}^{TE}(\omega) =  -\frac{1}{2k_0^2}- \frac{a^2}{8}\left(2 \gamma - 1 - i \pi + 2\log\frac{k_0 a}{2}\right) + \mathcal{O}(k_0^2a^2).
\end{align}
Once again, we can write the dressed polarizability of a single scatterer as
\begin{align}
    \alpha_d^{TE} &= \frac{\alpha_0(\omega)}{1 - k_0^2 \alpha_0(\omega) S_{2d}^{TE} / \Sigma },
\end{align}
which, to leading order, yields
\begin{align}
    \alpha_d^{TE} &\approx \frac{\alpha_r^{TE}(\omega)}{1 - \frac{i k_0^2}{8} \alpha_r^{TE}(\omega)}, \label{eq:Tayloralpha2d_TE}
\end{align}
where
\begin{align}
    \alpha_r^{TE}(\omega) &= \frac{\alpha_0 }{1 + \frac{\alpha_0}{2 \Sigma} + \frac{\alpha_0 k_0^2}{8\pi}\left( 2\gamma -1 + 2 \log\frac{k_0 a}{2}\right)} \underset{k_0 \to 0}{\to} 2 \Sigma \frac{\delta\varepsilon}{2 + \delta\varepsilon} = 2 \Sigma \frac{\varepsilon - 1 }{\varepsilon + 1}. \label{eq:ClausiusMossotti2dTE}
\end{align}
The latter quasi-static limit is a known variation of the Clausius-Mossotti relation for $2d$ vector waves~\cite{Jackson,Silveirinha2006}.

Note that the expression of $\alpha_d$ as a function of $S$ is different for $2d$ waves.
Since the extinction cross-section $\sigma_e$ is always defined as $\sigma_e = k_0 \textrm{Im}\left[ \alpha_d \right]$, this implies that the scattering cross-section has a dimensionality-dependent definition (a conclusion that may also be reached by dimensional analysis).
In $2d$, the definition in the limit $a \to 0$ becomes~\cite{Leseur2016,Leseur2016b}
\begin{align}
    \sigma_s^{TM} = \frac{k_0^3}{4} \left| \alpha_d^{TM} \right|^2, \\
    \sigma_s^{TE} = \frac{k_0^3}{8} \left| \alpha_d^{TE} \right|^2.
\end{align}
This implies that the optical theorem being valid requires
\begin{align}
    Im[S^{TM}] = \frac{\Sigma}{4} = \frac{\pi a^2}{4}, \\
    Im[S^{TE}] = \frac{\Sigma}{8} = \frac{\pi a^2}{8}.
\end{align}

Finally, let us compute the LDOS of vacuum for TE and TM waves, starting with the latter,
\begin{align}
    \varrho_0^{TM}(\omega) = \frac{2 \omega}{\pi c^2} \textrm{Im}\left[ G_0^{TM}(\bm{\rho},\bm{\rho}; \omega)\right]. \label{eq:LDOS_vacuum_TM}
\end{align}
Using the expression of the Green's function, this equation becomes
\begin{align}
    \varrho_0^{TM}(\omega) = \frac{ \omega}{2 \pi c^2} \lim_{\rho \to 0^{+}} \textrm{Re}\left[ H_0^{(1)}(k_0 \rho)\right]. 
\end{align}
To find an explicit expression, it is useful to write the Taylor expansion
\begin{align}
    H_0^{(1)}(x) = 1 + \frac{i}{\pi}\left(2 \gamma - 2 \log \frac{x}{2}\right) + \mathcal{O}(x^2),
\end{align}
leading to 
\begin{align}
    \varrho_0^{TM}(\omega) = \frac{ \omega}{2 \pi c^2}.
\end{align}
As a result, the $TM$ LDOS, analogous to Eq.~\ref{eq:LDOS_explicit}, reads
\begin{align}
    \varrho(\bm{\rho}, \omega)^{TM} = \varrho_0(\omega)^{TM} + \frac{2 \omega^3}{\pi c^4} \textrm{Im}\left[ \alpha(\omega) \left[\mathcal{G}_0^{(1),TM}(\bm{\rho}; \left\{\bm{\rho}_{i}\right\}; \omega)\cdot\mathcal{W}^{TM}(\left\{\bm{\rho}_{ij}\right\};\omega)\cdot {^t}\mathcal{G}_0^{(1),TM}(\bm{\rho}; \left\{\bm{\rho}_{j}\right\}; \omega) \right]\right], \label{eq:LDOS_explicit_TM}
\end{align}
where every matrix with exponent $TM$ is similar to that of the $3d case$ but with the $3\times 3$ Green's tensor elements replaced by scalar TM Green's functions, can be written with the vacuum contribution factored out
\begin{align}
    \delta \varrho^{TM}(\bm{\rho}, \omega) \equiv \frac{\varrho^{TM}(\bm{\rho}, \omega)}{\varrho_0^{TM}(\omega)} - 1 = 4 k_0^2 \textrm{Im}\left[ \alpha(\omega) \left[\mathcal{G}_0^{(1),TM}(\bm{\rho}; \left\{\bm{\rho}_{i}\right\}; \omega)\cdot\mathcal{W}^{TM}(\left\{\bm{\rho}_{ij}\right\};\omega)\cdot {^t}\mathcal{G}_0^{(1),TM}(\bm{\rho}; \left\{\bm{\rho}_{j}\right\}; \omega) \right]\right]. \label{eq:LDOS_explicit_2d_TM}
\end{align}

Likewise, the vacuum LDOS for the TE mode reads
\begin{align}
    \varrho_0^{TE}(\omega) = \frac{2 \omega}{\pi c^2} \textrm{Im}\left[ \textrm{Tr}_2\overline{\overline{G}}_0^{TE}(\bm{\rho},\bm{\rho}; \omega)\right]. \label{eq:LDOS_vacuum_TE}
\end{align}
To compute it, it is useful to first write the trace of the Green's function for any pair of points. 
Making use of $\textrm{Tr}_2 \hat{\bm{R}}\otimes\hat{\bm{R}} = 1$, one gets
\begin{align}
    \textrm{Tr}_2 \overline{\overline{G}}^{TE}_0(\bm{\rho},\bm{\rho}'; \omega) = \frac{i}{4} PV\left[ H_0^{(1)}(k_0 R)\right] - \frac{\delta(\bm{R})}{ k_0^2},
\end{align}
so that
\begin{align}
    \textrm{Im}\left[\textrm{Tr}_2 \overline{\overline{G}}^{TE}_0(\bm{\rho},\bm{\rho}'; \omega) \right] = \frac{1}{4} \textrm{Re}\left[ H_0^{(1)}(k_0 R)\right] \underset{R\to 0}{\to} \frac{1}{4},
\end{align}
and
\begin{align}
    \varrho_0^{TE}(\omega) = \frac{ \omega}{2\pi c^2}. \label{eq:LDOS_vacuum_explicit_TE}
\end{align}
Consequently, using the definition
\begin{align}
    \varrho^{TE}(\bm{r}, \omega) = \varrho_0^{TE}(\omega) + \frac{2 \omega^3}{\pi c^4} \textrm{Im}\left[ \alpha(\omega) \textrm{Tr}_2 \left[\mathcal{G}_0^{(1),TE}(\bm{\rho}; \left\{\bm{\rho}_{i}\right\}; \omega)\cdot\mathcal{W}^{TE}(\left\{\bm{\rho}_{ij}\right\};\omega)\cdot {^t}\mathcal{G}_0^{(1),TE}(\bm{\rho}; \left\{\bm{\rho}_{j}\right\}; \omega) \right]\right], \label{eq:LDOS_explicit_TE}
\end{align}
one may write
\begin{align}
    \delta \varrho^{TE}(\bm{\rho}, \omega) \equiv \frac{\varrho^{TE}(\bm{\rho}, \omega)}{\varrho_0^{TE}(\omega)} - 1 = 4 k_0^2 \textrm{Im}\left[ \alpha(\omega) \textrm{Tr}_2 \left[\mathcal{G}_0^{(1),TE}(\bm{\rho}; \left\{\bm{\rho}_{i}\right\}; \omega)\cdot\mathcal{W}^{TE}(\left\{\bm{\rho}_{ij}\right\};\omega)\cdot {^t}\mathcal{G}_0^{(1),TE}(\bm{\rho}; \left\{\bm{\rho}_{j}\right\}; \omega) \right]\right]. \label{eq:LDOS_explicit_2d_TE}
\end{align}

Instead of the LDOS of vacuum, one may instead use the LDOS of an infinite medium with dielectric constant $\varepsilon_m$ as a normalization, which would read, for both TE and TM,
\begin{align}
    \varrho_0(\omega; \varepsilon_m) = \frac{\omega \varepsilon_m}{2 \pi c^2}. \label{eq:LDOS_infinitemedium_2d}
\end{align}

Note that one may write the full LDOS in this geometry using the same definition as in $3d$, which, due to the linearity of the trace, leads to
\begin{align}
    \varrho^{2d}(\bm{r}, \omega) = \varrho^{TM}(\bm{r}, \omega) + \varrho^{TE}(\bm{r}, \omega).
\end{align}
Combining previous expressions, the final result for the relative change of this object compared to vacuum reads
\begin{align}
    \delta \varrho^{2d}(\bm{\rho}, \omega) \equiv \frac{\varrho^{2d}(\bm{\rho}, \omega)}{\varrho_0^{2d}(\omega)} - 1 = 4 k_0^2 \textrm{Im}\left[ \alpha(\omega) \textrm{Tr}_3 \left[\mathcal{G}_0^{(1),2d}(\bm{\rho}; \left\{\bm{\rho}_{i}\right\}; \omega)\cdot\mathcal{W}^{2d}(\left\{\bm{\rho}_{ij}\right\};\omega)\cdot {^t}\mathcal{G}_0^{(1),2d}(\bm{\rho}; \left\{\bm{\rho}_{j}\right\}; \omega) \right]\right], \label{eq:LDOS_explicit_2d_all}
\end{align}
where this time matrices with exponent $2d$ refer to matrices of $3\times 3$ block-diagonal tensors containing TE and TM modes.

\subsection{Validation with Finite-Difference methods}

\begin{figure}
    \centering
    \includegraphics[width=0.9\linewidth]{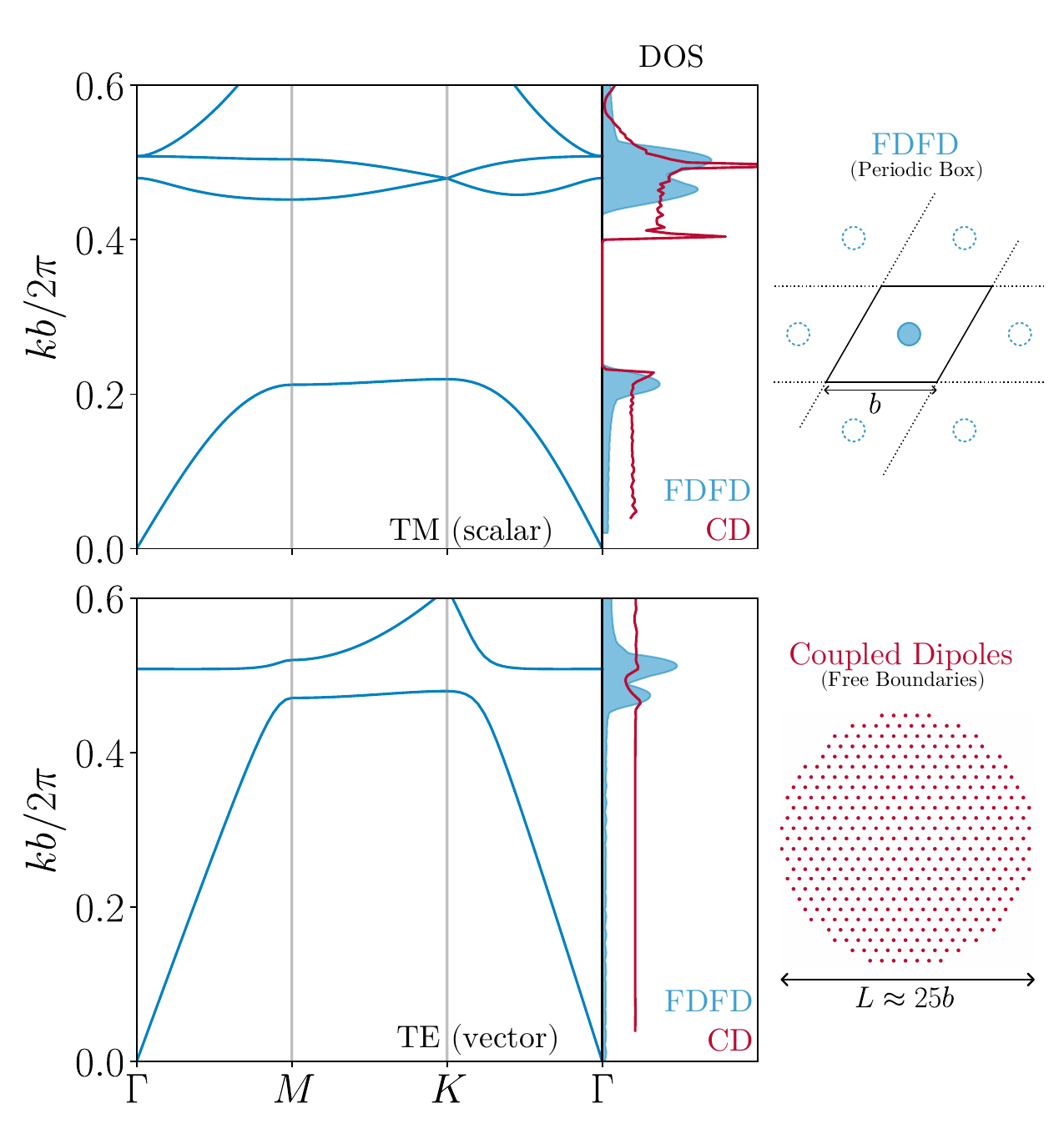}
    \caption{\textbf{FDFD comparison.}
    FDFD calculation of the low-lying band structure of a triangular lattice for the TM (top) and TE (bottom) polarizations.
    The band structure is shown (left), and used to compute a DOS (middle, shaded blue curves), which is compared to our coupled dipoles (CD) DOS estimates (red lines).
    In the right-most column, we remind the geometries considered in both approaches.}
    \label{fig:FDFD}
\end{figure}
In this section, we briefly show that the results we obtain by CDM are consistent with commonly used DOS measurements using finite-difference methods in the frequency domain (FDFD)~\cite{Johnson2001}.
In short, the method consists in discretizing a configuration into pixels (or voxels in $3d$) of material, each with a homogeneous dielectric constant, then in solving for eigenvalues of the discretized Maxwell equations in Fourier space.
This method relies on a periodicization of real space, so that it is particularly well suited to study crystalline systems.
It however scales very badly with the number of pixels needed to describe the system.
As a result, its use is limited to rather small systems of scatterers, and it is not clear how periodicized results relate to the properties of finite-size materials with real boundaries.
This is precisely why, like a large number of past works on optics in disordered materials~\cite{Carminati2021}, we decided to use a coupled dipoles approach.

Since FDFD measurements are still common in the field of disordered materials, see \textit{e.g.} Refs.~\onlinecite{Froufe-Perez2017,Klatt2022}, we however give a minimal proof that both approaches give compatible results in a simple example in which FDFD is tractable.
We consider a single dielectric rod with $n = 6$ with radius $a$, within a rhombic unit cell such that periodicity produces an infinite triangular lattice with lattice spacing $b$, and with a filling fraction $\phi = 5\%$.
The system is discretized onto a grid of $256^2$ pixels.
We then perform an FDFD analysis using the standard MPB package~\cite{Johnson2001}, and compare the output with a coupled dipoles approach for a disk-shaped chunk of a triangular lattice with a diameter $L \approx 25 b$.
The results are reported in Fig.~\ref{fig:FDFD}, for both TM and TE polarizations.
We show band structures in the usual angle-modulus representation.
We extract a DOS from these bands in a standard way, by broadening bands by some smooth finite-width kernel (here a quadratic polynomial kernel with finite support, or ``Epanechnikov'').
As expected, a gap develops in the region $k b / 2\pi \lesssim 1$, and it is much wider in TM than TE.
We show that our coupled dipoles calculations reproduce similar features.
Note that this is a result, more than a validity check: FDFD approaches essentially consider the infinite perfect lattice for crystals (or introduce costly Perfectly Matched Layers that also stray from experimental conditions~\cite{Oskooi2010}), while coupled dipoles consider a finite chunk with an actual boundary in free boundary conditions.
As such, one may expect coupled dipoles to follow more closely experimental conditions.

\section{Effective medium theory}

We here further discuss the expressions used in the main text as an effective medium theory.
To do so, it is useful to introduce the so-called $T$-matrix formalism~\cite{Carminati2021,Lagendijk1996}.
Its definition stems from the Born series, obtained by iterating the Lippman-Schwinger equation, Eq.~\ref{eq:Lippman-Schwinger}, so as to get rid of $\bm{E}$ in the right-hand side.
Adopting short-hand operator notations for integrals and variables, the Born series reads
\begin{align}
    \bm{E} = \bm{E}_{inc} + \overline{\overline{G}}_0 U\bm{E}_{inc} + \overline{\overline{G}}_0 U \overline{\overline{G}}_0 U\bm{E}_{inc} + \overline{\overline{G}}_0 U \overline{\overline{G}}_0 U \overline{\overline{G}}_0 U\bm{E}_{inc} + \ldots \label{eq:Born_series}
\end{align}
where we introduce the scattering potential 
\begin{align}
    U (\bm{r};\omega) = k_0^2 \delta\varepsilon(\bm{r,\omega}),
\end{align}
and each occurrence of $\overline{\overline{G}}_0$ is implicitly accompanied by an integral over a spatial variable.
From the Born series, the $T$-matrix is an operator defined, in reduced notations, as
\begin{align}
   T \equiv U + U \overline{\overline{G}}_0U + U \overline{\overline{G}}_0U \overline{\overline{G}}_0U + \ldots \label{eq:T_matrix},
\end{align}
which is essentially a rewriting of the Dyson equation~\cite{Carminati2021}, such that Eq.~\ref{eq:Born_series} may be rewritten in operator form as
\begin{align}
    \bm{E} = \bm{E}_{inc} + \overline{\overline{G}}_0 T\bm{E}_{inc} \label{eq:Born_series_T}
\end{align}
or, writing all operations and variables explicitly again,
\begin{align}
    \bm{E}(\bm{r};\omega) = \bm{E}_{inc}(\bm{r};\omega) + \int d^d \bm{r}' d^d \bm{r}''  \overline{\overline{G}}_0 (\bm{r},\bm{r}') T(\bm{r', \bm{r}''}; \omega)\bm{E}_{inc} (\bm{r}'')\label{eq:Born_series_T_explicit},
\end{align}
where the explicit expression of the $T$ function that appeared in the last equation is
\begin{align}
    T(\bm{r}_1,\bm{r}_2; \omega) = &U(\bm{r}_1; \omega) \delta(\bm{r}_1 - \bm{r}_2) \overline{\overline{I}} + U(\bm{r}_1; \omega) \overline{\overline{G}}_0(\bm{r}_1, \bm{r}_2; \omega) U(\bm{r}_2; \omega)  \nonumber \\ 
    &+ \int d^d\bm{r}_3 U(\bm{r}_1; \omega) \overline{\overline{G}}_0(\bm{r}_1, \bm{r}_3; \omega) U(\bm{r}_3; \omega) \overline{\overline{G}}_0(\bm{r}_3, \bm{r}_2; \omega) U(\bm{r}_2; \omega) + \ldots
\end{align}
Note that, to avoid more cumbersome notations, we here choose not to write $T$ and $t_i$ as $\overline{\overline{T}}$ and $\overline{\overline{t}}_i$, even though they are technically tensors for vector waves.
Also note that the series that defines $T$ may be rewritten in operator notations as
\begin{align}
    T = U \sum\limits_{n=0}^{\infty} (\overline{\overline{G}}_0 U)^n = \frac{U}{\mathbb{1} - \overline{\overline{G}}_0 U}. \label{eq:Tasinverse}
\end{align}

In the context of coupled dipoles, the scattering potential breaks down into a sum over individual scatterers,
\begin{align}
   U (\bm{r};\omega) = \sum\limits_{i=1}^{N} U_i (\bm{r};\omega) = \sum\limits_{i=1}^{N} k_0^2 \delta\varepsilon \, \mathbb{1}(||\bm{r} - \bm{r}_i || < a ),
\end{align}
so that
\begin{align}
    T(\bm{r}_1,\bm{r}_2; \omega) = &\sum\limits_{i=1}^{N}U_i(\bm{r}_1; \omega) \delta(\bm{r}_1 - \bm{r}_2) \overline{\overline{I}} + \sum\limits_{i,j=1}^{N}U_i(\bm{r}_1; \omega) \overline{\overline{G}}_0(\bm{r}_1, \bm{r}_2; \omega) U_j(\bm{r}_2; \omega)  \nonumber \\ 
    &+ \sum\limits_{i,j,k=1}^{N}\int d^d\bm{r}_3 U_i(\bm{r}_1; \omega) \overline{\overline{G}}_0(\bm{r}_1, \bm{r}_3; \omega) U_j(\bm{r}_3; \omega) \overline{\overline{G}}_0(\bm{r}_3, \bm{r}_2; \omega) U_k(\bm{r}_2; \omega) + \ldots
\end{align}
To make this expression simpler, it is useful to introduce the $T$-matrix $t_i$ of scatterer $i$, 
\begin{align}
    t_i(\bm{r}_1,\bm{r}_2; \omega) = &U_i(\bm{r}_1; \omega) \delta(\bm{r}_1 - \bm{r}_2) \overline{\overline{I}} + U_i(\bm{r}_1; \omega) \overline{\overline{G}}_0(\bm{r}_1, \bm{r}_2; \omega) U_i(\bm{r}_2; \omega)  \nonumber \\ 
    &+ \int d^d\bm{r}_3 U_i(\bm{r}_1; \omega) \overline{\overline{G}}_0(\bm{r}_1, \bm{r}_3; \omega) U_i(\bm{r}_3; \omega) \overline{\overline{G}}_0(\bm{r}_3, \bm{r}_2; \omega) U_i(\bm{r}_2; \omega) + \ldots
\end{align}
For a small scatterer within the coupled dipoles approximation, this $t_i$ can be computed analytically.
Indeed, by definition, for a single scatterer at $\bm{r}_0$,
\begin{align}
    \bm{E}(\bm{r}; \omega) = \bm{E}_{inc}(\bm{r}; \omega) + \int d^d\bm{r}_1 \overline{\overline{G}}_0(\bm{r}, \bm{r}_1; \omega) t_i(\bm{r}_1,\bm{r}_0; \omega) \bm{E}_{inc}(\bm{r}_0; \omega). \label{eq:tmatrix_def}
\end{align}
Working within the coupled dipoles approximation, one may establish, as in Eq.~\ref{eq:outsideE} that, outside of the scatterer,
\begin{align}
    \bm{E}(\bm{r}\neq \bm{r}_0; \omega) = \bm{E}_{inc}(\bm{r}; \omega) +  k_0^2 \alpha_0(\omega) \overline{\overline{G}}_0(\bm{r},\bm{r_0}; \omega) \bm{E}(\bm{r}_0; \omega)
\end{align}
with $\alpha_0(\omega) = V \delta\varepsilon$ the bare polarizability of the scatterer.
Furthermore, inside the scatterer, Eq.~\ref{eq:linearsystem}
\begin{align}
    \bm{E}(\bm{r}_0; \omega) = \bm{E}_{inc}(\bm{r}_0; \omega) +  k_0^2 \delta\varepsilon S(\omega) \bm{E}(\bm{r}_0; \omega).
\end{align}
Combining the last two equations leads to
\begin{align}
    \bm{E}(\bm{r}\neq \bm{r}_0; \omega) = \bm{E}_{inc}(\bm{r}; \omega) +  k_0^2 \alpha_d(\omega)\overline{\overline{G}}_0(\bm{r},\bm{r_0}; \omega) \bm{E}_{inc}(\bm{r}_0; \omega)\label{eq:CD_singlescat}
\end{align}
where we recall that the dressed polarizability is defined by
\begin{align}
    \alpha_d(\omega) \equiv \frac{\alpha_0(\omega)}{1-k_0^2 \delta\varepsilon S(\omega)}. \label{eq:alphad_CD}
\end{align}
Altogether, identifying Eqs~\ref{eq:CD_singlescat} and~\ref{eq:tmatrix_def} leads to the expression of the $t_i$ of a single scatterer within the coupled dipoles approximation,
\begin{align}
    t_i(\bm{r}, \bm{r}'; \omega) = k_0^2 \alpha_d(\omega) \delta(\bm{r}- \bm{r}_i) \delta(\bm{r}'- \bm{r}_i) \overline{\overline{I}}.\label{eq:t_matrix_element}
\end{align}

Introducing these single-scatterer objects, the full $T$-matrix can then be written as a resummation over the $t_i$'s,
\begin{align}
    T(\bm{r}_1,\bm{r}_2; \omega) = &\sum\limits_{i=1}^{N}t_i(\bm{r}_1, \bm{r}_2) + \sum\limits_{j\neq i = 1}^{N} \int d^d \bm{r}_3  d^d \bm{r}_4\, t_i(\bm{r}_1, \bm{r}_3) \overline{\overline{G}}_0(\bm{r}_3, \bm{r}_4; \omega) t_j(\bm{r}_4; \bm{r}_2)  \nonumber \\ 
    &+ \sum\limits_{k\neq j \neq i = 1}^{N}\int d^d\bm{r}_3 d^d\bm{r}_4 d^d\bm{r}_5 d^d\bm{r}_6  t_i(\bm{r}_1; \bm{r}_3) \overline{\overline{G}}_0(\bm{r}_3, \bm{r}_4; \omega) t_j(\bm{r}_4; \bm{r}_5) \overline{\overline{G}}_0(\bm{r}_5, \bm{r}_6; \omega) t_k(\bm{r}_6; \bm{r}_2) + \ldots
\end{align}
which may be written in condensed form as
\begin{align}
    T = &\sum\limits_{i=1}^{N}t_i+ \sum\limits_{j\neq i = 1}^{N}  t_i \overline{\overline{G}}_0 t_j + \sum\limits_{k\neq j \neq i = 1}^{N}  t_i \overline{\overline{G}}_0 t_j \overline{\overline{G}}_0 t_k + \ldots \label{eq:T_from_t}
\end{align}
In particular, Eq.~\ref{eq:Born_series_T} may be written as a function of individual $t_i$,
\begin{align}
    \bm{E} = \bm{E}_{inc} +  \sum\limits_{i=1}^N\overline{\overline{G}}_0 t_i\bm{E}_{inc} + \sum\limits_{j\neq i = 1}^{N}  \overline{\overline{G}}_0  t_i \overline{\overline{G}}_0 t_j \bm{E}_{inc} + \sum\limits_{k\neq j \neq i = 1}^{N} \overline{\overline{G}}_0   t_i \overline{\overline{G}}_0 t_j \overline{\overline{G}}_0 t_k \bm{E}_{inc} + \ldots \label{eq:Born_series_ti},
\end{align}
which is a resummed version of the Born series written in terms of dressed polarizabilities instead of bare ones.

For future purposes, it is also useful to introduce the Fourier transform of dyadic operators (either propagators or $T$-matrices), here generically represented by the symbol $A$,
\begin{align}
    \widehat{A}(\bm{k}_1,\bm{k}_2) \equiv \int d^d\bm{r}_1 d^{d}\bm{r}_2 A(\bm{r}_1, \bm{r}_2) e^{i \bm{k}_2\cdot\bm{r}_2 - i \bm{k}_1 \cdot \bm{r}_1}.
\end{align}

\subsection{Irreducible vertex and $T$-matrix}

One may show~\cite{Carminati2021} that disorder-averaged field correlations in disordered media are related to the cross-correlation of the propagator of the medium, $\langle \overline{\overline{G}} \, \overline{\overline{G}} {}^{\star} \rangle$, where the average is performed over realizations of the disorder.
This cross-correlation is related to the irreducible vertex $\Gamma$ of the field correlations by:
\begin{align}
   &\left\langle \overline{\overline{G}}(\bm{r}_1, \bm{r}_2;\omega) \, \overline{\overline{G}} {}^{\star} (\bm{r}_1', \bm{r}_2';\omega) \right\rangle = \left\langle \overline{\overline{G}}(\bm{r}_1, \bm{r}_2;\omega) \right\rangle \, \left\langle\overline{\overline{G}} {}^{\star} (\bm{r}_1', \bm{r}_2';\omega) \right\rangle \nonumber \\ &+ \int d^{d} \bm{R}_1 d^{d} \bm{R}_2 d^{d} \bm{R}_1' d^{d} \bm{R}_2' 
 \left\langle \overline{\overline{G}}(\bm{r}_1, \bm{R}_1;\omega) \right\rangle \, \left\langle\overline{\overline{G}} {}^{\star} (\bm{r}_1', \bm{R}_1';\omega) \right\rangle\Gamma(\bm{R}_1, \bm{R}_2; \bm{R}_1', \bm{R}_2') \left\langle \overline{\overline{G}}(\bm{R}_2, \bm{r}_2;\omega)  \, \overline{\overline{G}} {}^{\star} (\bm{R}_2', \bm{r}_2';\omega) \right\rangle.
\end{align}
One may also show that the irreducible vertex can be written as a diagrammatic expansion.
In most cases, it is truncated to its leading order, by the ladder (or Bethe-Salpeter) approximation.
In statistically homogeneous scattering media, translational invariance leads to large constraints on its Fourier transform, defined through
\begin{align}
    \widehat{\Gamma}(\bm{k}_1, \bm{k}_2; \bm{k}_1', \bm{k}_2') \equiv \int  d^d\bm{r}_1 d^d\bm{r}_2 d^d\bm{r}_1' d^d\bm{r}_2' \Gamma(\bm{r}_1, \bm{r}_2; \bm{r}_1', \bm{r_2}') e^{i \bm{k}_2\cdot \bm{r}_2 - i \bm{k}_1\cdot \bm{r}_1 + i \bm{k}_1'\cdot \bm{r}_1' - i \bm{k}_2'\cdot \bm{r}_2'},
\end{align}
which is then given by
\begin{align}
    \widehat{\Gamma}(\bm{k}_1, \bm{k}_2; \bm{k}_1', \bm{k}_2') = (2\pi)^3 \delta(\bm{k}_1 - \bm{k}_2 - \bm{k}_1' + \bm{k}_2') \gamma(\bm{k}_1, \bm{k}_2; \bm{k}_1', \bm{k}_2').
\end{align}
To leading order in $T$ and within the Bethe-Salpeter approximation, one may show that
\begin{align}
    \Gamma(\bm{r}_1,\bm{r}_2; \bm{r}_1', \bm{r}_2') = \left\langle T(\bm{r}_1,\bm{r}_2)T^{\star}(\bm{r}_1', \bm{r}_2') \right\rangle - \left\langle T(\bm{r}_1,\bm{r}_2) \right\rangle \left\langle T^{\star}(\bm{r}_1', \bm{r}_2') \right\rangle
\end{align}
which, after some algebra~\cite{Carminati2021}, leads, in the case of coupled dipoles, to
\begin{align}
    \gamma(\bm{k}_1, \bm{k}_2; \bm{k}_1', \bm{k}_2') = \rho 
 \, t(\bm{k}_1, \bm{k}_2) t^\star(\bm{k}_1', \bm{k}_2') + \rho^2 t(\bm{k}_1, \bm{k}_2) t^\star(\bm{k}_1', \bm{k}_2') \widehat{h}(\bm{k}_2 - \bm{k}_1), \label{eq:irrvertex_bethesalpeter}
\end{align}
with $h(\bm{r}) = g(\bm{r}) - 1$ the total correlation function of the point pattern formed by the set of scatterer centers, and $t$ the $T$-matrix of a single scatterer placed at the origin.

\subsection{Self-energy}

From the quantities introduced in previous subsection, one may evaluate (within the Bethe-Salpeter approximation and assuming translational invariance of the medium) a number of effective scattering properties that may then be used to model the medium as a continuum.
The most common property to compute is the scattering mean free path $\ell_s$, or its inverse the scattering coefficient $\mu_s$, which may be written, in $3d$ scalar waves~\cite{Carminati2021}, as
\begin{align}
    \mu_s = \frac{1}{\ell_s} = \frac{1}{16\pi^2} \int d^2 \hat{\bm{k}}' \gamma(k_R \hat{\bm{k}}, k_R \hat{\bm{k}}'; k_R \hat{\bm{k}}, k_R \hat{\bm{k}}'),
\end{align}
where the integral runs over the surface of a sphere and $k_R$ is such that
\begin{align}
    k_R^2 = k_0^2 + \text{Re}\,\Sigma(k_0), \label{eq:keff_real}
\end{align}
with $\Sigma$ the self-energy of the medium.
Since we do not want to justify the expression of $\ell_s$ but only how we intend to use it, in this section we discuss the definition and value of $\Sigma$.
It is defined via the Dyson equation for the disorder-averaged medium which, in reduced operator notations, reads, for scalar waves
\begin{align}
    \langle G \rangle = G_0 + G_0 \Sigma \langle G \rangle.
\end{align}
Note in particular that for translationally invariant media, all quantities in this equation reduce to a simple Fourier-space kernel when performing spatial Fourier transforms,
\begin{align}
    \widehat{A}(\bm{k},\bm{k}') = (2\pi)^d \widehat{A}(\bm{k})  \delta(\bm{k} - \bm{k}'),
\end{align}
so that, using the regularized expression for the Fourier transform of the propagator,
\begin{align}
    \widehat{G}_0(\bm{k}) = \frac{1}{k^2 - k_0^2 - i \epsilon},
\end{align}
one finds
\begin{align}
    \langle\widehat{G}\rangle(\bm{k}) = \frac{1}{k^2 - k_0^2 - \Sigma(\bm{k}) - i \epsilon}.
\end{align}
This last expression justifies Eq.~\ref{eq:keff_real}: the presence of the medium shifts the wave number to an effective value
\begin{align}
    k_{\text{eff}}^2 = k_0^2 + \Sigma(\bm{k}),
\end{align}
and $k_R$ is its real part that encodes the effective wavelength in the medium, as one may define an effective dielectric constant and an effective refractive index through
\begin{align}
    \varepsilon_{\text{eff}} \equiv \frac{k_\text{eff}^2}{k_0^2} = n_\text{eff}^2.
\end{align}
The choice of evaluating $\Sigma$ at $k_0$ in Eq.~\ref{eq:keff_real} is known as the on-shell approximation, which is tantamount to assuming $| \Sigma (\bm{k})| \ll k_0^2$ in the studied regime, so that the propagator is still peaked around $k_0$, and, assuming that the medium is isotropic, 
\begin{align}
    k_{\text{eff}}^2 \approx k_0^2 + \Sigma(k_0).
\end{align}
In other words, $k_R \approx k_0$ and $k_I = \text{Im} k_{\text{eff}} \ll k_0$, so that
\begin{align}
k_{\text{eff}} \approx k_R + i \frac{k_I^2}{2 k_R}.
\end{align}

It is common to write the self-energy directly as a function of the $T$-matrix~\cite{Carminati2021,Vynck2023}, using the fact that, by definition
\begin{align}
    \Sigma(\bm{r},\bm{r}') = \int d^d\bm{r}_1 d^d\bm{r}_2 \overline{\overline{G}}_0^{-1}(\bm{r}, \bm{r}_1) \left(\langle \overline{\overline{G}}\rangle (\bm{r}_1, \bm{r}_2) - \overline{\overline{G}}_0 (\bm{r}_1, \bm{r}_2)\right) \langle \overline{\overline{G}}\rangle^{-1}(\bm{r}_2, \bm{r}')
\end{align}
where we introduced the inverse propagators defined by
\begin{align}
    \int d^d \bm{r_1} \overline{\overline{G}}^{-1}(\bm{r}, \bm{r}_1) \overline{\overline{G}}(\bm{r}_1, \bm{r}')= \delta(\bm{r} - \bm{r}') \overline{\overline{I}},
\end{align}
together with the definition of the averaged $T$ matrix in the Dyson equation,
\begin{align}
    \langle \overline{\overline{G}} \rangle(\bm{r}, \bm{r}') = \overline{\overline{G}}_0(\bm{r}, \bm{r}') + \int d^d\bm{r}_1 d^d \bm{r}_2 \overline{\overline{G}}_0(\bm{r}, \bm{r}_1) \langle T \rangle(\bm{r}_1, \bm{r}_2) \overline{\overline{G}}_0(\bm{r}_2, \bm{r}').
\end{align}
Putting both definitions together yields, in short-hand notations (which makes the inverse easier to write),
\begin{align}
    \Sigma = \langle T \rangle \left( \mathbb{1} +  \overline{\overline{G}}_0 \langle T \rangle  \right)^{-1}.
\end{align}
This expression may also be expanded into powers of $\langle T \rangle$, leading, in short-hand notations, to
\begin{align}
    \Sigma = \langle T \rangle -  \langle T \rangle G_0 \langle T \rangle + \langle T \rangle G_0 \langle T \rangle G_0 \langle T \rangle   - \langle T \rangle G_0 \langle T \rangle G_0 \langle T \rangle G_0 \langle T \rangle + \ldots \label{eq:self_from_avgT}
\end{align}

To use this expression, recall, Eq.~\ref{eq:T_from_t},
\begin{align}
    \langle T \rangle = & \sum\limits_{i=1}^{N}\langle t_i \rangle+ \sum\limits_{j\neq i = 1}^{N}  \langle t_i \overline{\overline{G}}_0 t_j \rangle + \sum\limits_{k\neq j \neq i = 1}^{N} \langle  t_i \overline{\overline{G}}_0 t_j \overline{\overline{G}}_0 t_k \rangle+ \ldots \label{eq:avgT_from_t}
\end{align}
where the sums over more than 2 indices are only forbidden from having equal successive indices (for instance, the sum over $3$ indices verifies $j \neq i$ and $k \neq j$ but one may have $i = k$, and such ``recurrent scattering'' terms may play important roles in expansions~\cite{Monsarrat2022}).
For coupled dipoles, using Eq.~\ref{eq:t_matrix_element},
\begin{align}
    \langle T \rangle (\bm{r}, \bm{r}') = & k_0^2 \alpha_d(\omega)\left\langle\sum\limits_{i=1}^{N} \delta(\bm{r}- \bm{r}_i) \delta(\bm{r}'- \bm{r}_i) \right\rangle \overline{\overline{I}} \\ &+ k_0^4 \alpha_d(\omega)^2    \left\langle  \sum\limits_{j\neq i = 1}^{N} \delta(\bm{r}- \bm{r}_i)  \overline{\overline{G}}_0 (\bm{r}_i, \bm{r}_j)  \delta(\bm{r}'- \bm{r}_j)  \right\rangle + \ldots\label{eq:avgT_from_t_explicit}
\end{align}
The ensemble averages over realizations of the medium may then be computed.
The first one reads
\begin{align}
    \left\langle \sum\limits_{i=1}^{N}\delta(\bm{r}- \bm{r}_i) \delta(\bm{r}'- \bm{r}_i) \right\rangle &= \left\langle \sum\limits_{i=1}^{N} \delta(\bm{r}- \bm{r}_i) \right\rangle \delta(\bm{r}- \bm{r}') \\ 
    &= \left\langle \rho(\bm{r}) \right\rangle \delta(\bm{r}- \bm{r}')\\ 
    &= \rho_0 \delta(\bm{r}- \bm{r}'),
\end{align}
while the second one reads
\begin{align}
    \left\langle  \sum\limits_{j\neq i = 1}^{N} \delta(\bm{r}- \bm{r}_i)  \overline{\overline{G}}_0 (\bm{r}_i, \bm{r}_j)  \delta(\bm{r}'- \bm{r}_j)  \right\rangle &= \overline{\overline{G}}_0 (\bm{r}, \bm{r}')  \left\langle  \sum\limits_{j\neq i = 1}^{N} \delta(\bm{r}- \bm{r}_i) \delta(\bm{r}'- \bm{r}_j)  \right\rangle \\
    &= \overline{\overline{G}}_0 (\bm{r}, \bm{r}') \left( \left\langle \rho(\bm{r}) \rho(\bm{r}') \right\rangle - \rho_0 \delta(\bm{r} - \bm{r}')\right) \\
    &= \rho_0^2 \overline{\overline{G}}_0 (\bm{r}, \bm{r}') g(\bm{r}, \bm{r}')
\end{align}
where we used the fact that the disorder-averaged joint distribution of finding a point at both $\bm{r}_1$ and $\bm{r}_2$ enters the definition of the pair distribution function via~\cite{Hansen2006}
\begin{align}
    \langle \rho(\bm{r}_1) \rho(\bm{r}_2)\rangle &\equiv \langle \rho(\bm{r}_1)\rangle \langle \rho(\bm{r}_2)\rangle g(\bm{r}_1, \bm{r}_2) + \langle\rho (\bm{r}_1) \rangle \delta(\bm{r}_1 - \bm{r}_2)  \\
    &= \rho_0^2 g(\bm{r}_1, \bm{r}_2) + \rho_0 \delta(\bm{r}_1 - \bm{r}_2) \label{eq:homo_g}
\end{align}
where the $\delta(\bm{r}_1 - \bm{r}_2)$ accounts for the $i=j$ term of the product of the two density fields, which is usually excluded from the definition of $g$.
One may then write the average $T$ matrix up to double non-recurrent scattering as
\begin{align}
    \langle T \rangle (\bm{r}, \bm{r}') \approx & \rho_0 k_0^2 \alpha_d(\omega) \delta(\bm{r} - \bm{r}') \overline{\overline{I}} + \rho_0^2 k_0^4 \alpha_d(\omega)^2  \overline{\overline{G}}_0 (\bm{r}, \bm{r}') g(\bm{r}, \bm{r}').  \label{eq:avgT_paircorrelations}
\end{align}
Going to higher-order terms generally requires being able to write $n$-body correlations in the point pattern explicitly.
For general models of disorder, this is intractable, although one may write higher-order terms for some specific types of disorder (\textit{e.g.} Gaussian fields~\cite{Carminati2021} for which one may write Ward's identity to write every moment as a function of the $1$- and $2$-point correlations, or Poisson point patterns for which $\langle \rho(\bm{r}_1) \ldots \rho(\bm{r}_n)\rangle = \rho_0^n$).
Thus, to keep this discussion general, we will truncate $\langle T \rangle$ at the level of pair correlations, an approximation that is valid in systems with $\rho_0 \alpha_d (\omega) \ll 1$.
Note that for coupled dipoles, Eq.~\ref{eq:alphad_CD},
\begin{align}
    \rho_0 \alpha_d(\omega) = \phi \frac{\delta\varepsilon (\omega)}{1 - k_0^2 \delta\varepsilon (\omega) S(a,\omega)}
\end{align}
so that $\rho_0 \alpha_d (\omega) \ll 1$ is more restrictive than just $\phi \delta\varepsilon \ll 1$: one must also avoid strong resonances.

We here explicitly exclude recurrent scattering, as Refs.~\cite{Monsarrat2022,MonsarratThesis} showed that they are mostly relevant to analytical DOS estimates. 
Stopping at that order yields, for the self-energy
\begin{align}
    \Sigma(\bm{r}, \bm{r}') &\approx \rho_0 k_0^2 \alpha_d  \delta(\bm{r} - \bm{r}') \overline{\overline{I}} + \rho_0^2 k_0^4 \alpha_d^2  \overline{\overline{G}}_0 (\bm{r}, \bm{r}') h(\bm{r}, \bm{r}')
\end{align}
where $h = g-1$ was introduced.

\subsection{Effective medium properties without recurrent terms}

\subsubsection{$3d$ scalar waves}

Going back to effective medium properties, in $3d$ scalar waves, the scattering coefficient $\mu_s$ and the scattering mean free path $\ell_s$ can be written as
\begin{align}
    \mu_s = \frac{1}{\ell_s} = \frac{1}{16\pi^2} \int d^2 \hat{\bm{k}}' \gamma(k_R \hat{\bm{k}}, k_R \hat{\bm{k}}'; k_R \hat{\bm{k}}, k_R \hat{\bm{k}}') \label{eq:mu_s}.
\end{align}
Using Eq.~\ref{eq:keff_real} and evaluating it to leading order in density yields
\begin{align}
    k_R^2 \approx k_0^2 + \text{Re}\, \Sigma (k_0) \approx k_0^2 (1 + \rho_0 \text{Re}[\alpha_d(\omega)]) 
\end{align}
and, from Eq.~\ref{eq:irrvertex_bethesalpeter}
\begin{align}
    \gamma(\bm{k}_1, \bm{k}_2; \bm{k}_1', \bm{k}_2') = \rho 
 \, t(\bm{k}_1, \bm{k}_2) t^\star(\bm{k}_1', \bm{k}_2') + \rho^2 t(\bm{k}_1, \bm{k}_2) t^\star(\bm{k}_1', \bm{k}_2') \widehat{h}(\bm{k}_2 - \bm{k}_1),
\end{align}
with the $t$-matrix of a single scatterer placed at the origin given by Eq.~\ref{eq:t_matrix_element}
\begin{align}
    t(\bm{k}_1, \bm{k}_2) = k_0^2 \alpha_d(\omega),
\end{align}
and where the Fourier transform of the total correlation function verifies
\begin{align}
   1 + \rho \widehat{h}(\bm{k}) = \widetilde{S}(\bm{k}).
\end{align}
All in all, at this level of approximation and for scalar $3d$ waves
\begin{align}
    \gamma(\bm{k}, \bm{k}'; \bm{k}, \bm{k}') &= \rho | t(\bm{k}, \bm{k}')|^2 \widetilde{S}(\bm{k} - \bm{k'}) \label{eq:generic_gamma_ells}\\ &= \rho k_0^4 |\alpha_d(\omega)|^2 \widetilde{S}(\bm{k} - \bm{k}' ),
\end{align}
so that, to leading order,
\begin{align}
    \mu_s (k_0 \hat{\bm{k}}) &= \frac{1}{16\pi^2} \int d^2 \hat{\bm{k}}' \rho k_0^4 |\alpha_d(\omega)|^2 \widetilde{S}(k_R(\hat{\bm{k}} - \hat{\bm{k}}')).
\end{align}

It is useful to perform a change of variables to spherical coordinates using the $\bm{k}$ direction as the one containing the poles:
\begin{align}
    d^2\hat{\bm{k}}' &= \sin\theta d\theta d\phi
\end{align}
instead of $\hat{\bm{k}}'$.
For isotropic scatterers and isotropic structure factors, $t$ and $S$ are both only functions of $q \equiv k_R |\hat{\bm{k}} - \hat{\bm{k}} '|$, which verifies
\begin{align}
    q^2 = 2 k_R^2 (1 - \cos\theta) = 4 k_R^2 \sin^2 \frac{\theta}{2},
\end{align}
so that
\begin{align}
    q dq = k_R^2 \sin\theta d\theta.
\end{align}
All in all, for isotropic scatterers and isotropic media, the generic expression in Eq.~\ref{eq:generic_gamma_ells} can be used in Eq.~\ref{eq:mu_s} to write
\begin{align}
    \mu_s &= \frac{1}{16\pi^2} \int d^2 \hat{\bm{k}}' \rho | t(k_R\hat{\bm{k}}, k_R \hat{\bm{k}}')|^2 \widetilde{S}(k_R (\hat{\bm{k}}  - \hat{\bm{k}}')) \\
    &= \frac{1}{16\pi^2} \int d^2 \hat{\bm{k}}' \rho | t(q)|^2 \widetilde{S}(q) \\ 
    &= \frac{1}{8\pi} \int\limits_{0}^\pi d \theta \sin\theta \rho | t(q)|^2 \widetilde{S}(q) \\
    &= \frac{1}{8\pi k_R^2} \int\limits_{0}^{2 k_R} d q \, q \, \rho | t(q)|^2 \widetilde{S}(q).
\end{align}
One then often defines the form factor $F(q)$ as~\cite{Carminati2021,Vynck2023} 
\begin{align}
    F(q) \equiv \frac{k_R^2}{16\pi^2} |t(q)|^2,
\end{align}
so that
\begin{align}
    \mu_s &= \frac{2\pi \rho}{k_R^4} \int\limits_{0}^{2 k_R} d q \, q \, F(q) \widetilde{S}(q). \label{eq:mus_3dscalar}
\end{align}
This expression is useful as it is applicable even for $t$-matrices more complicated than those of coupled dipoles, and may for instance include Mie scattering terms.

In the special case of coupled dipoles, since $t$ is a constant, one finds
\begin{align}
    \mu_s &= \frac{\rho k_0^4 |\alpha_d(\omega)|^2}{8 \pi k_R^2} \int\limits_{0}^{2 k_R} d q \, q \, \widetilde{S}(q) \\
    &= 2\pi \rho \sigma_s \frac{1}{k_R^2}  \int\limits_{0}^{2 k_R} d q \, q \, \widetilde{S}(q).
\end{align}
where we introduced the scattering cross-section for a small scatterer in $3d$ scalar waves,
\begin{align}
    \sigma_s \equiv \frac{1}{16\pi^2} \int d^2\hat{\bm{k}} | t(k_R\hat{\bm{k}}, k_R\hat{\bm{k}}')|^2 = \frac{k_0^4}{4 \pi} |\alpha_d(\omega)|^2.
\end{align}
It is then interesting to consider the leading order approximation for $k_R$, namely, assuming $\delta\varepsilon\in\mathbb{R}$,
\begin{align}
    \mu_s = 2\pi \rho \sigma_s \frac{1}{k_0^2(1 + \rho_0 \text{Re}[\alpha_d(\omega)])}  \int\limits_{0}^{2 k_0\sqrt{1 + \rho_0 \text{Re}[\alpha_d(\omega)]}} d q \, q \, \widetilde{S}(q).
\end{align}
At this level, one finds that a collection of scatterers, compared to a single one, has an enhanced or suppressed scattering coefficient purely based on the radial integral of $\widetilde{S}$ below an effective frequency.
Notably, a SHU system ($\widetilde{S}(q) = 0$ for $q < K$) is transparent until $K = k_R/2$ at this level of approximation (and at the next~\cite{Leseur2016}), while a phase-separated system has enhanced scattering.

Another quantity of interest is the phase function,
\begin{align}
    p(\hat{\bm{k}}, \hat{\bm{k}}') &= \frac{\ell_s}{4\pi}  \gamma(k_R \hat{\bm{k}}, k_R \hat{\bm{k}}'; k_R \hat{\bm{k}}, k_R \hat{\bm{k}}') \\
    &=  \frac{\ell_s}{4\pi}  \rho | t(k_R\hat{\bm{k}}, k_R \hat{\bm{k}}')|^2 \widetilde{S}(k_R (\hat{\bm{k}}  - \hat{\bm{k}}')) \\
    & = \frac{\ell_s}{4\pi} \rho k_0^4 |\alpha_d(\omega)|^2 \widetilde{S}(k_R (\hat{\bm{k}}  - \hat{\bm{k}}')), \\
    &= \frac{\rho \sigma_s}{ \mu_s} \widetilde{S}(k_R (\hat{\bm{k}}  - \hat{\bm{k}}')),
\end{align}
a similar expression to that reported in $2d$ in Ref.~\cite{Leseur2016}.
This function is homogeneous in the limit of uncorrelated scatterer positions.
In the context of gyromorphs, the ring of peaks technically couples a finite set of $\bm{k}$-vectors to each other in the phase function at the value $k_R = k_p/2$.
However, in the limit $G \gg 1$, the discrete aspect of this set of peaks is expected to become less important.
Likewise, one may define the anisotropy factor
\begin{align}
    g \equiv \frac{1}{4\pi} \int d^2 \hat{\bm{k}}' \hat{\bm{k}}\cdot \hat{\bm{k}}' p(\hat{\bm{k}}, \hat{\bm{k}}')= 1 - \ell_s \frac{\pi \rho}{k_R^6} \int\limits_{0}^{2k_R} q^3 F(q) \widetilde{S}(q) dq
\end{align}
the transport mean free path and the transport coefficient,
\begin{align}
    \mu_t = \frac{1}{\ell_t} = \frac{1-g}{\ell_s} = \frac{\pi \rho}{k_R^6} \int\limits_{0}^{2k_R} q^3 F(q) \widetilde{S}(q) dq
\end{align}
and the extinction mean free path $\ell_e$, that verifies
\begin{align}
    \mu_e = \frac{1}{\ell_e} = 2 \text{Im} k_{\text{eff}}.
\end{align}
The latter always verifies $\mu_e = \mu_s + \mu_a$ with $\mu_a$ the absorption coefficient.
Furthermore, note that within the on-shell approximation,
\begin{align}
    k_{\text{eff}} &\approx k_0 \sqrt{1 + \frac{\Sigma(k_0)}{k_0^2}} \\
    &\approx k_0 + \frac{\Sigma(k_0)}{2k_0}
\end{align}
so that
\begin{align}
    \mu_e \approx \frac{\text{Im}\Sigma(k_0)}{k_0}.
\end{align}

In particular, one may check that for $\widetilde{S} = 1$, $g = 0$, $\mu_s = 4\pi \rho_0 \sigma_s$, $p(\hat{\bm{k}},\hat{\bm{k}}') = 1/(4\pi)$.

\subsubsection{$2d$ scalar waves}

For $2d$ scalar waves, the expressions are modified and follow~\cite{Leseur2016b}
\begin{align}
    \mu_s &= \frac{1}{8\pi k_R} \int d^1 \hat{\bm{k}}' \gamma(k_R \hat{\bm{k}}, k_R \hat{\bm{k}}; k_R \hat{\bm{k}}', k_R \hat{\bm{k}}'),\\
    p(\theta, \theta') &= \frac{\ell_s}{8 \pi k_R} \gamma(k_R \hat{\bm{k}}, k_R \hat{\bm{k}}; k_R \hat{\bm{k}}', k_R \hat{\bm{k}}'),\\
    g&\equiv \frac{1}{2\pi} \int\limits_{-\pi}^{\pi} d\theta' \cos (\theta - \theta') p(\theta,\theta').
\end{align}
The expressions are fairly similar to the $3d$ ones, however their simplified forms differ.
Take the expression for $\mu_s$,
\begin{align}
    \mu_s &= \frac{1}{8\pi k_R} \int d^1 \hat{\bm{k}}' \rho | t(k_R\hat{\bm{k}}, k_R \hat{\bm{k}}')|^2 \widetilde{S}(k_R (\hat{\bm{k}}  - \hat{\bm{k}}')).
\end{align}
In $2d$, one simply has $d^1 \hat{\bm{k}}' = d\theta$, and the integral runs over the circle, $\theta \in [-\pi;\pi)$.
Like in $3d$, the integrand only depends on $q = k_R|\hat{\bm{k}}  - \hat{\bm{k}}'| = k_R \sqrt{2-2\cos\theta}$, but now
\begin{align}
    \mu_s &= \frac{1}{8\pi k_R} \int\limits_{-\pi}^{\pi} d\theta \rho | t(q)|^2 \widetilde{S}(q).
\end{align}
As a result, the change of variable is not as immediate as in $3d$~\cite{Conley2014}, one needs to write, for $\theta \in [0; \pi]$
\begin{align}
    \theta = \arccos \left(1 - \frac{q^2}{2 k_R^2} \right)
\end{align}
so that
\begin{align}
    \frac{d\theta}{dq} &= \frac{q/k_R^2}{\sqrt{1 - \left(1 - q^2/(2k_R^2)\right)^2}}, \\
    &=\frac{1/k_R}{\sqrt{1 - (q/2k_R)^2}}
\end{align}
leading to
\begin{align}
    \mu_s &= \frac{\rho_0}{4\pi k_R^2} \int\limits_{0}^{2 k_R} dq \frac{| t(q)|^2 \widetilde{S}(q)}{{\sqrt{1 - (q/2k_R)^2}}}.
\end{align}

In the coupled dipoles limit, like in $3d$, one may introduce the scattering cross-section of a single scatterer,
\begin{align}
    \sigma_s \equiv \frac{k_0^3 |\alpha_d (\omega)|^2}{4} = \frac{|t(\bm{k}_1, \bm{k}_2)|^2}{4k_0},
\end{align}
so that
\begin{align}
    \frac{\mu_s}{\rho_0 \sigma_s} &= \frac{k_0}{\pi k_R^2} \int\limits_{0}^{2 k_R} dq \frac{\widetilde{S}(q)}{{\sqrt{1 - (q/2k_R)^2}}}.
\end{align}

The phase function can similarly be written as~\cite{Leseur2016b} 
\begin{align}
    p(\theta, \theta') &= \frac{\rho_0 k_0 \sigma_s}{8 \pi k_R\mu_s} \widetilde{S}(k_R (\hat{\bm{k}}  - \hat{\bm{k}}')) \\
    &= \frac{\rho_0 k_0 \sigma_s}{8 \pi  k_R \mu_s} \widetilde{S}(k_R (\hat{\bm{k}}  - \hat{\bm{k}}')).
\end{align}
Finally, the anisotropy factor can be written as
\begin{align}
    g &= \frac{\rho_0 \sigma_s}{8\pi \mu_s} \int\limits_{-\pi}^{\pi} d\theta \cos\theta \widetilde{S}(q = k_R\sqrt{2 - 2\cos\theta}),
\end{align}
and 
\begin{align}
    \cos\theta d\theta = \frac{1 - q^2/(2k_R^2)}{\sqrt{1 - (q/2k_R)^2}} \frac{dq}{k_R}
\end{align}
so that
\begin{align}
    g &= \frac{\rho_0 \sigma_s}{4\pi k_R \mu_s} \int\limits_{0}^{2k_R} dq \frac{1 - q^2/2k_R^2}{\sqrt{1 - (q/2k_R)^2}}\widetilde{S}(q).
\end{align}
From the previous expressions, the transport coefficient reads
\begin{align}
    \mu_t = (1-g)\mu_s &= \frac{\rho_0 \sigma_s}{\pi k_R} \int\limits_{0}^{2 k_R} dq \frac{\widetilde{S}(q)}{\sqrt{1 - (q/2k_R)^2}}\left(\frac{q^2}{8k_R^2} + \frac{k_0 }{ k_R}  - \frac{1}{4}\right).
\end{align}
In particular, one may check that for $\widetilde{S} = 1$, $g = 0$, $\mu_s = \rho_0 \sigma_s k_0/4 k_R$, $p(\theta,\theta') = 1/(2\pi)$.

\subsubsection{$3d$ vector waves}

For vector waves, the calculations become more intricate due to the presence of several polarization components.
In $3d$, results can be found in Refs.~\cite{VanTiggelen1994,Cherroret2016,Carminati2021,Vynck2023}.
In short, the main change is that the irreducible vertex $\Gamma$ is a tensor acting on polarization components.
However, in the radiative transfer limit, only the (real-space) transverse component participates in energy transfer, so that one may show that, in $3d$~\cite{Vynck2023}
\begin{align}
    \frac{1}{\ell_s}p(\hat{\bm{k}},\hat{\bm{k}}') = \frac{1}{32\pi^2} \text{Tr}\left[ \widehat{P}_{k,\perp} \otimes \widehat{P}_{k,\perp} \cdot \Gamma(k_R \hat{\bm{k}}, k_R\hat{\bm{k}}'; k_R\hat{\bm{k}}, k_R\hat{\bm{k}}' ) \cdot \widehat{P}_{k',\perp} \otimes \widehat{P}_{k',\perp} \right],
\end{align}
where we introduced the projectors
\begin{align}
    \widehat{P}_{q,\perp} \equiv \overline{\overline{I}} - \widehat{\bm{q}} \otimes\widehat{\bm{q}}
\end{align}
and $k_R^2 = k_0^2 + \text{Re} \Sigma_\perp(k_0)$ to leading order within the on-shell approximation.
Since the phase function is normalized,
\begin{align}
    \int d^{d-1}\hat{k}'p(\hat{\bm{k}},\hat{\bm{k}}') = 1,
\end{align}
this implies\footnote{The prefactor is here introduced like in Ref.~\cite{Vynck2023}, Ref.~\cite{Cherroret2016} instead used a prefactor $1/8\pi$ but absorbed a $1/(2\pi)^2$ into the definition of an angular average.}
\begin{align}
    \frac{1}{\ell_s} = \frac{1}{32\pi^2} \int d^{d-1}\hat{k}'\text{Tr}\left[ \widehat{P}_{k,\perp} \otimes \widehat{P}_{k,\perp} \cdot \widehat{\Gamma}(k_R \hat{\bm{k}}, k_R\hat{\bm{k}}'; k_R\hat{\bm{k}}, k_R\hat{\bm{k}}' ) \cdot \widehat{P}_{k',\perp} \otimes \widehat{P}_{k',\perp} \right].
\end{align}
In that expression, note that (not including two-scatterer recurrent scattering terms), the derivation can be performed fairly easily since, for coupled dipoles, the single-scatterer $T$-matrix $t$ is simply proportional to $\overline{\overline{I}}$, so that 
\begin{align}
    \widehat{\Gamma}(\bm{k}_1, \bm{k}_2; \bm{k}_1', \bm{k}_2') \propto \overline{\overline{I}} \otimes \overline{\overline{I}}.
\end{align}
As a result, the irreducible vertex has equal transverse and longitudinal projections, with a kernel given by the same equation as in the scalar wave case, leading to~\cite{Vynck2023}
\begin{align}
    \mu_s &= \frac{1}{16\pi^2} \int d^2 \hat{\bm{k}}' \rho | t(k_R\hat{\bm{k}}, k_R \hat{\bm{k}}')|^2 \widetilde{S}(k_R (\hat{\bm{k}}  - \hat{\bm{k}}')).
\end{align}
Thus, when recurrent terms are not considered, the expression of $\mu_s$ is exactly analogous to that of scalar waves.
The same can be said of $g$ and $\mu_e$, with the usual expression for the extinction mean free path~\cite{Vynck2023}
\begin{align}
    \frac{1}{\ell_e} \approx \frac{\text{Im} \Sigma_\perp(k_0)}{k_0}.
\end{align}

\subsubsection{$2d$ vector waves}

In $2d$, results for vector waves can be found in Refs.~\cite{Monsarrat2022,MonsarratThesis}.
The scattering coefficient can be written as
\begin{align}
    \frac{1}{\ell_s} = \frac{1}{16\pi^2 k_R} \int d^{d-1}\hat{k}'\text{Tr}\left[ \widehat{P}_{k,\perp} \otimes \widehat{P}_{k,\perp} \cdot \widehat{\Gamma}(k_R \hat{\bm{k}}, k_R\hat{\bm{k}}'; k_R\hat{\bm{k}}, k_R\hat{\bm{k}}' ) \cdot \widehat{P}_{k',\perp} \otimes \widehat{P}_{k',\perp} \right].
\end{align}
Like for the $3d$ case, when recurrent terms are not considered, the expression is the same for vector waves and scalar waves,
\begin{align}
    \mu_s &= \frac{1}{8\pi k_R} \int d^1 \hat{\bm{k}}' \rho | t(k_R\hat{\bm{k}}, k_R \hat{\bm{k}}')|^2 \widetilde{S}(k_R (\hat{\bm{k}}  - \hat{\bm{k}}')).
\end{align}

\subsection{Effective medium properties through simple integrals}

\begin{figure}
    \centering
    \includegraphics[width=0.48\linewidth]{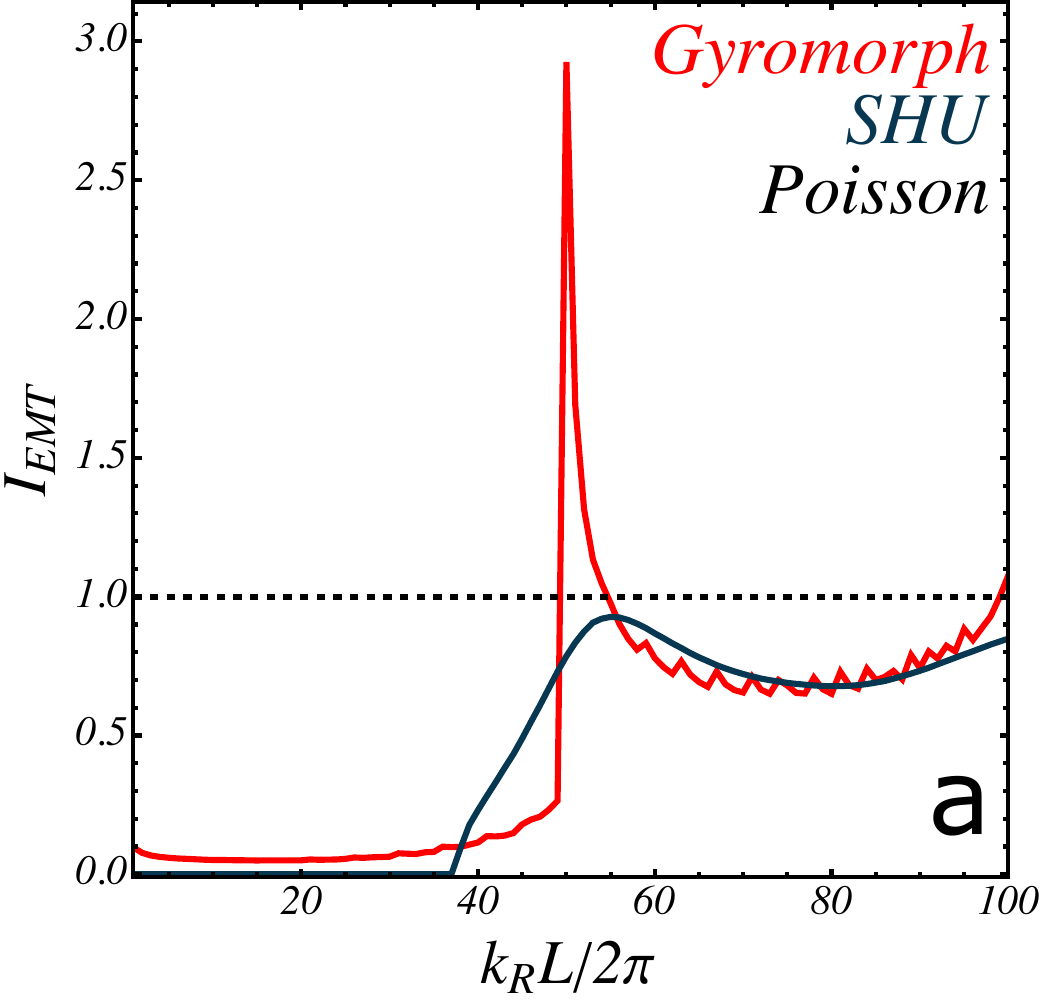}
    \includegraphics[width=0.48\linewidth]{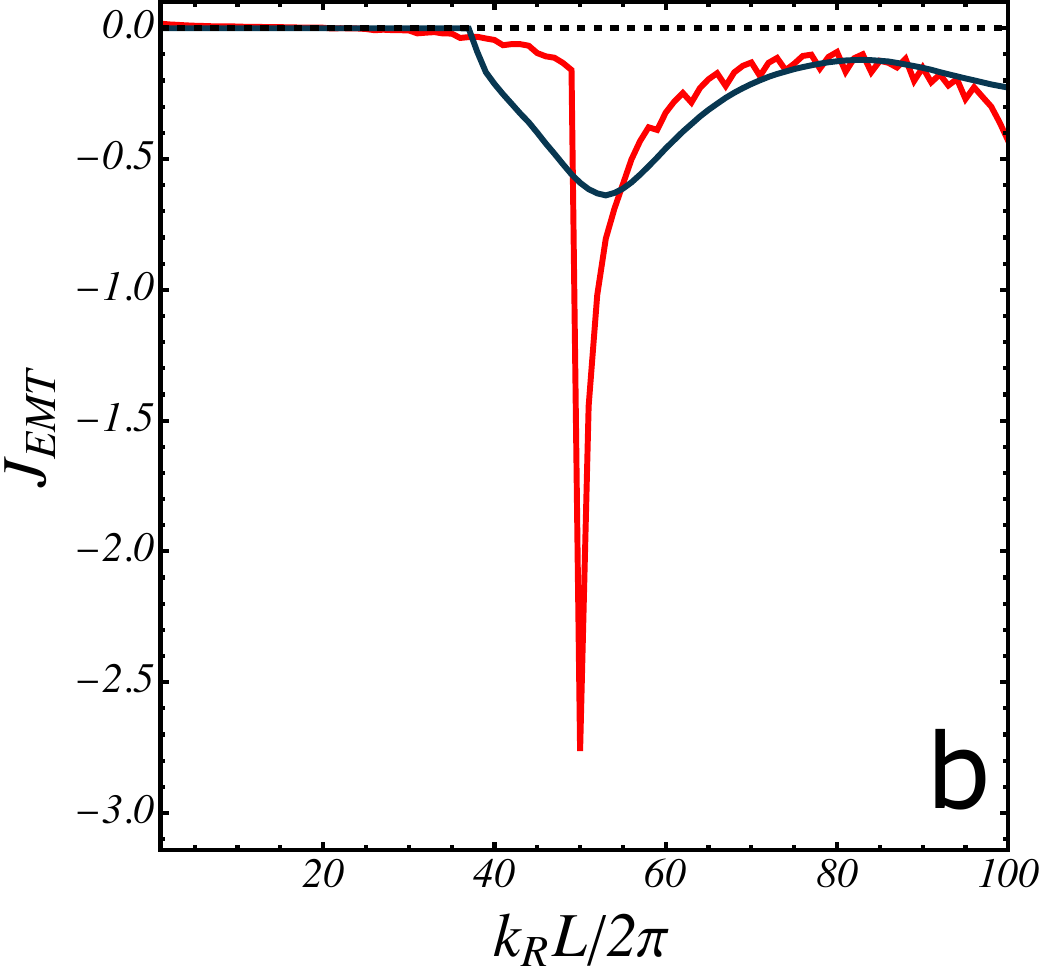}
    \caption{\textbf{Effective Medium Integral.}
    $(a)$ $I_{EMT}$ and $(b)$ $J_{EMT}$ as functions of $k_R L / 2 \pi$ for gyromorphs (red), SHU systems (dark blue) and Poisson point patterns (dashed black line).
    }
    \label{fig:EMT_int}
\end{figure}

At the level of approximation of the theory we use, the effective medium properties are proportional to an integral that is only a function of $k_R$ and $S$.
As a result, one may understand effective medium properties looking at an integral that is independent of optical parameters.
For $2d$ systems, $\ell_s$ and $g$ only depends on 
\begin{align}
    I_{EMT}(k_R) &= \frac{1}{\pi}\int\limits_{0}^{2 k_R} dq \frac{ \widetilde{S}(q)}{{\sqrt{1 - (q/2k_R)^2}}}, \\
    J_{EMT}(k_R) &= \frac{1}{\pi}\int\limits_{0}^{2 k_R} dq \frac{ \left(1 - q^2/k_R^2\right)\widetilde{S}(q)}{{\sqrt{1 - (q/2k_R)^2}}},
\end{align}
where the prefactors ensure that $I_{EMT} = 1$ for a Poisson point pattern.
These functions are computed for the average $S(k)$ across $30$ realizations of gyromorphs and SHU systems, and compared to the Poisson case, in Fig.~\ref{fig:EMT_int}.
Since $\ell_s$ is obtained from dilations of the $x$ axis of  Fig.~\ref{fig:EMT_int}$(a)$ and a multiplication by a prefactor, one can immediately see that gyromorphs display a much more prominent feature at $K/2$ (associated with a reduction in $\ell_s$) compared to both Poisson and SHU systems.
In addition, the dominant effect of stealthiness is clearly visible as enhanced transparency at low $K$, consistent with expectations.
Finally,  Fig.~\ref{fig:EMT_int}$(b)$, shows that both systems clearly favor backscattering, $g<0$. 
Based on this analysis alone, the structure factor $S(k)$ of gyromorphs is expected to be more effective than that of SHU systems at producing short scattering mean free paths---and thus, potentially, bandgaps.

\subsection{Transport mean-free paths}

\begin{figure}
    \centering
    \includegraphics[width=0.5\linewidth]{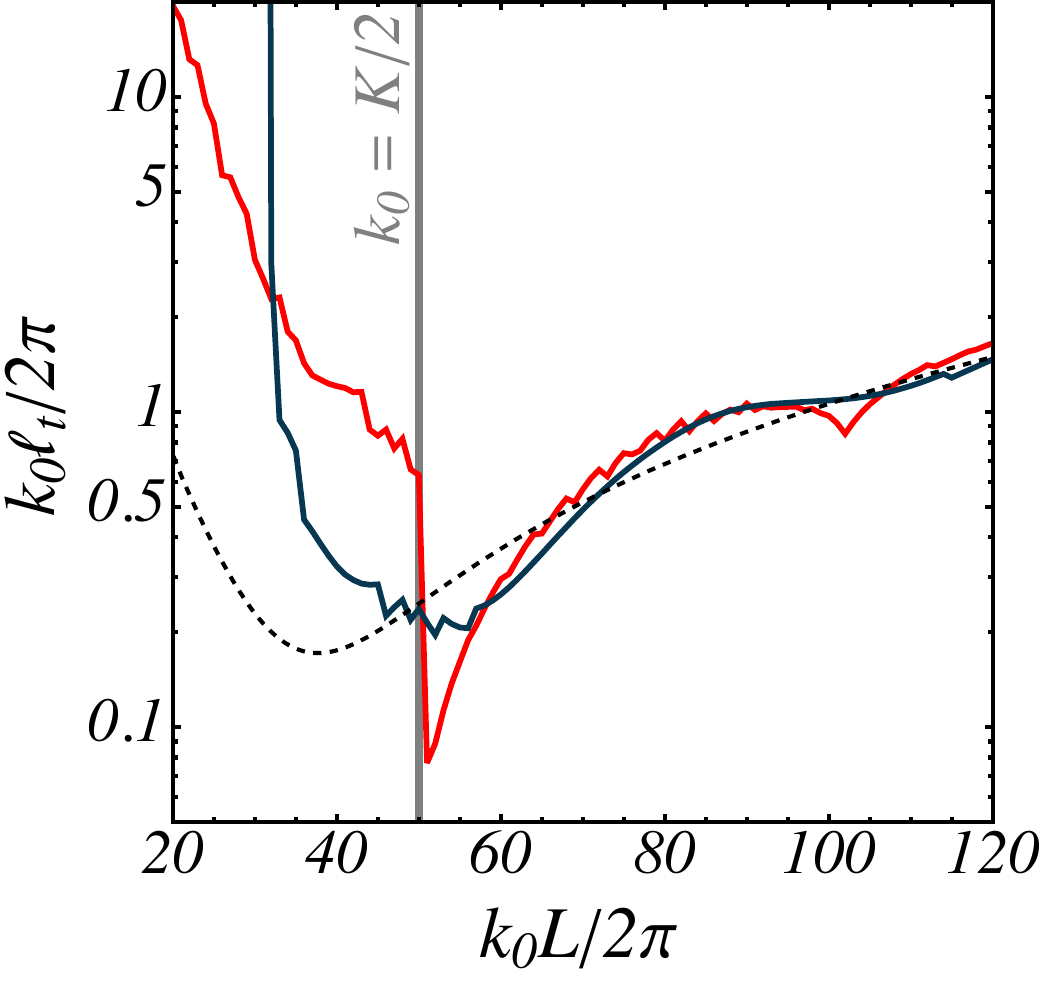}
    \caption{\textbf{Transport mean-free paths.}
    Rescaled transport mean-free path $k_0 \ell_t /2\pi$ against the rescaled frequency $k_0 L/2\pi$ for $G=60$ gyromorphs (red), SHU (dark blue) and Poisson point patterns (dashed black line), computed for $n=3$, $\phi = 0.05$ for scalar $2d$ waves.
    A gray line indicates $k_0 = K/2$.}
    \label{fig:ellt}
\end{figure}

In the main text we discuss the estimates of the anisotropy factor $g$ and of the scattering mean-free path $\ell_s$ for gyromorphs, SHU systems and Poisson point patterns.
We here provide the corresponding plot of the transport mean-free path, $\ell_t = \frac{\ell_s}{1-g}$, as estimated from our effective medium theory, in Fig.~\ref{fig:ellt}.
As claimed in the main text, the transport mean-free path of gyromorphs is estimated to be significantly lower than that of SHU systems at the target frequency.

\bibliography{PostDoc-StefanoMartiniani}